\documentclass{ws-ijmpd}
\usepackage[numbers, compress]{natbib}
\usepackage{color}
\usepackage{cancel}
\usepackage{wrapfig}
\usepackage{ulem}
\usepackage{float}
\usepackage{empheq}

\newcommand{\varv}{{\rm v}}

\newcommand{\mpl}{M_{\rm pl}}

\newcommand{\calO}{{\cal O}}

\newcommand{\calH}{{\cal H}}
\newcommand{\calP}{{\cal P}}
\newcommand{\calB}{{\cal B}}
\newcommand{\calR}{{\mathcal R}}
\newcommand{\calS}{{\mathcal S}}
\newcommand{\eq}[2]{\begin{equation}\label{#1}{#2}\end{equation}}

\newcommand{\dint}{\mathrm{d}}
\newcommand*\widefbox[1]{\fbox{#1}}

\usepackage{hyperref}

\newcommand{\Mpc}{{\rm Mpc}}
\newcommand{\GHz}{{\rm GHz}}

\newcommand{\expf}[1]{{{\rm e}^{#1}}}
\newcommand{\Jbb}{\mathcal{J}}

\newcommand{\zh}{{z_{\rm h}}}

\newcommand{\zmudc}{{z_{\rm dc}}}

\newcommand{\nS}{n_{\rm S}}

\newcommand{\nrun}{n_{\rm run}}

\newcommand{\taudot}{\dot{\tau}}
\newcommand{\kD}{k_{\rm D}}
\newcommand{\rs}{r_{\rm s}}
\newcommand{\cs}{c_{\rm s}}

\newcommand{\id}{{\,\rm d}}

\newcommand{\beq}{\begin{equation}}   %

\newcommand{\eeq}{\end{equation}}   %

\newcommand{\beqa}{\begin{eqnarray}}   %

\newcommand{\eeqa}{\end{eqnarray}}   %

\newcommand{\beal}{\begin{align}}

\newcommand{\enal}{\end{align}}

\newcommand{\bspl}{\begin{split}}

\newcommand{\espl}{\end{split}}

\newcommand{\bsub}{\begin{subequations}}

\newcommand{\esub}{\end{subequations}}

\newcommand{\bmulti}{\begin{multline}}   %

\newcommand{\beqm}{\begin{mathletters}}   %

\newcommand{\eeqm}{\end{mathletters}}   %

\newcommand{\Ne}{N_{\rm e}}

\newcommand{\The}{\theta_{\rm e}}

\newcommand{\sigT}{\sigma_{\rm T}}

\newcommand{\vek} [1]{\mbox{\boldmath${#1}$\unboldmath}}

\newcommand{\pot}[2]{#1 \times 10^{#2}}

\begin{document}
\title{Features and New Physical Scales in Primordial Observables:\\ Theory and Observation}

\author{Jens Chluba$^{\,a, b)}$\footnote{jchluba@ast.cam.ac.uk}}
\author{Jan Hamann$^{\,c,e)}$\footnote{jan.hamann@sydney.edu.au}}
\author{Subodh P. Patil$^{\,d,e)}$\footnote{subodh.patil@unige.ch (corresponding author)}}

\address{\vspace{15pt}$^{a)}$Kavli Institute for Cosmology Cambridge,\\ Madingley Road, \\Cambridge CB3 0HA, \\U.K.}
\address{$^{b)}$Department of Physics and Astronomy,\\ Johns Hopkins University,\\ 3400 N. Charles St.,
Baltimore, MD 21218,\\USA}
\address{$^{c)}$Sydney Institute for Astronomy,\\ School of Physics A28, \\The University of Sydney, NSW 2006,\\ AUSTRALIA}
\address{$^{d)}$Department of Theoretical Physics,\\ University of Geneva,\\ 24 Quai Ansermet, Geneve-4, CH-1211,\\ SWITZERLAND}
\address{$^{e)}$CERN Theory Division,\\ PH-TH, Case C01600,\\
Geneva-23, CH-1211,\\ SWITZERLAND}

\maketitle

\ccode{CERN-PH-TH-2015-096}
\ccode{PACS numbers:}
\pagebreak
\begin{abstract}
{
All cosmological observations to date are consistent with adiabatic, Gaussian and nearly scale invariant initial conditions. These findings provide strong evidence for a particular symmetry breaking pattern in the very early universe (with a close to vanishing order parameter, $\epsilon$), widely accepted as conforming to the predictions of the simplest realizations of the inflationary paradigm. However, given that our observations are only privy to perturbations, in inferring something about the background that gave rise to them, it should be clear that many different underlying constructions project onto the same set of cosmological observables. Features in the primordial correlation functions, if present, would offer a unique and discriminating window onto the parent theory in which the mechanism that generated the initial conditions is embedded. In certain contexts, simple linear response theory allows us to infer new characteristic scales from the presence of features that can break the aforementioned degeneracies among different background models, and in some cases can even offer a limited spectroscopy of the heavier degrees of freedom that couple to the inflaton. In this review, we offer a pedagogical survey of the diverse, theoretically well grounded mechanisms which can imprint features into primordial correlation functions in addition to reviewing the techniques one can employ to probe observations. These observations include cosmic microwave background anisotropies and spectral distortions as well as the matter two and three point functions as inferred from large-scale structure and potentially, 21 cm surveys.
}
\end{abstract}
\tableofcontents

\newpage

\section{Introduction \label{sec:introduction}}
{

Recent cosmological observations suggest that our universe is rather accurately modelled by the simplest version of a $\Lambda$-CDM cosmology \cite{Planck:2015xua,Sievers:2013ica,Hou:2012xq,Hinshaw:2012aka}
\footnote{Although interesting deviations from its predictions might be present at longer wavelengths with marginal significance \cite{Adam:2015rua}.}. The remarkable fact that one can model the large-scale evolution of the universe from recombination (redshift $z\simeq 10^3$) all the way through to the present epoch with just six adjustable parameters speaks of an underlying simplicity and elegance that begs for a deeper explanation, possibly from some fundamental physics setup. 

It is commonly accepted at present that the inflation paradigm provides such an explanation \cite{Starobinsky:1980te,Guth:1980zm,Linde:1981mu}. However, of the parameters of the $\Lambda$-CDM model-- to wit, the baryon density $\Omega_b$, the density of cold dark matter $\Omega_c$, the dark energy density $\Omega_\Lambda$, the reionization optical depth $\tau$, the scalar spectral index $n_\mathrm{s}-1$ and the amplitude of the primordial power spectrum $\Delta^2_{\mathcal R}$-- only the scalar tilt is typically {\it predicted} by any given inflationary construction\footnote{The amplitude is typically tuned to fit with observations and the other parameters are put in by hand a posteriori.}. Although one can invoke some selective criteria of `inference to best theory'  \cite{Wells:2012dx} to take the current set of observations as confirmation of the inflationary paradigm, it needn't be too much of a stretch to probe the available observations for further evidence before declaring a particular class of inflationary models as the true description of the very early universe. Furthermore, given the large degeneracies that exist between the predictions of different models \cite{Easson:2010zy, Easson:2010uw, Easson:2012id} (even drastically different cosmologies \cite{Wands:1998yp}) at the level of the primordial power spectrum, one might ask, are there any other handles on the data that increase our ability to discriminate between these, or even test the very inflationary paradigm itself?

In part due to such considerations, a fair amount of theoretical effort has been devoted recently to understanding the consequences of various inflationary models with respect to their predictions for primordial non-Gaussianity \cite{Bartolo:2004if}. If observed, the interaction statistics of the inflaton field offers us a window onto its self-couplings, its couplings to other fields and appropriately limited information about its embedding in some parent theory. Although many types of primordial non-Gaussianity could plausibly yet turn up in observations of the cosmic microwave background (CMB), the preliminary conclusion from the recent {\it Planck} mission \cite{Ade:2015ava} is that all observations are consistent with the presence of only those non-Gaussianities generated by foregrounds and non-linearities during and after recombination \cite{Creminelli:2004pv,Novosyadlyj2006,Pitrou:2010sn,Huang:2012ub,Su:2012gt}. It thus remains a pertinent question to ask whether or not we have exhausted all the information content that is potentially available in the primordial two-point correlation functions. Even if primordial non-Gaussianity is ever detected, the question takes on added significance in light of the vast amount of data due to become available in the near to long term through large-scale structure \cite{Laureijs:2011gra, Schlegel:2011zz, Tyson:2003kb}, 21~cm observations \cite{Loeb:2003ya, Furlanetto:2006jb, vanHaarlem:2013dsa, Rawlings:2004wk} and observations of the spectral distortion of the CMB \cite{Kogut2011PIXIE, PRISM2013WPII}. In particular, the potential gain in the number of modes available to us -- roughly, having access to information in a three-dimensional volume as opposed to a surface (the last scattering surface) -- and potentially over a greater range of comoving scales will enable us to ask new questions, and old questions much more meaningfully of the data \cite{Huang:2012mr}.

One particular class of questions that has occupied a great deal of attention in the literature over the years has been concerned with whether or not there are {\it features}, or localized deviations from scale invariance in the primordial power spectrum. From the perspective of linear response theory, one can rephrase certain aspects of this in terms of the more basic question: are there any new characteristic scales present in the primordial data? Clearly when we observe the cosmic microwave background we see evidence for the scale of gravity ($M_\mathrm{pl}$) and the characteristic mass scale of the mechanism that generated primordial structure (for example $\Delta^2_{\mathcal R} \sim V_0/(\epsilon M_\mathrm{pl}^4)$ during single field slow roll)\footnote{In putative models where the big bang is emergent or the result of a bounce from a prior contracting phase, the amplitude is predicted by the characteristic scale of the underlying dynamics. In the context of a particular string construction for example, $\Delta^2_{\mathcal R} \sim m^4_s/M_\mathrm{pl}^4$ \cite{Brandenberger:2006vv, Brandenberger:2014faa}. An analogous conclusion also follows in the context of cyclic universe models \cite{Lehners:2010ug}.}, up to the order parameter $\epsilon$ which would be fixed by a positive detection of primordial tensor modes\footnote{Although this too makes assumptions of the field content of the universe between laboratory and inflationary scales even if these fields are completely decoupled from the inflaton sector \cite{Antoniadis:2014xva}.}. The observed deviation from scale invariance in the spectral index \cite{Sievers:2013ica,Hou:2012xq,Hinshaw:2012aka,Ade:2015lrj} indicates a dynamical origin for the adiabatic perturbations, though directly inferring a mass scale from these depends on a host of model dependent priors. So we ask: might there be other scales lurking in the data that, if probed, could offer us hints of the scales that parametrize the parent theory or model in which the mechanism that generated primordial structure is embedded? 

At this point the reader might well be wondering whether or not state of the art CMB observations haven't already confirmed a featureless, red tilted primordial power spectrum to a large degree of accuracy \cite{Planck:2015xua,Sievers:2013ica,Hou:2012xq,Hinshaw:2012aka} thus rendering the questions at hand irrelevant? In actual fact, what has been confirmed is that in a priori parametrizing the initial conditions of the $\Lambda$-CDM universe with an almost scale free initial power spectrum (normalized at a fixed comoving scale) with the spectral tilt and a possible logarithmic running as parameters to scan over (among others), one arrives at a best fit value $n_\mathrm{s} \approx 0.96$ \cite{Planck:2015xua,Sievers:2013ica,Hou:2012xq,Hinshaw:2012aka} with no significant evidence for running. This is qualitatively different from saying that such a spectrum has been {\it observed}. CMB observations are intrinsically limited not just by cosmic variance but also by the {\it theoretical priors} that one has to invoke in order to analyse the data at all, in addition to having to contend with the unavoidable compression inherent in the projection of spacetime dependent inhomogeneities onto a surface of last scattering\footnote{\label{gain}Roughly, since foregrounds begin to dominate the CMB sky at multipoles around $\ell \sim 2000$, one has at most access to approximately $10^6$ pixels of data in the CMB, evidently a compression of the total information content of the inhomogeneities at last scattering by a factor of about $10^3$.}. This intrinsically limits our ability to directly reconstruct the primordial correlators from any given set of CMB observables, perhaps evident in the observation that features could already be present in currently available data, but not with any notable significance \cite{Ade:2015lrj}. Given that future large-scale structure (LSS) data promises to recover some subset of the (volume) information content of the primordial power spectrum projected out by the CMB onto the surface of last scattering, one might hope to look for the presence of features in the primordial correlators with much greater significance. The prospect of extracting 3-d information on the primordial power spectrum at hitherto unseen scales through observations of the 21 cm background and spectral distortions of the CMB only makes this endeavour even more pertinent.

In the context of inflationary cosmology, transient stronger couplings in the effective theory of the adiabatic mode, the presence of initial particles or modified vacua, particle production, phase transitions and modulations of the effective potential during inflation all furnish examples where localized features and ringing will be sourced in the power spectrum and higher order correlators. From these features, one can in principle infer characteristic mass scales which offer a window (with appropriate degeneracies from the perspective of the effective theory) onto the couplings and field content of the parent theory in which inflation is embedded. Characteristic scales that parametrize relaxation to the attractor, or the decay of any initially present particles could offer us a glimpse onto pre-inflationary dynamics or excited initial states if inflation lasted not much longer than required by the solution to the horizon problem. Additionally, the presence of localized violations of slow roll and features in the potential will also source features in the power spectrum that can in principle, be distinguished from the previously mentioned possibilities.

In what follows, we will review the theory, the motivation and the observational prospects for detecting features and new characteristic scales in the primordial correlation functions. As mentioned, the enterprise is especially relevant given the wealth of data from LSS and 21 cm observations that are due or envisaged in the near future. The latter promises to give us access to comoving scales far beyond those available to us in CMB and LSS at redshifts where they remain in the linear regime. In combination with measurements of spectral distortions of the CMB, we are thus being offered a much larger window onto inflation as it progressed. 

Although the following treatment won't be exhaustive, we hope to cover the basic aspects of those mechanisms that are well motivated theoretically and whose predictions might conceivably be tested against current and future cosmological data. The first half of our treatment begins in the next section by formulating a perturbative understanding of the origin of features in the comoving curvature power spectrum and bispectrum (to be precisely defined further), denoted $\calP_\calR$ and $\calB_\calR$, respectively, that model many physical situations of interest. From these, we hope to gain an understanding of which signatures to expect from a given context and whether they can be distinguished from each other. We then review some of the more widely studied mechanisms that could generate features in primordial correlators and how these relate to their underlying fundamental physics realizations. The second half of the treatment commences with a review of how features in the primordial correlators of $\calR$ imprint on the temperature and polarization anisotropies of the CMB as well as in distortions of its spectrum, after which we turn our attention towards observational aspects of searching for such features. Certain details of our treatment that require more involved elaborations are covered in the appendices.  

}

\section{Preliminaries \label{sec:preliminaries}}
{

To a large degree of accuracy, the cosmological perturbations observed in the CMB are adiabatic in nature. One can interpret this as an indication for them having been generated by (i) a single degree of freedom, or (ii) through some mechanism that converts entropy perturbations associated with multiple degrees of freedom into adiabatic perturbations shortly before or (iii) after horizon exit \cite{Lyth:2001nq}. For simplicity and predictability, we will focus on the former scenario in what follows, however, it should be clear that the basic examples that we'll cover in the following generalize straightforwardly.  

\subsection{The generation of curvature perturbations}
In what follows, we will set up a simple framework for understanding how features can be generated and imprinted in the primordial power spectrum in any given context. We begin with the most general quadratic action for cosmological perturbations in terms of the so-called Mukhanov-Sasaki (MS) variable \cite{Mukhanov:1990me,Seery:2005wm}:
\eq{ms}{S_2 = \frac{1}{2}\int \dint^4x~\left(v'^2 - c_\mathrm{s}^2(\nabla v)^2 + \frac{z''}{z}v^2\right)} 
where the above is defined on conformal coordinates and where
\eq{zdef}{z := a \frac{\phi_0'}{\mathcal H c_\mathrm{s}}.}
In the above, primes represent derivatives with respect to conformal time $\tau$, $\phi_0$ the background inflaton field, $\calH := a'/a$ and $c_\mathrm{s}$ the adiabatic sound speed. There are multiple advantages with working with the MS variable. Chief among them is the fact the action for the MS variable (in conformal coordinates) is that of a canonically normalized scalar field on Minkowski space, and hence quantization and specifying the initial vacuum state is straightforward. In addition, all information about the background solution which we perturb around to leading order is contained in $c_\mathrm{s}$ and $z$. The relationship with the comoving curvature perturbation\footnote{In this review, we adopt the notation $\calR$ for the \textit{comoving curvature perturbation}, that is the curvature purturbation on a foliation where the perturbed momentum flux $\delta T^i_0$ vanishes. In single field inflation, this is equivalent to the foliation where the inflaton perturbation $\delta\phi$ vanishes -- a gauge choice also known as `unitary' gauge, whose significance we shall discuss shortly. Recently, the notation $\zeta$, traditionally reserved for the curvature perturbation on \textit{constant energy density hypersurfaces} has become popular notation for the comoving curvature perturbation. The two variables are equivalent in the context of single field inflation but will differ in general. However this switch in notation -- popularized by \cite{Maldacena:2002vr} -- differs from historical use and the convention in which observations are reported \cite{Planck:2015xua, Hinshaw:2012aka}, to which we defer.} is given by
\eq{ccp}{v = z\calR,} 
in terms of which the perturbed quadratic action (reverting to cosmological time $t$) reads
\eq{ccpa}{S_2 = \int \dint^4x~a^3\epsilon M^2_\mathrm{pl}\left(\frac{\dot \calR^2}{c_\mathrm{s}^2} - \frac{(\nabla\calR)^2}{a^2}\right),}
where dots refer to cosmological time derivatives and $\epsilon := -\dot H/H^2 = \dot\phi_0^2/(2M^2_\mathrm{pl}H^2)$. The advantage of working with the comoving curvature perturbation is that in the absence of isocurvature perturbations, it is conserved at super-horizon scales to all orders in perturbation theory \cite{Salopek:1990jq} (to arbitrary order in loops) \cite{Pimentel:2012tw, Senatore:2012ya, Assassi:2012et}, which allows us to relate its value at horizon re-entry directly to its value at horizon exit during inflation.

Since temperature and polarization anisotropies directly relate to $\calR$, all late time observables will be expressed in terms of correlators of these. However for purposes of disentangling the various sources of features that could imprint on observables during inflation, it is useful for the moment to continue working with the MS variable. It should be emphasized that (\ref{ms}) and (\ref{ccpa}) only assume the monotonicity of $\phi_0$ (i.e. that it is a good physical clock) and that the background deviates from exact de Sitter space\footnote{Otherwise the scalar perturbations would be pure gauge.}. The primary object of interest to compute in determining the CMB temperature and polarization correlation functions is the so-called dimensionless power spectrum, which for the comoving curvature preturbation is defined as
\eq{2pf}{2\pi^2\delta^3(\vec k + \vec q)\calP_{\calR,\psi_\mathrm{I}}(k):= k^3\langle \psi_\mathrm{I}|\widehat \calR_{\vec k} \widehat \calR_{\vec q} |\psi_\mathrm{I}\rangle|_\mathrm{in~in},}
with $k:=|\vec k|$ and where the subscripts indicate that we are interested in computing finite time correlation functions, requiring us to work in the Schwinger-Bakshi-Mahanthappa-Keldysh or `in-in' formalism for short\footnote{For a review of this formalism in the context of computing cosmological correlators, see \cite{Weinberg:2005vy}. For certain caveats regarding its implementation in computing higher order cosmological correlators, see  \cite{Adshead:2009cb}.}, which we assume to always be the case, henceforth dropping the subscripts indicating so. We review the basics of the in-in formalism tailored to our purposes in \ref{in-in}. 

For the purposes of the discussion to follow, we allow for the perturbations to begin at some fixed (or asymptotic) initial time in an arbitrary initial state $|\psi_\mathrm{I}\rangle$, which in most cases we will take as the Bunch-Davies state. As we shall see in Sect.~\ref{sec:observation}, the TT angular power spectrum can be expressed in terms of its multipole coefficients from (\ref{2pf}) for the low multipoles ($\ell \lesssim 30$) \cite{Mukhanov:2005xxx,Mukhanov:2003xr} as
\eq{aps}{C_\ell = \frac{2}{9\pi}\int \frac{\dint k}{k}\calP_{\calR,\psi_\mathrm{I}}(k)j_\ell^2[k(\tau_0-\tau_r)],}
where from (\ref{ccp}) $\calP_{\calR,\psi_\mathrm{I}} = \calP_{v,\psi_\mathrm{I}}/z^2$, where $j_\ell$ is the spherical Bessel function of order $\ell$ and where $\tau_0-\tau_r$ represents the comoving distance to the last scattering surface. 

Before we proceed to discuss the generation of features in power spectrum, it is useful to remind ourselves how a scale invariant spectrum is generated in the standard slow roll inflationary scenario, the very definition of which requires that
\eq{sreps}{\epsilon = -\dot H/H^2 \equiv \frac{\dot\phi_0^2}{2M^2_\mathrm{pl}H^2}}
be small (the latter equivalence follows from the Friedmann equations). Assuming $\epsilon$ to be constant (certainly accurate to first order in slow roll parameters), one can integrate the above in cosmological time to yield $H = \frac{1}{\epsilon t + c}$, with $c$ some integration constant that we define as $c := H^{-1}_0$. Integrating once more and expressing the new integration constant that appears in terms of the scale factor $a_0$ at $t=0$ we find
\eq{asol}{a(t) = a_0(1 + \epsilon H_0 t)^{1/\epsilon}}
Clearly when $\epsilon \to 0$ the above is the very definition of the exponential:
\eq{asolas}{\lim_{\epsilon\to 0} a(t) = a_0~e^{H_0 t}.}
We stress however that (\ref{asol}) is exact for any finite value of $\epsilon$ that is constant. Converting to conformal time and dropping the subscript on $H$, the scale factor (\ref{asol}) becomes
\eq{asolc}{a = \frac{1}{[-\tau H(1-\epsilon)]^{\nu-1/2}},}
where we have defined
\eq{nudef}{\nu := \frac{3-\epsilon}{2(1-\epsilon)}.}
We now presume a phase of slow roll with $c_\mathrm{s} \equiv 1$ and with $\epsilon$ constant or small enough so that we can neglect higher order corrections. In this limit, the equations of motion that result from (\ref{ms}) per Fourier mode are given by
\eq{msk}{v''_k + \left(k^2 - \frac{\nu^2-\frac{1}{4}}{\tau^2}\right)v_k = 0}
which follows from re-expressing (\ref{zdef}) as $z = a\sqrt{2\epsilon}M_\mathrm{pl}$ and using (\ref{asolc}). The solution for the mode function requires the additional input of boundary conditions, which we define in the asymptotic past. Specifically, at early enough times ($\tau \to -\infty$), a particular comoving scale will be deep within the horizon and for which spacetime will look like it effectively has the properties of Minkowski space. Requiring that all such modes begin in their corresponding Minkowski space vacuum means that the solutions must asymptote to the plane wave solution $v_k \sim e^{-i k \tau}/\sqrt{2k}$, which fixes the normalization and phase of the two independent solutions of (\ref{msk}) to be such that
\eq{vsol}{v_k = \frac{\sqrt\pi}{2}e^{i\left(\nu + \frac{1}{2}\right)\frac{\pi}{2}}\sqrt{-\tau}H^{(1)}_\nu(-k\tau)} 
corresponds to the mode functions of the so called Bunch-Davies vacuum state \cite{Birrell:1984xxx,Bunch:1978yq}. From the above and (\ref{ccp}), one finds that the mode functions for the comoving curvature perturbation are given by
\eq{mfdef}{\calR_k(\tau) = \frac{e^{i\left(\nu + \frac{1}{2}\right)\frac{\pi}{2}}}{2}\sqrt{\frac{\pi}{2\epsilon}}\frac{[H(1-\epsilon)]^{\nu-1/2}}{M_\mathrm{pl}}(-\tau)^\nu H_\nu^{(1)}(-k \tau),}
which straightforwardly reduces to the familiar expression in the limit $\nu \to 3/2$:
\eq{mf3/2}{\calR_k(\tau) = i \frac{H}{M_\mathrm{pl}}\frac{1}{\sqrt{4\epsilon k^3}}(1 + i k \tau)e^{-i k\tau}.}
Given that we expand the comoving curvature perturbation around the Bunch-Davies vacuum $|\psi_\mathrm{I}\rangle \equiv |0\rangle_{BD}$ in terms of the Fourier modes (\ref{mfdef}) as
\eq{oexp}{\widehat\calR_{\vec k} = \calR_k(\tau)\widehat a_{\vec k} + \calR^*_k(\tau){\widehat a^\dag}_{-\vec k},}
where the interaction picture creation and annihilation operators satisfy the usual commutation relations $[\widehat a_{\vec k}, {\widehat a^\dag}_{\vec q}] = (2\pi)^3 \delta^{(3)}(\vec k - \vec q)$, one can straightforwardly evaluate the power spectrum (\ref{2pf}) at late times ($k\tau \to 0^{-}$) as\footnote{Where we make use of the asymptotic expansion of the Hankel function $\displaystyle\lim_{z\to 0}H^{(1)}_\nu(z) \sim -i/\pi\Gamma(\nu)(2/z)^\nu$.}
\eq{ltps}{\calP_\calR(k) = \frac{H^2}{8\pi^2\epsilon M^2_\mathrm{pl}}\left(\frac{k}{H}\right)^{n_\mathrm{s}-1}}
where the so called spectral tilt is given by
\eq{tilt}{n_\mathrm{s}-1 = 2\eta - 4\epsilon}
with $\eta := -\ddot\phi/(H\dot\phi)$ being the familiar `eta' parameter\footnote{The $\eta$ contribution to the tilt arises from the fact that once one allows for time dependence of the slow roll parameters, the tilt gets extra contributions that can be obtained by to first approximation through the mnemonic of simply evaluating the power spectrum at the time the {\it comoving} wave number $k$ crosses the horizon, inducing an implicit $k$ dependence in the $\epsilon$ factor in the denominator of (\ref{ltps}) which can be determined from the identity $\dot\epsilon = 2H\epsilon(\epsilon-\eta)$.}. In fact, since the time variation of the slow roll parameters are themselves second order in slow roll quantities (more generally, $\epsilon$ and $\eta$ are a subset of an infinite hierarchy of `slow roll' parameters \cite{Liddle:1994dx,Kinney:2002qn}) one can understand the right hand side of (\ref{tilt}) as depending adiabatically on $k$, implying a possible running of the tilt, a situation for which there exists no decisive evidence to date \cite{Ade:2015lrj}. Thus for small enough values of the slow roll parameters, the power spectrum is almost scale invariant. We are now equipped to turn our attention towards the main topic of this section-- how the various mechanisms that could superpose features on this spectrum do so. 

\subsection{Features and characteristic scales in $\calP_\calR(k)$}
We first recall the action for the MS variable (\ref{ms})
\eq{msf}{\nonumber S_2 = \frac{1}{2}\int \dint^4x~\left(v'^2 - c_\mathrm{s}^2(\nabla v)^2 + \frac{z''}{z}v^2\right)} 
where we take note of the fact that the above describes the action for an adiabatic perturbation around {\it any} cosmological background (not just those that are slow roll inflating) provided that the field $\phi_0 $ furnishes a good physical clock (i.e. so that $z = a\frac{\phi_0'}{c_\mathrm{s} \calH}$ never vanishes). Furthermore, we take note of the fact that to leading order in the effective theory (see the following subsection) all information of the background solution is contained in the functions $c_\mathrm{s}$ and $z$, and all information of the boundary conditions are contained in the initial state $|\psi_\mathrm{I}\rangle$, or more generally in the initial density matrix $\widehat\rho_\mathrm{I}$, usually taken to represent the Bunch-Davies vacuum. 

We now imagine a fiducial slow roll attractor solution with $c_\mathrm{s}\equiv c_0$, $z = z_0(\tau)$. Retracing our steps in the last section results in the power spectrum
\eq{psact}{\calP_\calR(k) = \frac{H^2}{8\pi^2\epsilon  M^2_\mathrm{pl}}\frac{1}{c_0}\left(\frac{c_0 k}{H}\right)^{3-2\nu_0}}
with the spectral index 
\eq{si}{n_0-1 = 3-2\nu_0}
now corresponding to that generated by the background trajectory defined by $z_0$. We assert that a wide range of situations where localised features and oscillations can be superposed onto such an attractor spectrum can be understood in a `perturbative' context\footnote{With the exception of situations where slow roll is interrupted at intermediate epochs, see Sect. \ref{sec:theory} for a discussion of this possibility.}. That is to say, all observable correlators can be computed as perturbative corrections to the correlators generated on the background solution defined by $c_0$ and $z_0$. The underlying reasoning is simple -- we generically expect any features that might be present to be smaller in magnitude than the amplitude of the spectrum locally. To flesh this out, in what follows, we take $c_0 = 1$ without loss of generality, and imagine a localized drop in the speed of sound or localized deviations from the fiducial attractor solution. As we shall show in the next section, this is the resultant in many situations of interest (including those incorporating loop effects) that generate localized features. 

Consider the time dependence of the background quantities $c_\mathrm{s}(\tau)$ and $z(\tau)$ as perturbative deviations from the fiducial quantities $c \equiv c_0$ and $z_0(\tau)$ by defining
\eq{wdef}{w(\tau):= c_0^2 - c_\mathrm{s}^2(\tau),} 
\eq{Wdef}{W(\tau):= \frac{z''}{z} - \frac{z_0''}{z_0}.}
Doing so allows us to write the quadratic action (\ref{ms}) in terms of a `free' part and an `interacting' part as
\eq{S}{S_2 = \frac{1}{2}\int \dint^4x~\left(v'^2 - c_0^2(\nabla v)^2 + \frac{z_0''}{z_0}v^2\right) + S_{2,\mathrm{int}}} 
with
\eq{Sint}{S_{2,\mathrm{int}}:= \frac{1}{2}\int \dint^4x~\left[w(\tau)(\nabla v)^2 + W(\tau)v^2\right].}
The presumption is that both $w(\tau)$ and $W(\tau)$ differ significantly from zero only over a finite interval, and remain perturbatively small uniformly in time as we'll explicitly show to be the case over several examples in the following section. Even though $w$ and $W$ both implicitly depend on the adiabatic sound speed, for the purposes of computing corrections to the attractor power spectrum, we can treat them as independent perturbations.  

Given that the fiducial attractor solution admits the mode expansion  
\eq{vfid}{v^0_k = \frac{\sqrt\pi}{2}e^{i\left(\nu_0 + \frac{1}{2}\right)\frac{\pi}{2}}\sqrt{-\tau}H^{(1)}_{\nu_0}(-c_0 k\tau)}
with the index $\nu_0$ now characterizing the attractor defined by $z_0$, we can expand the `free' part of the MS operator $\widehat v$ in the interaction picture in terms of its mode decomposition as
\eq{oexpv}{\widehat v_{\vec k} = v^0_k(\tau)\widehat a_{\vec k} + v^{0*}_k(\tau){\widehat a^\dag}_{-\vec k},}
where the interaction picture creation and annihilation operators satisfy the usual relations $[\widehat a_{\vec k}, {\widehat a^\dag}_{\vec q}] = (2\pi)^3 \delta^{(3)}(\vec k - \vec q)$. From the results of \ref{in-in}, we can readily infer that the two point correlation function of the MS variable is corrected by the perturbation $W(\tau)$ as
\begin{equation}
\delta_W \langle {\widehat v}_{{k}_1}(\tau)\, {\widehat v}_{{k}_2}(\tau) \rangle \ =  (2\pi)^3\, \delta^3({\vec k}_1 + {\vec k}_2)\int^\tau_{\tau_0}\dint {\tau'}~ 2W({\tau'})\Im \left\{G^0_{{k}_1}(\tau,{\tau'})\, G^0_{{k}_2}(\tau,\tau')\right\} \,\label{npansW}
\end{equation}
and by the perturbation $w(\tau)$ as
\begin{equation}
\delta_w \langle {\widehat v}_{{k}_1}(\tau)\, {\widehat v}_{{k}_2}(\tau) \rangle \ =  (2\pi)^3\, k_1^2 \delta^3({\vec k}_1 + {\vec k}_2)\int^\tau_{\tau_0}\dint {\tau'}~ 2w({\tau'}) \Im \left\{G^0_{{k}_1}(\tau,{\tau'})\, G^0_{{k}_2}(\tau,\tau')\right\} \,\label{npansw}
\end{equation}
where $\Im\{...\}$ denotes the imaginary part of, and where the `free field' Green's functions are given by
\eq{gfsol}{G^0_{k}(\tau,\tau')\ = \ \frac{\pi}{4}\ \sqrt{\tau\,\tau'}\ H_{\nu_0}^{(1)}(- c_0 k\,\tau)\, H_{\nu_0}^{(2)}(- c_0k\,\tau') \ .}
For the special case of a fiducial de Sitter background, this is given by
\eq{dg}{G^0_k(\tau,\tau') = \frac{1}{2c_0k}\Bigl(1 - \frac{i}{c_0 k\tau}\Bigr)\Bigl(1 + \frac{i}{c_0 k\tau'}\Bigr)e^{-ic_0 k(\tau-\tau')}.}
Converting to the power spectrum for the comoving curvature perturbation at late times\footnote{One must take care in evaluating the power spectrum at late enough times ($\tau \to 0^-$) such that the residual evolution of quantities that are ordinarily conserved on super-horizon scales subsides sufficiently. Failing to do so (for example, by evaluating quantities at horizon crossing) can result in errors as large as the quantities we are attempting to compute, such as the spectral tilt \cite{Nalson:2011gc, Nalson:2013tm}.}, at which point by assumption the background trajectory will have reverted to the fiducial attractor solution parametrized by $z_0$ and $c_0$, results in the following correction to the power spectrum
\eq{psc}{\Delta \calP_\calR(k) = -\calP_\calR(k) 4\pi^4\int^0_{\tau_0} \dint\tau  \left[\begin{matrix} W(\tau)\\ k^2 w(\tau) \end{matrix}\right]\Im\left\{ (-\tau)H_{\nu_0}^{(2)^2}(-c_0 k \tau)\right\} }
with the attractor power spectrum $\calP_\calR(k)$ given by the expression (\ref{psact}), and where the features are induced either by the perturbations $w$ or $W$ (given by (\ref{wdef}) and (\ref{Wdef}) respectively), or both, as will generically be the case when the speed of sound transiently changes during inflation. Hence, corrections to the attractor power spectrum manifest as the integral transform of the interaction potentials $W(\tau)$ and $w(\tau)$ with the kernel
\eq{kernel}{K_{\nu_0}(\tau, k) := \Im\left\{ (-\tau)H_{\nu_0}^{(2)^2}(-c_0 k \tau)\right\}.}
Since the fiducial background solution is presumed to be that of a slow rolling attractor, one can readily expand the kernel in powers of slow roll parameters such that to leading order one has:
\eq{kernelds}{K_{3/2}(\tau, k) := \frac{2}{\pi}\frac{1}{c_0 k}\Im\left\{e^{2i c_0 k \tau} \left(1 + \frac{i}{c_0 k \tau}\right)^2\right\},}
where $K_{3/2}$ is the kernel for a fiducial de Sitter attractor and subleading corrections can readily be computed by the expansion
\eq{kerexp}{H^{(2)}_{\nu_0}(z) = H^{(2)}_{3/2}(z) + \frac{1- n_0}{2}\frac{\partial H^{(2)}_{3/2}}{\partial \nu}(z)}
where the first Taylor coefficient $\nu_0-3/2$ can be expressed in terms of the attractor tilt by (\ref{si}). For completeness, we note that 
\eq{od}{\frac{\partial H^{(2)}_{3/2}}{\partial \nu}(z) = i\pi J_{3/2}(z) + [Ci(2z) - i Si(2z)]H^{(1)}_{3/2}(z) + \frac{2}{z}H^{(2)}_{1/2}(z),}
though it is evident that to leading order (\ref{kernelds}) suffices. Thus we arrive at the key expression
\eq{psd}{\boxed{\frac{\Delta \calP_\calR}{\calP_\calR}(k) = -\frac{8\pi^3}{c_0 k}\int^0_{\tau_0} \dint\tau  \left[\begin{matrix} W(\tau)\\ k^2 w(\tau) \end{matrix}\right] \Im\left\{e^{2i c_0 k \tau} \left(1 + \frac{i}{c_0 k \tau}\right)^2\right\}}}
Some remarks are in order at this point. It should be clear that the features induced in the power spectrum only have finite support in $k$ space if the interaction potentials $W(\tau)$ and $w(\tau)$ do not contain arbitrarily fast variations. That is, unless the potentials have a non-zero projection onto the kernel (\ref{kernelds}) for arbitrarily high momenta, the features induced in the power spectrum will cut off. One can expect reasonable bounds on how fast the interaction potentials can vary from the requirement that our background solution derive consistently from an underlying low energy effective description. Specifically, sudden changes in the inflationary potential or in the speed of sound are limited by the requirements that the inflaton embeds itself consistently in some UV completion, which implies the validity of a consistent derivative expansion for the inflaton field and its fluctuations, precluding arbitrarily fast variations in $W(\tau)$ and $w(\tau)$\footnote{Conversely, any `sharp' features in the inflaton potential or sudden drops in $c_s$ (defined as those for whose derivatives become comparable to the cut-off of the theory) are unlikely to be radiatively stable, as quantum corrections tend to smear these out in field space or in time, respectively.}. For example, as we shall review in the next section, adiabaticity requirements bound variations in the speed of sound as \cite{Cespedes:2012hu, Achucarro:2012yr}
\eq{adc}{|\dot c_s| \ll M |1-c_s^2|,}  
where $M$ parametrizes the scale of heavy physics that we've integrated out in obtaining our low energy effective theory (EFT). 

We also wish to remark that changes in the speed of sound can reasonably imprint at much smaller scales in the CMB relative to localized features in the potential (or localized deviations off the attractor). This is evident from the $k^2$ factor that dresses the kernel (\ref{kernelds}) for the interaction potential $w(\tau)$ relative to the interaction potential $W(\tau)$ (recalling (\ref{wdef}) and (\ref{Wdef})). This will prove important not only due to the fact that a locally varying sound speed is strongly motivated from underlying EFT considerations, but also as localized features are much easier to constrain at shorter comoving scales in the CMB, as we shall see shortly\footnote{This is due to the same volume vs surface area gain in the number of modes available as noted in footnote \ref{gain}.}. The tantalizing prospects of being able to discern the matter power spectrum at scales far beyond those accessible in the CMB through future 21 cm observations is not to be taken lightly in this regard.
\begin{center}
\begin{figure}[h]
\includegraphics[width=12.5cm]{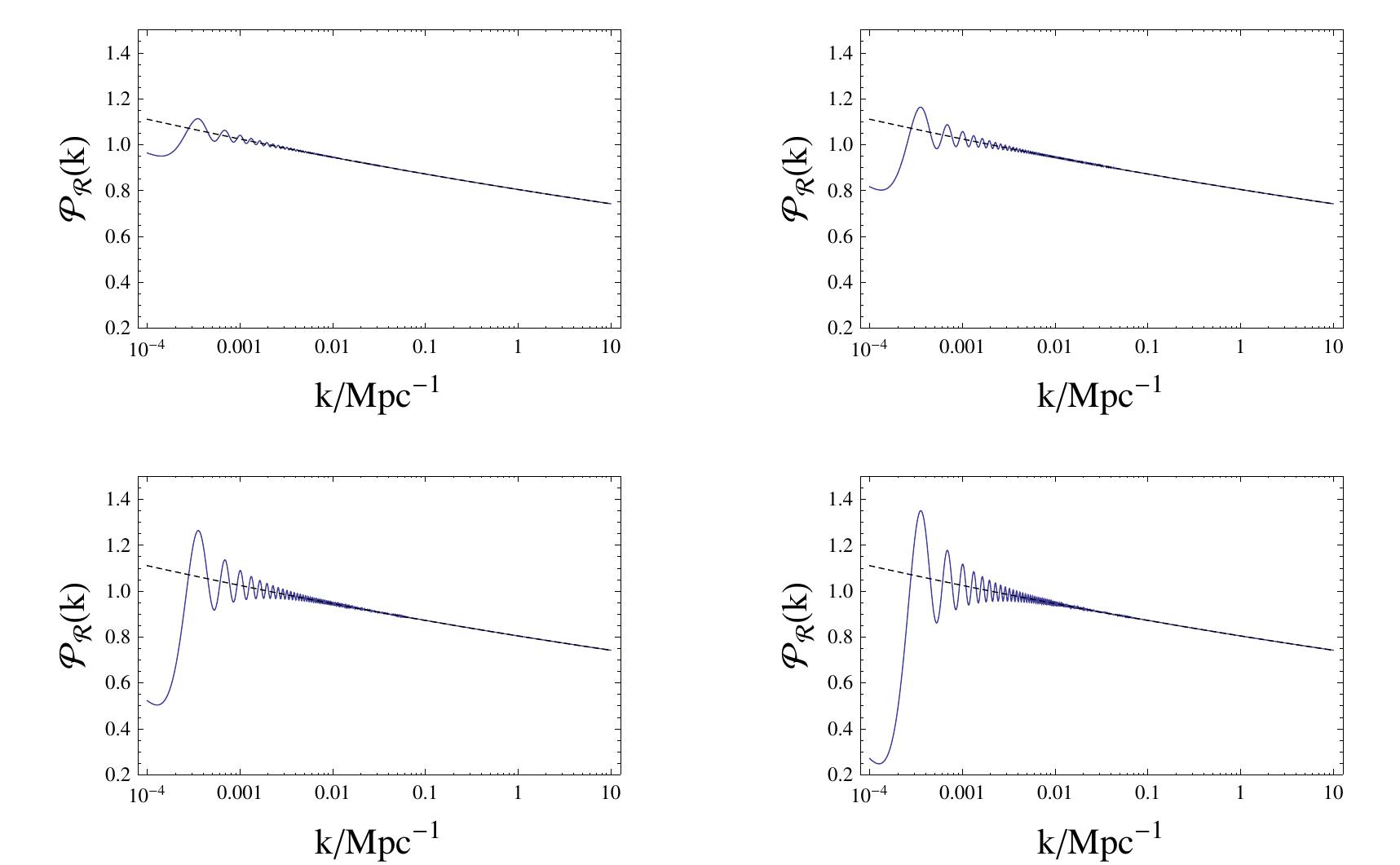}
\caption{\label{fig:relax} Relaxation to the attractor with $W(\tau) = \lambda e^{-(\tau-\tau_0)\mu}$, with $\lambda = 5\times 10^{-5}/(4\pi^4), \tau_0 = -10^4$ and with $\mu$ running from $2,1,0.5$ and $0.35$ in the upper left, upper right, lower left and lower right panels, respectively. The fiducial attractor has spectral index $n_s = 0.965$.}
\end{figure}
\end{center}
\begin{figure}[h]
\includegraphics[width=12.5cm]{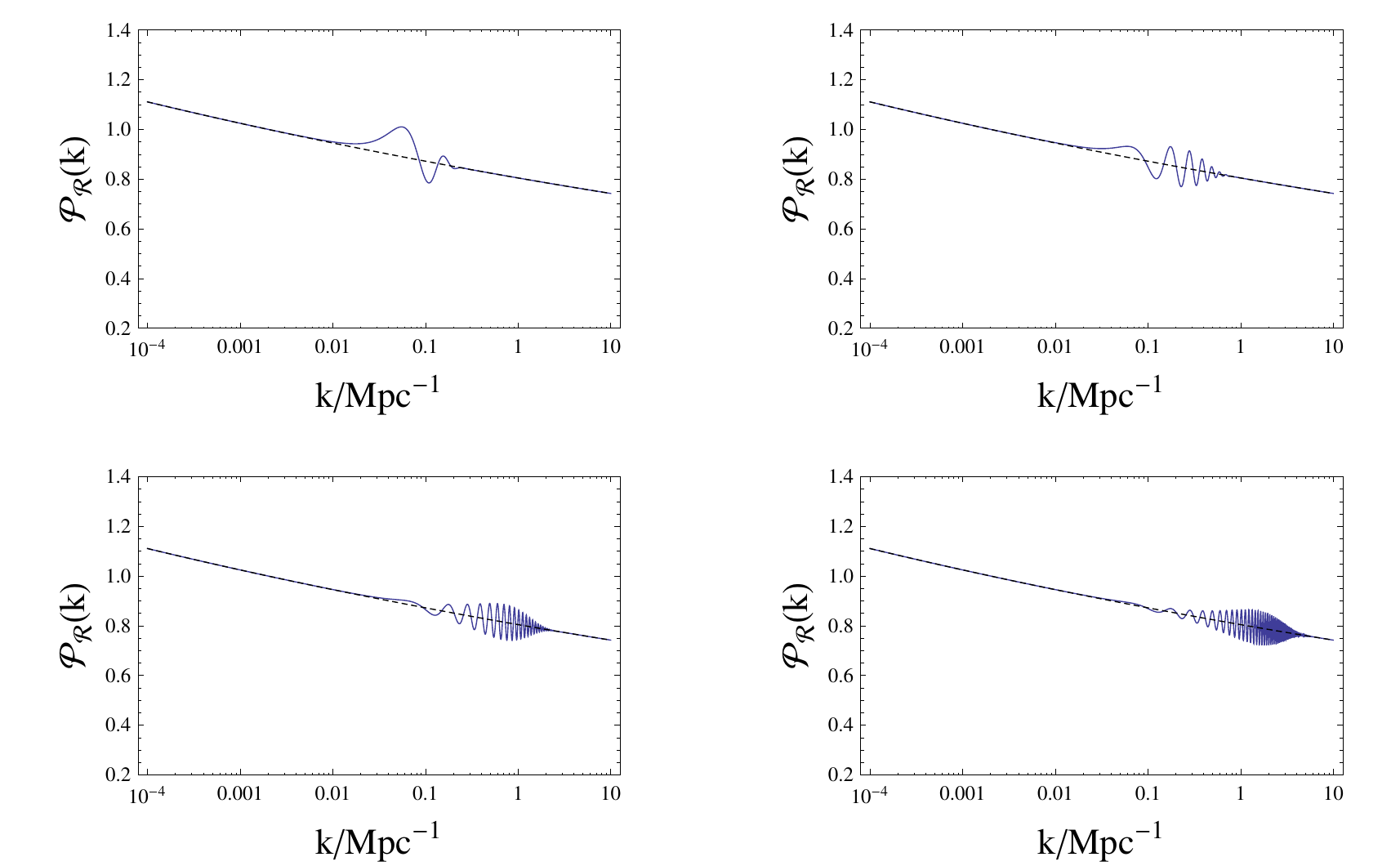}
\caption{Transient drop in $c_s$, modelled by $w(\tau) =  \lambda \tau^2e^{-(\tau-\tau_0)^2\mu}$, with  $\lambda = 2\times 10^{-4}/(4\pi^4), \tau_0 = -30$ and with $\mu$ running from $0.01,0.1,1$ and $5$ in the upper left, upper right, lower left and lower right panels, respectively. The fiducial attractor has spectral index $n_s = 0.965$. \label{fig:pulse_cs}}
\end{figure}
For illustrative purposes, we consider the effects of localized variations in the potentials $W(\tau)$ and $w(\tau)$ that qualitatively model a background that relaxes to the attractor (Fig. \ref{fig:relax}) and a transient drop in $c_s$ (Figs. \ref{fig:pulse_cs} and \ref{fig:long_cs}) respectively. We take note of the salient differences between these examples viz. the evident suppression of power at large comoving scales and the ease with which varying speeds of sound can imprint features at shorter comoving scales without violating adiabaticity. We stress that all physical examples that locally imprint features (with all modes initially in their adiabatic vacuum state) are captured to leading order in the variations of the potentials $w(\tau)$ and $W(\tau)$. These examples include certain models that locally interrupt slow roll, models with localized features in the potential, situations where localized particle production occurs either at tree or loop level, among others. We will be surveying the possibilities in the following section. 

\begin{figure}[h]
\includegraphics[width=12.5cm]{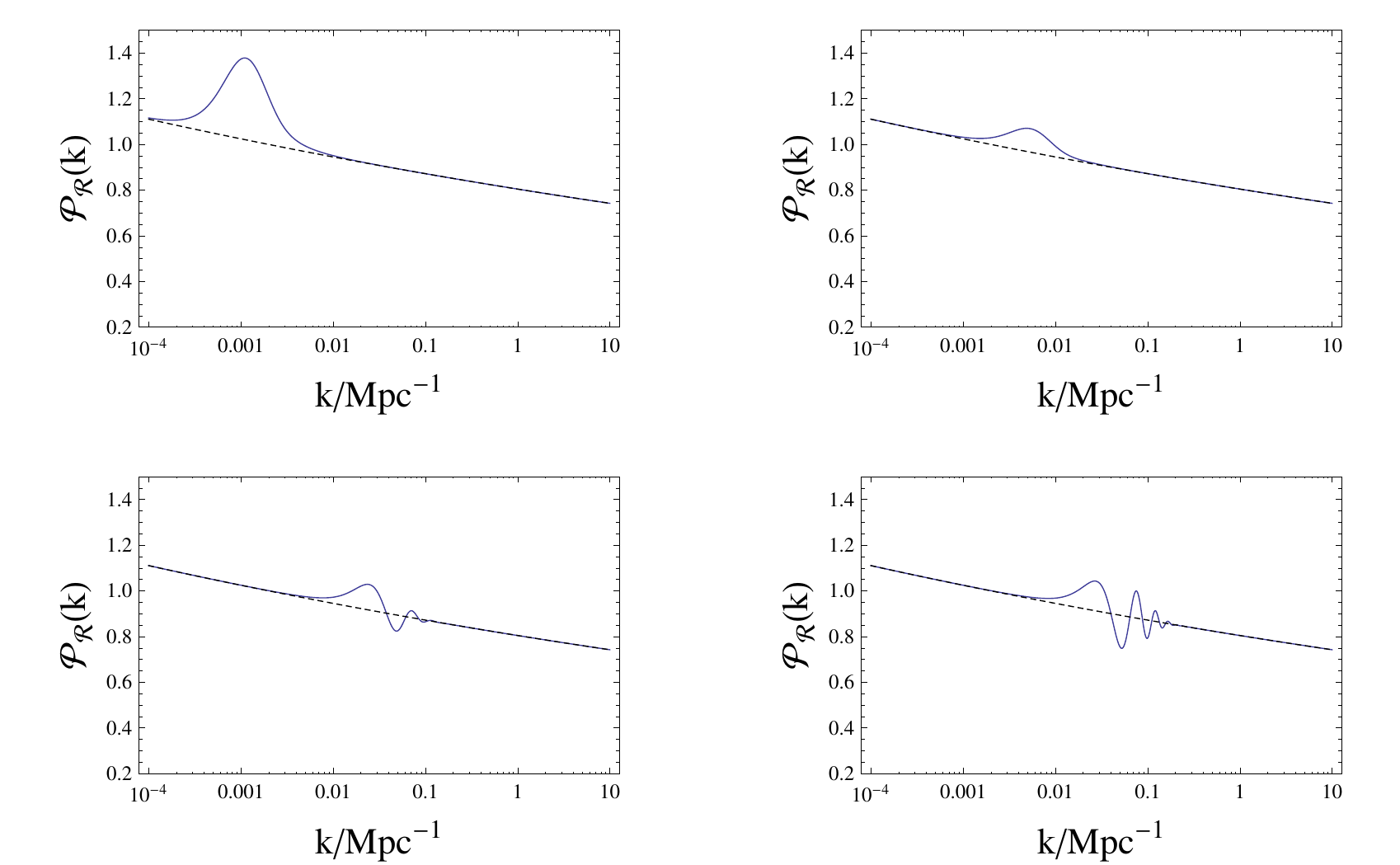}
\caption{\label{fig:long_cs} Transient drop in $c_s$, modelling a so called `ultra-slow turn',  represented by $w(\tau) =  \lambda \tau^2 ({\rm  Tanh}[(\tau-\tau_i)\mu] - {\rm Tanh}[(\tau-\tau_f)\mu])$, with  $\lambda = 5\times 10^{-5}/(4\pi^4), \tau_i = -75, \tau_f = -60$ and with $\mu$ running from $0.001, 0.005, 0.05$ and $0.1$ in the upper left, upper right, lower left and lower right panels, respectively. The fiducial attractor has spectral index $n_s = 0.965$.}
\end{figure}

Situations in which the perturbation modes begin in states that differ from the adiabatic vacuum (in the case of de Sitter space, this is the Bunch-Davies vacuum) can further modulate features on the power spectrum. Reconsider (\ref{2pf}) with an arbitrary initial state $|\psi_I\rangle$   
\eq{mic}{ 2\pi^2\delta^3(\vec k + \vec q)\calP_{\calR,\psi_\mathrm{I}}(k):= k^3\langle \psi_\mathrm{I}|\widehat \calR_{\vec k} \widehat \calR_{\vec q} |\psi_\mathrm{I}\rangle|_\mathrm{in~in}.} 
Were $|\psi_I\rangle$ to correspond to an eigenstate of the number operator of the curvature quanta (or were it the result of tracing over a thermal density matrix), the result would be a straightforward modulation of the power spectrum 
\eq{dprm}{\calP_{\calR,\psi_I}(k) = \calP_{\calR,0}(k)[1 + 2n(k)] \  ,}
where $n(k)$ is the occupation number of the $k^{th}$ mode. In general, any state for which $\langle {\widehat a}_{\vec k} \rangle = \langle \widehat a^\dag_{\vec k} \rangle = 0$ while $\langle \widehat a_{\vec k} \widehat a^\dag_{\vec p}\rangle$ is $\delta$--function correlated\footnote{\label{flabel} Such states are particular examples of a more general class of states-- the so--called ``random phase'' states-- for which the simple expression (\ref{dprm}) results.  \cite{Kandrup:1988vg,Kandrup:1988sc}.} results in this phase independent modulation. More generally, we might be interested in situations where the initial state does not satisfy these properties. Consider for example that at $t_I$, the initial state corresponded to some state for which mode by mode, the mode functions were rotated as 
\begin{equation}
\begin{split}
\label{fexp}
\calR'_{\vec k} &= \cosh\theta_{\vec k}\, \calR_{\vec k} \ +\ e^{-i\delta_{\vec k}}\, \sinh\theta_{\vec k}\, \calR^\star_{\vec k} \ , \\ \calR'^\star_{\vec k} &= \cosh\theta_{\vec k}\,
\calR^\star_{\vec k} \ + \ e^{i\delta_{\vec k}}\, \sinh\theta_{\vec k}\, \calR_{\vec k} \ ,
\end{split}
\end{equation}
where the transformation is defined on the hypersurface at the time $t_I$. In this case, one can write the initial state as $|\psi_I\rangle = {\widehat U}(\Theta) |0\rangle$, where ${\widehat U}(\Theta)$ is a unitary operator that we will explicitly construct shortly. The correlator we're interested in computing (\ref{mic}) can therefore be rewritten as $\langle 0 | {\widehat U}^{\dagger } {\widehat \calR}_{\vec k}(t) {\widehat U} \,
{\widehat U}^{\dagger } {\widehat \calR}_{\vec q}(t) {\widehat U} \, |0 \rangle$, so that it is equivalent to calculating it in the original adiabatic vacuum after having performed a unitary transformation on the $\widehat \calR$ operators. Evidently, the interaction picture field that produces quanta with the wavefunctions $R'_{\vec k}$ is given by
\begin{eqnarray}
{\widehat \calR}' (t,{\vec x}) \ = \ \ {\widehat U}^{\dagger} \ {\widehat \calR} (t,{\vec x}) \ {\widehat U} && =\ \int \frac{\dint^3k}{(2 \pi)^{3/2}} \ \Bigl[ \calR'_{\vec k}(t)\, e^{i {\vec k} \cdot {\vec x}}\, {\widehat a}_{\vec k} \ + \calR'^\star_{\vec k}(t)\, e^{-i {\vec k} \cdot {\vec x}}\, {\widehat a}^\dag_{\vec k}\Bigr] \ \\ \nonumber &&= \ \int \frac{ \dint^3k}{(2 \pi)^{3/2}} \ \Bigl[ \calR_{\vec k}(t)\, e^{i {\vec k} \cdot {\vec x}}\, {\widehat b}_{\vec k} \ +\calR^\star_{\vec k}(t)\, e^{-i {\vec k} \cdot {\vec x}}\, {\widehat b}^\dag_{\vec k}\Bigr] \ ,
\end{eqnarray}
where the two sets of creation and annihilation operators relate as
\begin{equation}
\begin{split}
\label{car} {\widehat b}_{\vec k} &= \cosh\theta_{\vec k}\, {\widehat a}_{\vec k} \ + \ e^{i\delta_{\vec k}}\, \sinh\theta_{\vec k}\, {\widehat a}^\dag_{- \vec k} \ , \\
{\widehat b}^\dag_{\vec k} &= \ e^{-i\delta_{\vec k}}\, \sinh\theta_{\vec k}\, {\widehat a}_{-\vec k} \ + \ \cosh\theta_{\vec k}\, {\widehat a}^\dag_{\vec k} \ .
\end{split}
\end{equation}
If the transformation (\ref{fexp}) only rotates modes up to some fixed comoving momentum scale, then the transformation is unitary\footnote{The Hilbert spaces generated mode by mode by the Heisenberg algebras related by the transformation (\ref{car}) are only unitarily equivalent if $\theta_{k}$ has finite support. That is, we require the transformation (\ref{car}) to tend to the identity sufficiently fast as $|\vec k| \to \infty$. If $\theta_{k}$ had support for arbitrarily large values of $\vec k$ (e.g. when $\theta_{k}$ is a constant as is the case for the so called $\alpha-$vacua \cite{Chernikov:1968zm, Mottola:1984ar, Allen:1985ux}), the two Hilbert spaces would be unitarily inequivalent \cite{Strocchi:1985cf}.} and can explicitly constructed:
\begin{equation}
{\widehat U} (\Theta) \ = \ {\rm Exp}\left\{-\frac{1}{2}\, \int \dint^3k~[\Theta^\star_{\vec k} \, {\widehat a}_{\vec k}  {\widehat a}_{- \vec k} \, -
\, \Theta_{\vec k}\, {\widehat a}^{\dag }_{\vec k} {\widehat a}^{\dag }_{- \vec k} ]\right\} \ ; \label{UT2}\, \Theta_{\vec k} := \theta_{\vec k} \, e^{i\delta_{\vec k}} \ ,
\end{equation}
and where one can always choose $\theta_k > 0$. One can readily check that this transformation effects (\ref{car}):
\begin{equation}
{\widehat U}^{\dagger}  \, {\widehat a}_{\vec k} \, {\widehat U}\ = \ {\widehat b}_{\vec k}.\
\end{equation}
Relating the transformation (\ref{fexp}) in terms of the more familiar Bogoliubov coefficients
\begin{equation}
\label{Bog}
\alpha_{\vec k} = \cosh\theta_{\vec k} \ , \qquad \beta_{\vec k} = e^{-i \delta_{\vec k}}\sinh\theta_{\vec k} \ ,
\end{equation}
one can evaluate the in-in correlator of interest as
\begin{eqnarray}
&& \langle \psi_I| {\widehat \calR}_{\vec k}(t)\, {\widehat \calR}_{\vec q}(t) \, |\psi_I\rangle \ \ = \
\langle \, 0| {\widehat \calR}'_{\vec k}(t)\, {\widehat \calR}'_{\vec q}(t) \,
|0 \rangle  =  \  |\calR'_{\vec k}|^2 \delta^3 ({\vec k} + {\vec q}) \nonumber \\
&&  = \, |\calR_{\vec k}|^2 \left\{ 1 + 2 |\beta_{\vec k}|^2 \ + \ 2 \ \cos\, (\delta_{\vec k} + 2 \Delta_{\vec k}) \ |\alpha_{\vec k} \beta_{\vec k}|  \right\}
\delta^3 ({\vec k} + {\vec q}), \label{psi1}
\end{eqnarray}
where we've defined the phase $\calR_{\vec k} = e^{i \Delta_{\vec k}}|\calR_{\vec k}|$. In position space, the power spectrum is obtained from
\begin{equation}
\langle \psi_I| {\widehat \calR}^2 (t,{\vec x}) \, |\psi_I\rangle \ = \ \int \frac{\dint^3 k}{(2 \pi)^3}\, |\calR_{\vec k}|^2
\left\{ 1 + 2 |\beta_{\vec k}|^2 + 2\cos\, (\delta_{\vec k} + 2 \Delta_{\vec k}) \ |\alpha_{\vec k} \beta_{\vec k}|  \right\} \ , \label{psi2}
\end{equation}
so that finally, we obtain the generalization of (\ref{dprm}) to states with potentially correlated phases
\begin{equation}
\calP_{\calR,\psi}(k) = \calP_{\calR,0}(k)\left\{ 1 + 2 |\beta_{\vec k}|^2 \ + 2 \cos \,(\delta_{\vec k} + 2 \Delta_{\vec k}) \ |\alpha_{\vec k} \beta_{\vec k}|  \right\} \ . \label{psB}
\end{equation}
Therefore, in addition to the contributions that result from localized variations of the potentials (\ref{wdef}) and (\ref{Wdef}) that result in (\ref{psd}), we have the following additional contributions from initial state modifications:
\eq{misps}{\boxed{\frac{\Delta \calP_\calR}{\calP_\calR}(k) = 2 |\beta_{\vec k}|^2 + 2 \cos\, (\delta_{\vec k} + 2 \Delta_{\vec k})|\alpha_{\vec k} \beta_{\vec k}|}}
where demanding time reversal invariance of the modified vacua imposes the condition $\delta_{\vec k} \equiv 0$ \cite{Allen:1985ux}. Similar to the adiabaticity conditions that restrict how much support (\ref{psd}) has in $k$--space, \textit{we realize that the above modulation of the power spectrum must also decay sufficiently fast}. This is because from the perspective of (\ref{Bog}), support for $\Delta \calP_\calR/\calP_\calR$ for arbitrarily large $k$ would imply an infinite energy density of particle excitations over the adiabatic vacuum. 

Having deduced the most general forms in which features can imprint on the primordial power spectrum in the context of single field models (\ref{psd})(\ref{misps}), it is useful (especially if we are also interested in how features may imprint in higher order correlation functions) to ground the rest of our treatment in the framework of the effective theory of the adiabatic mode, to which we presently turn our attention. 

\subsection{The effective theory of the adiabatic mode}
By treating all the degrees of freedom available at any given energy scale as {\it effective} degrees of freedom whose dynamics are governed by an {\it effective action}, we can parametrize our ignorance of the fundamental theory from which these degrees of freedom descend in a powerful and predictive manner \cite{Donoghue:1995cz, Manohar:1996cq, Burgess:2003jk}. In cosmology as in particle physics, an effective field theory (EFT) treatment is of particular use in searches for "beyond the standard model" physics. In the context of inflationary cosmology, an effective treatment proves indispensable given that we have little idea as of yet as to what the inflaton truly represents as a physical degree of freedom. 

There are two broad classes of questions one might be interested in-- those pertaining to the background and those pertaining to perturbations around this background (which may not even admit a perturbative description), each with its own relevant operator (i.e. derivative) expansion. Questions in the former class might concern the naturalness of obtaining enough inflation in a given context, that is to say concerning loop corrections to the inflationary effective potential and kinetic terms, whether reheating occurs efficiently enough, questions concerning the initial onset of inflation and so on. For such questions, a covariant derivative expansion-- one that is not premised on a particular background solution (though it is premised on expanding in a particular basis of operators)-- is of utility \cite{Burgess:2003zw, Weinberg:2008hq}. If we presume that such questions have been addressed, thus taking the background solution for granted, questions concerning perturbations around this background and in particular of the adiabatic perturbations, are usefully addressed in the EFT expansion of the adiabatic mode \cite{Creminelli:2006xe, Cheung:2007st}. The virtue of expanding in this particular operator basis is that it exploits the properties of the adiabatic mode as the Goldstone boson associated with spontaneously breaking time translation invariance through cosmological evolution \cite{Low:2001bw}, providing a useful book keeping device on the expansion.

The analysis commences with the observation that in any given spacetime where cosmological evolution is driven by a single effective degree of freedom, there exists a preferred foliation defined by surfaces of constant field value, \textit{i.e. that this degree of freedom serves as a good clock}. Denoting this field $\phi$, this is simply the observation that in any arbitrary foliation where the field and its fluctuations around some homogeneous background solution decompose as
\eq{pbg}{\phi(x,t) = \phi_0(t) + \delta\phi(t,x),}
one can make an infinitesimal coordinate reparametrization that gauges away the fluctuations completely. Namely, the transformation $t \to t  + \pi$ such that $\pi = \delta\phi/\dot\phi_0$\footnote{Where the dot denotes the derivative w.r.t. the time variable that defines the particular foliation.} gauges away the $\phi$ fluctuations as \cite{Mukhanov:1990me}
\eq{cmg}{\phi(t,x) = \phi_0 + \delta\phi(t,x) \to \phi_0 + \delta\phi(t,x) - \dot\phi_0\pi \equiv \phi_0(t).}

In this foliation, surfaces of constant $\phi$ coincide with surfaces of constant time, and the only fluctuating degrees of freedom are evidently in the metric. Thus it appears as if what was initially a fluctuating scalar degree of freedom has been ``eaten'' by the graviton, which acquires a propagating longitudinal polarization. This is made manifest in the ADM parametrization of the metric \cite{Arnowitt:1962hi}
\eq{adm}{\dint s^2 = -N^2\dint t^2 + h_{ij}(\dint x^i + N^i \dint t)(\dint x^j + N^j \dint t)} 
where $N, N^i$ are respectively the lapse function and shift vector that define the foliation and where $h_{ij}$ is the induced three metric on the spatial hypersurfaces. The utility of this decomposition lies in the fact that no time derivatives of the $N$ or $N^i$ appear in the action, and hence correspond to non-propagating fields that enforce constraints, i.e. as Lagrange multipliers. Therefore all the dynamics of the gravitational field is contained in the 3-metric $h_{ij}$, which we parametrize as
\eq{3m}{h_{ij} = a^2(t)e^{2\calR}\delta_{ij},}     
where we have neglected tensor perturbations, and where $\calR$ is the non-linear generalization of the variable introduced by Bardeen, Steinhardt and Turner \cite{Bardeen:1983qw} in linear perturbation theory, which can be shown (in the absence of non-adiabatic perturbations) to be conserved on super-horizon scales to all orders in perturbations \cite{Salopek:1990jq} and to arbitrary loop order \cite{Pimentel:2012tw, Senatore:2012ya, Assassi:2012et}.
The variable $\calR$ is the conformal mode of the 3-metric of the spatial hypersurfaces and corresponds to a local rescaling of the scale factor. In pure gravity, this mode is constrained not to propagate \cite{York:1972sj}, however this is not so once coupled to a scalar degree of freedom. In comoving gauge (the slicing where $\phi(t,x) \equiv \phi_0(t)$), the conformal mode has evidently eaten the scalar fluctuation mode and induces a longitudinal polarization for the graviton\footnote{This is analogous to the process that occurs in gauge theories when the gauge symmetry is broken. For this reason, in this context, comoving gauge is often also referred to as \it{unitary gauge} \cite{Cheung:2007st}.} that we identify with $\calR$-- the local rescaling of the scale factor that is equivalent to a local time reparametrization of the background evolution. 

Effective field theory consists of writing down all operators consistent with the symmetries operative at the energy and curvature scales of interest, organized as an expansion in operator dimensions, suppressed by powers of the scale\footnote{Which counts powers of canonically normalized fields as well as derivatives.} that defines when new physical degrees of freedom become relevant. To any given order, the coefficients of these operators are to be fixed by a finite number of observations. In the present context, we note that the foliation defined by (\ref{cmg}) implies that all scalar metric and matter fluctuations are encoded in the dynamics of the 3-metric (\ref{3m}) through the adiabatic mode $\calR$. Thus it must be that any action for the perturbations be organized as an operator expansion in powers of $\calR$ and its derivatives that preserves this privileged foliation. Clearly, any combination of operators that describe perturbations around the background solution that is invariant under spatial diffeomorphisms of the hypersurfaces are permitted to appear in the effective action. Cheung \textit{et al} \cite{Cheung:2007st} have shown that in the absence of strong gravitational effects, a convenient operator basis is provided by powers of $(g^{00} + 1) := \delta g^{00} = 1 - 1/N^2$ and $\delta E_{ij} := E_{ij} - E^0_{ij}$, where the superscripts denote unperturbed quantities and where
\eq{edef}{E_{ij} = NK_{ij} := \frac{1}{2}\left(\dot h_{ij} - \nabla_iN_j - \nabla_j N_i\right),}
with $K_{ij}$ being the extrinsic curvature of the foliation and where $\nabla$ denotes covariant derivatives w.r.t. the 3-metric $h_{ij}$. That is, to the action that describes the background solution, we are free to add arbitrary combinations of $\delta g^{00}$ and $\delta E_{ij}$ that are scalars under spatial diffeomorphisms, which we subsequently organize in powers of $\calR$ and its derivatives. The latter results after we solve for the $N$ and $N^i$ constraints in terms of $\calR$, substitute the result back into the action, and expand in orders of the perturbation. The action for the background is given by \cite{Cheung:2007st}
\eq{bgact}{S_{bg} = \int \dint^4x\sqrt{-g}\left[\frac{\mpl^2}{2}R -\mpl^2\left(\frac{1}{N^2}\dot H + 3H^2 + \dot H \right)\right].} 
This can either be viewed as the result of having re-expressed the matter content that induced the cosmological evolution in terms of geometrical quantities through the background equations of motion $G_{\mu\nu} = \mpl^{-2}T_{\mu\nu}$, or simply as the unique form that enforces tadpole cancellation once we perturb around the background solution. The operators that we can add to this background action encode the specific details of the matter sector that induced the background cosmology. Given that $\delta g^{00}$ and $\delta E_{ij}$ are zeroth and first order in derivatives respectively though both at least first order in perturbations, up to quadratic order, we can add any combination of the following operators 
\eq{qapt}{\mathcal L_{(2)} \sim (\delta g^{00})^2,~ \delta g^{00}\delta E^i_i,~(\delta E^i_i)^2,~\delta E^{ij}\delta E_{ij},}
each of which can appear with arbitrary time dependent coefficients. Therefore, it might naively appear that the resulting quadratic action for $\calR$ might depend on four arbitrary functions\footnote{Although second order in perturbations, the $\delta E_{ij}^2$ operators generate terms up to quartic order in derivatives once the constraints have been solved for. One can in fact act on the operators (\ref{qapt}) with arbitrary contractions of spatial covariant derivatives to generate more higher derivative terms at any given order in perturbations \cite{Gwyn:2012mw}, although these will be relatively suppressed by powers of the high energy scale that limits the validity of the theory. We do not consider this possibility in the counting that follows, although one can readily generalize the argument to do so.}. However, up to quadratic order, the last two terms in (\ref{qapt}) result in contributions that differ only by an integration by parts.  

Hence only three independent functions will serve to specify the quadratic action once the constraints have been solved for to express everything in terms of $\calR$. One of these functions, $\lambda(t)$, will be the functional coefficient of the quartic operator $(\nabla^2\calR)^2$ generated by the $(\delta E^i_i)^2$ term  \cite{Baumann:2011dt}, whereas the other two are evidently given by $c_s(t)$ and $\epsilon(t)$ (\ref{ccpa}):
\begin{equation}
\nonumber
 S_2 = \int \dint^4x~a^3\epsilon M^2_\mathrm{pl}\left(\frac{\dot \calR^2}{c_\mathrm{s}^2} - \frac{(\partial\calR)^2}{a^2} + \lambda\frac{(\partial^2\calR)^2}{a^4}\right) 
\end{equation}
In the discussion to follow, we only consider the first two terms in the above to leading order. 

Similar reasoning at the level of the cubic action, one can add any of the following operators 
\eq{capt}{\mathcal L_{(3)} \sim (\delta g^{00})^3,~ (\delta g^{00})^2 \delta E^i_i,~\delta g^{00}(\delta E^i_i)^2,~\delta g^{00} \delta E^{ij}\delta E_{ij},~\delta E^k_k \delta E^{ij}\delta E_{ij},~ (\delta E^i_i)^3}
which naively implies that the cubic action can depend on six arbitrary functions. However just as before, up to cubic order in perturbations, the last two terms in the above give contributions that differ only by an integration by parts. Hence the cubic action depends on an additional five functions on top of cubic terms generated by the four independent operators listed in (\ref{qapt}) (see \cite{Bartolo:2010bj, Bartolo:2010di, Bartolo:2010im} for a detailed study of the effects of these operators on the CMB three and four point functions).

A great simplification results if the matter sector that induced cosmological evolution does not couple to the shift vector $N^i$ in any way (or only beyond some order in a derivative expansion), such as would be the case if the Lagrangian of the matter sector had the form
\eq{sms}{\mathcal L_m = \mathcal L_m(\phi,\nabla\phi),} 
where $\phi$ is some scalar degree of freedom. That is, the Lagrangian depends on a single scalar degree of freedom and its first derivatives alone or up to some order in the derivative expansion. Lagrangians that fall in to this class include the so called $k-$inflationary, or derivatively coupled scalar models \cite{ArmendarizPicon:1999rj, Dubovsky:2005xd} and the leading terms of any effectively single field model up to quartic order  \cite{Burgess:2012dz}. In this case, all but the first operator of (\ref{capt}) contribute vanishingly\footnote{Note that this would not be true if $\mathcal L_m$ depended on higher order derivatives of $\phi$ (such as powers of $\square \phi$ that enter non-redundantly) as would typically be the case for an effective action with terms suppressed by factors greater than $1/M^4$ where $M$ defines the heavy mass scale that limits the validity of the theory  \cite{Burgess:2012dz}.} to the cubic action, which hence only introduces one more function-- the coefficient of $(\delta g^{00})^3$. Denoting the coefficient of the $(\delta g^{00})^2$ and $(\delta g^{00})^3$ operators $M_2^4$ and $M_3^4$ respectively, we see that for a matter sector of the form (\ref{sms}) the action up to cubic order could only depend on only three functions and their derivatives-- $\epsilon, M_2^4$ and $M_3^4$. For this restricted subclass, beginning with the action
\eq{bgact}{S = \int \dint^4x\sqrt{-g}\left[\mpl^2\frac{R^{(4)}}{2} -\mpl^2\left(\frac{\dot H}{N^2} -3H^2 - \dot H\right) + \frac{M_2^4}{2!}(\delta g^{00})^2 + \frac{M_3^4}{3!}(\delta g^{00})^3\right],}
we solve for the lapse function and the shift vectors $N$ and $N^i$ respectively and substitute them back into the action to yield the quadratic and cubic actions: 
\eq{qactfin}{S_2 = \mpl^2\int \dint^4x~a^3\epsilon\left(\frac{\dot \calR^2}{c_\mathrm{s}^2} - \frac{(\partial\calR)^2}{a^2}\right)} 
\begin{eqnarray}
\label{cubactfinr}
S_3 = \mpl^2\int \dint^4 x\Biggl[&-&\epsilon a\calR (\partial\calR)^2 + \frac{3\epsilon}{c_s^2} a^3 \calR \dot\calR^2 - \epsilon a^3 \frac{\dot\calR^3}{H}\left(2c_s^{-2} - 1 + \frac{4}{3}\frac{M_3^4}{\mpl^2\dot H}\right) \\ \nonumber &-& 2 a^3 \partial_i\theta\partial^2\theta\partial_i\calR + \frac{a^3}{2}\left(3\calR - \frac{\dot\calR}{H}\right)(\partial_i\partial_j\theta\partial_i\partial_j\theta - \partial^2\theta\partial^2\theta)\Biggr]
\end{eqnarray}
where the adiabatic speed of sound is given by
\eq{csdef}{\frac{1}{c_s^2} = 1 - \frac{2 M_2^4}{\mpl^2\dot H},}
and where\footnote{We note that $\partial^2$ does not contain any factors of the scale factor}
\eq{sv}{\partial^2\theta = -\frac{\partial^2\calR}{a^2 H} + \frac{\epsilon}{c_s^2}\dot\calR~}
is the scalar part of the solution to the shift vector to linear order $N^i_T = \partial_i\theta$.\textit{ We see in particular that one function-- $M_2^4(t)$-- appears in front of several operators in both the quadratic and cubic actions through the precise combination that is $c_s$} (\ref{csdef}). The consequences of this simple fact for cosmological observables will occupy us in Sect. 3.1. 

Before closing this subsection, we wish to highlight the fact that although not immediately apparent, the cubic action (\ref{cubactfinr}) is exactly \textit{second order} in the slow roll parameters as an operator expansion\footnote{However, when calculating correlation functions, one typically resorts to an expansion of the operator $\widehat \calR$ in terms of normal modes, which themselves contain powers of the slow roll parameters. Ref.  \cite{Burrage:2011hd} proposes a redefined operator basis that consolidates this implicit dependence in powers of slow roll parameters.}. More specifically, each non-vanishing cubic contribution to the finite time correlators is suppressed by two powers of $\epsilon$ and/or its derivatives. One can see this at the level of the action after sufficient integrations by parts \cite{Maldacena:2002vr, Collins:2011mz}, or by explicit calculation of finite time correlators. This property persists to all orders-- the EFT of the adiabatic mode is an expansion in powers of $\epsilon$ and its derivatives, and is trivial in the limit $\epsilon \to 0$. It is in this precise sense that $\epsilon$ can be considered as the \textit{order parameter} associated with spontaneously breaking time translation invariance.

\subsection{Features and characteristic scales in $\calB_\calR(k)$}
Analogous to the power spectrum (\ref{2pf}), the object of interest to calculate for cubic correlation function is the so called bispectrum $\calB_\calR(\vec{k}_1,\vec{k}_2,\vec{k}_3)$, defined via
\eq{bsdef}{\left\langle \widehat \calR_{\vec k_1}(t)\widehat  \calR_{\vec k_2}(t)\widehat  \calR_{\vec k_3}(t) \right\rangle = 
(2\pi)^3\delta^{(3)}\left(\sum_i\vec{k}_{i}\right)\calB_\calR(\vec{k}_1,\vec{k}_2,\vec{k}_3).}
In a statistically isotropic universe, the bispectrum is a function of the shape and the size (but not orientation) of the triangle formed out of the three component wavevectors, so that
\eq{}{\calB_\calR(\vec{k}_1,\vec{k}_2,\vec{k}_3) \equiv \calB_\calR({k}_1,{k}_2,{k}_3)}
Extracting such a quantity out of the CMB is a very challenging endeavour, so it is convenient to further compress the information contained in the bispectrum through various diagnostics for which it is easier to construct efficient estimators. One such diagnostic is the dimensionless $f_{\rm NL}$ parameter, defined for a fixed shape configuration (denoted by the superscript $\triangle$) as
\begin{equation}\label{fNL-def}
f_\mathrm{NL}^\triangle(k_1,k_2,k_3) \equiv \frac{10}{3} \frac{(k_1k_2k_3)^3 \calB_\calR}{(2\pi)^4\calP_\calR^2\sum_i k_i^3} \, .
\end{equation}
The origin of this diagnostic follows from the fact that from a variable $\calR_{\rm G}(x)$ that obeys Gaussian statistics, one can construct a variable $\calR_{\rm NG}(x)$ with non-Gaussian statistics of the `local' type\footnote{`Local' here refers to the fact that the non-Gaussianities are mostly due to large deviations of the nature of the fluctuations \textit{at a given point} from that which would be expected for Gaussian statistics. The bispectrum in this case would peak for configurations where one of the wavevectors is very small.} (which has the same expectation value and variance as $\calR_{\rm G}(x)$) through
\eq{}{\calR_{\rm NG}(x) = \calR_{\rm G}(x) + \frac{3}{5}f^{\rm loc}_{\rm NL}\left(\calR_{\rm G}^2(x) - \langle\calR_{\rm G}^2(x)\rangle\right) }
where the factor $3/5$ is convention and follows from the fact that during matter domination, the Newtonian potential $\Phi$ and $\calR$ relate as $\calR = \frac{5}{3}\Phi$. Such a variable generates a non-zero bispectrum such that in the configuration $k_1 \to 0, k_2 \approx k_3$, the quantity on the right hand side of (\ref{fNL-def}) will evaluate to $f^{\rm loc}_{\rm NL}$. We could also consider (\ref{fNL-def}) for the configuration where $k_1 \approx k_2 \approx k_3$, which is particularly sensitive to non-Gaussianities of the so-called \textit{equilateral} type. This diagnostic is most sensitive to non-Gaussianities generated by three-point interactions between (non-collinear) modes that exit the horizon at roughly the same time, such as that which would be generated by reduced adiabatic sound speeds. The reason for this is that the former enhances (spatial) derivative interactions, which are suppressed on super-horizon scales, and therefore contribute the most for modes that cross the horizon simultaneously (i.e. in the equilateral configuration).

How then, does one generate non-Gaussianities in the first place? For economy of discussion, we will restrict the following discussion to the action (\ref{bgact}) defined only by the independent functions $\epsilon, M^4_2, M^4_3$\footnote{As we discuss in the following section, these include the leading order contributions obtained from integrating out heavy fields that couple to the inflaton.}. It should be clear that time variations in $M^4_2(t)$ generate time variations in the resulting quadratic (\ref{qactfin}), cubic (\ref{cubactfinr}) (and all higher order) contributions to the action whereas $M^4_3(t)$ only generates contributions to the cubic (and higher order) terms in the action. Therefore any features in the power spectrum induced by time variation of $M^4_2$ (i.e. through time variations of $c_s$ (\ref{csdef})), \textit{would induce correlated features in the bispectrum} in a manner than can be calculated, as we do in the following section. From our previous discussion, it should be clear that the expectation value (\ref{bsdef}) is a finite time correlation function, most conveniently evaluated in the in-in formalism. From (\ref{inin}) in \ref{in-in}, this is given by
\eq{inin0}{\left\langle \widehat \calR_{\vec k_1}(t)\widehat  \calR_{\vec k_2}(t)\widehat  \calR_{\vec k_3}(t) \right\rangle = i \int^{t}_{-\infty}\dint t_1\left\langle[H_I(t_1), \widehat \calR_{\vec k_1}(t)\widehat  \calR_{\vec k_2}(t)\widehat  \calR_{\vec k_3}(t)] \right\rangle + ...}
where $H_I(t)$ denotes the interaction Hamiltonian constructed from (\ref{cubactfinr}) given the free action (\ref{qactfin}), where in spite of the presence of derivative interactions, the interaction Hamiltonian is given in terms of the interaction Lagrangian simply as $H_I = -L_I$ (see \cite{Weinberg:2005vy} for a discussion of this subtlety). As discussed earlier, although not immediately evident from (\ref{cubactfinr}), the cubic action is in fact suppressed by one more order in slow roll parameter (i.e. $\epsilon$ and its derivatives) relative to the quadratic action so that \textit{to leading order in slow roll}, it is sufficient to consider only the first term (\ref{inin0}) in the general expression \ref{inin}. As we shall see shortly, although it is straightforward (if involved) to evaluate how features generated in the power spectrum correlated with features in the bispectrum extracted from (\ref{inin0}), there is a particular limit in which matters are greatly simplified. 

Since $\calR$ represents the non-linear realization of time translation invariance on a cosmological spacetime, a non-trivial relationship must exist between the two and three point correlation functions (since the action of the symmetry mixes terms of different orders). Consider the bispectrum in the so-called squeezed limit, where $k_1 \to 0, k_2 \approx k_3$. Since $k_1$ represents a much longer wavelength mode than $k_2$ and $k_3$ the long wavelength mode is indistinguishable from a rescaling of the background over which the other two modes represent perturbations as they exit the horizon. Therefore the result would be the same if long wavelength mode were simply factored out of the expectation value (now taken over the rescaled geometry) with small corrections that go as $k^2_1/k_3^2$. Consequently, it can be shown that the three point function relates to the power spectrum in a precise manner:
\begin{equation}\label{crelation}
\boxed{\lim_{k_1 \to 0} \calB_\calR({k}_1,{k}_2,{k}_3){\longrightarrow} - \calP_\calR(k_1)\calP_\calR(k_3)\frac{\dint \log \calP_\calR(k_3)}{\dint \log k_3} \, }
\end{equation}
which is the so-called single field consistency relation first derived in \cite{Maldacena:2002vr} and generalized through the `background wave' argument given above in \cite{Creminelli:2004yq}. From the presence of the logarithmic derivative of the last factor in (\ref{crelation}), we see that any localized features in $\calP_\calR$ directly sources features at commensurate wavelengths in $\calB_\calR$ in a simple manner in the squeezed limit. Although more involved computationally, later sections also will consider correlated features generated for more general shape configurations.

Having equipped ourselves with the requisite formalism, we now proceed to survey the diverse theoretical motivations for the appearance of features and new characteristic scales in cosmological observables. Our goal is to highlight those mechanisms for which a concrete rationale exists from the perspective of fundamental theory and to emphasize the role their detection might play in learning about the true nature of the degree(s) of freedom responsible for generating primordial structure. 

}

\section{Theoretical motivations \label{sec:theory}}
{
After decades of interplay between theory and observation, in cosmology as in particle physics, we have successfully arrived at a phenomenological model that accounts for most observations of the large scale structure of the universe. We've dubbed this the "Standard Model" of cosmology, which purports that the universe began in a phase of primordial inflation that eventually resulted in a hot, thermalized big bang parametrized by a $\Lambda$-CDM cosmology. However, similar to the situation in particle physics, we are left searching for any firm indications as to the deeper theory that explains the seemingly arbitrary parameters of the cosmological standard model. It is in this context that the search for features, if present, can be loosely likened to the search for physics beyond the standard model in particle physics -- searching for "bumps" in certain primordial correlators that might betray the existence of new physics through the new characteristic scales that they imply, which in some cases can even be traced to the existence of very massive particles that couple to the inflaton (see following subsection). The positive detection of any such features and the concomitant characteristic scales that are implied by them would in principle, allow us open the door further ajar on the physics underpinning the $\Lambda$-CDM model. 

Having oriented ourselves with an introduction to the physics of the adiabatic mode, we now turn to a review the diverse theoretical mechanisms that naturally generate features in the primordial correlators in particular realizations of inflation.

\subsection{The influence of heavy fields \label{sec:heavyfields}}
Scalar fields appear to proliferate in any attempt to reconcile cosmology with fundamental physics. In many string and super gravity theories, scalar moduli seem to create as many problems as they do potential solutions \cite{McAllister:2007bg, Burgess:2011fa} for the simple reason that a typical low energy description contains many of them that are massless. In order to obtain a low energy description that is consistent with the observed absence of any gravitational strength long range fifth forces, one is required to give these scalar fields large masses to avoid spoiling the predictions of big bang cosmology, as well as to render them consistent with current bounds on fifth force experiments \cite{Hoyle:2000cv, Kapner:2006si}. Once one identifies a mechanism for this to occur\footnote{See for instance ref.\cite{Grana:2005jc, Douglas:2006es, Blumenhagen:2006ci} in the context of `flux compactifications' in type II string theory, and refs.\cite{Patil:2004zp, Patil:2005fi} for a complimentary approach in the context of Heterotic string theory.}, one might consider looking for variants that leaves one, or more of these fields light enough to be a viable candidate for the inflaton. From this viewpoint, one typically expects that the inflaton is a (or possibly one of many) light degree of freedom in a multi-dimensional field space.

Having multiple light fields participate in the dynamics of inflation represents an important universality class of models, known simply as multi-field inflation \cite{Wands:2007bd, Langlois:2010xc}. In this section, we consider a particular limit of these models -- the effectively single field limit -- that represents the likeliest realization of all single field models when we consider inflation's embedding in a UV complete theory. That is, we consider the inflaton to be the lightest field in a multi-dimensional field space where all other fields have masses that are much greater than the scale of inflation\footnote{\label{heavdef}That is to say, $M^2 \gg H^2$, where $M$ represents the lightest mass of any of the heavy fields coupled to the inflaton.}. 

The standard lore would be that if the fields that couple to the inflaton are heavy enough, a truncation approximation suffices and we can simply ignore the effects of the heavy fields (defined as in footnote \ref{heavdef}). However, this fails to be true in many interesting contexts \cite{Tolley:2009fg, Achucarro:2010jv} because of the usual adage that \textit{truncating a degree of freedom is not always the same as integrating it out}. When the field undergoes non-geodesic motion in field space (i.e. turns), if the radius of the turn starts to compare with the mass scale associated with the orthogonal (i.e. heavy) directions, heavy \textit{fields} can become excited completely consistent with the fact that no \textit{heavy quanta} are created \cite{Achucarro:2010da, Cespedes:2012hu, Achucarro:2012sm, Achucarro:2012yr, Cespedes:2013rda}. This is not too different from the physics of bob-sledding -- a bob-sledder going around a bend in the track freely moves up and down the track without exciting any of the normal (i.e. vibrational) modes orthogonal to their trajectory. So it is possible for the inflaton to excite the \textit{heavy fields} orthogonal to it as the trajectory bends in field space without exciting the (Wilsonian) \textit{fast modes}, consistent with decoupling in the usual sense\footnote{The simple reason for this is that on time dependent backgrounds, there is a misalignment in field space between  heavy and light fields and the fast and slow modes of the theory as defined by the Hamiltonian \cite{Achucarro:2012yr, Burgess:2012dz}.}. It is possible to derive a single field effective description when the effective mass of the normal excitations to the trajectory satisfies:
\eq{eme}{H^2 \ll M^2_{\rm eff};~~M^2_{\rm eff} = M^2 - \dot\theta^2,}
and where $\dot\theta$ represents the instantaneous angular velocity of the background trajectory in field space. In this effective single field description, the net effect of the interactions of the heavy fields with the adiabatic mode are encapsulated (to leading order) in a reduced and potentially varying adiabatic speed of sound given by \cite{Achucarro:2010jv}
\eq{csred}{\boxed{\frac{1}{c_s^{2}} = 1 + \frac{4\dot\phi^2_0}{M^2_{\rm eff}\kappa^2} \equiv 1 + \frac{4\dot\theta^2}{M^2_{\rm eff}}}}
where $\kappa$ is the radius of curvature of the background trajectory $\dot\phi_0 = \kappa\dot\theta$, and $\dot\phi_0$ its velocity\footnote{Which in a multi-field context is defined as $\dot\phi_0^2 := \dot\phi_a\dot\phi^a$, where $\phi^a$ are the field co-ordinates with indices raised and lowered by the sigma model metric.}. 

In the language of the EFT of the adiabatic mode, one can understand this as having generated the leading order operator $(\delta g^{00})^2$ after having integrated out the heavy degrees of freedom, with the coefficient
\eq{m2coff}{M^4_{2} = \frac{\dot\phi_0^4}{\kappa^2M^2_{\rm eff}}.} 
Higher order operators are generated, but these are typically subleading when the speed of sound remains close to unity. For example, it can be shown on general grounds \cite{Senatore:2009gt} (and explicitly \cite{Achucarro:2012sm}) that the coefficient of the $(\delta g^{00})^3$ term $M^4_3$ is related to $M_2^4$ as 
\eq{m3m2}{M^4_3 \sim M^4_2(1-c_s^{-2})}
where the exact order unity constant of proportionality depends on the precise context. Hence, one can understand the effects of heavy fields that interact with the inflaton in the \textit{effectively single field limit} as inducing a reduced and varying speed of sound. Recalling that the speed of sound represents an independent parameter in the effective action for the adiabatic mode (\ref{qactfin}) (\ref{cubactfinr}), it shouldn't be surprising that it is possible to effect this without violating slow roll \cite{Achucarro:2010da} and without creating any heavy quanta \cite{Cespedes:2012hu, Achucarro:2012yr}, thus providing an EFT rationale for considering varying speeds of sound during single field slow roll inflation \cite{Tolley:2009fg, Cremonini:2010ua, Shiu:2011qw, Park:2012rh, Pi:2012gf}. The effectively single field regime is valid so long as the parameters defining the background, namely $\epsilon$ and $c_s$ remain small in the former case and vary slowly enough in the latter. The precise condition for $c_s$ can be phrased as \cite{Cespedes:2012hu, Achucarro:2012yr}
\eq{adc2}{|\dot c_s| \ll M |1-c_s^2|,}   
which serves to ensure that no heavy quanta are produced as inflation progresses\footnote{The condition (\ref{adc2}) is equivalent to the requirement that in the parent theory (i.e. before integrating out the heavy degrees of freedom), the frequencies of the heavy quanta $\omega_+$ satisfy the usual adiabaticity condition $\dot\omega_+/\omega^2_+ \ll 1$, ensuring that none will be created if they weren't present in the first place \cite{Achucarro:2012yr}. For a discussion of the conditions under which these heavy quanta do not align with the usual isocurvature quanta see refs \cite{Burgess:2012dz, Gao:2012uq}. See also ref.\cite{Castillo:2013sfa} for a detailed discussion of integrating out the heavy quanta when they do align with the isocurvature directions.}. The discussion so far has been restricted to the tree-level effective action. Although one can simply posit the UV completion of the theory to be such that reduced and varying speeds of sound result in the low energy EFT, this will require additional fine tuning unless we are guaranteed radiative stability in the presence of large derivative interactions, as is the case in contexts discussed in \cite{Nicolis:2004qq, Brouzakis:2014bwa, deRham:2014wfa}. 

One can of course, move away from the effectively single field regime, transiting through the `quasi-single field' regime \cite{Chen:2009we, Chen:2009zp, Noumi:2012vr, Gong:2013sma} (which we return to shortly) -- where the mass of the heavy field approaches that of the Hubble scale during inflation -- through to a regime where the heavy fields can be excited by sudden turns in the field trajectory (thus departing from adiabaticity) \cite{Gao:2012uq, Gao:2013ota, Noumi:2013cfa, Konieczka:2014zja}, or by being initially excited \cite{Jain:2015jpa}, or to situations where slow roll is temporarily violated \cite{Avgoustidis:2011em, Ribeiro:2012ar, Adshead:2013zfa} (see \cite{Battefeld:2014aea} for bounds on excitations of heavy fields of \textit{any} variety from recent data). However in the latter context, one has to take care that too rigorous violations of slow roll inflation not violate the validity of the single field effective description \cite{Adshead:2014sga, Cannone:2014qna}\footnote{A situation that requiring that $\epsilon$ and $\eta$ remain bounded throughout in combination with (\ref{adc2}) ensures won't be the case.}. In any of these cases, one has to account for the isocurvature modes that are generated and track their decay back into adiabatic perturbations, implying a significant model dependence to the generation of features unlike in the strictly adiabatic case, where due to the nature of the EFT expansion of the adiabatic mode, the features are governed at leading order by at most two functions $\epsilon$ and $c_s$. Thus we restrict ourselves for the rest of the discussion in this section to this case. 

One immediately realizes from (\ref{qactfin}) and (\ref{cubactfinr}) that any features in the power spectrum will correlate with features in the higher point correlation functions as well. This relationship was made precise in ref.\cite{Achucarro:2012fd} in the context where inflation is punctuated by transient drops in $c_s$ encoding the effect of heavy fields\footnote{Equivalently, turns in the inflaton trajectory in the parent theory.}. When the features generated by transient drops in the speed of sound remain perturbatively small (in the sense that $\Delta\calP(k)/\calP \ll 1$ uniformly) the relationship (\ref{psd}) can be inverted to express the drop in the speed of sound in terms of the features generated in the power spectrum\footnote{We note that this inversion is in general possible around any given background, and not just one corresponding to slow roll inflation. What is important is that the feature is generated by the variation of one parameter alone to make the inversion possible.} --
\begin{equation}
\frac{1}{c_s^2} = 1 - \frac{2i}{\pi} \int^\infty_{-\infty} \frac{\dint k}{k} \frac{\Delta\calP_\calR}{\calP_\calR}(k) e^{-2i k \tau} \, .
\end{equation}
In a variety of contexts, the cubic action depends only on $c_s^2$ and the function $M^4_3$ which itself depends on $c^2_s$ through (\ref{m3m2}) and an order unity coefficient that depends on the heavy physics that one has integrated out to arrive at the effective theory. Therefore it shouldn't be surprising that changes to the bispectrum -- defined in (\ref{bsdef}) --
\eq{}{\nonumber \left\langle \widehat \calR_{\vec k_1}(t)\widehat  \calR_{\vec k_2}(t)\widehat  \calR_{\vec k_3}(t) \right\rangle = 
(2\pi)^3\delta^{(3)}\left(\sum_i\vec{k}_{i}\right)B_\calR({k}_1,{k}_2,{k}_3),}
induced by the variation of $c_s$, can evidently be expressed entirely in terms of the features induced in the power spectrum. For example, when one heavy field has been integrated out in a two-field parent theory \cite{Achucarro:2012sm}, we have $M_3^4 = 3/4(1-c_s^{-2})M_2^4$, whence
\begin{eqnarray}
\label{bispectrumf}
\Delta B_\calR &=& \frac{(2\pi)^4\calP_\calR^2}{(k_1k_2k_3)^2} \left\{ -\frac{3}{2} \frac{k_1k_2}{k_3} \left[ \frac{1}{2k} \left( 1 + \frac{k_3}{2k} \right) \frac{\Delta\calP_\calR}{\calP_\calR}( k) - \frac{k_3}{4k^2} \frac{\dint}{\dint\log k} \left( \frac{\Delta\calP_\calR}{\calP_\calR} \right) \right] + \text{2 perm} \right. \nonumber\\
&+& \frac{1}{4} \frac{k_1^2+k_2^2+k_3^2}{k_1k_2k_3} \left[ \frac{1}{2k} \left( 4k^2 - k_1k_2 - k_2k_3 - k_3k_2 - \frac{k_1k_2k_3}{2k} \right) \frac{\Delta\calP_\calR}{\calP_\calR}( k) \right.\\
&-& \left.\left. \frac{k_1k_2+k_2k_3+k_3k_1}{2k} \frac{\dint}{\dint\log k} \left( \frac{\Delta\calP_\calR}{\calP_\calR} \right) + \frac{k_1k_2k_3}{4k^2} \frac{\dint^2}{\dint\log k^2} \left( \frac{\Delta\calP_\calR}{\calP_\calR} \right) \right] \right\}\biggl|_{k = (k_1 + k_2 + k_3)/2}\nonumber
\end{eqnarray}
A commonly used diagnostic for non-Gaussianity is provided by the so called $f_{\rm NL}$ parameters, defined for a particular shape configuration as (\ref{fNL-def})
\begin{equation}\label{fNL-B}
\Delta f_\mathrm{NL}^\triangle(k_1,k_2,k_3) \equiv \frac{10}{3} \frac{(k_1k_2k_3)^3 \Delta \calB_\calR}{(2\pi)^4\calP_\calR^2\sum_i k_i^3} \, .
\end{equation}
We thus understand the physical content of (\ref{bispectrumf}) as implying that any features in the primordial power spectrum generate correlated features, i.e. a scale dependence to the bispectrum that can be expressed as 
\begin{equation}\label{fnlgen}\boxed{
f^\triangle_{\rm NL} \sim c^\triangle_0(\vec{k})\frac{\Delta\calP_\calR}{\calP_\calR} + c^\triangle_1(\vec{k})\left(\frac{\Delta\calP_\calR}{\calP_\calR}\right)' + c^\triangle_2(\vec{k})\left(\frac{\Delta\calP_\calR}{\calP_\calR}\right)''}
\end{equation}
where primes denote logarithmic derivatives with respect to the comoving scale $k$. All model dependence and dependence of the particular shape configuration (but not its scale) are encoded in the $c_i^\triangle(\vec{k})$ coefficients \cite{Achucarro:2012fd}. In particular, we note that the single field consistency relation \cite{Maldacena:2002vr, Creminelli:2004yq} implies that $c^\triangle_1(\vec{k}) \to 0$ for the squeezed configuration, which is easily be verified to be the case in (\ref{bispectrumf}).
\begin{figure}[t]
\begin{center}
\epsfig{file=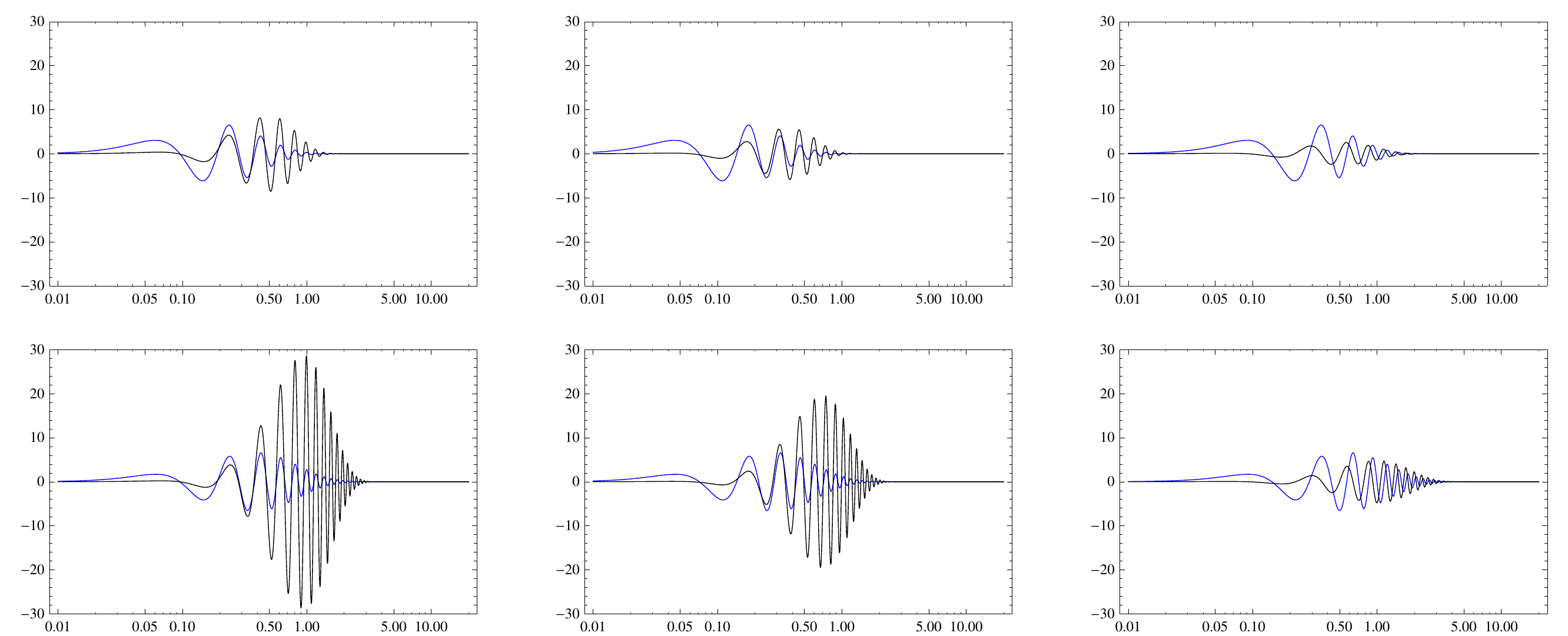, height=2.5in, width=5in}
\caption{\small{$f_{\rm NL}/\Delta_{\rm max}$ (Black lines) vs $\frac{\Delta \calP}{\calP}/\Delta_{\rm max}$ (Blue lines) for the equilateral (right), folded (middle) and squeezed shapes (left) for $\tau_0 k_* = -11$, $c =0.8$ (top) and $\tau_0 k_* = -11$, $c =1.5$ (bottom) respectively for the `cosh' drop in the speed of sound, given by $c_s^{2} =1 -\frac{\Delta_{\rm max}}{{\rm cosh}[c(\tau - \tau_0)]}$.}\label{correlated}}
\end{center}
\end{figure}
In figure \ref{correlated} we illustrate the correlation between features in the power spectrum and the correlated features in the bispectrum for three shape configurations, induced by a transient drop in the adiabatic speed of sound. Although the $f_{\rm NL}$ diagnostic defined in (\ref{fNL-B}) is not optimized for oscillating features, we simply use it as a useful heuristic for the time being, postponing a discussion of more appropriate estimators for features in the bispectrum to Sect. \ref{sec:observation}. One can readily generalize this analysis to higher order and the case where the slow roll parameters are taken to vary in addition \cite{Gong:2014spa, Gong:2014rna, Palma:2014hra}\footnote{See \cite{Konieczka:2014zja} for a study of the analagous situation where the turns are so fast such that (\ref{adc2}) is no longer satisfied and adiabaticity is violated.}. Clearly, seeing any such correlation between features in the power spectrum and the bispectrum would offer compelling evidence of the nature of the inflaton as an \textit{effectively single, weakly coupled} degree of freedom independent of the details of its UV completion\footnote{See ref.\cite{Achucarro:2013cva} for preliminary indications for such a correlation in first year Planck data.}. It should also be stressed that a positive detection of any such features would also permit a \textit{primitive spectroscopy} of the parent theory in that the overall envelope of the features that can be read off from Fig. \ref{correlated} relate directly to the scales $M$ and $\kappa$ given in Eq.~(\ref{eme}) and (\ref{csred}). 

In concluding this subsection we return to an important class of models -- quasi-single field inflation -- that sit at the limit of the scope of this review in that they interpolate between the single and multiple field regimes. Proposed in \cite{Chen:2009we, Chen:2009zp, Chen:2012ge}, it was observed that unless forbidden by a shift symmetry, loop corrections generate masses of order $m^2 \sim H^2$ for all otherwise light fields during inflation\footnote{This lends itself an interpretation in terms of thermal masses generated for all scalar fields with the identification (in natural units) of $m \sim T = H/(2\pi)$ being the temperature of de Sitter space \cite{Spradlin:2001pw}.} implying that the contribution of isocurvature modes \textit{during inflation} also has to be factored into the computation of late time correlation functions of $\calR$, typically generating oscillatory features in the bispectrum but preserving scale invariance of the power spectrum (see \cite{Arkani-Hamed:2015bza} for a recent investigation into this class of models from a more formal point of view, and \cite{Dimastrogiovanni:2015pla} for an investigation into its signatures for LSS). It was further suggested \cite{Chen:2011zf, Chen:2011tu} that excited isocurvature modes during quasi-single field inflation could be used as independent clocks other than that provided by the background inflaton (since the setup is intrinsically non-adiabatic) to measure primordial expansion history. See \cite{Sefusatti:2012ye} for the prospects of detecting these in LSS and CMB measurements.

\subsection{Light particle content and particle production \label{sec:partprod}}
An important generalization of effectively single field models motivated by various string theoretic constructions \cite{Green:1995ga, Kofman:2004yc}, are those for which the single field nature of the model is punctuated by regions in field space where new light degrees of freedom appear along the inflaton trajectory. This can happen in two ways -- when the window in which these new degrees of freedom appear is small or large relative to a Hubble time. The former case is quantified by when
\eq{sudden}{\frac{\dot\phi_0}{H\Delta_\phi} \gg 1,} 
where $\Delta_\phi$ is the range in field space where the additional massless degrees of freedom appear. The opposite case, corresponding to when the window is large relative to a Hubble time, is when the system essentially transits through a \textit{multifield} regime -- where the production of isocurvature modes is energetically favoured. We consider this regime to fall within the ambit of multi-field models (where we have to track the precise ways in which the isocurvature modes subsequently decay) and is not within the scope of this review, although it is straightforward though more involved to generalize the results to this case.
 
When the inflaton traverses a region where the once massive degrees of freedom it couples to suddenly become massless, a burst of particle production takes place. This is most easily inferred from the expression for the frequencies associated with the energy eigenstates of the isocurvature quanta, which we label $\psi$, given by
\eq{iso}{\omega^2_\psi(k) = k^2 + m^2_\psi(\phi),}
where prime denotes derivative w.r.t. $\phi$ and where $m^2_\psi(\phi)$ is the effective mass \cite{Achucarro:2010jv} of the heavy field (whose perturbations are the isocurvature modes) which depends on expectation value of the inflaton field\footnote{We take note here that in general, the notion of `heavy' and `light' can differ from the naive basis of fields in which the action is expressed \cite{Burgess:2012dz}. Here, and in all that follows, `heavy' and `light' distinguishes degrees of freedom whose contribution to the Hamiltonian are separated by a mass gap. This mass gap defines the cut-off of the low energy effective theory once the heavy fields have been integrated out.}. The adiabatic theorem states that were we to begin in a state where no heavy quanta are present, then provided
\eq{adbc}{\frac{\dot\omega_\psi}{\omega_\psi^2} \ll 1,}
them no $\psi$ quanta are created through the time evolution of the background. This condition implies
\eq{adbcexp}{ \frac{{m^2_\psi}'}{2\omega_\psi^3}\dot\phi_0 \ll 1.} 
Therefore the necessary condition for any heavy quanta to be produced is that their eigenfrequencies must change as inflation progresses, and that this change is either (i) fast enough (so that the numerator in the above becomes large), or (ii) that the background transits through a point where the heavy quanta become very light (so that the denominator becomes small). Examples typically considered in the literature focus on the latter possibility, although the former remains an interesting avenue which to date hasn't been as widely discussed in the literature. For this reason, we presently elaborate on a simple example that encompasses both cases in particular limits\footnote{See \cite{Adshead:2014sga} for theoretical bounds placed on the amount of adiabaticity violation during inflation from the requirement that the underlying effective theory remain weakly coupled, a situation which we stay on the right side of in the examples to follow.}.
\begin{wrapfigure}{r}{0.45\textwidth}
\vspace{-10pt}
  \begin{center}
    \includegraphics[width=0.45\textwidth]{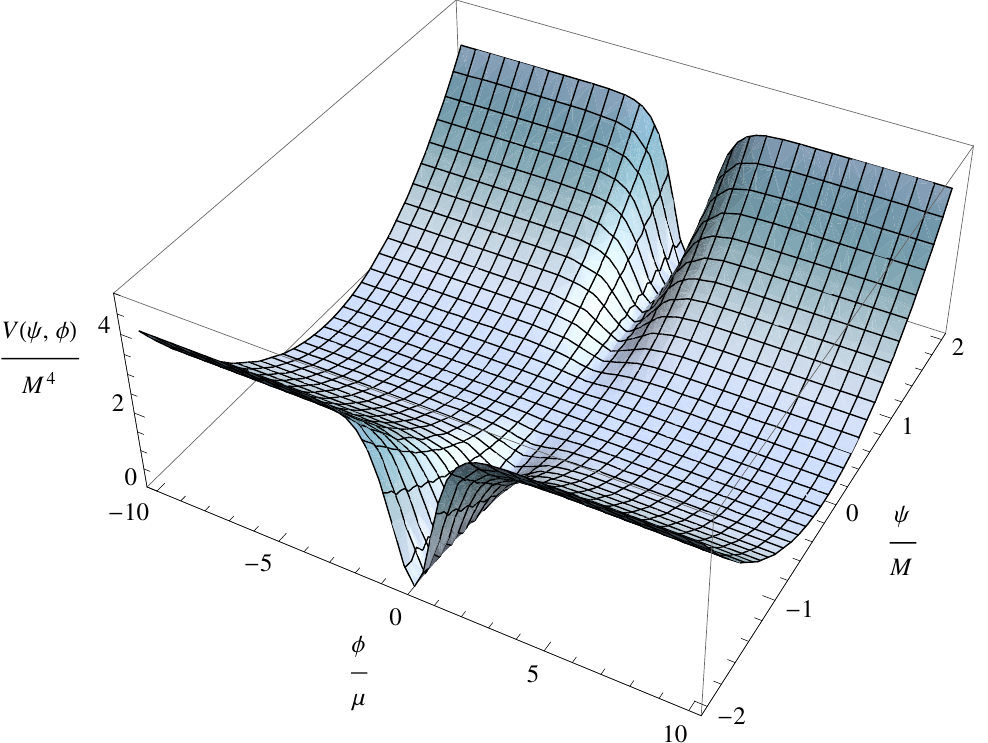}
  \end{center}
  \caption{$m^2_\psi = M^2\left(1 - {\rm Sech}^2[\phi/\mu]\right)$}
  \label{twofield}
\end{wrapfigure}

A prototypical potential that manifests a transient drop in the mass of the heavy field is illustrated in fig \ref{twofield}. As the inflaton rolls through the point where the field $\psi$ becomes light, isocurvature quanta production becomes favored. Since these quanta are only transiently light, they decay back into the adiabatic quanta upon becoming massive again. This initial burst of isocurvature production (followed by intermediate rescatterings) and subsequent decay can occur either with, or without significant energy transfer from the background inflaton field. If the backreaction is significant, the slow roll parameters characterising the background evolution can change and in some cases can even become transiently large -- a scenario known as interrupted slow roll \cite{Lesgourgues:1999uc, Hunt:2004vt, Hunt:2007dn, Jain:2008dw}. This effect superposes on to the effects of adiabatic-isocurvature scattering which occurs even when backreaction is not significant, provided only that the condition (\ref{adbc}) is violated. In order to illustrate both these effects, we consider a system of two fields $\phi$ (the inflaton) and $\psi$ with a potential given by
\eq{ep2}{V(\phi,\psi)= V_{\rm inf}(\phi) + M^2\psi^2\Bigl[1 - \lambda^2 {\rm sech}^2[(\phi - \phi_*)/\mu]\Bigr],}
with $V_{\rm inf}(\phi)$ being a potential that is capable of sustaining slow-roll. The mass term for the $\psi$ field is illustrated in fig \ref{twofield} for the case $\lambda = 1, \phi_* = 0$. A simple approximation scheme when dealing with dynamical situations in an interacting quantum field theory is given by the so called mean field approximation, which in the current context simply replaces the classical field product $\psi^2$ with the time dependent expectation value $\langle \psi^2\rangle$ (suitably renormalized). Therefore the \textit{effective potential} seen by the inflaton is given by 
\eq{infpotc}{V(\phi) = V_{\rm inf}(\phi) + M^2\langle\psi^2\rangle_\phi\Bigl[1 - \lambda^2 {\rm sech}^2[(\phi - \phi_*)/\mu]\Bigr],}
where the subscript on the expectation value denotes that in general, it is a functional of the background trajectory. In \ref{1l}, we derive this correction to the effective potential from a functional approach and relate it to the usual Coleman-Weinberg correction in the limit where the backreaction on the inflaton trajectory is negligible. 

From (\ref{infpotc}) one can readily compute the correction to the slow roll parameters, and hence the background solution induced by the correction to the potential
\eq{Vcorr}{\delta V(\phi) = M^2\langle\psi^2\rangle_\phi\Bigl[1 - \lambda^2 {\rm sech}^2[(\phi - \phi_*)/\mu]\Bigr].}
With the definition $\epsilon = \dot\phi_0^2/(2\mpl^2H^2)$, we can straightforwardly infer after some manipulations that to leading order:
\eq{deps}{\delta \epsilon = 2 \epsilon \frac{\delta V'}{V'}.}
From (\ref{zdef}), we find that this results in modulations of the function $z = a \mpl\sqrt{2\epsilon}$ with $c_s \equiv 1$. In terms of the effect on the perturbations (even though the system remains effectively single field throughout its evolution) it is informative to see how this is so in the regime we're interested in from the perspective of the parent two-field theory. We begin with the generalization of the MS equations (\ref{msf}) to a two field system \cite{Lee:2005bb} 
\eq{}{\nonumber v^{1''} + \Bigl(k^2 - \frac{a''}{a}\Bigr)v^1 + a^2\Bigl([V_{\phi\psi} + \sqrt\epsilon V_\psi/M_{\rm pl}]v^2 + [V_{\phi\phi} + 2\epsilon H^2(3-\epsilon) + 2\sqrt{2\epsilon}V_\psi] v^1\Bigr) = 0,} 
\eq{iso3}{v^{2''} +  \Bigl(k^2 - \frac{a''}{a}\Bigr)v^2 + a^2\Bigl([V_{\phi\psi} + \sqrt\epsilon V_\psi/M_{\rm pl}]v^1 + [V_{\psi\psi} + 2\epsilon M^2_{\rm pl}H^2 R]v^2\Bigr) = 0,}
where $v^1 := a\delta\phi$ and $v^2 := a\delta\psi$, and where $R$ is the Ricci scalar associated with the field space metric. We recall that in applying the mean field approximation in the above, \textit{we simply take expectation values of all powers of $\psi$}. Since $\langle\psi\rangle = 0$ by requiring the background to be on the minimum of the potential \footnote{In general this is not a consistent assumption to make if the background trajectory is turning in field space \cite{Achucarro:2010da}, however given  (\ref{ep2}), it can be shown that this is a consistent assumption to make even at the level of the effective action \cite{Burgess:2012dz}.}, only derivatives of (\ref{ep2}) involving only no, or two derivatives of $\psi$ survive. Hence the mean field approximation translates (\ref{iso3}) into:
\eq{curv4}{v^{1''} + \Bigl(k^2 - \frac{a''}{a}\Bigr)v^1 + a^2[\langle V_{\phi\phi}\rangle + 2\epsilon H^2(3-\epsilon)] v^1 = 0,} 
\eq{iso4}{v^{2''} +  \Bigl(k^2 - \frac{a''}{a}\Bigr)v^2 + a^2[ \langle V_{\psi\psi}\rangle + 2\epsilon M^2_{\rm pl}H^2 R]v^2 = 0,}
\textit{Thus we see that the two perturbations effectively decouple}, justifying the notion that this is an effectively single field model (through satisfying (\ref{sudden})) although like the examples from the previous section, not without the influence of the heavy fields present. Thus, we see that particle production influences the effective potential seen by the curvature perturbation through the correction to $V_{\phi\phi}$ arising from the second term in (\ref{infpotc}). From (\ref{curv4}), we find:
\eq{curv5}{v^{1''} + \Bigl(k^2 - \frac{a''}{a}\Bigr)v^1 + a^2[V^{\rm inf}_{\phi\phi} + 2\epsilon H^2(3-\epsilon)] v^1 + a^2\langle\psi^2\rangle_\phi f(\phi_0)v^1 = 0,}
where the $\epsilon$ and $f(\phi)$ depend on the background solution $\phi_0$, and 
\eq{fp}{f(\phi):= M^2\Bigl[1 - \lambda^2 {\rm sech}^2[(\phi - \phi_*)/\mu]\Bigr]'' = \frac{2\lambda^2 M^2}{\mu^2}\frac{(1 - 3\,{\rm tanh}^2[(\phi-\phi_*)/\mu])}{{\rm cosh}^2[(\phi-\phi_*)/\mu]}.}
The last term in (\ref{curv5}) can be interpreted as the correction induced by a transient burst of particle production, which also modifies the background solution parametrized by $\epsilon$. Evidently, with the identification
\begin{eqnarray}
\label{zpp}\frac{z_0''}{z_0} &=& \frac{a_0''}{a_0} - a_0^2[V^{\rm inf}_{\phi\phi} + 2\epsilon_0 H_0^2(3-\epsilon_0)]\nonumber \\  \frac{z''}{z} &=& \frac{a''}{a} - a^2[V^{\rm inf}_{\phi\phi} + 2\epsilon H^2(3-\epsilon)] -a^2\langle\psi^2\rangle_\phi f(\phi_0)
\end{eqnarray}
One can infer the features generated by constructing the potential (\ref{Wdef}). It is straightforward to check that to leading order in slow roll parameters, this is simply 
\eq{Wpp}{W(\tau) = -a^2\langle\psi^2\rangle_\phi f(\phi_0),}
\begin{figure}[t]
\begin{center}
\epsfig{file=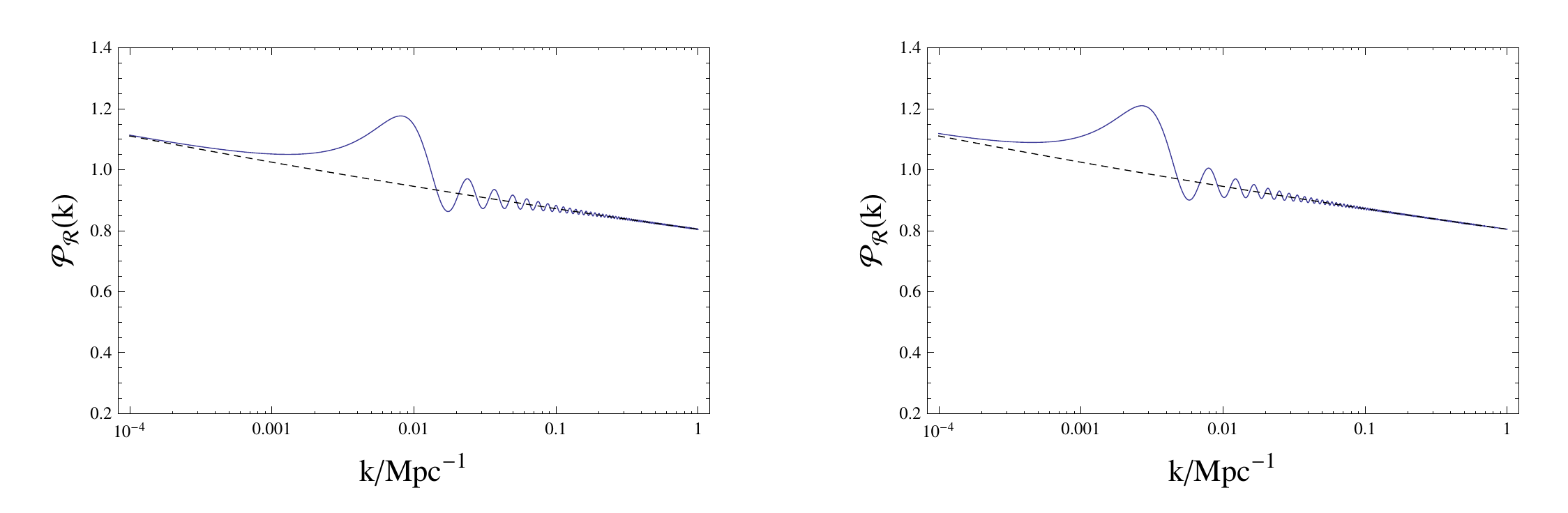, height=1.8in, width=5in}
\caption{\label{particle_prod} \small{$\frac{\Delta \calP}{\calP}$ induced by the particle production for the potential (\ref{ep2})}, with representative parameters $\Delta := \mu/\dot\phi_0 = H^{-1}, \lambda = 0.9, M = 10^{-3}\mpl, \epsilon = 0.01$ with $\phi = \phi_*$ approximately $4$ and $3$ $e$-folds (left and right, respectively) before the current horizon scale exited the Hubble radius during inflation.}
\end{center}
\end{figure}where one assumes that slow roll hasn't been interrupted (so that higher order corrections, neglected in the above, become relevant -- a situation which we will consider in the next subsection). The detailed evaluation of (\ref{Wpp}) is presented in \ref{aexample}, and we plot its effect on the power spectrum in figure \ref{particle_prod}. We note that in our derivation, the $\psi$ quanta \textit{do not need to become exactly massless\footnote{Which would correspond to the case $\lambda = 1$ -- see the figure in \ref{aexample} for observationally relevant particle production for smaller values of $\lambda$.} at $\phi_*$}, as has typically been considered in the literature, only that the scale $\mu$ be such that the adiabatic condition (\ref{adbcexp}) be violated, which can happen for reasonable values consistent with slow roll and the existence of the appropriate hierarchies between the mass of the heavy quanta and $H$. Of course, particle production is parametrically enhanced when the $\psi$ quanta become exactly massless, as considered in \cite{Chung:1999ve, Barnaby:2009mc, Barnaby:2009dd, Barnaby:2010sq, Barnaby:2010ke, Fedderke:2014ura}, whose results can be viewed as a particular case of the treatment elaborated upon here. 

An interesting generalization has also been considered in \cite{Battefeld:2010sw, Battefeld:2011yj, Battefeld:2012wa, Battefeld:2013bfl}\footnote{See also \cite{Battefeld:2010vr} for a related study in a fully multi-field context.}, where motivated by string theoretic considerations, the effects of traversing multiple intervals in field space where massless fields appear along (or nearby) the inflaton trajectory was considered. In the limiting case where these regions appear frequently (while satisfying (\ref{sudden}) throughout), a limiting velocity results for the inflaton due to continual particle production as inflation progresses. The resulting effective potential for the inflaton is consequently `flattened' even in cases where the tree level potential would not ordinarily sustain slow roll. This is but one example of potentially observable physical consequences the inflaton's embedding in its (multiple field) parent theory. In the next subsection we review other such examples that can result in the appearance of new characteristic scales in cosmological observables.  

\subsection{Effective potential modulations and phase transitions \label{sec:effpot}}

Perhaps the most direct manner in which one could generate features in the power spectrum would be to explicitly modulate the part of the potential responsible for slow roll inflation itself. As we have seen in the previous section, although this can easily happen due to couplings of the inflaton to other fields at one loop (even those with masses far greater than the Hubble scale) one could also entertain such modulations purely from a phenomenological perspective or motivated by a particular class of constructions. It is useful in this context to differentiate those potentials that cause localized `interruptions' in slow roll (in a sense that will be made precise) from those for which the slow roll parameters remain small and bounded. For this, we first need to distinguish the relative significance of the different slow roll parameters for the background evolution. Consider the definitions
\eq{}{\epsilon := \frac{\dot\phi^2}{2\mpl^2H^2},~\eta := -\frac{\ddot\phi}{\dot\phi H},~ \xi := \frac{\dddot\phi}{H^2\dot\phi}}
which are equivalent to the first three terms of the so-called `Hubble hierarchy' \cite{Kinney:2002qn, Hoffman:2000ue} of slow roll parameters. The term hierarchy follows from the fact that 
\begin{eqnarray}
\label{hhier}
\dot\epsilon &=& 2H\epsilon(\epsilon - \eta)\\ \nonumber
\dot\eta &=& H(\eta^2 + \eta\epsilon - \xi),
\end{eqnarray}
and similarly for $\dot\xi$, which would involve the next term in the hierarchy. In a precise sense $\epsilon$ is at the top of the Hubble hierarchy in that the condition $\epsilon < 1$ determines the very existence of a background that is undergoing accelerated expansion. Although $\eta$ being small is colloquially considered one of the requisites of `slow roll' inflation, this is not strictly true as $\eta \ll 1$ merely guarantees a sufficiently long period of slow roll if it and $\epsilon$ are both roughly constant (see \cite{Chen:2013aj} for an illustration of a background where $\epsilon$ is small and fast decreasing with $\eta$ of order unity). In other words, $\epsilon$ indicates whether inflation is happening at a given instant whereas $\eta$ indicates whether inflation will continue to happen in the future if both slow roll parameters remain constant. As we go further down the Hubble hierarchy, it should be clear that the time variation of a term is partly dependent on the next order term. 

With these distinctions as to the significance of the terms of the Hubble hierarchy in mind, we define `interrupted' slow-roll as realizations of inflation punctuated with periods where $\epsilon \gtrsim 1$. That interrupted inflation should generate features in the primordial correlators shouldn't come as a surprise given that by construction, it breaks time translation even more strongly than the quasi de Sitter background that gives rise to `almost' scale invariant perturbations. The first instance of interrupted slow roll in the literature is the reference \cite{Adams:1997de}, where it was realized that attempts to embed inflation in grand unified models could result in a series of phase transitions during inflation \cite{Vilenkin:1982wt}, generating an effective potential that has a series of breaks separating different phases of slow roll with differing Hubble rates (see \cite{Silk:1986vc, Salopek:1988qh, Hodges:1989dw, Mukhanov:1991rp, Polarski:1992dq, Starobinsky:1992ts} for earlier proposals that did not necessarily invoke intermediate phase transitions to obtain `multiple inflation', and \cite{Burgess:2005sb} for a subsequent justification from the string perspective). Such models would predict primordial power spectra with sequential steps at those comoving scales that are exiting the horizon at the time of each phase transition, with oscillatory transients (see \cite{Barriga:2000nk, Barriga:2000ma} for early attempts to probe galaxy surveys in combination with pre-WMAP CMB observations for evidence of these). It has been argued that evidence for such features exists in subsequent CMB observations approaching three sigma significance \cite{Hunt:2004vt, Hunt:2007dn}. Interestingly, these transient violations of slow roll also allow for detectably large local non-Gaussianities \cite{Hotchkiss:2009pj} ($f^{\rm loc}_{\rm NL} \sim 5 - 20$) as implied by (\ref{crelation}), although these will require the implementation of estimators that are sensitive to scale dependent local non-Gaussianity, a topic which we will return to in the next section. The authors of \cite{Jain:2008dw} furthermore considered interrupted slow roll as a mechanism through which to suppress primordial power at large angular scales, which as a corollary also generates oscillatory features that might account for certain `outliers' in the angular power spectrum \cite{Hazra:2010ve, Aich:2011qv}\footnote{In addition to generating potentially observable tensor non-Gaussianties over a limited range of scales \cite{Jain:2009pm, Martin:2014kja, Sreenath:2014nca}.}.

Within the context of standard, uninterrupted inflation (where $\epsilon$ remains small and uniformly bounded throughout) one can envisage situations where higher order terms in the Hubble hierarchy (e.g. $\eta, \xi$ in  (\ref{hhier})) can become transiently larger, consequently generating features. In order to make quantitative predictions in this context, one has to account for corrections to the usual formulae for the spectral properties of the CMB that ordinarily assume the smallness of the higher terms in the Hubble hierarchy. This was first investigated in \cite{Stewart:2001cd}, and has come to be known as the generalized slow roll (GSR) formalism. This formalism is necessary when features in the potential or the derivative couplings of the model (e.g. through non-canonical kinetic terms) themselves cause large jumps and deviations of the higher order Hubble slow roll parameters (unlike the examples studied in the previous two sub-sections), which might even be necessitated in certain multi-field set-ups \cite{Avgoustidis:2011em}. Features from steps in the potential or in non-canonical kinetic terms that necessitate the use of the GSR formalism have been investigated in \cite{Dvorkin:2009ne, Hu:2011vr, Miranda:2013wxa, Motohashi:2015hpa} for the power spectrum in addition to the bispectrum in \cite{Burrage:2011hd, Ribeiro:2012ar, Adshead:2011bw, Adshead:2011jq, Adshead:2012xz}. The effects on the polarization power spectra (which as we shall see in the next section, can resolve features with finer resolution) was investigated in \cite{Mortonson:2009qv, Miranda:2014fwa}.

Among the more traditional genre of models where the slow roll parameters (and their derivatives) remain bounded and small throughout inflation, a particular class of models involving the use of axions has received a large amount of attention as means of realizing large field inflation in the context of grand unified theories and string theory. Whenever a theory admits one or more gauge symmetries ($SU(3)_c$ in the case of QCD for example), one must also a priori allow for the possibility of dimension four operators such as
\eq{axionterm}{\theta\frac{g^2}{32\pi^2} G^a_{\mu\nu} \widetilde G^{a\,\mu\nu} \subset \mathcal L }
to enter the low energy effective Lagrangian. In the above, $G^a_{\mu\nu}$ is the field strength associated with the gauge group with gauge coupling $g$, with $\widetilde G^a_{\mu\nu}$ its dual field strength, and $\theta$ a constant parameter. Although classically (\ref{axionterm}) is a total derivative for constant $\theta$, such a term violates CP invariance in the quantum theory unless $\theta \equiv 0$. In the context of QCD for example, the absence of an electric dipole moment for the neutron bounds $\theta$ to be less than $10^{-11}$ (see \cite{Kawasaki:2013ae} for a general review of axions in cosmology, particularly as dark matter candidates). Quite why $\theta$ is observed to be so small when it could `naturally' be as large as order unity is known as the \textit{strong CP problem}. Peccei and Quinn \cite{Peccei:1977hh, Peccei:1977ur} proposed a dynamical solution to this problem by introducing a new spontaneously broken $U(1)$ symmetry, where the associated (pseudo) Nambu-Goldstone boson $\phi$ couples to the CP violating term through the dimension five interaction 
\eq{}{-\frac{\phi}{f}\frac{g^2}{32\pi^2} G^a_{\mu\nu} \widetilde G^{a\,\mu\nu} \subset \mathcal L,}
where $f$ is the so-called \textit{axion decay constant}. The action for the nominally CP violating sector then becomes
\eq{}{-\frac{1}{2}\partial_\mu\phi\partial^\mu\phi + \left(\theta - \frac{\phi}{f}\right)\frac{g^2}{32\pi^2} G^a_{\mu\nu} \widetilde G^{a\,\mu\nu} \subset \mathcal L.}
By a well known result non-perturbative result \cite{Coleman:1978ae}, instanton effects then induce an effective potential for the axion of the form
\eq{nppert}{V(\phi) = -\Lambda^4\,{\rm cos}\left(\theta - \frac{\phi}{f}\right)} 
where $\Lambda$ is a non-perturbatively generated mass scale. Such a potential has $\phi = f\theta$ as a minimum, cancelling the CP violating term completely, and thus solving the strong CP problem. Clearly $\phi \to \phi + c$ is a classical shift symmetry of the theory (broken only by non-perturbative effects to a discrete subgroup) since ${\rm Tr\,}G_{\mu\nu} \widetilde G^{\mu\nu}$ is a purely topological term, and so it appears natural to entertain $\phi$ as a candidate inflaton\footnote{Which through expanding (\ref{nppert}) around its minimum, can also be viewed as offering a concrete realization of $m^2\phi^2$ inflation.}. An immediate problem in doing so is that in order to satisfy the slow roll conditions, one requires the axion decay constant to be such that
\eq{}{f \gg \mpl}
which is hard to construct in a controlled setting (e.g. string theory \cite{Westphal:2014ana}) in addition to being suspect as a hierarchy at the quantum level. This motivated the authors of \cite{Kim:2004rp} to consider multiple axion fields with decay constants that are nearly (though not precisely) identical. In this case, the multi-field potential admits a nearly flat direction in the eigenbasis of the mass matrix for decay constants $f_1 \approx f_2$, and that these can readily be sub-Planckian \cite{Choi:2014rja, Shiu:2015xda}\footnote{The authors of \cite{Burgess:2014oma} noted that this is a special case of a more general `alignment' mechanism.}. 

An entirely different approach for obtaining inflation through axions was proposed by the authors of \cite{Silverstein:2008sg}. In the context of type IIB string theory, compactifying extra dimensions using fluxes generated by 2-forms (see \cite{Grana:2005jc} for a review) introduces axions to the 4-d effective theory. The presence of (NS5) branes wrapping around 2-cycles of the compact (Calabi-Yau) sub-manifold explicitly breaks the shift symmetry of the corresponding axion and introduces the linear term to (\ref{nppert}):
\eq{resonanceP}{V(\phi) = \mu^3\phi -\Lambda^4\,{\rm cos}\left(\theta - \frac{\phi}{f}\right)}
with $\mu$ a mass scale derived from the details of the flux compactification, the string coupling and the brane tension. It is the linear part of the potential that is responsible for inflation, with the cosine representing small, superposed periodic modulations. For large enough field values, the slow roll conditions are satisfied and a sizeable tensor mode background is generated. In addition, the non-perturbatively generated part of the potential induces periodic features in the power spectrum as inflation progresses \cite{Flauger:2009ab, Flauger:2014ana}\footnote{See \cite{Conlon:2011qp} for a criticism of this construction for having neglected the effects of brane-antibrane backreaction which induces corrections that spoil the conditions for inflation. See however \cite{Retolaza:2015sta} for a construction where these corrections are under control (\cite{Marchesano:2014mla} attempts to sidestep this issue completely by generating the linear term from an F-term rather than brane monodromy).}, which gives rise to a novel mechanism through which sizeable non-Gaussianities can be generated in single field inflation, sourced by a resonance between the perturbations and the oscillations of the background inflaton\footnote{In addition, in certain other regimes sizeable equilateral non-Gaussianities can be generated even in the absence of sizeable features generated in the power spectrum \cite{Hannestad:2009yx}.}. \textit{Resonant non-Gaussianity}, first observed in \cite{Chen:2008wn} and subsequently in the context brane \cite{Bean:2008na} and axion monodromy inflation \cite{Flauger:2010ja, Chen:2010bka} is in fact a general feature of any model where a smooth slow-rolling potential is modulated by sinusoidal oscillations (cf. \cite{Behbahani:2011it} for a general study of signatures of discrete shift symmetries during inflation, and \cite{Battefeld:2013xka} for studies of resonance in a multi-field context). As was first noticed in \cite{Chen:2006xjb}, the equation of motion for the mode functions of $\calR$ that follow from (\ref{ccpa})
\eq{}{\ddot\calR + \left(3H + \frac{\dot\epsilon}{\epsilon} - 2\frac{\dot c_s}{c_s} \right)\dot\calR - \frac{c_s^2}{a^2}\nabla^2\calR = 0}
can transiently exhibit parametric resonance for a range of modes (that fall within bands where the so called \textit{Floquet index} is briefly positive \cite{Bassett:2005xm}) if either $\epsilon$ or $c_s$ vary sinusoidally in time. In plainer terms, oscillating background quantities can generate resonant particle production of $\calR$ quanta over the brief window where the physical frequency of the mode $\omega_{\rm phys} = c_s k/a$ is commensurate with that of the oscillating background quantity. For the model defined by the potential (\ref{resonanceP}), we presume $c_s \equiv 1$ and that the evolution is monotonic, which is guaranteed if
\eq{}{b := \frac{\Lambda^4}{\mu^3 f} < 1}
It can be shown \cite{Flauger:2009ab, Flauger:2010ja} that to leading order, superposed oscillatory features (logarithmic in $k$) are generated of the form
\eq{}{\frac{\Delta\calP_\calR}{\calP_\calR}(k) = \kappa\,{\rm cos}\left[\frac{\phi_*}{f} - \frac{{\rm log}(k/k_*)}{\widetilde f}\right]}
where $\phi_*$ denotes the value of the background $\phi(t)$ when the comoving scale $k_*$ exits the horizon, and where we have defined 
\eq{dimdef}{\kappa := 3b\,(2\pi \widetilde f)^{1/2},~~~~ \widetilde f:= f\phi_*/\mpl^2.}
Presumably, terms subleading in slow roll neglected in the treatment of \cite{Flauger:2009ab, Flauger:2010ja} will also cause $n_s$ to decrease logarithmically in $k$, as required by the discussion following (\ref{psd}). We note in passing that it was observed in \cite{Meerburg:2014bpa} that long wavelength features of the sort that can occur in axion monodromy models can alleviate the tension with the observed lack of power at large angular scales.

Similarly, the bispectrum can be calculated as
\eq{ambs}{\calB_\calR(k_1,k_2,k_3) = \frac{(2\pi)^4 \Delta^4_\calR(k_*)}{(k_1k_2k_3)^2}\frac{\kappa}{8\widetilde f^2}\left[{\rm sin}\left(\frac{{\rm log}(K/k_*)}{\widetilde f} \right) + \widetilde f \sum_{i\neq j}\frac{k_i}{k_j} {\rm cos}\left(\frac{{\rm log}(K/k_*)}{\widetilde f}\right)\right] }
where $\Delta^2_\calR(k_*)$ is the amplitude of the power spectrum at the pivot scale $k_*$ usually corresponding to COBE normalization. The expression (\ref{ambs}) represents the leading order expression in slow roll and the dimensionless parameter $\widetilde f$ defined in (\ref{dimdef}), which according to (\ref{ambs}) has to be small if large enough non-Gaussianity is to be observed. The second term in the above is consequently suppressed relative to the first, except in the squeezed limit (\ref{crelation}) where its contribution ensures that the consistency relation is satisfied. As has been noted in \cite{Flauger:2010ja} although the overlaps\footnote{We shall discuss what we mean by overlaps between different shape configurations in more precise terms in Sect. \ref{sec:observation}.} between (\ref{ambs}) and the standard squeezed, equilateral and orthogonal shape configurations are small, significant non-Gaussianity is present in other configurations, calling for more refined estimators for this genre of bispectra \cite{Munchmeyer:2014cca}. We plot various profiles of the bispectrum in the figure below.
\begin{figure}[t]
\includegraphics[width=0.49\textwidth]{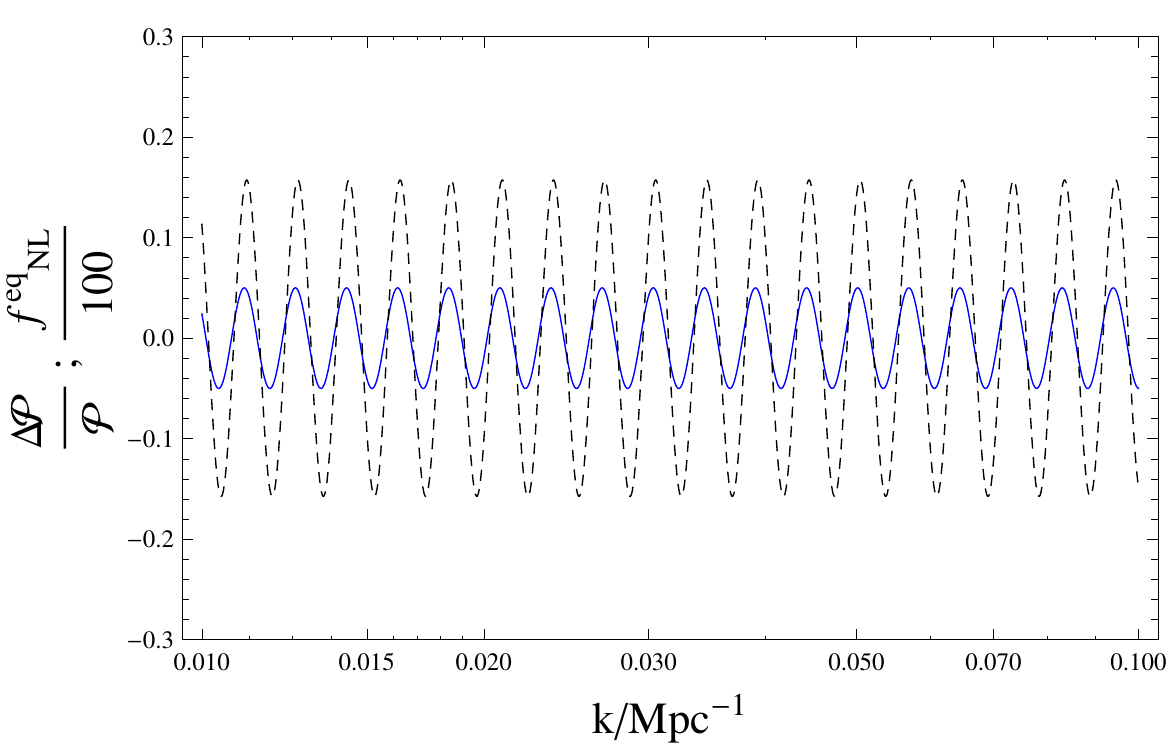}
\includegraphics[width=0.49\textwidth]{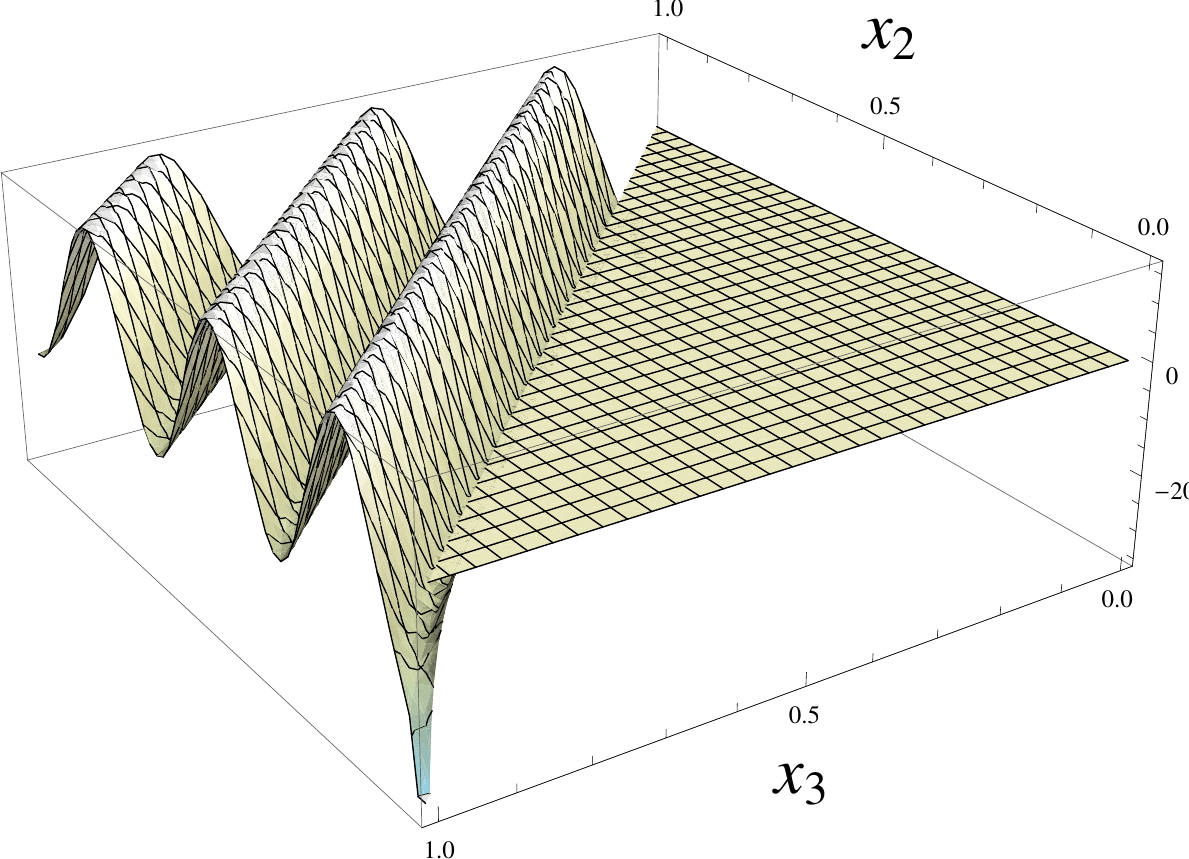}
\caption{Left plot: $\Delta \calP_\calR/\calP$ vs. $f_{\rm NL}^{\rm eq}/100$ for $\kappa = 0.05$, $\tilde f = 0.02$, and similarly for the right plot: $\calB_\calR(k_1,k_2,k_3)\frac{(k_1k_2k_3)^2}{(2\pi)^4 \Delta^4_\calR(k_*)}$, where we fix $k_1 = k_* = 0.002 {\rm Mpc}^{-1}$ and where we define $x_2 := k_2/k_1$ and $x_3 := k_3/k_1$. The restriction of the domain follows from the triangle inequality $1 \leq x_2 + x_3$. \label{fig:mondromy}}
\end{figure}

Searches for oscillatory features in the CMB two and three point functions have been extensive \cite{Ade:2015ava, Ade:2015lrj, Aich:2011qv} with no significant evidence in favor of them. It has however been argued that the use of improved estimators is required to be really sensitive to the bispectra generated by resonance models \cite{Munchmeyer:2014cca}. In addition, one can expect improved sensitivities to result from cross correlating with near term LSS surveys \cite{Huang:2012mr, Hazra:2012vs}, the gain in sensitivity coming from the expected gain in the number of modes observable at commensurate comoving scales. We shall return to the important question of improved sensitivity to features in the primordial two and three point functions through future LSS and 21 cm surveys in the next section.

\subsection{Pre-inflationary dynamics \label{sec:preinf}}
One of inflation's key motivations is that it purportedly accounts for the initial conditions of the hot big bang with few traces of the pre-inflationary state from which it emerged. That is, given a sufficiently long period of inflation, there will be no memory of the initial density matrix even if it corresponded to a state vastly different from the homogeneous, isotropic, thermalized initial plasma of the hot big bang\footnote{Although it has been argued that inflation requires homogeneity on scales larger than the horizon at the time of its onset, thus assuming one of the conditions one might seek to explain with it \cite{Vachaspati:1998dy} (see also \cite{Goldwirth:1991rj, Kung:1989xz, Berezhiani:2015ola}).}. This feature was initially coveted for its help in rendering models of grand unification (which generically overproduce cosmologically dangerous relics such as monopoles) consistent with observations, provided the reheat temperature was below the scale of any dangerous phase transitions. Some have even gone so far as to argue (in the context of chaotic, self reproducing inflation) that any given observer is `likeliest' to find themselves in a universe that inflated for a very large number of $e$-folds, rendering the question of the initial state of the universe moot\footnote{Cf. \cite{Mukhanov:2005sc} see \cite{Gibbons:2006pa} for a criticism of this reasoning. In general, all such arguments depend on a meaningful addressing of the so called measure problem of cosmology and should be taken with the appropriate disclaimers.}. 

However, as another illustration of a recurring theme of this review, these considerations take for granted the manner in which inflation embeds in some candidate parent theory. In the context of string compactifications for example \cite{Burgess:2011fa} it is rather delicate to arrange for potential that can sustain enough inflation (given arbitrary initial conditions) without overshooting the region where inflation is possible \cite{Brustein:1992nk}. From the effective field theory perspective, it is only a restricted class of models that can sustain enough inflation (i.e. undergo enough excursion in field space without encountering regions of the effective potential that spoil inflation) \cite{Burgess:2003zw}, the most common manifestation of which is the so-called eta problem (see \cite{Cicoli:2011zz} for a review). Such considerations have led some authors to argue that if it is so hard to arrange for enough inflation to occur, perhaps it makes sense to conclude that it didn't last much longer than is strictly necessary to solve the horizon problem \cite{Schwarz:2009sj, Ramirez:2011kk, Ramirez:2012gt}. A generic consequence of this, is that we'll be witness to the dynamics of the inflaton close to its onset in addition to the density matrix of the universe at that epoch\footnote{That is, any initial particle content and correlations before a long period of inflation dilutes them away.}.

It has been noted multiple times, in differing contexts that a generic suppression of power results for the scales that are exiting the horizon at the time of any pre-inflationary dynamics where slow roll has yet to be attained. This can be readily appreciated through the expression for the tilt of the spectrum for a solution \textit{that has settled onto an inflationary attractor} (\ref{tilt}):
\eq{}{\nonumber n_\mathrm{s}-1 = 2\eta - 4\epsilon.}
Recalling the definitions $\eta := -\ddot\phi/(H\dot\phi)$ and $\epsilon = -\dot H/H^2$, one can readily show that (\ref{hhier}):
\eq{epsdot}{\dot\epsilon = 2H\epsilon(\epsilon - \eta)}
\eq{etadot}{\dot\eta = H(\eta^2 + \eta\epsilon - \xi);~~\xi := \dddot\phi/(H^2\dot\phi)}
so that to first order in the slow roll parameters, one can get a good first estimate of the envelope of the power spectrum (that is, its overall shape neglecting oscillations) by simply replacing in (\ref{tilt}) the values of the slow roll parameters at the time $t_k$ at which the comoving scale $k$ is exiting the horizon:
\eq{tilt2}{n_\mathrm{s}(k)-1 \approx 2\eta(t_k) - 4\epsilon(t_k) + ...}
To see how any pre-inflationary (fast-roll) dynamics can suppress power for the scales which are exiting the horizon, we note that for \textit{decelerating} solutions in an expanding universe, $\eta > 0$. Furthermore we note that although $\epsilon$ is by definition positive, $\eta$ can transiently be large (i.e. much larger than $\epsilon$) without spoiling slow roll if $\epsilon$ is small enough\footnote{This can be inferred from the formulae (\ref{epsdot}) and (\ref{etadot}.}. Therefore a background solution that is decelerating fast enough (but is still within the slow roll regime) (\ref{tilt2}) implies a positive tilt to the spectrum for the modes exiting the horizon at that time, eventually settling onto an attractor spectrum that characterizes the attractor, favored to be slightly red by observations. This implies a suppression of power at longer comoving scales. Any spectrum with a positive tilt that settles onto a flat, or red tilt for shorter comoving scales (which exit the horizon once the background has settled onto its attractor) necessarily has its power suppressed at longer wavelengths relative to shorter wavelengths. For the tensor modes, one finds that $n_T(k) = -2\epsilon(t_k)$, so that conversely the envelope of the spectrum\footnote{If it is ever measurable -- see \cite{Song:2003ca} for cosmic variance limits on the latter that depend on the eventual tensor to scalar ratio once primordial gravity waves are positively detected.} is going to be enhanced at long wavelengths (up until scales at which higher order corrections to (\ref{tilt2}) become necessary and this first order formula can no longer be trusted \cite{Gong:2014qga}). We plot an illustrative example in fig. \ref{fig:low_supp}. 
\begin{center}
\begin{figure}[h]
\includegraphics[width=12.5cm]{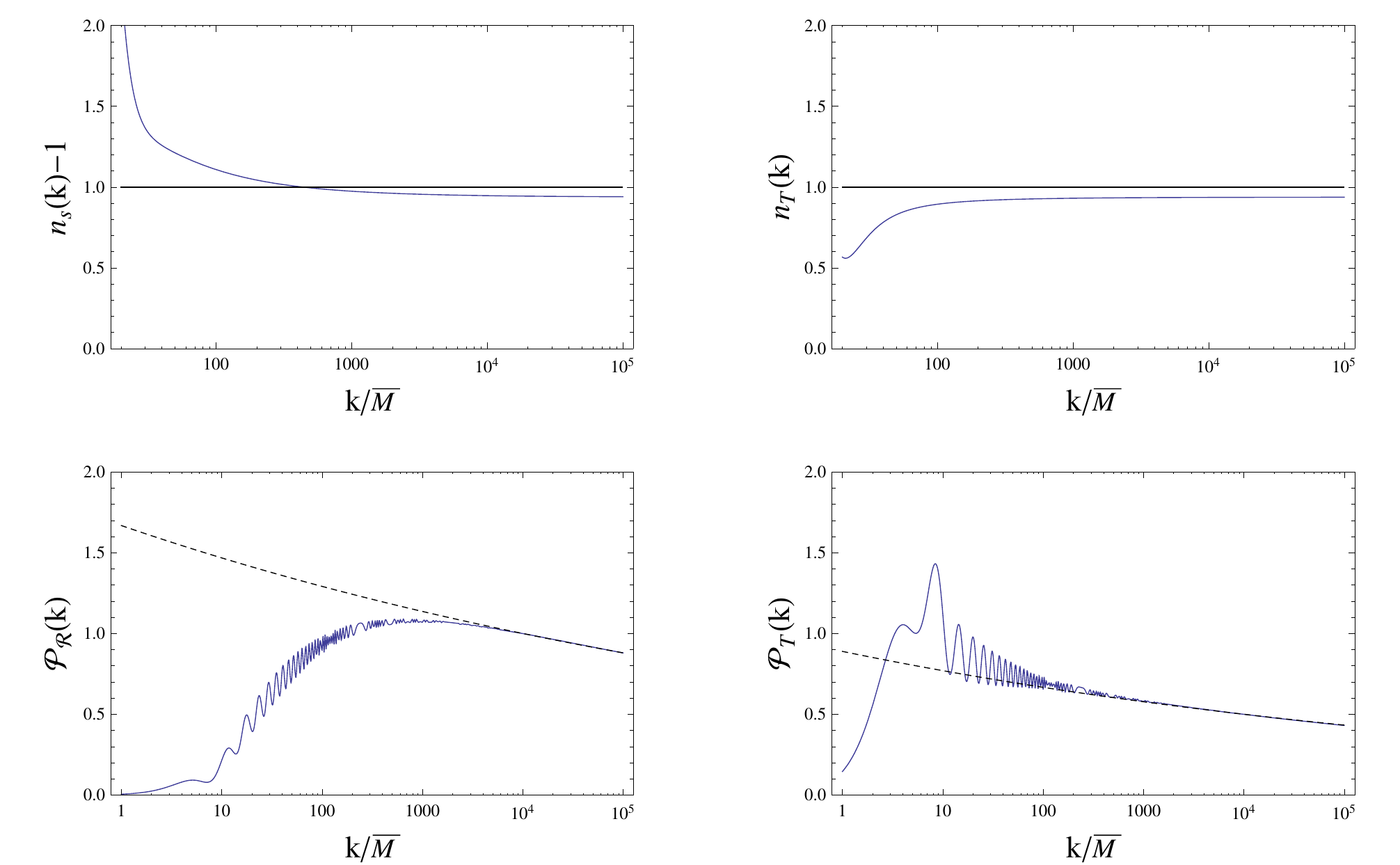}
\caption{The running of the spectral indices and the exact power spectrum for motion in the potential $V(\phi) = \bar M^4(e^{\sqrt 6 \phi/\bar M} + e^{\phi/(4\bar M)})$ \label{fig:low_supp}, which represents a potential that can sustain slow roll augmented with a contribution that results in fast roll if the field is far enough to the right, where we begin tracking the evolution (taken from a larger class of examples studied in \cite{Dudas:2012vv}). Exactly which of these modes we observe depends on the relative normalization of $\bar M$ w.r.t. $k_*$, the scale at which we COBE normalize.}
\end{figure}
\end{center}
We stress that although the primordial power spectrum is suppressed at the largest scales, exactly which part of it we are privy to through our observations depends on when the comoving scale $k_0$ corresponding to the present day Hubble horizon $H_0$ had exited the horizon during inflation. This is because the window functions that project the spatial variation of $\calR$ onto the temperature anisotropies on the surface of last scattering (depicted in fig. \ref{fig:temp_delta}) peak at a scale whose normalization relative to the modes exiting the horizon during this pre-inflationary dynamics could in principle, be arbitrary. That is, were we to observe the effects of any pre-inflationary dynamics, we would need this phase to have occurred within some small window (typically 6-8 $e$-folds) of the exit the comoving scale corresponding to $k_0$. This has to be conceded as a tuning, or at the very least, another cosmological coincidence although the latter has some motivation in the `just enough' inflation scenario. We also wish to stress that the only observable trace of a modification to the power spectrum at the largest scales typified by the example of fig. \ref{fig:low_supp} would be in the suppression of the coefficients $C_\ell$ of the angular power spectrum. Any oscillatory features are typically too rapid to be resolved by the broadness of the window function at such large scales (see fig. \ref{fig:temp_delta}).  

Following up on an observation first made by the authors of \cite{Contaldi:2003zv}, pre-inflationary `fast-roll' has been considered in various studies \cite{Destri:2009hn, Destri:2008fj} as a means to account for the statistically marginally low quadrupole seemingly apparent in the WMAP data. Motivated by related (but statistically much more significant) large angle anomalies in the CMB \cite{Copi:2005ff, Copi:2006tu, Copi:2008hw}\footnote{See \cite{Yoho:2015bla} for forecast gains in the statistical significance of the large angle anomalies from CMB polarization measurements.}, the authors of \cite{Schwarz:2009sj, Ramirez:2011kk, Ramirez:2012gt, Cicoli:2014bja} argued for a scenario where inflation ought to have lasted not much more than necessary and again studied the resulting suppression at large angular scales. 

Pre-inflationary fast roll has been justified from a variety of microphysical considerations that include the preference for inflation to occur around inflection points in a single-field context in \cite{Downes:2012gu} and in a string theoretic (multli-field) context in \cite{Cicoli:2013oba}. Notably, the authors of \cite{Dudas:2010gi} observed that the sort of potentials generated when supersymmetry is partly broken by the presence of D-branes and ortientifold planes in type II string theory, necessarily imply that in order to connect a late time power low inflating phase with the initial curvature singularity of the big bang, the scalar field must emerge from the big bang climbing up the potential\footnote{Thus the authors of \cite{Dudas:2010gi} provide an answer to the question (often taken for granted) as to why the inflaton begins up the potential in the first place.}. Subsequent investigations \cite{Dudas:2012vv, Kitazawa:2014dya, Kitazawa:2014mca, Kitazawa:2015uda} have shown that such a pre-inflationary `climbing' phase has a qualitatively much more pronounced effect on suppressing power at the largest scales than monotonic relaxation to slow roll from fast roll. Further motivation for a phase of pre-inflationary fast-roll was provided by the authors of \cite{Bousso:2013uia, Bousso:2014jca}, who argued that the generic initial conditions for a rolling scalar field after it nucleates from a false vacuum generically suppress power at large scales for entirely analogous reasons\footnote{See \cite{Kamenshchik:2014kpa, Kamenshchik:2015gua} also for a study of inflationary dynamics in the context of the mini-superspace approximation to the Wheeler-DeWitt equation where large scale power is similarly suppressed.}.

\subsection{Initial state effects? \label{sec:initstate}}

Having focussed on pre-inflationary \textit{dynamics} in the previous subsection, it seems only consistent to also consider pre-inflationary \textit{initial conditions} that differ from the usual Bunch-Davies (BD) vacuum state. Before we discuss the various theoretical motivations for doing so, we first address several important theoretical restrictions on modifying initial states that arise from considerations of obtaining a consistent perturbative quantization. We begin by first recalling the expression for the power spectrum (\ref{2pf}) with an arbitrary initial state $|\psi_I\rangle$ (\ref{mic}):
\eq{mic2}{ 2\pi^2\delta^3(\vec k + \vec q)\calP_{\calR,\psi_\mathrm{I}}(k):= k^3\langle \psi_\mathrm{I}|\widehat \calR_{\vec k} \widehat \calR_{\vec q} |\psi_\mathrm{I}\rangle|_\mathrm{in~in}.} 
Were we recall that were $|\psi_I\rangle$ to correspond to a so-called `random phase' state (such as an eigenstate of the number operator of the curvature quanta or a thermal state cf. the discussion in footnote \ref{flabel}), the result would simply be
\eq{dprm2}{\calP_{\calR,\psi_I}(k) = \calP_{\calR,0}(k)[1 + 2n(k)] \  ,}
where $n(k)$ corresponds to the occupation number of the $k^{th}$ mode and where $ \calP_{\calR,0}$ corresponds to the power spectrum obtained from the usual BD vacuum. In general, any state which can be obtained from the BD vacuum by a mode by mode transformation (\ref{fexp}) 
\begin{equation}
\begin{split}
\label{fexp2}
\calR'_{\vec k} &= \cosh\theta_{\vec k}\, \calR_{\vec k} \ +\ e^{-i\delta_{\vec k}}\, \sinh\theta_{\vec k}\, \calR^\star_{\vec k} \ , \\ \calR'^\star_{\vec k} &= \cosh\theta_{\vec k}\,
\calR^\star_{\vec k} \ + \ e^{i\delta_{\vec k}}\, \sinh\theta_{\vec k}\, \calR_{\vec k} \ ,
\end{split}
\end{equation}
corresponds to a rotation of the creation and annihilation operators into the new canonical pair:
\begin{equation}
\begin{split}
\label{car2} {\widehat b}_{\vec k} &= \cosh\theta_{\vec k}\, {\widehat a}_{\vec k} \ + \ e^{i\delta_{\vec k}}\, \sinh\theta_{\vec k}\, {\widehat a}^\dag_{- \vec k} \ , \\
{\widehat b}^\dag_{\vec k} &= \ e^{-i\delta_{\vec k}}\, \sinh\theta_{\vec k}\, {\widehat a}_{-\vec k} \ + \ \cosh\theta_{\vec k}\, {\widehat a}^\dag_{\vec k} \ ,
\end{split}
\end{equation}
which are given in terms of the familiar Bogoliubov coefficients
\begin{equation}
\label{Bog}
\alpha_{\vec k} = \cosh\theta_{\vec k} \ , \qquad \beta_{\vec k} = e^{-i \delta_{\vec k}}\sinh\theta_{\vec k} \ .
\end{equation}
The result is the generalization of (\ref{dprm2}) to states with potentially correlated phases\footnote{See the passage between eqs. (\ref{2pf}) and (\ref{misps}) for more complete details of the derivation.}:
\begin{equation}
\calP_{\calR,\psi}(k) = \calP_{\calR,0}(k)\left\{ 1 + 2 |\beta_{\vec k}|^2 \ + 2 \cos (\delta_{\vec k} + 2 \Delta_{\vec k}) \ |\alpha_{\vec k} \beta_{\vec k}|  \right\} \ , \label{psB2}
\end{equation}
where we define the phase $\Delta_{\vec k}$ via $\calR_{\vec k} = e^{i \Delta_{\vec k}}|\calR_{\vec k}|$, generating superimposed features on the power spectrum of the form 
\eq{misps2}{\frac{\Delta \calP_\calR}{\calP_\calR}(k) = 2 |\beta_{\vec k}|^2 + 2 \cos (\delta_{\vec k} + 2 \Delta_{\vec k})|\alpha_{\vec k} \beta_{\vec k}|}
which represents an expression that is without loss of generality (by the completeness of Hilbert space), the precise details of the initial state and its phase correlations being encoded in the various coefficients\footnote{Given a concrete scenario for pre-inflationary dynamics, these can be explicitly calculated.} in (\ref{psB2}). However, some important caveats are in order here.

The first caveat is a well known feature of quantum field theory. We first note that the formal transformation that effects the mode by mode rotation (\ref{car2}) that describes the modified vacuum state $|\psi_I\rangle$ is given by (\ref{UT2})
\begin{equation}
{\widehat U} (\Theta) \ = \ {\rm Exp}\left\{-\frac{1}{2}\, \int \dint^3k~[\Theta^\star_{\vec k} \, {\widehat a}_{\vec k}  {\widehat a}_{- \vec k} \, -
\, \Theta_{\vec k}\, {\widehat a}^{\dag }_{\vec k} {\widehat a}^{\dag }_{- \vec k} ]\right\} \ ; \label{UT3}\, \Theta_{\vec k} := \theta_{\vec k} \, e^{i\delta_{\vec k}} \ .
\end{equation}
So that 
\eq{psid}{|\psi_I\rangle = {\widehat U} (\Theta)|0_{BD}\rangle,}
where the subscript $BD$ is to emphasize that the original vacuum is the usual Bunch-Davies vacuum state. One is entitled to ask what the overlap of the modified vacuum state is with the original vacuum, given by
\eq{}{ \langle\psi_I|0_{BD}\rangle = \langle 0_{BD}| {\widehat U} (\Theta) |0_{BD}\rangle }
Requiring that the modified vacuum states be invariant under time reversal imposes the condition $\delta_{\vec k} \equiv 0$ \cite{Allen:1985ux}, so that $\Theta_{\vec k} \equiv \theta_{\vec k}$. One can then show that\footnote{See \cite{Umezawa:1993yq}, in particular eqs. (2.39) - (2.39).} 
\eq{uzstate}{ |\psi_I\rangle = {\rm Exp}\left[-\frac{1}{2}\int \dint^3k\,{\rm ln~ cosh}\,\theta_{\vec k} \right]{\rm Exp}\left[\frac{1}{2}\int \dint^3k\,{\rm ~ tanh}\,\theta_{\vec k}{a^\dag_{\vec k}}^2 \right]|0_{BD}\rangle}
so that the overlap between the modified vacuum state and the original BD vacuum is given by
\eq{finoverlap}{ \langle\psi_I|0_{BD}\rangle =  {\rm Exp}\left[-\frac{1}{2}\int \dint^3k\,{\rm ln~ cosh}\,\theta_{\vec k} \right].}
It is also straightforward to show that the state (\ref{uzstate}) has unit norm provided that
\eq{boundintbd}{\int \dint^3k\,{\rm ln~ cosh}\,\theta_{\vec k} < \infty,}
otherwise it would not be possible to describe $|\psi_I \rangle$ as a normalizable excitation over the vacuum $|0_{BD}\rangle$. This would imply that $|\psi_I\rangle$ exists in a different (unitarily inequivalent) Fock space that the original Bunch-Davies vacuum. 

The class of vacua that remain invariant under action of the de Sitter symmetry group are parametrized by the mode by mode constant rotation $\theta_{\vec k} \equiv \alpha$ where $\alpha$ is some constant. These are the so-called `alpha-vacua' \cite{Allen:1985ux} which have been entertained as plausible modifications to the usual BD vacuum for inflationary cosmology. However, it is immediately apparent that the bound (\ref{boundintbd}) is violated for any constant rotation, implying that the differing $\alpha$-vacua are unitarily inequivalent, wherein $\alpha$ serves as a super-selection parameter between them \cite{deBoer:2004nd}. So far this is merely a formal statement, and one might wonder why the BD vacuum might be preferred at all rather than any other $\alpha$-vacuum around which one can perturbatively quantize the fluctuations of the inflaton? It turns out that an interacting field theory defined over a generic $\alpha$-vacuum ground state is intrinsically \textit{non-local} in that one generates divergences that require counter-terms involving products of local fields with fields at the antipodal point in de Sitter space\footnote{Recalling that the usual FRW coordinate system used to describe de Sitter space only covers half of the spacetime manifold (see \cite{Spradlin:2001pw} for a discussion of various aspects of physics on de Sitter spacetimes).} \cite{Banks:2002nv}. The only vacuum for which this is not true is that state that is invariant under the antipodal map, which corresponds to $\alpha = 0$, or usual Bunch-Davies state.

Therefore, a reasonable restriction on the class of modified initial states appear to be those for which the bound (\ref{boundintbd}) is satisfied, which describes an initial state corresponding to a finite energy excitations over the BD vacuum. A corollary of (\ref{boundintbd}) is that any features generated by modified initial states must have finite support in $k$-space (i.e. the modulations to the power spectrum (\ref{misps}) must eventually decay). Physically, this is because there can only be a finite number of excited quanta up to a finite energy. Thought of another way, inflation by its very nature dilutes the universe and works to erase any memories of the pre-inflationary initial state by construction. Accordingly, any modified initial state (corresponding to the initial presence of particles with a range of momenta) will see these particles exponentially diluted and can only imprint in late time observables if inflation did not last too long (cf. the previous section), or if quanta of high enough momenta were initially present. The presence of the latter, clearly being bounded by the fact that one would like to remain in a regime where one is describing perturbative excitations around some slow-rolling background\footnote{This has frequently been phrased as requiring that the modified initial states do not `back-react' substantially on the background geometry to spoil its quasi de Sitter nature \cite{Porrati:2004gz, Brandenberger:2004kx, Greene:2004np, LopezNacir:2007jx}.}.   

With these restrictions in mind, we can survey the various theoretical motivations for considering modified initial states. The question rose to prominence primarily in the context of the so-called `trans-Planckian' problem of inflationary cosmology \cite{Martin:2000xs, Brandenberger:2000wr}. It was noted there that if one were to extrapolate the physical scales that we observe in the CMB back through 60 $e$-folds worth of expansion, a large fraction of the modes observed today would have once been in the super-Planckian regime. It then seems a relevant question to ask, could these modes be sensitive to physics above the Planck scale? After a spirited discussion in the literature \cite{Brandenberger:2002hs, Easther:2001fz, Kaloper:2002uj, Kaloper:2002cs, Danielsson:2002kx, Easther:2002xe, Goldstein:2002fc, Burgess:2002ub, Sriramkumar:2004pj, Greene:2005wk, Boyanovsky:2006qi, Collins:2009pf, Jackson:2010cw}, how best to answer this question crystallized around the issue of how one chooses to parametrize our ignorance of the evolution of the fluctuation modes the trans-Planckian regime\footnote{A analogue of this question had previously appeared in the context of Hawking radiation, where an observer at asymptotic infinity will observe photons from the tail of the black body distribution that will have been redshifted from initially trans-Planckian values. In this context, thermalization proves to be efficient in erasing any traces of non-adiabatic evolution in the trans-Planckian regime \cite{Easther:2001fz} (modelled by modified dispersion relations) \cite{Corley:1996ar}.}. 

One approach (borrowing from analogous situations in condensed matter systems) would be to model evolution in the regime where unknown physics completes the description of the dynamics with modified dispersion relations that encodes this physics. The motivation is straightforward -- massless quanta in a continuous medium obey the linear dispersion relation
\eq{lindisp}{\omega^2(k) = c^2 k^2}
whereas once modes start to approach the scale of where the continuum description stars to break down (such as approaching the inter site separation in the context of a lattice) the linear relationship acquires a convexity, 
\eq{lindisp2}{\omega^2(k) = c^2 \,{\rm sin}^2\,(k/k_c).}
where $k_c$ denotes the characteristic inverse wavelength associated with the microphysics of the system (e.g. a lattice spacing). Therefore one might phenomenologically consider different dispersion relations to see what their effects might be on CMB observables.  
\begin{figure}[t]
\begin{center}
\epsfig{file=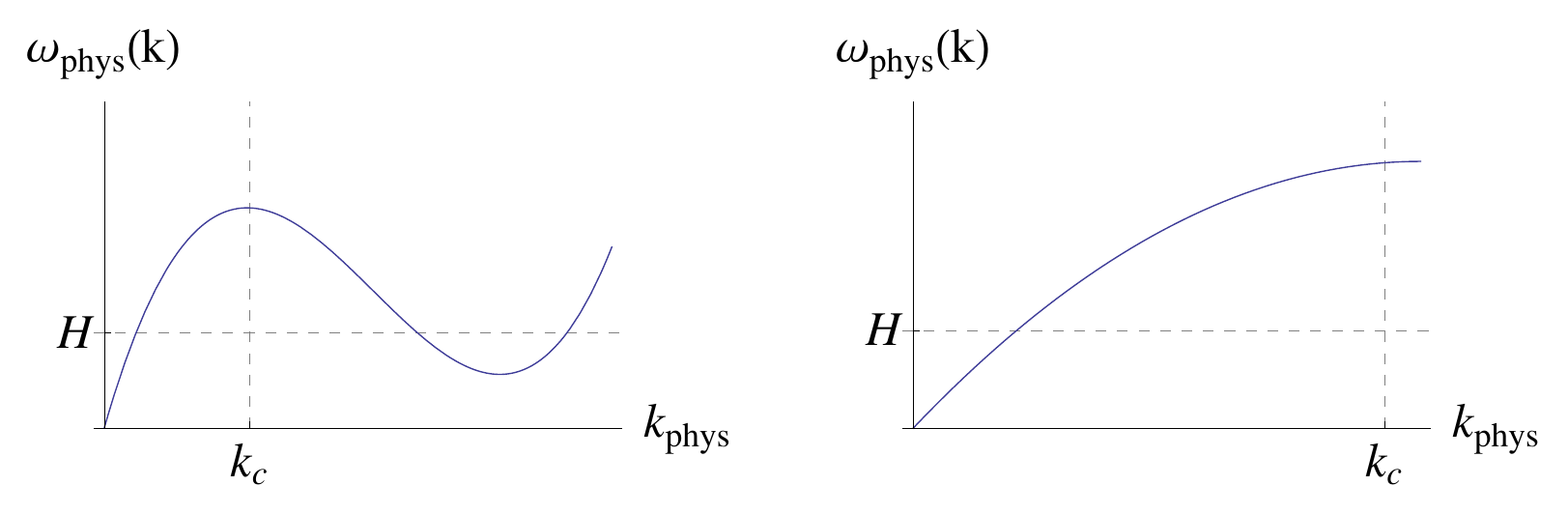, height=1.65in, width=5in}
\caption{\label{dispersion_figs2} Examples of phenomenological modified dispersion relations that violate the adiabatic approximation above $k_c$ (left) and those that don't (right). All wave vectors are in physical units $k_{\rm phys} := k/a$, where $k$ is the comoving wave vector.}
\end{center}
\end{figure}

It turns out, perhaps unsurprisingly, that any modified dispersion relations that do not violate adiabiticty (defined as $\dot\omega/\omega^2 \ll 1$) in the super-Planckian regime renders the modes as they emerge into the sub-Planckian regime in their usual adiabatic (BD) vacuum state \cite{Brandenberger:2002hs, Easther:2001fz, Danielsson:2002kx}, and thus do not affect CMB observables in any significant way\footnote{The arguments presented in these references trace the standard logic for when one can construct the adiabatic vacuum on a time dependent background, the details for which can be found in \cite{Birrell:1984xxx}.}. Specifically, modulated corrections to the power spectrum appear but are suppressed by a factor of $H^2/M_c^2$, where $M_c$ is the characteristic scale of the trans-Planckian physics, and is thus most likely to remain unobservably small. However dispersion relations that violate adiabaticity in the trans-Planckian regime can result in calculable (logarithmic) oscillations in the power spectrum. One conceivable manner in which the latter is possible is for dispersion relations of the type illustrated in the left hand plot of fig. \ref{dispersion_figs2}, which requires a concave regime for the dispersion relation \cite{Martin:2003kp} (at horizon crossing, $k_{\rm phys} = H$ so that adiabaticity is typically violated at $\omega \approx H$). However, absent any concrete model where this occurs in a controlled manner, one must take such dispersion relations with a grain of salt as they typically signal a transition among the propagating quanta rather than a bona fide modification to the dispersion relation in itself\footnote{For instance, liquid helium-4 in a supercooled phase exhibits such a dispersion relation, although its interpretation is in terms of a `second sound' where one transitions from a phase where standard phonons propagate to one where hypothesized `rotons' are the dominant propagating modes \cite{roton}.}. A general context in which modified dispersion relations finds theoretical grounding is in the of high energy Lorentz invariance violation, either considered phenomenologically \cite{Shankaranarayanan:2004iq, Collins:2007jc, Jacobson:2005bg} or in the context of a specific construction (such as Ho\v rava gravity \cite{Horava:2009uw, Blas:2009qj, Koh:2009cy, Ferreira:2012xa}). Related contexts where short distance modifications to mode propagation lead to Lorentz invariance violation as a corollary includes non-commutative geometry  \cite{Lizzi:1995kq, Chu:2000ww, Alexander:2001dr, Lizzi:2002ib, Huang:2003zp, Fukuma:2003pi, Tsujikawa:2003gh, Huang:2003fw, Kim:2004zca, Alavi:2004aq, Palma:2009hs, Rinaldi:2009ba, Koivisto:2010fk, Machado:2011fq, Perrier:2012nr} and inflation in the presence of a minimal length scale \cite{Kempf:2000ac, Kempf:2001fa, Easther:2001fi, Ashoorioon:2004vm, Niemeyer:2001qe, Danielsson:2004xw, Shankaranarayanan:2002ax, Schalm:2004xg, Greene:2005aj, Collins:2005nu, Alberghi:2003am, Hassan:2002qk, Starobinsky:2002rp, Tanaka:2000jw, Palma:2008tx}.
\begin{figure}[t]
  \begin{center}
    \includegraphics[width=0.7\textwidth]{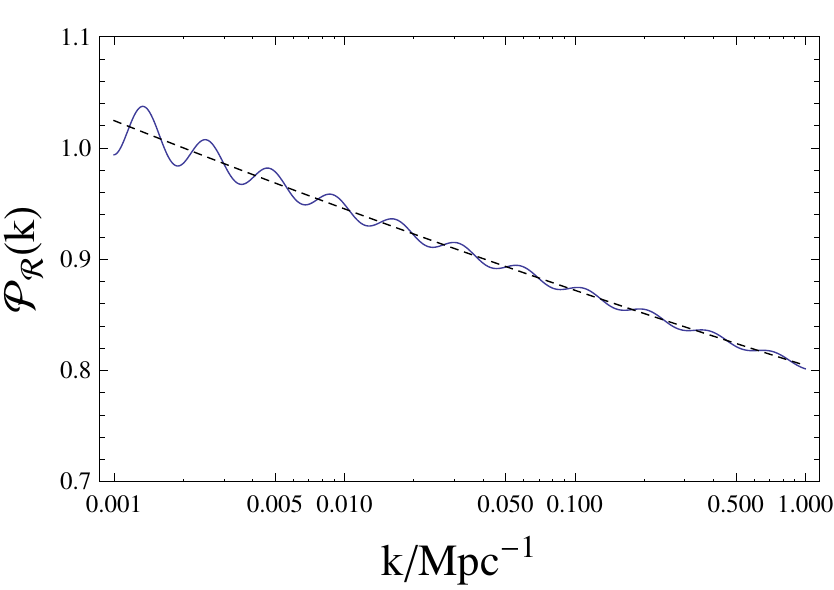}
  \caption{\label{logktrans} Features generated by a model where $H(k_0)/M_c \sim \mathcal O(10^{-2})$, with $k_0 = 10^{-3} {\rm Mpc}^{-1}$ \cite{Easther:2002xe}. }
  \end{center}
\end{figure}

Another approach one could take to parametrize our ignorance of how modes propagate in the trans-Planckian regime appeals to the tried and tested workhorse of effective field theory \cite{Kaloper:2002uj, Kaloper:2002cs, Danielsson:2002kx, Burgess:2002ub, Greene:2005wk}. Ordinarily, one might expect the effects of any physics beyond the cutoff of our description to be encoded in terms of higher dimensional operators expressed in the usual derivative expansion \cite{Kaloper:2002uj, Kaloper:2002cs, Burgess:2002ub}, seemingly drawing into question the need for any novel formalism. However, this ends up shuffling the issue from modifying the dynamics of mode propagation into the initial conditions, wherein the trans-Planckian question is now phrased in terms of requiring a specification of initial conditions with super-Planckian resolution. An EFT approach that seeks to address this invokes the so-called boundary effective field theory formalism \cite{Schalm:2004xg, Greene:2005aj, Schalm:2004qk}, whereby \textit{an arbitrary} set of initial conditions are parametrized by a boundary action defined on a so-called `New Physics Hyper-surface' (NPH) \cite{Bozza:2003pr} that consists of the usual derivative expansion but now with fields restricted to the boundary. These boundary operators can be used to construct arbitrary initial states, and allows one to systematically calculate the effects on CMB observables. It should be stressed that the NPH need not be a hyper-surface at some fixed initial time and could even correspond to one for which different comoving modes are `initialized' at differing coordinate times. The net result in this context as well is that the effects of trans-Planckian physics \textit{generically} results in modulations of (\ref{misps2}) where to leading order
\eq{}{|\beta_k|^2\sim \frac{H^2}{M_c^2},~ |\alpha_k|^2\sim 1 + \frac{H^2}{M_c^2},}
with phases that vary logarithmically in $k$ due to the fact that the mode functions always depend on the combination $k/(aH)$ and that $H$ decreases during slow roll (see \cite{Easther:2002xe} for details). Therefore, we find that the second term in (\ref{misps2}) contributes a logarithmically oscillating contribution to the power spectrum with an overall amplitude of $|\beta_k|\sim H/M_c$ (also decreasing logarithmically during slow roll) which can realistically result in percent level modulations (Fig. \ref{logktrans}). These appear to be on the threshold of detectability (in a sense that will be made precise in the following sections), although no significant evidence for these have appeared so far in WMAP \cite{Easther:2005yr, Meerburg:2010rp} and Planck data \cite{Planck:2015xua}\footnote{We should also highlight several works that further consider features and enhancements to the three point function of the adiabatic mode from modified initial states \cite{Meerburg:2009ys, Meerburg:2009fi, Chialva:2011iz, Chialva:2011hc} that in certain contexts, can also be arranged to preserve the scale invariance of the two point function \cite{Ashoorioon:2010xg, Ashoorioon:2011eg}.}.  

}

\section{Observables: measuring spatial fluctuations \label{sec:observation}}
{
In the previous section, we saw that there are a wide range of theoretical scenarios that can lead to features in the primordial power spectrum, which begs the question: are any of these realised in nature?  Addressing this requires us to confront the theoretical predictions with experimental evidence. However, our quest for features is not completely straightforward -- nature has endeavoured to muddle the waters in several nefarious ways.

First of all, the quantities predicted by theory, i.e. the statistical properties of the primordial curvature perturbation are not directly observable;  we can neither access anything primordial, nor can we ``see'' curvature perturbations.
At best, we can gain indirect information by measuring derived quantities, such as matter density perturbations or anisotropies in the cosmic microwave background that have evolved from the initial state.  In general, the relation of the primordial spectrum to actual observables depends on additional cosmological parameters unrelated to inflation (e.g. the baryon and dark matter densities, the Hubble parameter, the optical depth at recombination etc.) whose effect may, in some cases, be confused with a feature.

Secondly, the generation of initial perturbations is a stochastic process, and quantities like the primordial curvature perturbation or the matter density fluctuation are random fields. Exact, deterministic theoretical predictions of, for example, the power spectrum of these fields can only be made for ensemble averages (or equivalently, their average over an infinite volume).   When it comes to observing the universe however, we are limited to finite volumes. In this case, the theoretical power spectrum of the field over this volume becomes a random variable: its expectation value is given by the ensemble average, but it is subject to a sampling variance (in this context also known as {\it cosmic variance}) which depends on the size and shape of the volume observed. Note that sampling variance is an intrinsic theoretical uncertainty; the only way to reduce it is by increasing the observed volume.  This means that even if we had a perfect measurement, the power spectrum constructed from these data will be subject to a scatter which can obscure or even mimic the presence of features.

Thirdly, the scatter is compounded by statistical uncertainties that all realistic measurements are subject to (e.g., due to finite detector resolution, detector noise, shot noise, etc.). This problem can at least in principle be overcome with better technological means; once these uncertainties are reduced to below the level of the sampling variance, they cease to be relevant.

Finally, there is only a finite window of wavenumbers accessible to observation. The size of the universe's particle horizon today sets a lower limit on the wavenumber of observable perturbations $k \gtrsim 2 \times 10^{-4}$~Mpc$^{-1}$. The upper limit depends on the probe: structures at low redshifts ($z \lesssim 1$) have undergone non-linear evolution on scales $k \gtrsim \mathcal{O}(10^{-1})$~Mpc$^{-1}$, where mode-coupling has erased all but the broadest features.  At higher redshifts non-linear evolution becomes less of an issue, though in the case of the CMB, the original signal is significantly suppressed by Silk-damping from photon diffusion \cite{Silk1968} and tends to be swamped by foregrounds beyond $k \gtrsim \mathcal{O}(10^{-1})$~Mpc$^{-1}$. For other probes of the (matter) power spectrum such as 21~cm tomography, it may in principle be possible to see linearly coupled modes out to $k \approx \mathcal{O}(10^{2})$~Mpc$^{-1}$ \cite{Loeb:2003ya, Furlanetto:2006jb, vanHaarlem:2013dsa, Rawlings:2004wk}. Therefore, given presently available data and being generous, the observable window spans a good four orders of magnitude in $k$, corresponding to roughly 10 $e$-foldings of inflation; features produced outside this range are hidden from our direct view. However, as previously mentioned and as we shall elaborate upon further below, it will soon be possible to open the window on the shorter wavelength modes by tapping the information contained in spectral distortions and 21~cm surveys, the former potentially allowing us a window onto comoving scales up to  $k \approx \mathcal{O}(10^{4})$~Mpc$^{-1}$ \cite{PRISM2013WPII, Chluba2012inflaton}. This would add another 5 $e$-foldings of inflation to the picture.

So what options do we have to get a handle on the primordial power spectrum?  In this section, we will cover the most direct approach:  measuring inhomogeneities (or anisotropies) of the universe's constituents and relating them to primordial perturbations. All energy components of the  $\Lambda$CDM model except for dark energy (i.e., photons, neutrinos, baryons and cold dark matter), have inhomogeneities, making them potentially useful sources of information about the initial fluctuations. Out of these, we can exclude the perturbations of the neutrino background: while in principle a good probe of the primordial power spectrum~\cite{Hannestad:2009xu}, cosmic background neutrinos are elusive creatures and although their direct detection may be within reach with current technology \cite{Ringwald:2009bg, Betts:2013uya, Kaboth:2010kf}, a detection of their anisotropies is unlikely in the foreseeable future. This leaves us with photon and matter perturbations. The latter can be revealed either via visible tracers such as galaxies and neutral hydrogen gas, or through their weak gravitational lensing effect on background sources.  The perturbations of the photons (or, more precisely, of the baryon-photon fluid around decoupling) can be directly observed in the form of temperature and polarisation anisotropies of the cosmic microwave background. Furthermore, looking beyond anisotropies, one may potentially be able to constrain or even detect primordial features with just the monopole through distortions to the CMB spectrum. We elaborate upon this intriguing idea in detail in Section~\ref{sec:spectraldistortions}.

\subsection{Cosmic microwave background anisotropies}

Currently, the by far most sensitive probe for features, and the only one able to reach the lower limit of the observable wavenumber window, are the anisotropy measurements of the cosmic microwave background radiation. From the observed maps of the CMB's temperature fluctuations and polarisation, one can extract the \textit{angular power spectra} $\widehat{C}_\ell^{XY}$, where $X,Y \in \left\{\mathrm{T,E} \right\}$.\footnote{The $B$-component of the CMB polarisation is not sourced by scalar perturbations at linear order and hence not helpful in probing $\mathcal{P_R}(k)$.} These are defined as follows-- first consider expanding the observed temperature anisotropies in the sky in terms of spherical harmonics
\eq{dtot}{\frac{\Delta \widehat T}{\widehat T}(\vec n) = \sum_{\ell m} a^T_{\ell m} Y_{\ell m}(\vec n) }
where $\vec n$ is a unit vector on the sphere and where the $a^T_{\ell m}$ are stochastic variables. Were we to presume statistical isotropy, it can be shown that the average over an ensemble of universes satisfies
\eq{angdef}{\langle a^X_{\ell m} a^{Y*}_{\ell' m'} \rangle = C^{XY}_\ell \delta_{\ell\ell'}\delta_{m m'}}
which defines the $C^{XY}_\ell$'s as the angular auto/cross correlation function for the variables X/Y in question. The observed $\widehat{C}_\ell^{XY}$ are to be compared to the theoretical prediction, $C_\ell^{XY} $.  In the absence of initial vector and tensor perturbations, to linear order, the $C_\ell^{XY} $ are determined by the scalar contributions to the temperature and E-polarisation auto- and cross-correlation angular power spectra, which can be expressed in terms of the power spectrum of the comoving curvature mode as \cite{Lesgourgues:2013qba}
\eq{eq:cls}{\boxed{C_\ell^{XY} = \frac{1}{2\pi^2} \int \mathrm{d}\ln k \; \Delta^X_\ell(k,\tau_0) \Delta^Y_\ell(k,\tau_0) \mathcal{P_R}(k).}}
The transfer functions $\Delta^X_\ell$ are integrals over comoving time $\tau$,
\eq{eq:transfer}{\Delta^X_\ell(k,\tau_0) = \int_{\tau_{\mathrm{ini}}}^{\tau_0} \mathrm{d}\tau \; S^X(k,\tau) j_\ell(k(\tau-\tau_0)),}
with a geometrical part given by $j_\ell$, the spherical Bessel functions of order $\ell$, and the source functions $S^X$ which contain the dependence on non-primordial cosmological parameters such as the matter and baryon densities. The source functions are typically evaluated using a Boltzmann code (see \cite{Durrer:1127831} for a detailed derivation and \cite{Mukhanov:2003xr} for a reasonably accurate analytic understanding up to $\ell \sim 200$). The $C_\ell^{XY}$ thus computed are ensemble averages; for observations of the full sky, 
we can estimate the angular cross/auto power spectra by constructing the following \textit{estimator}:
\eq{}{\widehat C^{XY}_\ell = \frac{1}{2\ell + 1}\sum_m a_{\ell m}^X a_{\ell m}^{Y*}.}
From (\ref{angdef}) we see that although this estimator is \textit{unbiased} in the sense that
\eq{}{\langle \widehat C_\ell^{XY}\rangle = C_\ell^{XY},}
it has a non-zero (cosmic) variance~\cite{Martinez:1339567},
\begin{eqnarray}\label{cosvar}
{\rm var} ~\widehat C^{XY}_\ell &:=& \langle \widehat C_\ell^{XY} \widehat C_\ell^{XY} \rangle - \langle \widehat C_\ell^{XY} \rangle\langle\widehat C_\ell^{XY} \rangle\\
\nonumber &=& \frac{1}{(2\ell +1)^2} \sum_{m m'} \langle a^X_{\ell m}a^{Y*}_{\ell m} a^X_{\ell m'}a^{Y*}_{\ell m'} \rangle - {C_\ell^{XY}}^2\\ &=& \nonumber \frac{1}{(2\ell +1)^2} \sum_{m m'} \left[\langle a^X_{\ell m}a^{X}_{\ell m'}\rangle\langle a^{Y*}_{\ell m}a^{Y*}_{\ell m'} \rangle + \langle a^X_{\ell m}a^{Y*}_{\ell m'}\rangle\langle a^{X}_{\ell m'}a^{Y*}_{\ell m} \rangle \right]\\ &=& \nonumber \frac{1}{2\ell + 1}\left[C_\ell^{XX}C_\ell^{YY} + {C_\ell^{XY}}^2\right],
\end{eqnarray}
where the third line follows from the second by presuming that the $a_{\ell m}$'s satisfy almost Gaussian statistics and that they correspond to real variables (so that $a^{X*}_{\ell m} = a^X_{\ell -m}$). Therefore for the auto-correlation functions we have
\eq{}{\boxed{ {\rm var} ~\widehat C_\ell^{XX} = \frac{2}{2\ell + 1} {C_\ell^{XX}}^2}}
That is to say, the $\widehat{C}_\ell^{XY}$ are random variables with mean $C_\ell^{XY}$ and subject to a cosmic (sampling) variance given by (\ref{cosvar}). Realistic experiments will not be able to observe the CMB over the full sky, either due to experimental design (e.g., for ground-based or balloon experiments) or the need to mask areas where foregrounds cannot be removed, such as the galactic plane.
If only a fraction $f_\mathrm{sky}$ of the sky is covered, the cosmic variance will scale roughly with a factor $1/f_\mathrm{sky}$, and the inferred $\widehat{C}_\ell^{XY}$ are no longer uncorrelated.

Looking at equation~(\ref{eq:cls}), schematically, the angular power spectrum is a convolution of the primordial power spectrum with a window function, given by the product of two transfer functions. Invariably, this convolution leads to a loss of information: features in $\mathcal{P_R}(k)$ will be smeared out to a certain degree and show up less prominently in ${C}_\ell$.\footnote{Besides the smearing due to the geometrical projection, the observed angular power spectra are themselves subject to further convolutions due to (i) the effects of weak gravitational lensing on the CMB~\cite{Lewis:2006fu}, and (ii) the aberration caused by the proper motion of the observer with respect to the CMB rest frame~\cite{Jeong:2013sxy}.}   If the characteristic width of a feature in $k$-space exceeds the width of the window function on the corresponding scales, its amplitude in the angular power spectrum will be approximately suppressed by the inverse ratio between the two.   Thus, provided a sufficiently high frequency, even a relative modulation of the primordial spectrum of $\mathcal{O}(1)$ could be hidden in the angular power spectrum-- though the deviation from linear evolution required to produce such a large modulation is likely going to induce a strong signal in higher order correlations (see Section~\ref{sec:hoc}).

\subsubsection{CMB temperature anisotropies}

\begin{figure}[t]
\center
\includegraphics[height=.90\textwidth,angle=270]{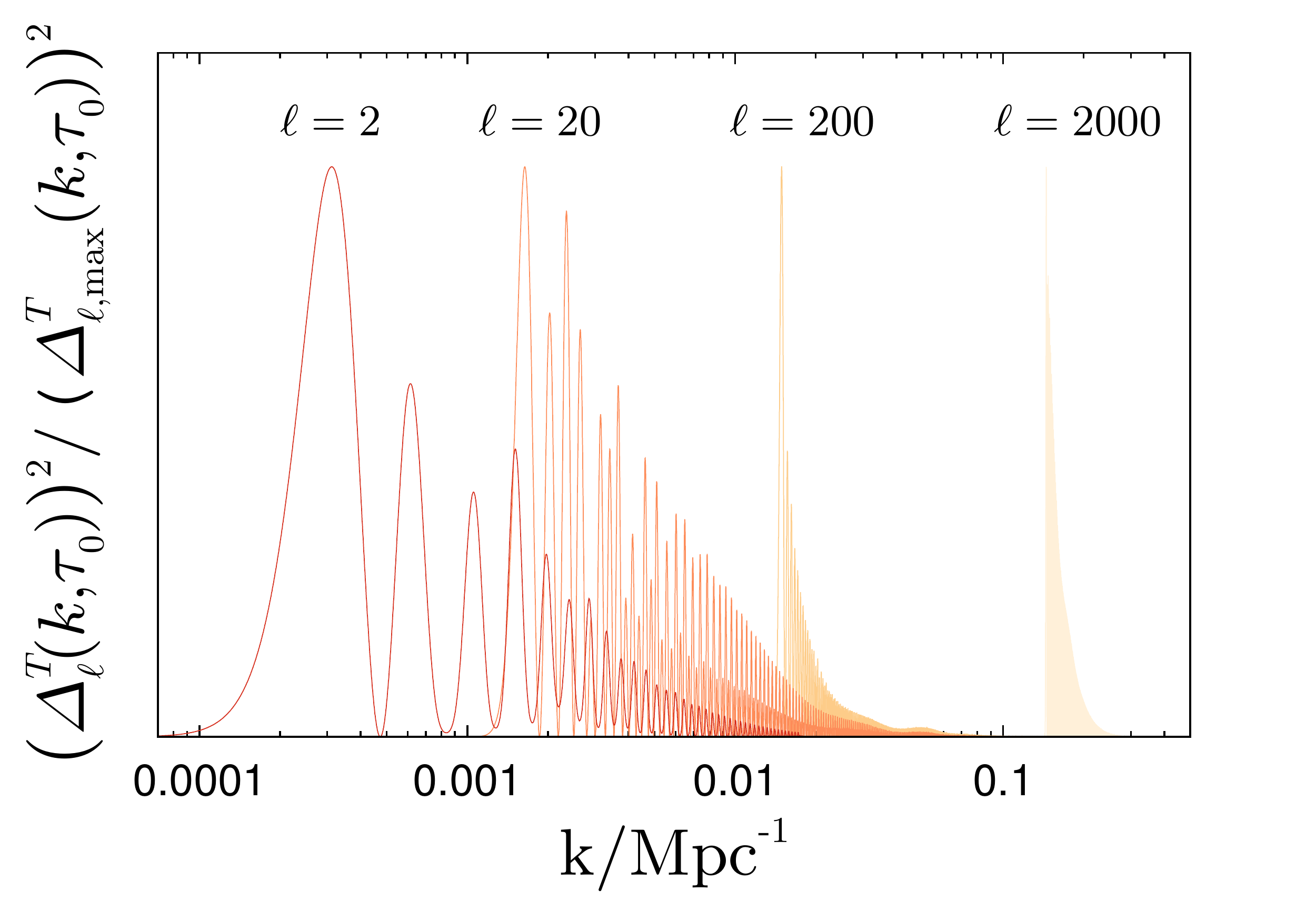}
\caption{Temperature auto-correlation angular power spectrum window functions for multipoles \mbox{$\ell = 2,20,200$}~and~$2000$, normalised to their respective maxima.\label{fig:temp_delta}}
\end{figure}

We show the window functions of the temperature anisotropies for multipoles \mbox{$\ell = 2, 20, 200$}  and $2000$ in Figure~\ref{fig:temp_delta}.  Their rapid oscillations are due to the Bessel functions, and changing the later-time cosmology will modify the envelopes.  
As one goes to higher multipoles, the log-width of the window functions decreases from about 0.4 at $\ell = 2$ to 0.045 at $\ell = 2000$ (assuming the Planck best-fit $\Lambda$CDM late-time cosmology). At multipoles $\ell \gtrsim 2000$, the primary CMB temperature signal becomes rapidly subdominant to secondary microwave anisotropies, e.g., from point sources, the cosmic infrared background or the Sunyaev-Zeldovich effects \cite{Zeldovich1969, Sunyaev1980}, which are which are much less sensitive to the presence of features.  Thus the range of wavenumbers over which the CMB temperature spectrum is sensitive to features is given roughly by the interval $0.0002~\mathrm{Mpc}^{-1} \lesssim k \lesssim 0.15~\mathrm{Mpc}^{-1}$. This range has in fact been almost entirely covered at the cosmic variance limit by the Planck mission~\cite{Adam:2015rua} (see Figure~\ref{fig:plonk_cls}).

\begin{figure}[t]
\center
\includegraphics[height=.90\textwidth,angle=270]{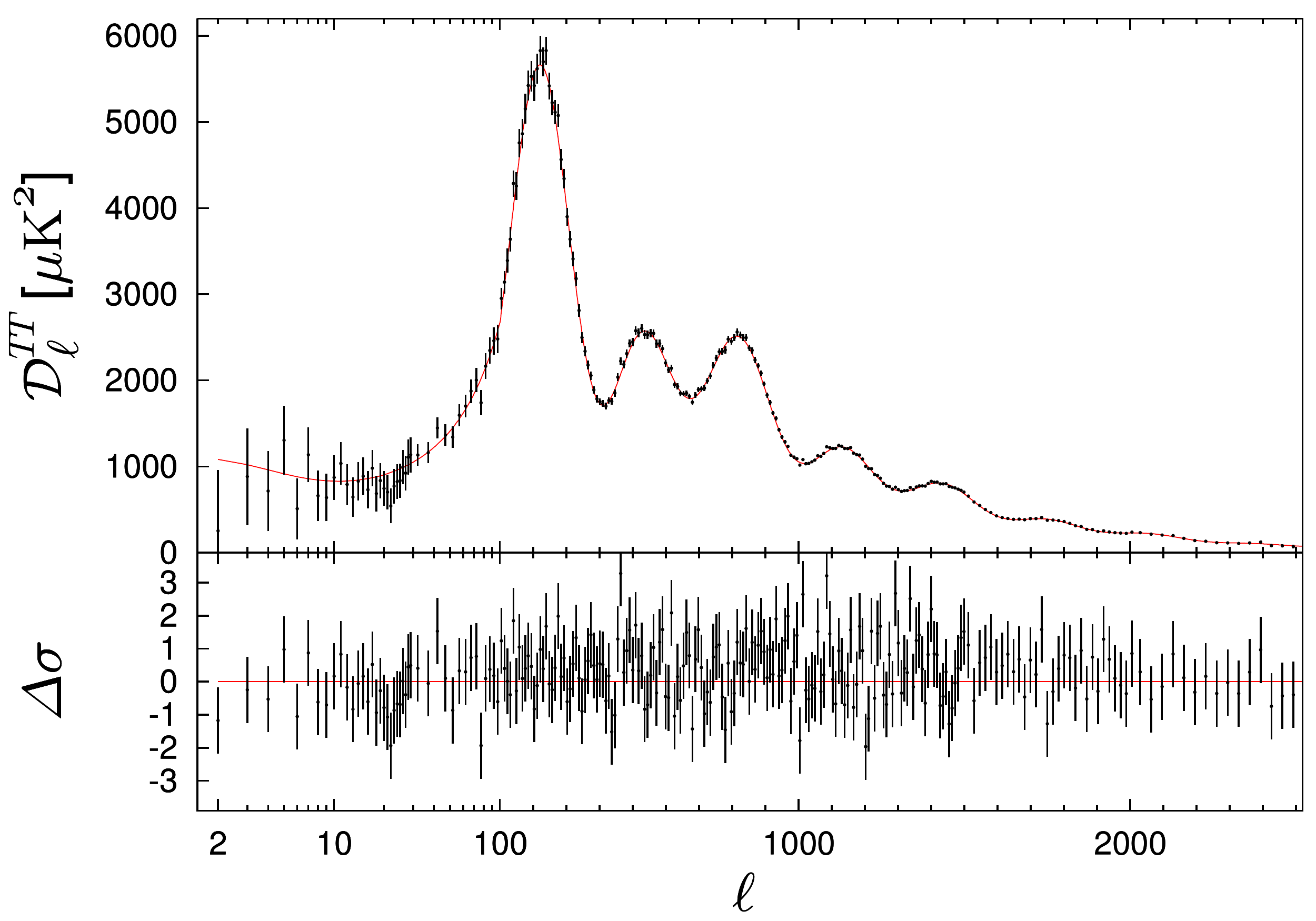}
\caption{\textit{Top:} Planck full mission measurement of the angular power spectrum of the temperature autocorrelation of the CMB~\cite{Adam:2015rua}. \textit{Bottom:} Residuals in units of standard deviations with respect to the best-fit $\Lambda$CDM model. For $\ell > 30$, the power spectrum is binned.  The error bars denote the variance due to sampling error and experimental uncertainties due to, e.g., detector noise, component separation or beam errors. \label{fig:plonk_cls}}
\end{figure}

\subsubsection{CMB polarisation anisotropies}

\begin{figure}[t]
\center
\includegraphics[height=.90\textwidth,angle=270]{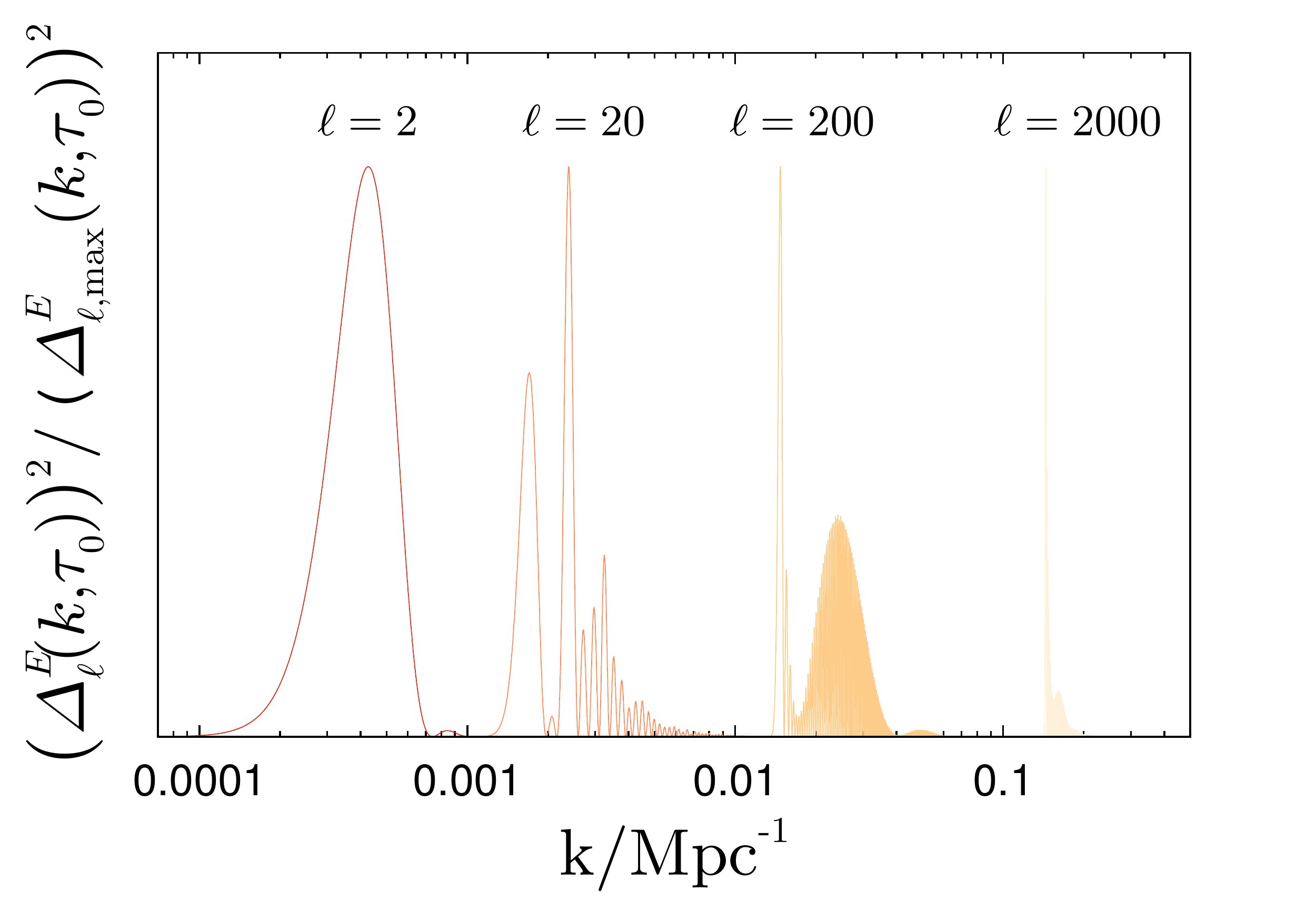}
\caption{Same as Fig.~\ref{fig:temp_delta}, for the $E$-polarisation autocorrelation.\label{fig:epol_delta}}
\end{figure}

The temperature anisotropies of the cosmic microwave background are not the only source of information one can obtain from the decoupling era.  The CMB is weakly linearly polarised due to Thomson scattering (the low energy limit of Compton scattering) of the CMB photons during decoupling~\cite{Bond:1984fp, Hu:1997hv}, sourced by the local quadrupole anisotropy in the incident radiation intensity field.  Measuring the angular power spectrum of the $E$-component of the CMB polarisation is a powerful complementary tool in the search for features in the primordial spectrum and potentially even surpasses the temperature in terms of sensitivity.  Not only is the small-scale polarisation signal to a lesser degree subject to foreground contamination, with the primary signal dominant up to $\ell \sim 4000$~\cite{Seiffert:2006vh,Bock:2008ww}, the polarisation window functions for a given multipole are also narrower than the corresponding temperature window functions (see Figure~\ref{fig:epol_delta})~\cite{Hu:2003vp}, leading to more sharply defined features in the observable $E$-autocorrelation spectrum. Here, the log-width of the window functions ranges from roughly 0.13 at $\ell = 2$ to 0.03 at $\ell = 2000$.
It should be noted however, that on the largest scales ($\ell \lesssim 20$), the polarisation power spectrum is dominated by the contribution generated during reionisation.  At those scales, features in the $EE$-spectrum could also imply a deviation from the commonly assumed nearly instantaneous reionisation model, rather than point to a primordial origin.

Mortonson et al.~\cite{Mortonson:2009qv}, for example, considered a putative oscillatory feature around $k = 2-5 \times 10^{-3}~\mathrm{Mpc}^{-1}$ (corresponding to $20 \lesssim \ell \lesssim 40$) which improves the fit to CMB temperature data by $2 \Delta \ln \mathcal{L}^{TT} \approx 10$ (see Sect. \ref{sec:statistics}) for a definition of the former quantity).  In a cosmic-variance limited measurement of the $E$-polarisation autocorrelation spectrum, such a feature can be expected to lead to $2 \Delta \ln \mathcal{L}^{EE} \approx 60$ when the reionisation history is known, and $\sim 30$ if it is allowed to freely vary. Since the $EE$-spectrum is suppressed by a few orders of magnitude with respect to the temperature spectrum, measuring it demands much higher instrumental sensitivity. Unlike Planck's temperature data, its polarisation data is not limited by cosmic variance (but nonetheless adds useful information at intermediate scales~\cite{Mortonson:2009qv,Ade:2015lrj}).  An all-sky cosmic variance-limited measurement to much smaller scales is feasible with current technology and several experimental designs have been proposed~\cite{Bock:2008ww,Bouchet:2011ck,Andre:2013afa}. Although none of the former have been selected for funding to date, ground based and sub-orbital CMB polarisation measurements (which either by themselves \cite{Lazear:2014bga} or taken in combination \cite{Ahmed:2014ixy, Niemack:2010wz, 2014arXiv1407.2973B, 2012SPIE.8452E..1EA, Kermish:2012eh, Keating:2011iq, Filippini:2011ds, Fraisse:2011xz, Eimer:2012ny, ReichbornKjennerud:2010ja, MacDermid:2014wca, Matsumura:2013aja,Kogut:2011xw, Ogburn:2012ma}, promise almost full sky coverage) remain one of the most promising future avenues for the detection of features, especially on the largest scales.

\subsection{The matter power spectrum}
Long after recombination, the initial inhomogeneities imprinted in the CMB evolved into the scaffolding upon which large scale structures formed. The central quantity to describe the statistical properties of these inhomogeneities (which are matter density fluctuations around the mean density) given by $\delta \rho_\mathrm{m}/\bar{\rho}_\mathrm{m} \equiv \delta_\mathrm{m}$, is the matter power spectrum $\mathcal{P}_\mathrm{m} (k,z)$, defined via
\begin{equation}
\langle \delta_\mathrm{m}(\mathbf{k},z) \delta_\mathrm{m}(\mathbf{k}',z) \rangle \equiv \frac{2\pi^2}{k^3}  \delta^3(\mathbf{k}-\mathbf{k}') \; \mathcal{P}_\mathrm{m}(k,z).
\end{equation}
As long as the fluctuation amplitude is small, $\delta_\mathrm{m} \ll 1$, matter fluctuations of different wavenumbers evolve independently and $\mathcal{P}_\mathrm{m}$ is directly related to the primordial power spectrum of curvature fluctuations via
\begin{equation}
\mathcal{P}_\mathrm{m} (k,z) = \int \mathrm{d}k' \; T_\mathrm{m}(k,k',z) \mathcal{P_R}(k) \approx T^\mathrm{lin}_\mathrm{m} (k,z) \mathcal{P_R}(k),
\end{equation}
where $T^\mathrm{(lin)}_\mathrm{m}$  is the (linear) matter transfer function containing the dependence on the intervening cosmology.  Compared to the CMB where information about the fluctuations is fundamentally limited to the two-dimensional surface of last scattering, the distribution of matter can be probed in three dimensions.  This has two important consequences: firstly, one gains sensitivity to the time evolution of the power spectrum which is very helpful in disentangling primordial features from non-primordial ones\footnote{Consider, e.g., the suppression of power on small scales due to the free-streaming effect of massive neutrinos.  This is a time-dependent effect that becomes more pronounced at later times, as opposed to a primordial power spectrum suppression, which will show up equally at all redshifts.}.  Secondly, the ability to detect density fluctuations along the line of sight results in an increase of the number of observable modes for a given wavenumber, greatly reducing the uncertainty due to sample variance, with $\mathrm{var}\, \mathcal{P}_\mathrm{m}(k) \propto V^{-1} k^{-2} \mathcal{P}_\mathrm{m}(k)$ for a given volume~$V$.

Despite the ability to recover some 3-dimensional information, a finite-volume survey will still be subject to a limited resolution in $k$.  This becomes apparent when one considers the spatial correlation function, which is related to the power spectrum by a Fourier transform. Take for instance a sinusoidal modulation of $\mathcal{P_R}(k)$ with frequency $\omega$: in the correlation function this corresponds to a spike at a length scale $l \propto \omega$.  If for a given frequency $\omega$ the survey volume is too small to accommodate this length scale, it will not be possible to see the spatial correlation and thus one also will not be able to resolve the modulation.
As a consequence, observations of the matter perturbations over a finite volume only allow us to construct an estimate of the power spectrum in band powers $\mathcal{P}_i$~\cite{Feldman:1993ky}, which are related to the full power spectrum in terms of a convolution with a set of window functions $\mathcal{W}_i(k)$:
\begin{equation}
\boxed{\mathcal{P}_i(k_i) = \int \mathrm{d}k' \; \mathcal{W}_i(k-k') \mathcal{P}(k').}
\end{equation}
Depending on the survey geometry, one can construct an optimal set of window functions that are as narrow and uncorrelated as possible without compromising the variance in $\mathcal{P}$~\cite{Tegmark:1995xxx}.  For a volume-limited all-sky survey out to a distance $R$, the optimal window function in the large-scale limit is given by 
\begin{equation}
\label{eq:window}
\mathcal{W}(k) = 4 \pi R \left( \frac{\sin k R}{\pi^2 - k^2 R^2} \right)^2,
\end{equation}
whose width scales like $R^{-1}$.  This effect is illustrated in the bottom panel of Fig.~\ref{fig:wiggle_delta}, for effective survey volumes $R^3 = 1 \; \mathrm{Gpc}^3 h^{-3}$, representative of current surveys (e.g., SDSS LRG~\cite{Reid:2009xm}), and $R^3 = 50 \; \mathrm{Gpc}^3 h^{-3}$, which is within reach of upcoming projects like the Large Synoptic Survey Telescope (LSST)~\cite{Abell:2009aa} or Euclid~\cite{Laureijs:2011gra}.  In the example here, the feature at large scales cannot be resolved.  The width of $\mathcal{W}$ for a $R \sim \mathcal{O}(1)$~Gpc is $\mathcal{O}(0.001) \; h\, \mathrm{Mpc}^{-1}$, so structures in the spectrum with a smaller characteristic size $\Delta k$ will be erased.

\begin{figure}[t]
\center
\includegraphics[height=.90\textwidth,angle=270]{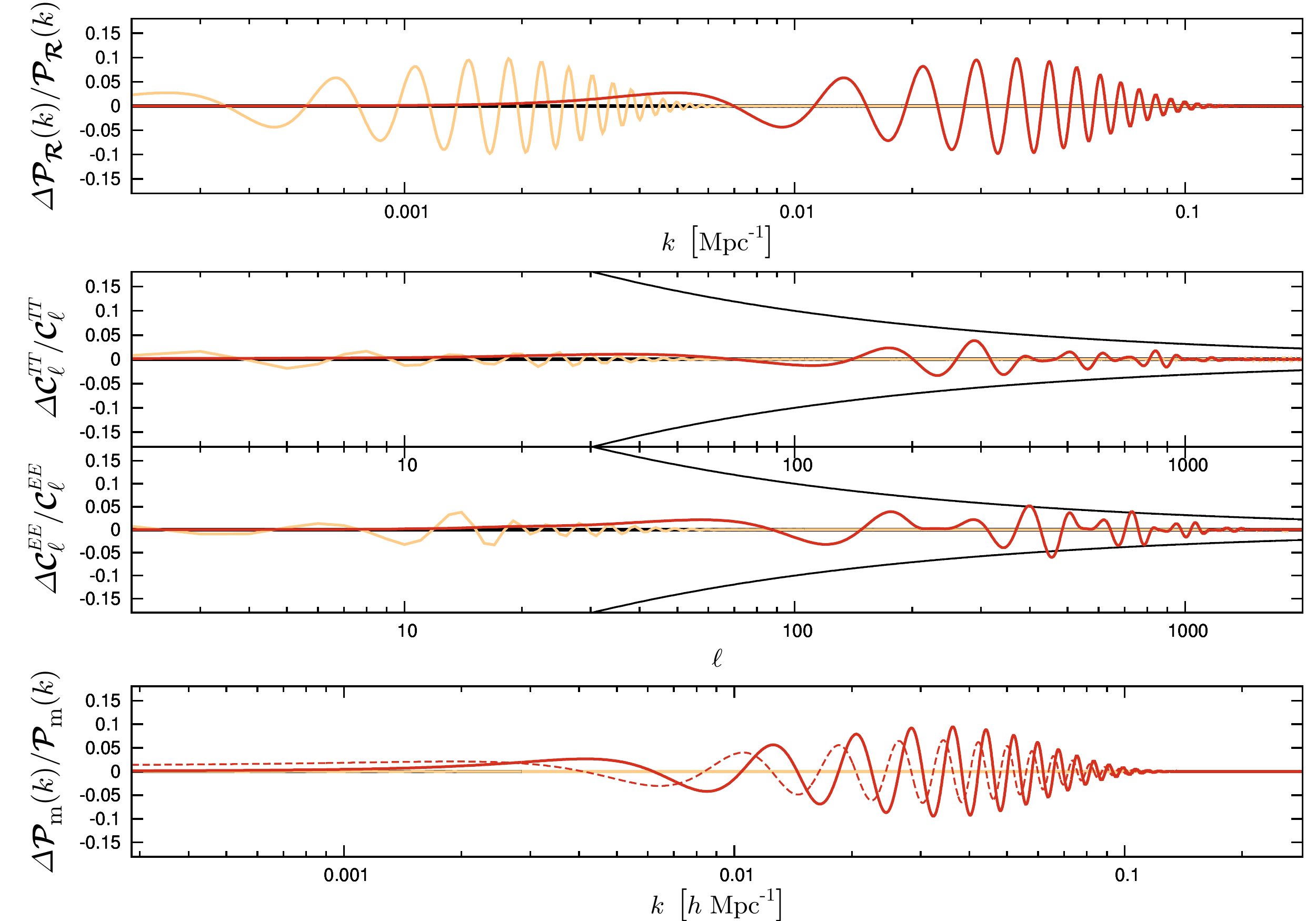}
\caption{Relative fluctuations for two example primordial features ({\it top}) and the corresponding modulation of the CMB temperature and $E$-polarisation angular power spectra ({\it centre}, thin black lines mark the cosmic variance).  {\it Bottom}: signature of the features in the matter power spectrum convolved with the window function of Eq.~\ref{eq:window} assuming effective survey volumes of $R^3 = 1$~Gpc$^3 h^{-3}$ (dotted lines) and $R^3 = 50$~Gpc$^3 h^{-3}$ (solid lines). \label{fig:wiggle_delta}}
\end{figure}

When it comes to mapping the distribution of matter in the universe, we can, depending on the circumstances, draw upon three different methods for the matter to reveal its presence: (i) gravitational lensing, (ii) absorption or (iii) emission of photons.  
The first category can be exploited by measuring the weak gravitational lensing of galaxies~\cite{Bartelmann:1999yn} or the CMB~\cite{Lewis:2006fu}.  While weak lensing has the great advantage of probing all forms of matter, it is, unfortunately not very sensitive at all to features in the spectrum.  The reason is that lensing is an integrated effect and depends on the matter distribution at all redshifts between source and observer; as a result, the lensing angular power spectrum for a given multipole has contributions from $\mathcal{P}_\mathrm{m}$ over a large range of wavenumbers, which leads to a very efficient smoothing of any but the most extreme primordial features.
Just like weak lensing, exploiting the second category needs a background light source in whose spectrum absorption due to absorbing matter can be detected.  In a cosmological context, the absorber would typically be neutral hydrogen due to its ubiquity in the early universe.  The distribution of matter along the line of sight can then be reconstructed by relating the redshift of particular absorption lines to the radial distance of the absorber and inferring its density at that distance from the absorption strength.  
The most straightforward probes of the matter power spectrum are based on emission, and in the following we will focus on the prospects of using galaxies and neutral hydrogen as tracers of $\mathcal{P}_\mathrm{m}$.

As long as the density fluctuations are small, i.e., $\delta \rho_\mathrm{m}/\rho_\mathrm{m} \ll 1$ the matter power spectrum is well approximated with linear perturbation theory.  At late times and small scales, however, the fluctuations are no longer small, linear theory breaks down\footnote{at redshift $z \sim 0$ this starts becoming relevant for wavenumbers $k \gtrsim 0.05 h$~Mpc$^{-1}$.} and a precise determination of the non-linear spectrum $\mathcal{P}_\mathrm{m}^\mathrm{nl}(k)$ becomes a rather difficult problem.  
From the perspective of looking for features, the breakdown of linear theory has another unwelcome side-effect: in the non-linear regime fluctuations of a given wavenumber $k$ do not evolve independently of neighbouring wavenumbers anymore (mode-coupling), which is going to further suppress fine structures in the spectrum.

\subsubsection{Galaxy surveys}
The use of galaxies as a tracer of the matter distribution has a long history in cosmology.  The measurement of their positions in the sky, along with their redshift, from which one can infer their radial distance, allows one to construct their three-dimensional power spectrum $\mathcal{P}_\mathrm{g}(k)$.  However, since the formation of galaxies depends on the density of matter in a non-trivial way, the galaxy power spectrum is not actually the same as the matter power spectrum which we can easily calculate from theory.  Rather, it is a biased tracer of $\mathcal{P}_\mathrm{m}(k)$.

On linear scales, the two are related by the simple relation $\mathcal{P}_\mathrm{g}(k) = b^2(z) \mathcal{P}_\mathrm{m}(k)$ with a redshift-dependent proportionality factor, the bias $b^2 > 1$, which in general also depends on the type of galaxies considered.  On non-linear scales, the bias also picks up a scale-dependence $\mathcal{P}_\mathrm{g}^\mathrm{nl}(k)(k) = b^2(k,z) \mathcal{P}_\mathrm{m}^\mathrm{nl}(k)$ which is even more challenging to predict than the non-linear corrections to the matter power spectrum.  The wavenumbers at which non-linear corrections and the scale-dependent bias for a given type of object start becoming important do increase as one goes to higher redshift, but at the same time accessing higher redshifts requires tracing intrinsically brighter objects which tend to be more strongly biased.  At the moment these effects are not even well enough understood for the standard cosmological model, let alone features models.  Conservatively speaking, it will be very difficult to obtain reliable information from measurements of galaxy clustering on scales beyond $k \approx 0.1 \; h \, \mathrm{Mpc}^{-1}$.

At present, galaxy power spectrum data are not as constraining as CMB temperature data which cover the same range, since only a small region of our cosmic neighbourhood is known to us, and the sample variance cannot compete with the CMB's cosmic variance.  Luckily, this is about to change though: within a decade from now, ongoing surveys such as the Dark Energy Survey (DES)~\cite{Abbott:2005bi} and next-generation large-volume surveys like DESI~\cite{Schlegel:2009uw}, the LSST~\cite{Abell:2009aa} or Euclid~\cite{Laureijs:2011gra} are going to map out a good fraction of our surroundings up to a redshift $z \approx 2$, allowing us to take advantage of the $V^{-1}$ scaling of the sample variance and eventually beat the current gold standard set by the CMB.

It is not only the lower sample variance, but also the possibility of obtaining 3-dimensional information that distinguishes probes of the matter power spectrum from the CMB.  As can be seen from the bottom panel of Figure~\ref{fig:wiggle_delta}, for a typical feature in this intermediate wavenumber range, the finite window functions of a survey with effective volume of $\mathcal{O}(1-10) \; \mathrm{Gpc}^3 \, h^{-3}$ 
do not lead to a significant degradation of the signal, unlike the CMB signal, which is markedly attenuated in this region.
Features with characteristic sizes smaller than $\Delta k \approx 0.001 \; h \, \mathrm{Mpc}^{-1}$ on the other hand, cannot be resolved, which essentially makes a detection of any but the broadest features on scales $k \lesssim 0.01$~Mpc$^{-1}$ unrealistic. Nonetheless, within an intermediate range of wavenumbers $\mathcal{O}(0.01) \; h \, \mathrm{Mpc}^{-1} \lesssim k \lesssim 0.1 \; h \, \mathrm{Mpc}^{-1}$, galaxy surveys are a very powerful resource and hold a lot of potential for the discovery of features.

\subsubsection{The Lyman-$\alpha$ forest}
Additional insights on features in the power spectrum may be gained at redshifts beyond $z \approx 2$ from the {\it Lyman-$\alpha$-forest}, i.e., absorption signatures of neutral hydrogen in the 
spectra of objects such as quasars or Lyman break galaxies caused by the Lyman-$\alpha$ transition~\cite{Croft:1997jf}, from which the flux power spectrum can be extracted. 
 While individual absorption spectra only yield an estimate of a one-dimensional power spectrum which would not be very helpful for the purpose of resolving features, combining the information of a sufficient number of them allows the 3-dimensional power spectrum (or its Fourier-equivalent, the correlation function) to be estimated~\cite{Slosar:2011mq,Slosar:2013fi,Delubac:2014aqe}.  

The Lyman-$\alpha$-forest covers a redshift range of $2 \lesssim z\lesssim 4$, and at these redshifts, non-linear evolution of the underlying matter power spectrum does not play as big a role as it does for the galaxy power spectrum, giving in principle access to fluctuations on smaller scales $\mathcal{O}(0.01) \; h \, \mathrm{Mpc}^{-1} \lesssim k \lesssim 1 \; h \, \mathrm{Mpc}^{-1}$.  However, the gas physics relating the flux power spectrum to the matter power spectrum is complicated and requires computationally expensive hydrodynamical simulations; to what extent this would affect features is an open question, but it does not seem unreasonable to assume that one can expect at least a qualitative sensitivity to deviations from smoothness from these data.

\subsubsection{21~cm surveys}
An even more radical idea to use neutral hydrogen as a tracer of the matter distribution and reconstruct its three-dimensional structure is by measuring emission or absorption due to the 21~cm hydrogen spin flip transition~\cite{Scott:1990bla}.  This could be done either via a measurement of the emission caused by reionisation~\cite{McQuinn:2005hk} in a redshift range of $6 \lesssim z \lesssim 15$, or via the absorption of CMB photons during the dark ages $15 \lesssim z \lesssim 200$.  Whereas the redshift range of galaxy surveys ($z < 2$) covers only about 5\% of the Hubble volume, 21~cm emission could in conceivably make an additional $\sim 20\%$ of the Hubble volume accessible, plus another $\sim 40\%$ from absorption~\cite{Wright:2006up} (assuming full sky coverage) and hence in principle allowing survey volumes up to $\mathcal{O}(1000\ \mathrm{Gpc}{^3} h^{-3})$.  With correspondingly narrower window functions, 21~cm surveys will be able to extend sensitivity to features down to wavenumbers $k \sim 0.001 h$~Mpc$^{-1}$, reaching almost the range of the CMB.  
Furthermore, the matter fluctuations at the redshifts targeted by 21~cm surveys are even smaller, and thus even less subject to non-linear effects, opening the window to features on scales as small as $k = \mathcal{O}(10^2) h$~Mpc$^{-1}$, thus allowing us to extend our search for features to scales inaccessible to other probes of spatial fluctuations.

While obtaining all the potential information from such a 21~cm survey would certainly be the ultimate dream of any cosmologist, it is at this point not clear whether this goal can ever be achieved.  The cosmological signal is extremely feeble, and disentangling it from astrophysical foregrounds is going to be a very challenging task~\cite{Pritchard:2011xb} for the reionisation era, and even more so for the dark ages. To date, no cosmological signal has been detected although experimental programmes are on the way to evaluate the potential of this observable. Upcoming radio-surveys such as the
Square Kilometre Array~\cite{Blake:2004pb} or the Tianlai project~\cite{Chen:2012xu} do have cosmology-related science goals, but a competitive sensitivity to features still lies in the more distant future, and will require an experiment such as the Fast Fourier Transform Telescope~\cite{Tegmark:2008au}.

\subsection{Higher order correlators \label{sec:hoc}}

As discussed in sections \ref{sec:preliminaries} and \ref{sec:theory}, any features in the two point correlation functions will correlate with features in higher order correlation functions for adiabatic fluctuations. In the CMB, one can thus entertain looking for such features in the bispectrum. Although detecting small amounts of non-Gaussianity with reasonable statistical significance remains the observational challenge of the day, the possibility that cross correlations (possibly even among different tracers) might exist with the power spectra enhances the prospects for their detection\footnote{Although we shall focus the discussion that follows on the bispectrum, as reviewed in \cite{Ade:2015ava}, one can look to quantify primordial non-Gaussianities in a number of other ways including making use of Minkowski functionals and wavelets to name but a few.}. 

How does one then go about searching for primordial non-Gaussianities? Recall first that (\ref{dtot}), the expression for the angular power spectrum (\ref{eq:cls}) follows from
\eq{almdef}{a^T_{\ell m} = 4\pi(-i)^\ell\int\,\frac{d^3k}{(2\pi)^3}\Delta_\ell(\vec k, \tau_0)\calR(\vec k) Y_{\ell m}(\hat k)}
where the $\Delta_\ell$'s are given by (\ref{eq:transfer}) and where $\hat k$ is the unit norm vector on the (momentum) 2-sphere. Analogous to the definition of the angular power spectrum (\ref{eq:cls}), we see that given the definition of the bispectrum (\ref{bsdef})
\eq{}{\nonumber \left\langle \widehat \calR_{\vec k_1}(t)\widehat  \calR_{\vec k_2}(t)\widehat  \calR_{\vec k_3}(t) \right\rangle = 
(2\pi)^3\delta^{(3)}\left(\sum_i\vec{k}_{i}\right)\calB_\calR(\vec{k}_1,\vec{k}_2,\vec{k}_3),}
we can define the temperature \textit{angular bispectrum} in analogy to (\ref{angdef}) as
\eq{}{\calB^{\,\ell_1\ell_2\ell_3}_{m_1m_2m_3} = \langle a^T_{\ell_1 m_1} a^T_{\ell_2 m_2} a^T_{\ell_3 m_3} \rangle.}
From (\ref{almdef}) and (\ref{bsdef}) it can be shown that the angular bispectrum relates to the momentum space bispectrum as \cite{Fergusson:2008ra, Fergusson:2009nv}
\begin{align}\nonumber
\calB^{\,\ell_1\ell_2\ell_3}_{m_1m_2m_3} =\left(\frac{2}{\pi}\right)^3\,\int y^2 dy\, &\prod_{i=1}^3 k_i^2dk_i\,\calB_\calR(k_1,k_2,k_3)\Delta_{\ell_1}(k_1,\tau_0)\Delta_{\ell_2}(k_2,\tau_0)\Delta_{\ell_3}(k_3,\tau_0)\\& \times j_{\ell_1}(k_1 y)j_{\ell_2}(k_2 y)j_{\ell_3}(k_3 y)\, G^{\ell_1\ell_2\ell_3}_{m_1m_2m_3}
\end{align}
where
\eq{}{G^{\ell_1\ell_2\ell_3}_{m_1m_2m_3}:= \sqrt{ \frac{(2\ell_1+1)(2\ell_2+1)(2\ell_3+1)}{4\pi}}{\begin{pmatrix}\ell_1 && \ell_2&& \ell_3\\0 && 0 && 0\end{pmatrix}}{\begin{pmatrix}\ell_1 && \ell_2&& \ell_3\\m_1 && m_2 && m_3\end{pmatrix}},}
results from the angular integration of the product of three spherical harmonics\footnote{The dummy integral over $y$ results from switching to the integral representation of $\delta^{(3)}(\sum k_i)$.}, and where the quantities in the parentheses are the Wigner 3-j symbols. Since isotropy remains a good approximation to the observed sky, it is also common to work with the \textit{averaged} bispectrum defined as
\eq{}{\calB^{\ell_1\ell_2\ell_3} = \sum_{\{m_i\}}{\begin{pmatrix}\ell_1 && \ell_2&& \ell_3\\m_1 && m_2 && m_3\end{pmatrix}} \langle a^T_{\ell_1 m_1} a^T_{\ell_2 m_2} a^T_{\ell_3 m_3} \rangle.}
\begin{wrapfigure}{r}{0.4\textwidth}
  \begin{center}
    \includegraphics[width=0.5\textwidth]{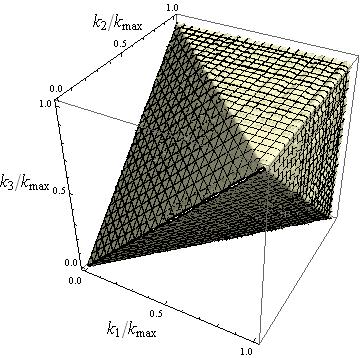}
  \end{center}
  \caption{The domain $\mathcal V$ of $\calS(k_1,k_2,k_3)$\label{domain2}}
\end{wrapfigure}

Given the inherent difficulties of extracting three point functions of putatively very weakly coupled perturbations from observations\footnote{The bispectrum signal being too weak to be able to extract individual multipoles-- the only practical option being to use an estimator that sums over all mulitipoles (i.e. sky averages).}, it is much more practical to ask the question of how significantly (in the statistical sense) a particular shape of the bispectrum is realized in the actual data. To this end, we need a useful way of characterizing a potentially arbitrary signal. Although the $f_{\rm NL}$ class of diagnostics (\ref{fNL-def}) are useful for the sort of non-Gaussianities expected from specific classes of almost scale invariant models of single field inflation, they are inadequate for searches for features, for more general models of inflation \cite{Ribeiro:2011ax, Battefeld:2011ut, Regan:2013wwa,  Byun:2013jba}, or for the type of non-Gaussian signals expected from topological defects. Shellard, Fergusson and Liguori have proposed just such a general scheme  \cite{Fergusson:2008ra, Fergusson:2009nv} which we review briefly in what follows. We first define the shape function 
\eq{}{\calS(k_1,k_2,k_3) = \frac{1}{N_\calB}\left(k_1k_2k_3\right)^2\calB_\calR(k_1,k_2,k_3)}
where $N_\calB$ is some suitable normalization that depends on the shape in question. Given that the observable number of modes is finite (the upper limit coming from the scale at which foregrounds start to dominate the CMB signal), the domain for the shape function is given by the tetrapyd defined by $k_1,k_2,k_3 \leq k_{\rm max}$ in conjunction with the triangle inequalities $k_1 \leq k_2 + k_3$ (+ permutations). We denote this domain $\mathcal V$, illustrated in Fig. \ref{domain2}.
We now ask, given an actual realization in the CMB of the bispectrum shape $A$, how well would an estimator based on some other shape $B$ pick up this signal? For this, we introduce the notion of a `cosine' between two shapes \cite{Babich:2004gb}, defined as
\eq{}{\mathcal C(A,B):=\frac{(\calS_A,\calS_B)}{(\calS_A,\calS_A)^{1/2}(\calS_B,\calS_B)^{1/2}}}
where the scalar product between two shapes $(A,B)$ is defined as
\eq{bssp}{(A,B):=\int_{\mathcal V}\,d\mathcal V\,\omega\, \calS_A^*\calS_B,}
where one is obliged to introduce a non-trivial measure $\omega(k_1,k_2,k_3)$ to weight for non-trivial scaling of the bispectrum. For the purposes of the present discussion we don't need to consider it further, referring the interested reader to \cite{Fergusson:2009nv, Battefeld:2011ut} for details. The size of this overlap indicates how much an estimator optimized for shape $B$ would detect the true signal from shape $A$. Clearly, if we had a \textit{basis} of estimators that could capture an arbitrary shape, we would in principle be sensitive to an arbitrary signal. With the scalar product (\ref{bssp}), one can indeed construct such an (orthonormal) basis of functions via the Gram-Schmidt procedure on $\mathcal V$. The precise basis one obtains depends on the root basis vector one starts the orthogonalization procedure with. A separable polynomial basis suitable for various scale invariant bispectra was elaborated upon in \cite{Fergusson:2008ra, Fergusson:2009nv, Liguori:2010hx, Fergusson:2010dm}. For sensitivity to oscillatory signatures in the bispectrum (as is expected in resonance models), Meerburg proposed a separable Fourier mode basis \cite{Meerburg:2010ks}\footnote{Expressions for the first few basis vectors in the polynomial and Fourier mode bases are detailed in the respective \cite{Fergusson:2009nv, Meerburg:2010ks}.}. More recently, Byun et al. proposed a basis of localized piecewise spline functions tailored towards non-separable bispectra \cite{Byun:2015rda}. Given a particular choice, one can then make a \textit{modal expansion} of a given shape function as
\eq{modalexp}{\calS(k_1,k_2,k_3) = \sum_{n}\alpha^R_n R_n(k_1,k_2,k_3)}
expanded in terms of the orthonormal basis vectors
\eq{}{(R_n,R_m) = \delta_{n m}}
where the $a^R_n$ coefficients are obtained as
\eq{}{\alpha^R_n = (\calS,R_n)}
In this way, one can in principle characterize an \textit{arbitrary} bispectrum shape (although the rapidity of convergence of the expansion (\ref{modalexp}) will depend on the compatibility of the basis chosen with the realized shape). To date, searches for features in the bispectrum alone have been inconclusive although recent efforts to cross correlate these with searches for features in the power spectrum have returned results of statistically marginal significance~\cite{Achucarro:2013cva, Achucarro:2014msa, Fergusson:2014hya, Fergusson:2014tza}. In the near future this enterprise will receive an added fillip from the results of a variety of LSS surveys (as discussed earlier) \cite{Huang:2012mr, CyrRacine:2011rx}. However, extracting information about primordial non-Gaussianity from LSS observables such as the galaxy bispectrum is still subject to host of theoretical uncertainties (concerning scale dependent biasing at shorter comoving scales in addition to requiring a better understanding the relationship between the observed matter power spectrum and bispectrum and their primordial counterparts at mildly non-linear scales) \cite{Wagner:2011wx, Agullo:2012cs, Leistedt:2014zqa}, although this promises to be an active field of research in the coming years.

}

\section{Observables: measuring CMB spectral distortions \label{sec:spectraldistortions}}
{
So far, we only discussed CMB observables related to the temperature and polarisation anisotropies. There is, however, another valuable piece of information which may become available in the future: the CMB {\it energy spectrum}. Since the measurements of COBE/{\it FIRAS} \cite{Mather1994, Fixsen1996}, the spectrum of the CMB is known to be extremely close to a perfect blackbody at a temperature of $T_0=(2.726\pm0.001)\,{\rm K}$ today \cite{Fixsen2009}. Even if no deviation from a blackbody spectrum --- commonly referred to as {\it spectral distortion} --- was detected, this remarkable observation places very tight constraints on the thermal history of our universe, ruling out cosmologies with extended periods of significant energy release that could have disturbed the equilibrium between matter and radiation in the post-BBN era. One source of energy release is due to the dissipation of small-scale perturbations in the photon-baryon fluid \citep{Sunyaev1970diss, Daly1991, Hu1994, Chluba2012, Chluba2012inflaton, Pajer2012b}, as we explain here.

Nearly 25 years have passed since the launch of COBE, and from the technological point of view already today it should be possible to improve the absolute spectral sensitivity by more than three orders of magnitude \cite{Kogut2011PIXIE, PRISM2013WPII}. This will open a new window to the early universe, on one hand allowing us to directly probe processes that are present within the standard cosmological paradigm (e.g., distortions from the reionisation and recombination epochs), but on the other hand also opening up a huge {\it discovery space} to unexplored non-standard physics (e.g., decaying/annihilating relic particles, evaporation black holes or cosmic strings) \citep[for overview see][]{Chluba2011therm, Sunyaev2013, Chluba2013fore, Tashiro2014, PRISM2013WPII}. At this stage, CMB spectral distortion measurements are furthermore only possible from space, so that in contrast to $B$-mode polarisation science competition from the ground is largely excluded, making CMB spectral distortions a unique target for future CMB space missions \citep{Chluba2014Science}.
As we highlight in this section, CMB spectral distortions can help us to constrain the {\it shape} and {\it amplitude} of the small-scale power spectrum at wavenumbers ${\rm few}\times\,\Mpc^{-1}\lesssim k \lesssim \pot{2}{4}\,\Mpc^{-1}$, scales that are inaccessible by other means and complement CMB anisotropy measurements at larger scales.

\subsection{Thermalisation physics primer}
The physics going into the cosmological {\it thermalisation} calculation --- the process that restores the pure blackbody spectrum after some departure from thermal equilibrium --- are pretty simple and well understood, allowing us to make precise predictions for different thermal histories and energy release scenarios. For primordial spectral distortions, we are mainly concerned with the average CMB spectrum, so that spatial perturbations can be neglected and the universe can be described as uniformly expanding, thermal plasma consisting of free electrons, hydrogen and helium atoms and their corresponding ions inside a uniform bath of CMB photons. We shall also restrict ourselves to redshifts $z\lesssim \pot{\rm few}{7}$, when electron-positron pairs already completely disappeared, since earlier thermalisation is perfect from any practical point of view and no observable distortion remains. 

Under these circumstances, any energy release inevitably causes a momentary distortion of the CMB spectrum. This can be understood with the following simple arguments: a pure blackbody spectrum, $B_\nu(T)$, is fully characterised by one number, its temperature $T$. Changing the energy density, $\rho_\gamma\propto T^4$, of the photon field by $\Delta \rho_\gamma/\rho_\gamma\ll 1$ (e.g., from some particle decay) means that the photon number also has to be readjusted by $\Delta N_\gamma/N_\gamma\approx (3/4)\Delta \rho_\gamma/\rho_\gamma$ to restore the blackbody relations $\rho_\gamma\propto T^4$ and $N_\gamma\propto T^3$. The photons, furthermore, have to be distributed according to $\simeq \partial B_\nu/\partial T$ in energy, to correctly shift the initial blackbody temperature from $T$ to $T'\approx T+(1/4) \Delta \rho_\gamma/\rho_\gamma$ without creating a distortion. Thus, the correct relation between photon number and energy density as well as the correct change of the distribution function are required to avoid a distortion. If one of these conditions is violated (e.g., if only energy is transferred) then a distortion is inevitable.

In the early universe, the double Compton (DC) and Bremsstrahlung (BR) processes are controlling the number of CMB photons, while Compton scattering (CS) allows photons to diffuse in energy. The interplay of these interactions between matter and radiation determines the precise shape of the CMB spectrum at any stage of its evolution. When studying different energy release mechanisms, one question thus is whether there was enough time between the energy release event and our observation to {\it produce} and {\it redistribute} those distortion photons\footnote{Energy release is the most common mechanism to produce distortions. However, the adiabatic cooling of matter in fact extracts energy from the CMB \citep{Chluba2005, Chluba2011therm}, so that an excess of photons is found in the CMB spectrum. In this case, DC and BR absorb photons.}, thereby fully recovering from the perturbation and completing the thermalisation process. 

\subsubsection{Distortion visibility function}
The thermalisation problem has been studied thoroughly both analytically \cite{Zeldovich1969, Sunyaev1970mu, Illarionov1975, Danese1982, Burigana1995, Chluba2005, Khatri2012b, Khatri2012mix, Chluba2013refmu} and numerically \cite{Burigana1991, Hu1993, Burigana2003, Procopio2009, Chluba2011therm, Chluba2013Green, Chluba2013fore}. From these studies, the following simplified picture can be drawn: at $z\gtrsim \pot{2}{6}$, when the universe is less than a few month old, the thermalisation process is extremely efficient and practically any distortion can be erased until today. At lower redshifts, the CMB spectrum becomes vulnerable to disturbances in the thermal history and only small amounts of energy can be ingested without violating the tight experimental bounds from COBE/{\it FIRAS} \cite{Mather1994, Fixsen1996,Fixsen2002} and other distortion measurements \cite{Kogut2006ARCADE, tris1, arcade2}. 

The transition from efficient to inefficient thermalisation is encoded by the {\it distortion visibility function}, $\Jbb(z, z')$, which  determines by how much the distortion amplitude (regardless of its shape) is suppressed between two redshifts $z$ and $z'<z$. Due to the huge entropy of the universe (there are $\simeq \pot{1.6}{9}$ times more photons than baryons), the DC process \citep{Lightman1981, Thorne1981, Gould1984, Chluba2011d} is the most important source of soft photons at high redshifts, so that the distortion visibility function is roughly given by $\mathcal{J}(z, z')\approx \expf{-(z/\zmudc)^{5/2}}\expf{(z'/\zmudc)^{5/2}}$, with thermalisation redshift $\zmudc\approx\pot{1.98}{6}$ which is determined by the efficiency of DC photon production and Compton redistribution \cite{Danese1982, Burigana1991, Hu1993}. Improved approximations for the visibility function exist \cite{Khatri2012b, Chluba2013refmu}, but for simple estimates the above expression suffices.
For our purpose, we also set $z'=0$ and then use $\mathcal{J}(z)=\mathcal{J}(z, 0)=\expf{-(z/\zmudc)^{5/2}}$. The distortion visibility function thus cuts off exponentially for $z\gtrsim \zmudc$ (see Fig.~\ref{fig:Jfunc}). How far into the {\it cosmic photosphere} \cite{Bond1996} --- the epoch of the universe during which $\Jbb\ll 1$ --- one could view, thus depends on the absolute sensitivity of the experiment and how much initial energy had to be thermalised. 

With a PIXIE-type experiment \cite{Kogut2011PIXIE}, one might be able to detect a distortion created as early as $z\simeq\pot{6}{6}$ if $\Delta \rho_\gamma/\rho_\gamma\simeq 0.01$ of energy were liberated by some process. At much later times ($z\lesssim \pot{\rm few}{5}$), the distortion visibility is very close to unity ($\leftrightarrow$ basically all injected energy will still be visible as a distortion today; see Fig.~\ref{fig:Jfunc}) and the upper limits from COBE/{\it FIRAS} imply $\Delta \rho_\gamma/\rho_\gamma\lesssim \pot{6}{-5}$ \citep{Fixsen1996}. With a PIXIE-type experiment, this could be improved to $\Delta \rho_\gamma/\rho_\gamma\lesssim \pot{8}{-9}$, allowing us to constrain tiny amounts of energy release related to standard processes occurring in our universe.

\subsubsection{Types of primordial distortions}
While the distortion visibility tells us how much of the released energy will still be visible as a spectral distortion today, it does not fix the {\it shape} of the distortion. Here, three regimes are most important: at $z\gg \pot{2}{6}$, thermalisation is extremely efficient (distortion visibility $\mathcal{J}\ll 1$) and CS, DC and BR are able to adjust the initial blackbody spectrum $B_\nu(T)$ to $B_\nu(T+\Delta T)\approx B_\nu(T)+\partial_T B_\nu(T) \Delta T+\mathcal{O}(\Delta T^2/T^2)$ with $\Delta T/T\approx (1/4)\Delta\rho_\gamma/\rho_\gamma$ assuming that a total energy of $\Delta\rho_\gamma/\rho_\gamma\ll 1$ was released. The spectral shape of a {\it temperature shift} [at lowest order in $\Delta T/T$], $G_\nu=(1/4) \,T \partial_T B_\nu(T)=\frac{1}{4} (2h\nu^3/c^2) x\expf{x}/(\expf{x}-1)^2$ with $x=h\nu/kT$ (see Fig.~\ref{fig:Greens}), is {\it not} a distortion but just describes the change of the initial spectrum required to restore full equilibrium. For this both scattering and number changing processes have to be very rapid. Since there is no prediction for the CMB monopole temperature, this part of the CMB spectrum can only be used to constrain changes in the CMB temperature; however, for this another method to determine the CMB temperature higher redshifts (e.g., recombination or BBN) is needed \citep{Simha2008, Jeong2014, Planck2015params}.

In the next regime, valid until $z\simeq \pot{3}{5}$, the efficiency of DC and BR gradually reduces while photons are still efficiently redistributed in energy by the Compton process. In this case, electrons and photons are in kinetic equilibrium with respect to CS, forming a {\it chemical potential} or $\mu$-distortion \cite{Sunyaev1970mu}, but thermalisation stops being complete ($\leftrightarrow$ the distortion visibility function $\mathcal{J}$  approaches unity). Thus, the change of the initial blackbody is given by the superposition of a temperature shift and a pure $\mu$-distortion, $M_\nu\approx 1.401 \,T \partial_T B_\nu(T)[0.4561 - kT/h\nu]$.

\begin{figure}[t] 
   \centering
   \includegraphics[width=0.96\columnwidth]{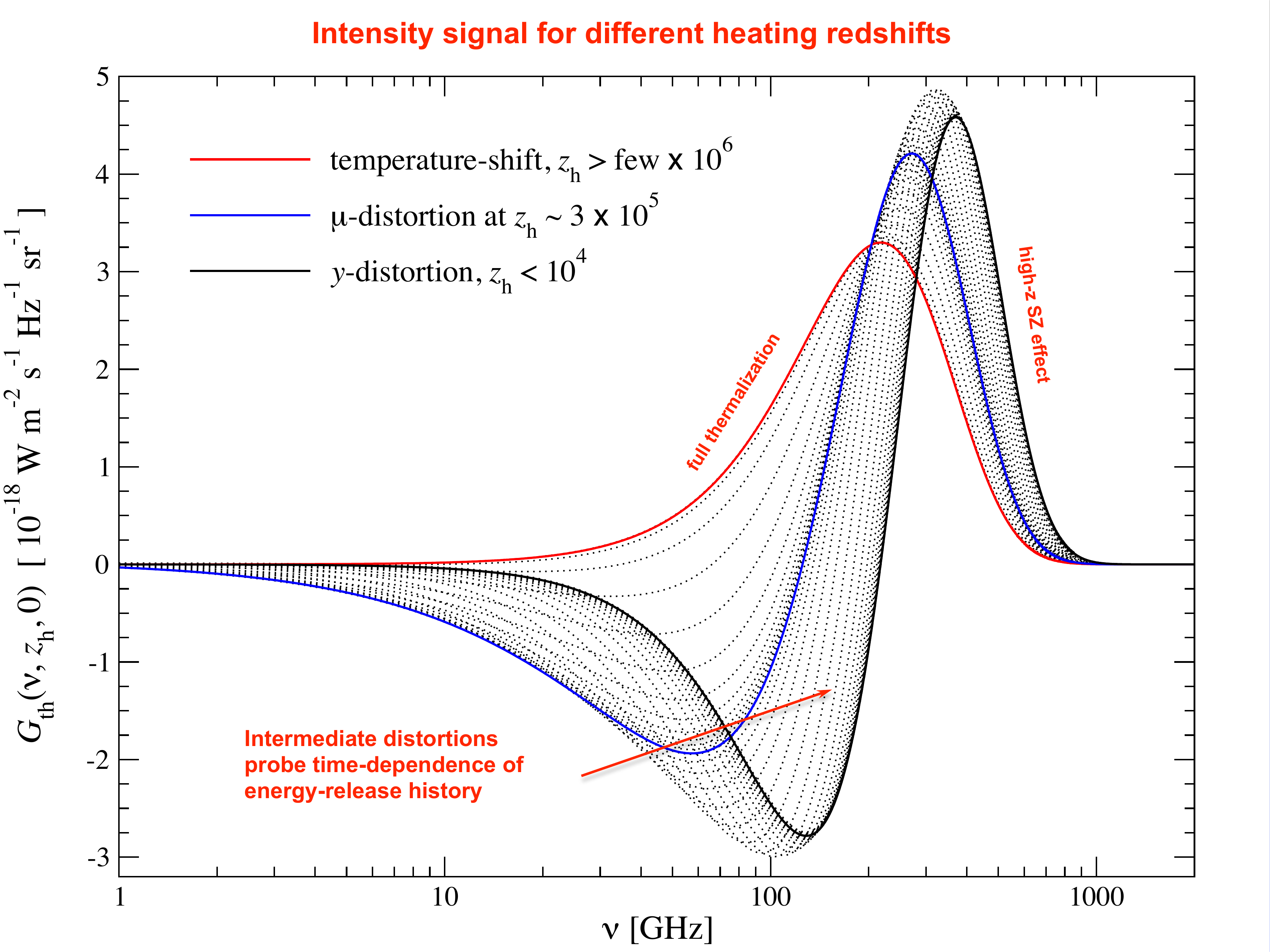}
   \caption{Change in the CMB spectrum after a single energy release at different heating redshifts, $z_{\rm h}$. For $\zh\gtrsim \pot{\rm few}{6}$, a temperature shift, $G_\nu$, is created. Around $\zh \simeq \pot{3}{5}$ a pure $\mu$-distortion, $M_\nu$, appears, while at $\zh\lesssim 10^4$ a pure $y$-distortion, $Y_\nu$, is formed. At all intermediate stages, the signal is given by a superposition of these extreme cases with a small (non-$\mu$/non-$y$ or $r$-type) residual distortion that contains valuable information about the time-dependence of the energy-release process (Figure adapted from \cite{Chluba2013Green}).}
   \label{fig:Greens}
\end{figure}
At $z\lesssim 10^4$, up-scattering of photons by electrons also becomes inefficient and photons diffuse only little in energy. In this era, a Compton-$y$ distortion is formed, also known in connection with the Sunyaev-Zeldovich effect of galaxy clusters \cite{Zeldovich1969}. The shape of a pure $y$-distortion is given by $Y_\nu\approx (1/4) \,T \partial_T B_\nu(T)[x\coth(x/2)-4]$, showing a characteristic decrease of the effective temperature at low frequencies and an increment in the CMB Wien-tail. 
The classical $\mu$- and $y$-distortion, have a slightly different shape (see Fig.~\ref{fig:Greens}). For a $\mu$-distortion, the deviation from the CMB blackbody vanishes at $\nu\simeq 124\,\GHz$, while the cross-over frequency for a $y$-distortion is $\nu\simeq 217\,\GHz$. With future experiments, one can thus hope to distinguish these two types of distortions, and since a $\mu$-distortion can only be formed in the very early stage of the universe, its amplitude directly constrains episodes of early energy release at $z\gtrsim \pot{5}{4}$ or until about 100 years after the Big Bang.

The shape of any primordial distortion caused by energy release is close to a superposition of the extreme cases described above. But the reader probably noticed a gap in our description between $10^4\lesssim z\lesssim \pot{3}{5}$. In this regime, scattering becomes inefficient in redistributing photons over frequency and the distortion morphs between a $\mu$- and $y$-distortion, but the transition is non-linear in the energy exchange and a smaller residual (non-$\mu$/non-$y$ or $r$-type) distortion is formed in addition. The sum of $y$-, $\mu$- and $r$-distortion is sometimes called {\it intermediate} or {\it hybrid} distortion, but additional information is only gained from the {\it residual} distortion.
Although in earlier numerical studies this regime was also mentioned \cite{Burigana1991, Hu1995PhD}, only in the past years it was stressed that the residual distortion contains valuable time-dependent information \cite{Chluba2011therm, Khatri2012mix, Chluba2013Green}, which allows us to distinguish different energy release scenarios \cite{Chluba2013fore, Chluba2013PCA}. This adds another dimension to the CMB, delivering more than just two numbers related to the  $\mu$- and $y$-distortion amplitudes. This is especially important since at late times ($z\lesssim 10-20$), the formation of structures reheats the medium to temperatures $T\simeq 10^4\,{\rm K}-10^5\,{\rm K}$. In this era, a large uniform $y$-distortion is formed with $y\simeq 10^{-7}-10^{-6}$ \cite{Sunyaev1972b, Hu1994pert, Cen1999, Miniati2000, Refregier2000, Oh2003, Zhang2004}, which will swamp any primordial $y$-signal. Without the $r$-distortion we were left only with the amplitude of the $\mu$-distortion and thus could just constrain the overall energy release!

\subsection{Computing and characterising the distortion}
\label{sec:Greens_meth}
It is straightforward to calculate the distortion for any energy release history directly integrating the corresponding Boltzmann equations \citep[e.g.,][]{Illarionov1975, Burigana1991, Hu1993, Procopio2009}. One flexible numerical approach is {\sc CosmoTherm} \cite{Chluba2011therm}, which for the first time allowed direct integration of the thermalisation problem explicitly including the time-dependence for a wide range of energy-release scenario. However, for case-by-case studies and parameter estimation full numerical approaches are too time-consuming. Fortunately, the problem can be simplified: generally we expect the distortion to be very small, so that the Boltzmann equations can be linearised. In this case, a Green's function approach can be used \cite{Chluba2013Green}. This allows us to precisely calculate the distortion for a wide range of energy release scenarios. 

Given an energy release history, $\id (Q/\rho_\gamma)/\id z$, which describes the fractional energy release relative to the photon energy density as a function of redshift, with the thermalisation Green's function, $G_{\rm th}(\nu, z)$, we can thus write \cite{Chluba2013Green}
\begin{align}
\label{eq:Greens}
\boxed{\Delta I_\nu\approx \int G_{\rm th}(\nu, z') \frac{\id (Q/\rho_\gamma)}{\id z'} \id z'}.
\end{align}
Here, $G_{\rm th}(\nu, z)$, encodes the relevant thermalisation physics which is independent of the energy release scenario (see Fig.~\ref{fig:Greens}), and $\Delta I_\nu$ is the final change of the initial blackbody after the energy release.
With Eq.~\eqref{eq:Greens}, one computation of $\Delta I_\nu$ only takes a tiny fraction of a second. This allows us to perform parameter estimations for different energy release scenarios, as first shown in \cite{Chluba2013fore}.
The Green's function approach is part of {\sc CosmoTherm} and available at \url{www.Chluba.de/CosmoTherm}. Alternative schemes were discussed in \citep{Khatri2012mix}, highlighting analytic approximations.

\subsubsection{Simple analytic estimates}
Ignoring the (small) correction from the $r$-distortion in the $\mu/y$-transition regime ($10^4\lesssim z\lesssim \pot{3}{5}$), we can estimate the shape of the intermediate distortion, describing it just as a superposition of $\mu$ and $y$ terms. In simple words, knowing the fractional energy that is released during the $y$-era and $\mu$-era we obtain an approximation for the distortion shape. 
For the classical approximation, one simply assumes that the transition between $\mu$ and $y$ is abrupt around $z\simeq z_{\mu y}\simeq \pot{5}{4}$ \cite{Hu1993}. Since at $z\lesssim z_{\mu y}$, the distortion visibility function is extremely close to unity ($\leftrightarrow$ all the released energy ends up as a distortion) and $G_{\rm th}(\nu, z)\approx Y_\nu$, we have 
\begin{align}
\label{eq:Y_def}
\Delta I_\nu\approx Y_\nu \int_0^{z_{\mu y}}\frac{\id (Q/\rho_\gamma)}{\id z'} \id z'=Y_\nu \left.\frac{\Delta \rho_\gamma}{\rho_\gamma}\right|_y.
\end{align}
Using instead the classical definition for the $y$-distortion, $Y_{\rm SZ}=4Y_\nu$ \cite{Zeldovich1969}, then implies $\Delta I_\nu\approx y \,Y_{\rm SZ}$ with Compton-$y$ parameter $y\approx (1/4) \Delta \rho_\gamma/\rho_\gamma|_y$. 

At $z\gtrsim z_{\mu y}$, the distortion visibility drops, such that the effective energy release in the $\mu$-era is $\Delta \rho_\gamma/\rho_\gamma |_\mu\approx \int_{z_{\mu y}}^\infty \mathcal{J}(z') [\id (Q/\rho_\gamma)/\id z'] \id z'$ with $\mathcal{J}(z)\approx \expf{-(z/\zmudc)^{5/2}}$. The $\mu$-distortion contribution can thus be estimated as
\begin{align}
\label{eq:M_def}
\Delta I_\nu\approx\ M_\nu \int_{z_{\mu y}}^\infty \mathcal{J}(z') \frac{\id (Q/\rho_\gamma)}{\id z'} \id z' = M_\nu \left.\frac{\Delta \rho_\gamma}{\rho_\gamma}\right|_\mu.
\end{align}
In earlier derivations, another normalisation for $\mu$ was used, $M_{\rm SZ}=M_\nu/1.401$  \cite{Sunyaev1970mu}, such that $\Delta I_\nu\approx \mu \, M_{\rm SZ}$ with $\mu$-parameter $\mu\approx 1.401\,\Delta \rho_\gamma/\rho_\gamma|_\mu$.

\begin{figure}[t] 
   \centering
   \includegraphics[width=\columnwidth]{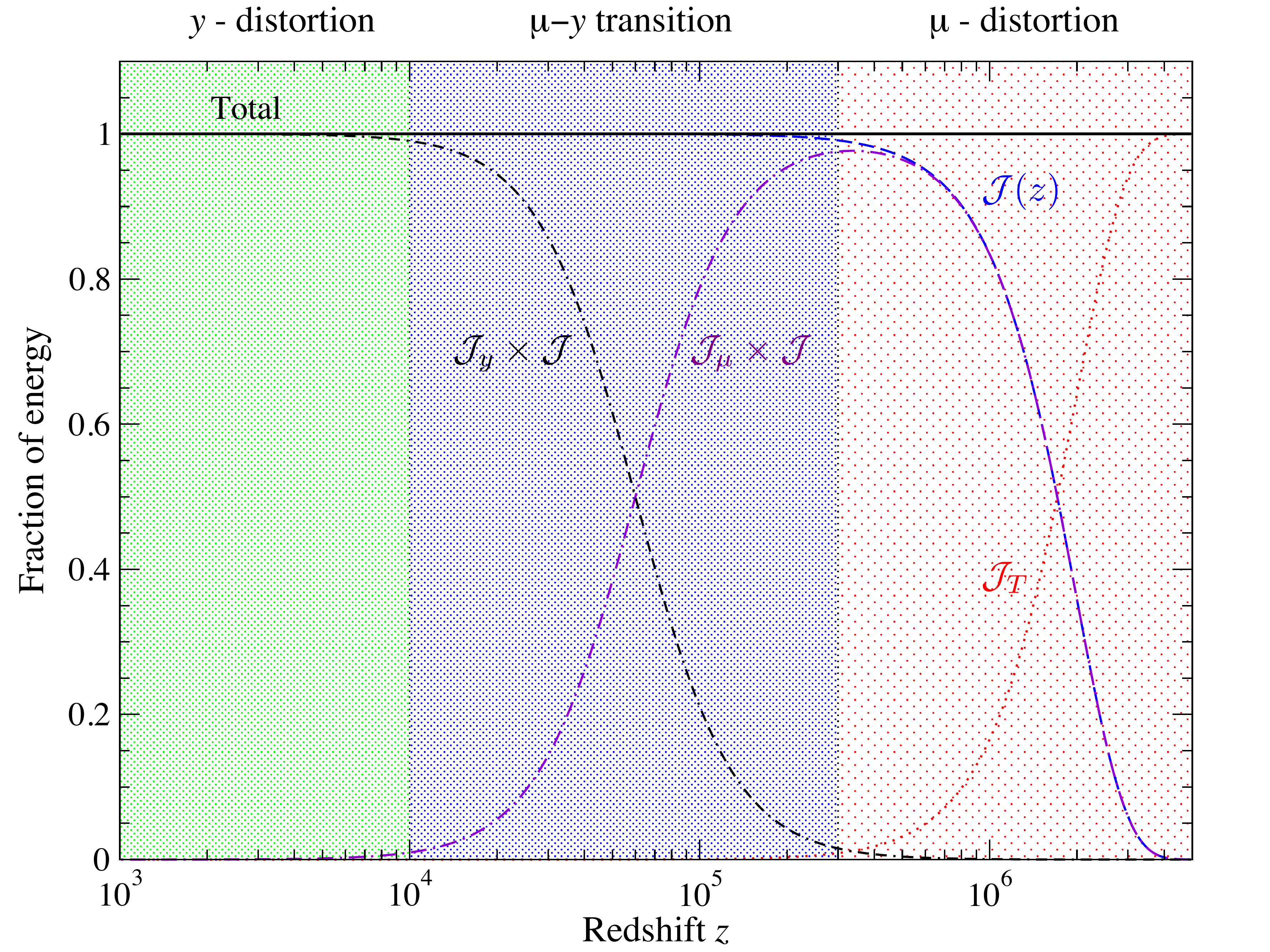}
   \caption{Simple approximations for the distortion visibility function and energy branching ratios in the different distortion eras. Here, $\mathcal{J}_T(z)\approx 1-\mathcal{J}(z)$, denotes the fraction of energy that is fully thermalised and we used $\mathcal{J}_\mu(z)\approx 1-\mathcal{J}_y(z)$, enforcing zero leakage of energy into the $r$-distortion.}
   \label{fig:Jfunc}
\end{figure}

In addition, we can estimate the total temperature shift caused by the energy release process by simply computing the amount of energy that actually thermalised: $\Delta \rho_\gamma/\rho_\gamma |_T\approx \int_{0}^\infty [1-\mathcal{J}(z')] [\id (Q/\rho_\gamma)/\id z'] \id z'$. Putting everything together, we therefore have
\begin{align}
\label{eq:Greens_approx}
\Delta I_\nu\approx Y_\nu\,\left.\frac{\Delta \rho_\gamma}{\rho_\gamma}\right|_y+M_\nu\,\left.\frac{\Delta \rho_\gamma}{\rho_\gamma}\right|_\mu+G_\nu\,\left.\frac{\Delta \rho_\gamma}{\rho_\gamma}\right|_T\equiv y\,Y_{\rm SZ}+\mu\,M_{\rm SZ}+G^\ast_\nu\,\Delta_T.
\end{align}
Here, we introduced $G^\ast_\nu=4 G_\nu$ and $\Delta_T=\Delta T/T\approx (1/4) \Delta \rho_\gamma/\rho_\gamma |_T$. The different definitions for the spectral functions are historical and for the Green's function the normalisation was fixed  energetically, $\int G_\nu \id\nu=\int M_\nu \id\nu=\int Y_\nu \id\nu\equiv \rho_\gamma(T)/4\pi$. 

The approximations for both the distortion visibility function and the hybrid regime between the $\mu$- and $y$-distortion can in principle be improved. One very simple extension is to estimate the branching ratios, $\mathcal{J}_y(z')$ and $\mathcal{J}_\mu(z')$, of energy into $y$ and $\mu$, respectively \citep{Chluba2013Green}. This gives the approximations
\bsub
\label{eq:Greens_approx_improved}
\begin{empheq}[box=\widefbox]{align}
\left.\frac{\Delta \rho_\gamma}{\rho_\gamma}\right|_y \approx \int_0^\infty \mathcal{J}_y(z')\,\mathcal{J}(z')\,\frac{\id (Q/\rho_\gamma)}{\id z'} \id z'
\\
\left.\frac{\Delta \rho_\gamma}{\rho_\gamma}\right|_\mu \approx \int_0^\infty \mathcal{J}_\mu(z')\,\mathcal{J}(z') \frac{\id (Q/\rho_\gamma)}{\id z'} \id z'
\end{empheq}
\esub
for the effective energy release in the $y$- and $\mu$-era, with \citep{Chluba2013Green}
\bsub
\label{eq:branching_approx_improved}
\begin{align}
\mathcal{J}_y(z)&\approx \left(1+\left[\frac{1+z}{\pot{6}{4}}\right]^{2.58}\right)^{-1}
\\
\mathcal{J}_\mu(z)&\approx 1-\exp\left(-\left[\frac{1+z}{\pot{5.8}{4}}\right]^{1.88}\right).
\end{align}
\esub
To ensure full energy conservation (no leakage of energy to the $r$-distortion), instead one can use $\mathcal{J}_\mu(z)\approx 1- \mathcal{J}_y(z)$ [see Fig.~\ref{fig:Jfunc}]. These expressions are accurate at the $10\%-30\%$ level for the concordance model. For higher precision, it is simpler to use the Green's function method, so that we do not go into more detail here.

\subsubsection{Information in the residual ($r$-type) distortion}
The approximation, Eq.~\eqref{eq:Greens_approx}, only captures part of the full information stored by the spectral distortion, since it neglects the non-$\mu$/non-$y$ contributions related to the $r$-distortion. This information can be accessed with future CMB spectrometers and provides valuable insight in the precise time-dependence of the energy release process, which then can used to constrain the scale-dependence of the small-scale power spectrum. But what is the new information that can be distilled from this? 
The total change of the CMB blackbody spectrum caused by a wide range of energy release scenarios can be expressed as
\begin{align}
\label{eq:distortion_full}
\boxed{\Delta I_\nu = y\,Y_{\rm SZ}+\mu\,M_{\rm SZ}+G^\ast_\nu\,\Delta_T + R(\nu)},
\end{align}
where $R(\nu)$ is the residual distortion contribution. Under real world conditions, this part of the distortion signal has no unique definition but depends on the experimental settings such as the frequency range, number of channels, bandpasses, and channel sensitivities \citep{Chluba2013PCA}. These parameters (as well as contaminations by foreground components and systematics) determine the {\it orthogonality} ($\leftrightarrow$ how well different signals can be distinguished) of different distortion shapes and are thus crucial for the definition of $R(\nu)$ with respect to $\mu$, $y$ and $T$.

\begin{figure}[t]
\centering
\includegraphics[width=\columnwidth]{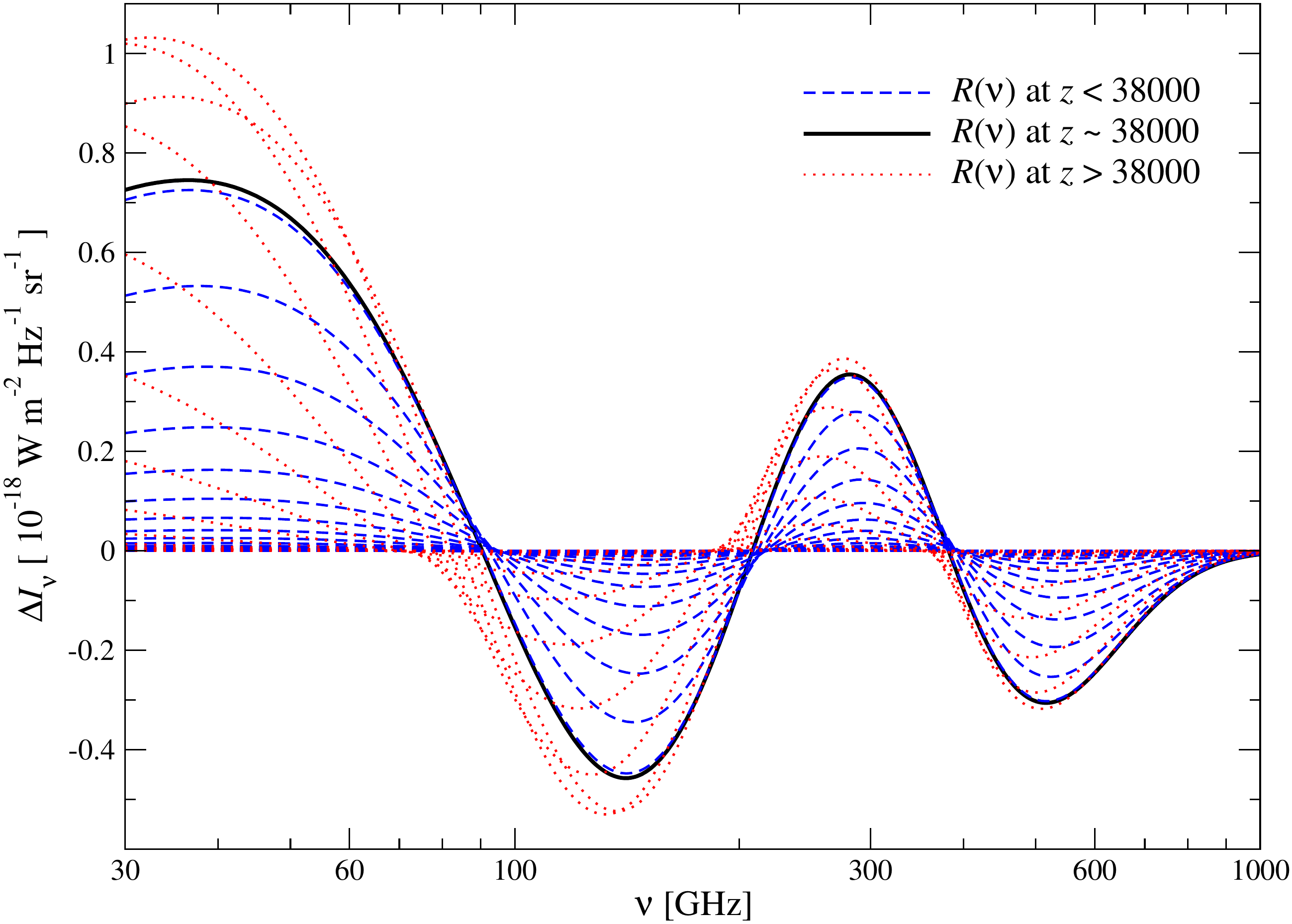}
\caption{Residual distortion for different heating redshifts in the range $10^4\lesssim z \lesssim \pot{3}{5}$. For the construction, we assumed instrumental parameters $\{\nu_{\rm min}, \nu_{\rm max},\Delta\nu\}=\{30, 1000, 1\}\,{\rm GHz}$ and diagonal noise covariance (the figure is taken from \citep{Chluba2013PCA}).}
\label{fig:R_z}
\end{figure}

For single energy release at different redshifts, channels with width $\Delta \nu=1\,\GHz$ in the range $30\,\GHz=\nu_{\rm min}\leq \nu \leq \nu_{\rm max}=1000\,\GHz$ and uncorrelated constant noise, $\Delta I_{\rm c}$, per channel, the $r$-distortion is shown in Fig.~\ref{fig:R_z}. The typical amplitude of the $r$-distortion signal is largest at low frequencies ($\nu\lesssim 60\,\GHz$), reaching $\simeq 10\%-30\%$ of the total distortion signal \citep{Chluba2013Green}. Thus, a lot of the additional information is contained in this frequency domain.

We now have a full description of the spectral signal for single energy release, consisting of specific parameters $p=\{\mu, y, \Delta_T\}$ and $R(\nu, z)$; however, what is the real information content of $R(\nu, z)$? Clearly, for a given experimental sensitivity, the differences between two cases $R(\nu, z_1)$ and $R(\nu, z_2)$ can only be explored if $\Delta R=R(\nu, z_2)-R(\nu, z_1)$ exceed the noise, $\Delta I_{\rm c}$, at a sufficient level. Furthermore, for smooth energy release histories, the $r$-distortion is a superposition of signals from different redshifts. Thus, different energy release scenarios can only be discerned to a limited degree. 

All this can be quantified more precisely using a signal eigenmode analysis [also known as {\it principal component analysis} (PCA)], which decomposes the residual distortion signal into orthogonal/independent spectral components, $S^{(k)}_i=S^{(k)}(\nu_i)$, in the different frequency channels, $\nu_i$, ranked by the degree to which they can be measured \citep{Chluba2013PCA}. The details are not important here, but the PCA delivers a {\it compression} of the information from the multi-frequency spectral data, $\Delta I_i=\Delta I(\nu_i)$, to a number of new parameters, $p=\{\mu, y, \Delta_T, \mu_k\}$, that characterise the distortion signal and its relation to the energy release history. 
The $\mu_k$ are the eigenmode amplitudes of the $r$-distortion signal, which can be expressed as 
\begin{align}
\label{eq:res_dist}
R(\nu_i) = \sum_k S^{(k)}_i \mu_k.
\end{align}
Typically one can expect to be sensitive to only a few of the $\mu_k$, depending on the sensitivity of the experiment and the type of energy release. By construction, the $\mu_k$ do not correlate with each other or any of the other spectral parameters and their error bars, $\Delta \mu_k$, can be estimated from the PCA. This is specific to the chosen experimental parameters, however, the constructed eigenmodes can also be used for other experimental settings although then the modes become correlated \citep{Chluba2013PCA}.

Thinking of the spectral signal, $\Delta I_i$, as a $N$-dimensional vector in frequency space, this implies that 
\begin{align}
\label{eq:mu_est}
\mu_k=\sum_i S^{(k)}_i \Delta I_i  \,\Big/ \sum_i (S^{(k)}_i)^2. 
\end{align}
For a given energy release scenario, the eigenmode amplitudes, $\mu_k$, can thus be computed as simple dot product once the signal eigenmodes are determined. This greatly simplifies the parameter estimation and model comparison process \citep{Chluba2013PCA}. For PIXIE-like settings, the eigensignal vectors, $S^{(k)}_i$, are part of {\sc CosmoTherm}. We will use this method below to forecast the sensitivity of CMB spectral distortion measurements to features in the primordial power spectrum.

\subsubsection{What can we learn from the distortion signal?}
Different energy release scenarios cause different sets of signal parameters, $p_i$. These {\it distortion eigenspectra} can be used for parameter estimation and model selection. To distinguish different models, their energy release histories have to differ at a sufficient level during the $\mu-y$ transition era ($10^4\lesssim z \lesssim \pot{3}{5}$), where the produced $r$-distortion signal is largest. Models with $m$ parameters can only be fully constrained if at least $m$ distortion parameters can be accessed spectroscopically. At least one additional parameter is needed to select between two $m$-parameter energy release scenarios.

In combination with measurements of the light element abundances, it is in principle possible to constrain $\Delta_T=\Delta T/T$ (or equivalently the change of the CMB monopole temperature $T_0$) and thus the total amount of entropy production between BBN and today. For PIXIE-type settings, one can expect to reduce the uncertainty of $T_0$ from $\simeq 1\,$mK \citep{Fixsen2009} to $\simeq 13\,$nK \citep[95\% c.l.,][]{Chluba2013PCA}. However, the constraints on post-BBN entropy production are currently limited to very large energy release at $z\gtrsim \pot{2}{6}$ with $\Delta \rho_\gamma/\rho_\gamma \simeq 0.01$ \citep[e.g.,][]{Simha2008, Jeong2014}, so that generally one cannot learn as much from more precise measurements of $T_0$. 

As mentioned above, the value of the $y$-parameter will be dominated by the late-time distortion signal created by reionisation and structure formation. Although for a PIXIE-like experiment the uncertainty for the average $y$-parameter may decrease from $\simeq\pot{1.5}{-5}$ \citep{Fixsen1996} to $\simeq \pot{4}{-9}$ \citep[95\% c.l.,][]{Kogut2011PIXIE}, the primordial contributions will be hard to separate in a model-independent way. Uncompensated atomic transitions in the recombination era may help us to refine the picture a little \citep{Chluba2008c, Sunyaev2009}, possibly allowing us to distinguish pre- from post-recombination $y$-distortions, but a more detailed discussion is beyond the scope of this review. For our purpose, the value of $y$ will be interpreted as a foreground parameter and shall not be used to distinguish various energy release scenarios.

We are therefore left with $p'=\{\mu, \mu_k\}$ to learn about different energy release scenarios. 
For a PIXIE-like experiment the uncertainty of $\mu$ could decrease from $\simeq\pot{9}{-5}$ \citep{Fixsen1996} for COBE/FIRAS to $\simeq \pot{3}{-8}$ \citep[95\% c.l.,][]{Kogut2011PIXIE, Chluba2013PCA}. The possible detection limits for $\mu_1$, $\mu_2$ and $\mu_3$ are $\simeq \pot{3}{-7}$, $\simeq \pot{2}{-6}$ and $\simeq \pot{7}{-6}$, respectively \citep{Chluba2013PCA}.
The value of $\mu$ can be used as a proxy for the total energy release, while the $\mu_k$ inform us about its time-dependence. It was, for instance, shown that decaying particle scenarios \citep[e.g.,][for early discussions of distortion constraints]{Sarkar1984, Kawasaki1986, Ellis1992, Hu1993b} give rise to a different distortion eigenspectrum than annihilating particles, so that these cases may be distinguished if a sufficient sensitivity is reached \citep{Chluba2013fore, Chluba2013PCA}. Similarly, one may be able to constrain the location and amplitude of features in the small-scale power spectrum, as we discuss now.

\subsection{Probing the small-scale power spectrum with distortions}
Spectral distortions could allow us to constrain the shape and amplitude of the small-scale power spectrum. The physics causing the distortion signal is very simple and just based on the fact that the mixing of blackbodies with different temperatures does {\it not} produce a pure blackbody spectrum. In our universe, blackbodies at different temperatures are set up on various scales by inflation and the photon mixing process is simply mediated by Thomson scattering (no energy exchange required!) at scales shorter than the photon diffusion length well within the horizon. 
In this section, we highlight how from the small-scale power spectrum of curvature perturbations, $P_\calR(k)=2\pi^2 k^{-3}\,\calP_\calR(k)$, we can obtain the spectral distortion and corresponding constraints.

\subsubsection{Damping of small-scale acoustic modes}
Measurements of the CMB anisotropies cover a wide range of physical scales, starting from our horizon scale (wavenumber $k\simeq \pot{2}{-4}\,\Mpc^{-1}$) all the way to $k\simeq 0.2\,\Mpc^{-1}$ deep inside the CMB damping tail (see Fig.~\ref{fig:plonk_cls}). 
Parts of the CMB sky that are separated by more than $\simeq 1^\circ$ (or multipole $\ell \simeq 200$) correspond to independent regions of our universe, which at the time of recombination have basically not been in causal contact since the end of inflation. This characteristic angular scale corresponds to the size of the sound horizon\footnote{The comoving distance  sound waves in the photon-baryon fluid traveled since inflation.} at recombination, $\rs\simeq 150\,\Mpc$, defining the location of the first peak in the CMB temperature power spectrum.  At much larger angular separation, the observed temperature anisotropies still represent the quasi-scale independent initial perturbations with small corrections due to the ISW effect. This is reflected in the nearly flat and featureless shape of the CMB power spectrum at $\ell < 50$ (see Fig.~\ref{fig:plonk_cls}). In contrast, at smaller scales, the fluctuation power exhibit oscillations ($\leftrightarrow$ acoustic peaks) which decrease in amplitude due to Silk damping caused by photon diffusion  \citep{Silk1968}. 

The CMB damping tail is now measured to high precision with ACT \citep{act, Calabrese2013}, SPT \citep{spt} and Planck \citep{Planck2013params, Planck2015params}. If we were able to {\it undo} the effect of Silk damping, we would find much larger fluctuations at small scales \citep[e.g., see Fig. 7 of][]{Hu1997}. The power stored by the initial perturbations was erased by the photon diffusion process and is equivalent to energy release \citep{Sunyaev1970diss, Daly1991, Hu1994, Chluba2011therm, Khatri2011BE}. Initially, this causes a $y$-type distortion which subsequently thermalises \citep{Chluba2012, Khatri2012short2x2, Pajer2012b}. 
Physically, photons are simply mixed together by the isotropising effect of Thomson scattering (and free streaming after recombination), so that no energy exchange between matter and radiation is actually required to produce the distortion. Energy is just directly transferred from the fluctuating part of the photon field (acoustic waves) to the uniform part, when comparing to an initial blackbody at the average CMB temperature.

\begin{figure}[t] 
   \centering
   \includegraphics[width=0.99\columnwidth]{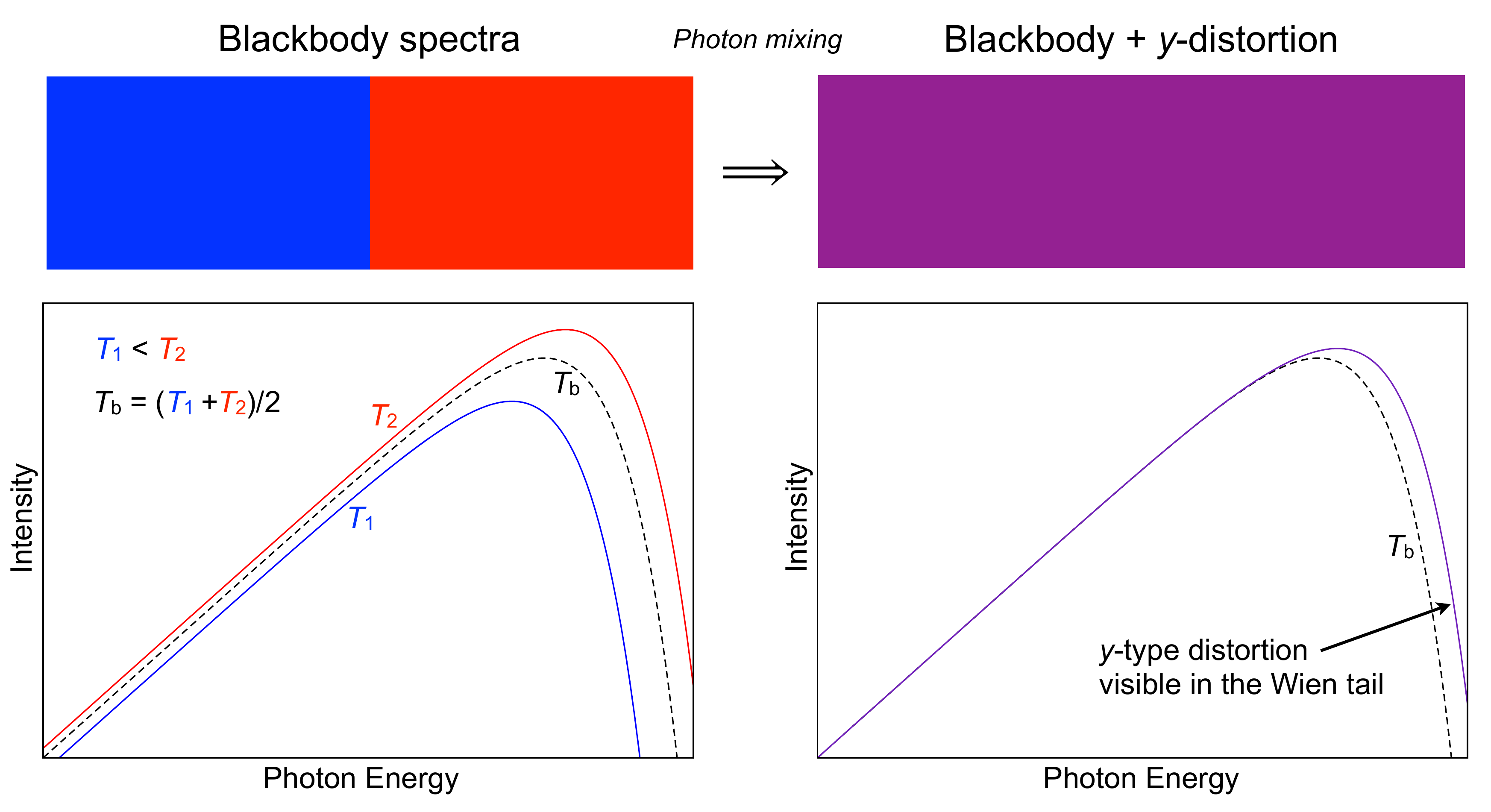}
   \caption{Illustration for the superposition of blackbodies. We envision blackbody photons inside a box at two temperatures $T_1$ and $T_2$, and mean temperature $T_{\rm b}=\frac{1}{2}(T_1+T_2)$ initially (left panel). Thomson scattering mixes the two photon distributions without changing the photon number or energy. The averaged distribution is not a pure blackbody but at second order in the temperature difference exhibits a $y$-type distortion in the Wien tail (right panel).}
   \label{fig:Superposition}
\end{figure}

Without any more detailed derivations, it is thus already clear that larger small-scale power gives rise to larger distortions \citep{Sunyaev1970diss, Daly1991, Hu1994, Chluba2011therm}. Furthermore, the shape of the small-scale power spectrum will determine the effective heating rate at different times and thus can be probed using spectral distortions \citep{Chluba2012, Chluba2012inflaton, Powell2012, Khatri2013forecast, Chluba2013fore, Chluba2013PCA, Clesse2014}. The distortion also depends on the type of perturbations (adiabatic $\leftrightarrow$ isocurvature) \citep{Daly1991, Hu1994isocurv, Dent2012, Chluba2013iso}, although here we shall focus on adiabatic modes.
If more small-scale perturbations are present in different directions, an anisotropic distortion could be created, allowing us to probe scale-dependent non-Gaussianity in the ultra-squeezed limit \citep{Pajer2012, Ganc2012, Biagetti2013, Razieh2015} or possibly cosmic bubble collisions \citep{Chluba2014Science}. Finally, also tensor and vector modes lead to dissipation; however, unless a very blue power spectrum is realised, the associated distortion signals should be sub-dominant \citep{Amin2014, Ota2014, Chluba2015}.

\subsubsection{Superposition of blackbodies and creation of distortions}
\label{sec:sup_disc}
To understand the physics a little better, in Fig.~\ref{fig:Superposition} we illustrate the superposition of blackbodies, $B_\nu(T)$, with two different temperatures, $T_1$ and $T_2$ (equal weight), simply thinking about photons trapped inside a box. Initially, the photon distribution at any location is a blackbody at a specified temperature. Thomson scattering converts the photon field to one uniform distribution, which afterwards is described by a blackbody at temperature $T_{\rm b}=(T_1+T_2)/2$ plus a $y$-type distortion \citep{Zeldovich1972, Chluba2004, Stebbins2007}
\begin{subequations}
\label{eq:superposition}
\begin{align}
\label{eq:superposition_a}
\left<I_\nu\right>=\frac{B_\nu(T_1)+B_\nu(T_2)}{2}
&\approx B_\nu(T_{\rm b})+2 G^\ast_\nu(T_{\rm b}) \, y_{\rm sup}+Y_{\rm SZ}(T_{\rm b}, \nu) \, y_{\rm sup}
\\[-0.3mm]
\label{eq:superposition_b}
&\approx B_\nu(T_{\rm b})+Y^\ast_{\rm SZ}(T_{\rm b}, \nu) \, y_{\rm sup}
\\[1mm]
\label{eq:superposition_c}
&\approx B_\nu(T_{\rm b}[1+2 y_{\rm sup}])+Y_{\rm SZ}(T_{\rm b}, \nu) \, y_{\rm sup}.
\end{align}
\end{subequations}
Here, we defined the $y$-type distortion function, $Y^\ast_{\rm SZ}(T_{\rm b}, \nu)=Y_{\rm SZ}(T_{\rm b}, \nu)+2\,G^\ast_\nu(T_{\rm b})$, and effective $y$-parameter, $y_{\rm sup}=\frac{1}{8} \,(T_2-T_1)^2/T^2_{\rm b}=\frac{1}{2}\Delta T^2/T^2_{\rm b}$, for the two temperatures $T_1=T_{\rm b}-\Delta T$ and $T_2=T_{\rm b}+\Delta T$. 
The three ways of writing the averaged distribution are equivalent, but conveniently isolate the different limiting behaviours at low and high frequencies and the effect on photon number and energy density.  

We can see from Fig.~\ref{fig:Superposition} that  in the Rayleigh-Jeans part of the spectrum the difference between $\left<I_\nu\right>$ and $B_\nu(T_{\rm b})$ is extremely small, while in the Wien tail a distortion is clearly visible. The function $Y^\ast_{\rm SZ}(T_{\rm b}, \nu)$ in Eq.~\eqref{eq:superposition_b} captures these two extremes, vanishing for $h\nu \ll kT_{\rm b}$, but exhibiting a $y$-type dependence, $Y^\ast_{\rm SZ}(T_{\rm b}, \nu)\approx Y_{\rm SZ}(T_{\rm b}, \nu)\approx G^\ast_\nu(T_{\rm b}) \frac{h\nu}{k T_{\rm b}}$ for $h\nu \gg k T_{\rm b}$. 
Computing the photon number and energy densities from Eq.~\eqref{eq:superposition_a}, we find 
\begin{align}
\label{eq:densities}
\left<N_\gamma\right>\approx N^{\rm bb}_\gamma(T_{\rm b})[1+ 6 y_{\rm sup}]
&=N^{\rm bb}_\gamma(T_{\rm b})[1+ 3\Delta T^2/T_{\rm b}^2],
\nonumber\\[1mm]
\left<\rho_\gamma\right>\approx \rho^{\rm bb}_\gamma(T_{\rm b})[1+12 y_{\rm sup}]
&=\rho^{\rm bb}_\gamma(T_{\rm b})[1+ 6\Delta T^2/T_{\rm b}^2],
\end{align}
respectively. Here, $N^{\rm bb}_\gamma(T) \propto T^3$ and $\rho^{\rm bb}_\gamma(T)\propto T^4$ denote the number and energy density of a blackbody at temperature $T$. We also used the integral relations $4\pi\int G^\ast_\nu(T_{\rm b})\id \nu = 4\pi\int Y_{\rm SZ}(T_{\rm b}, \nu)\id \nu =4\,\rho^{\rm bb}_\gamma(T_{\rm b})$, $4\pi\int G^\ast_\nu(T_{\rm b})\id \nu/h\nu = 3 N^{\rm bb}_\gamma(T_{\rm b})$ and $\int Y_{\rm SZ}(T_{\rm b}, \nu)\id \nu/h\nu =0$.

Equation~\eqref{eq:densities} shows that the number and energy densities of the averaged photon distribution are both a bit larger than expected from the mean temperature, $T_{\rm b}$. With respect to $B_\nu(T_{\rm b})$, an energy density of $\Delta \rho_\gamma/\rho^{\rm bb}_\gamma(T_{\rm b})\simeq 6\Delta T^2/T_{\rm b}^2$ was added, while the number density changed by $\Delta N_\gamma/N_\gamma(T_{\rm b})\approx 3 \Delta T^2/T_{\rm b}^2=(1/2) \Delta \rho_\gamma/\rho^{\rm bb}_\gamma(T_{\rm b})$. This violates the blackbody relations for the number and energy densities [i.e., $\Delta N_\gamma/N_\gamma(T_{\rm b})\approx (3/4) \Delta \rho_\gamma/\rho^{\rm bb}_\gamma(T_{\rm b})$], indicating that the isotropised spectrum is indeed distorted. The relative energy stored by the initial $y$-distortion is $\Delta \rho^y_\gamma/\rho^{\rm bb}_\gamma(T_{\rm b})\simeq 4y_{\rm sup} \simeq 2 \Delta T^2/T_{\rm b}^2$ [cf. Eq.~\eqref{eq:superposition_a}], which is only $1/3$ of the total energy contributed by the fluctuating part. The other $2/3$ of energy directly raise the temperature of the uniform blackbody part, i.e., $B_\nu(T_{\rm b})\rightarrow B_\nu(T_{\rm b}[1+2 y_{\rm sup}])$ in Eq.~\eqref{eq:superposition_c}, and thus do not source a distortion \citep{Chluba2012}.

Once the photon field isotropised by Thomson scattering, the usual thermalisation process starts, attempting to redistribute and replenish photons, converting $y\rightarrow\mu\rightarrow \Delta T$. After thermalisation completes, in total $\Delta N_\gamma/N_\gamma(T_{\rm b})\approx (1/4) \Delta \rho_\gamma/\rho^{\rm bb}_\gamma(T_{\rm b}) \approx (3/2)\Delta T^2/T_{\rm b}^2$ of photons were produced by DC and BR emission. This is exactly the right number of photons required to fully thermalise the $y$-distortion part of the uniform photon distribution and to obtain a final blackbody at temperature $T'=T_{\rm b}[1+(3/2) \Delta T^2/T_{\rm b}^2]$ following from the total energy conversion, $\Delta \rho_\gamma/\rho^{\rm bb}_\gamma(T_{\rm b})\simeq 6\Delta T^2/T_{\rm b}^2$.

\subsubsection{Computing the effective heating rate}
In our universe, a spectral distortion is created by the superposition of blackbodies of different temperature due to Thomson scattering. One additional aspect we need to consider more carefully is the scale at which dissipation occurs. 
At recombination, photons can freely travel over distances shorter than the Thomson scattering mean free path, $\lambda_{\rm T}\simeq 1/\sigT \Ne \simeq 0.013\,\Mpc$ (at $z\simeq 1100$), which is $\simeq 10^4$ smaller than the sound horizon at that time. At scales $\lambda_{\rm T}\lesssim \lambda\lesssim \lambda_{\rm H}$ (with Hubble horizon $\lambda_{\rm H}\simeq \sqrt{3}\, \rs\simeq 260\,\Mpc$), photons perform a {\it random walk}, traveling a typical distance $\lambda_{\rm D}\simeq \sqrt{N_{\rm sc}} \lambda_{\rm T}$. The number of scatterings, $N_{\rm sc}$, can be estimated by comparing the comoving distance a photon could have traveled in a Hubble time with the mean free path, such that at recombination the diffusion scale is $\lambda_{\rm D}\simeq \sqrt{\eta \lambda_{\rm T}}\simeq 2\,\Mpc$ \citep{Hu1997}; this is about $75$ times smaller than $\rs$. A more detailed estimate gives $\lambda_{\rm D}\simeq 7\,\Mpc$ or $\kD=\lambda_{\rm D}^{-1}\simeq 0.14\,\Mpc^{-1}$ (see below) for the damping scale at recombination, but the qualitative picture does not change.
At earlier times, the damping scale was even smaller and perturbations at $k\gtrsim \kD$ were efficiently erased sourcing CMB spectral distortions and starting the thermalisation process. 

A detailed treatment of the dissipation problem in second order perturbation theory was carried out by \cite{Chluba2012}. There the physical picture for the creation of the distortion was clarified, demonstrating that the total energy conversion is $9/4$ larger than from the classical estimates \citep{Sunyaev1970diss, Daly1991, Hu1994} and that only $1/3$ of this energy actually sources a distortion, initially appearing as $y$-type distortion. 
The explicit expression for the effective heating rate from scalar perturbations, including second order effects, gauge dependence and all photon multipoles, reads \citep{Chluba2012, Khatri2012short2x2, Chluba2013iso}:
\begin{empheq}[box=\widefbox]{align}
\label{eq:Sac_full}
\frac{\id (Q_{\rm ac}/\rho_\gamma)}{\id z}
\approx&
\frac{4 a \taudot}{H}\int \frac{k^2\id
  k}{2\pi^2}P_\calR(k)
  \left[\frac{\left(3\Theta_1-\varv\right)^2}{3}+\frac{9}{2}\Theta_2^2
  \nonumber\right.
  \\
  &\qquad\qquad\qquad
  \left.
  -\frac{1}{2}\Theta_2\left(\Theta_0^{\rm P}+\Theta_2^{\rm P}\right)+\sum_{\ell\ge 3}(2\ell+1)\Theta_{\ell}^2\right],
\end{empheq}
where $\taudot =\sigT\Ne c \approx \pot{4.4}{-21}(1+z)^{3}\,{\rm sec^{-1}}$ is the rate of Thomson scattering, $a=1/(1+z)$ the scale factor normalised to unity today and $H\approx \pot{2.1}{-20}\,(1+z)^2 {\rm sec^{-1}}$ the Hubble expansion rate\footnote{The approximations for $\taudot$ and $H$ are only valid during the radiation-dominated era.}. Here, $\Theta_\ell$ and $\Theta^{\rm P}_\ell$ denote the photon temperature and polarisation transfer functions and $\varv$ the one for the baryon velocity. Additional corrections from polarisation terms have been neglected, but do not contribute at any significant level during the tight-coupling era of interest to us \citep{Chluba2015}.

It was shown that in the tight-coupling regime ($z\gtrsim 1000$), the temperature quadrupole anisotropy is most important \citep{Chluba2012}. Effects of baryon loading, bulk flows, and photon multipoles with $\ell>2$ can be neglected. This greatly simplifies the expression for the effective heating rate, for which only the transfer functions for $\Theta_2$, $\Theta_0^{\rm P}$ and $\Theta_2^{\rm P}$, determining the {\it shear viscosity} of the photon fluid \citep{Weinberg1971, Kaiser1983}, really matter. These transfer functions can be computed using standard Boltzmann codes, like {\sc Camb} \citep{camb}, and are also included as part of {\sc CosmoTherm}. For a single mode, these oscillate rapidly, so that the heating rate is a function of time \citep[e.g., Fig.~1 of][]{Chluba2012inflaton}, reflecting the fact that dissipation is effective at different phases of the wave propagation. Dissipation is furthermore most effective where the gradients of the temperature field are steepest.

Using the tight coupling approximation \citep{Hu1996anasmall}, one can write\footnote{Note that in our definition $\Theta^{\rm Hu}_\ell =(2\ell+1)\Theta_\ell$.} $\tau'\Theta_2\simeq \frac{8}{15} k \Theta_1$, where $\tau'=\taudot \, a/c$ and $\Theta_0^{\rm P}+\Theta_2^{\rm P}\simeq \frac{3}{2}\,\Theta_2$, so that
\begin{align}
\label{eq:Sac_full_TC}
\frac{\!\id (Q_{\rm ac}/\rho_\gamma)}{\id z}
\approx&
\frac{4 a \taudot}{H}
\!\int \!\frac{k^2\id
  k}{2\pi^2}P_\calR(k) \frac{15}{4}\Theta_2^2
  \approx
\frac{4 c^2}{H a \taudot } \!\int\! \frac{\id
  k}{2\pi^2} k^4 P_\calR(k) \frac{16}{15}\Theta_1^2.
\end{align}
For adiabatic initial conditions, one has $\Theta_1(z, k)\approx A\,(\cs/c) \sin(k\rs)\,\expf{-k^2/\kD^2}$, where $\cs\approx c/\sqrt{3}$ is the sound speed, $\rs\simeq \int \cs\!\id \eta \approx \cs \eta\approx \pot{2.7}{5} (1+z)^{-1}$ the sound horizon and $A \simeq[1+(4/15)R_\nu]^{-1}\simeq 0.9$ with neutrino loading $R_\nu=\rho_\nu/(\rho_\gamma+\rho_\nu)\simeq 0.41$. Neglecting baryon loading, the damping scale, $\kD$, follows from first order perturbation theory and is determined by \citep[e.g., see][]{Hu1996anasmall, DodelsonBook}
\beal
\label{eq:k_diss}
\partial_t \kD^{-2}
\approx\frac{\cs^2}{2 a^2 \taudot}\frac{16}{15}
\approx   \frac{8 c^2}{45 a^2 \taudot}, 
\end{align}
where the factor $16/15$ includes polarisation effects \citep{Kaiser1983}. In the radiation dominated era, one has $\kD\approx \pot{4.0}{-6}(1+z)^{3/2}\,\Mpc^{-1}$. 
Inserting this into Eq.~\eqref{eq:Sac_full_TC}, we find
\begin{align}
\label{eq:Sac_full_TC_final}
\frac{\id (Q_{\rm ac}/\rho_\gamma)}{\id z}
&\approx
\frac{4 A^2}{H a } \, \frac{\cs^2}{\taudot}\,\frac{16}{15}\int \frac{\id
  k}{2\pi^2} k^4 P_\calR(k) \sin^2(k\rs)\,\expf{-2 k^2/\kD^2}
\nonumber
\\[2mm]
&\approx 
- 4A^2\int \frac{k^2 \id
  k}{2\pi^2} P_\calR(k) \sin^2(k\rs)\,\partial_z \expf{-2 k^2/\kD^2}.
\end{align}
From this it is straightforward to show that for a single mode, most of the energy is released at redshift $z_{\rm diss}\simeq \pot{4.5}{5}\left(k/10^3\Mpc^{-1}\right)^{2/3}$ \citep[see,][]{Chluba2012inflaton}. Thus, in the $y$-era ($z\lesssim 10^4$), modes with wavenumber $k\lesssim 3\,\Mpc^{-1}$ dissipate, while in the $\mu$-era ($\pot{3}{5}\lesssim z\lesssim \pot{\rm few}{6}$) we are sensitive to $544\,\Mpc^{-1}\lesssim k \lesssim \pot{\rm few}{4}\,\Mpc^{-1}$. For $3\,\Mpc^{-1}\lesssim k \lesssim 544\,\Mpc^{-1}$, the hybrid ($y+\mu\,+\,r$) distortion is produced, allowing us to probe the shape of the small-scale power spectrum. For modes with $k\simeq 45\,\Mpc^{-1}$ about half of the energy causes $y$ and the other half a $\mu$-distortion.

Assuming smooth $P_\calR(k)$, we can neglect the variation of $\sin^2(k\rs)$ and simply use the time-averaged value $\simeq 1/2$. In this case, the effective heating rate becomes
\bsub
\label{eq:Sac_full_TC_final_II}
\begin{empheq}[box=\widefbox]{align}
\label{eq:Sac_full_TC_final_II_a}
\frac{\id (Q_{\rm ac}/\rho_\gamma)}{\id z}
&\approx 
- 2A^2 \frac{\id}{\id z} \int \frac{k^2 \id k}{2\pi^2} P_\calR(k) \, \expf{-2 k^2/\kD^2}
\\[2mm]
\label{eq:Sac_full_TC_final_II_b}
&\approx
\frac{A^2}{H a} \, \frac{32 c^2}{45 \taudot} \int \frac{\id k}{2\pi^2} k^4 P_\calR(k) \, \expf{-2 k^2/\kD^2}.
\end{empheq}
\esub
The minus sign is due to the fact that for decreasing small-scale power energy is liberated. With this expression, we can approximate the effective heating rate for adiabatic perturbations given the curvature power spectrum $P_\calR(k)$. For isocurvature modes, a mode-dependent heating efficiency and $k$-space weighting, $A^2\rightarrow C^2(k)$, has to be added \citep{Chluba2013iso}, but here we do not consider isocurvature modes in more detail. Using the Green's function method (Sect.~\ref{sec:Greens_meth}), with Eq.~\eqref{eq:Sac_full_TC_final_II} we are thus able to compute the distortion and perform parameter estimation.

For simpler estimates, one can insert Eq.~\eqref{eq:Sac_full_TC_final_II} into Eq.~\eqref{eq:Greens_approx_improved} to compute the effective energy release in the $\mu$- and $y$-eras numerically. Approximating the energy branching ratios, $\mathcal{J}_i(z)$ in Eq.~\eqref{eq:branching_approx_improved}, as step-functions, one can introduce $k$-space window functions to directly approximate the $\mu$ and $y$-parameters. This approach was used by \cite{Chluba2012inflaton} and later refined in \cite{Chluba2013iso}, yielding
\bsub
\label{eq:mu_y}
\begin{empheq}[box=\widefbox]{align}
\mu_{\rm ac}&\approx  \int_{k_{\rm min}}^\infty \frac{k^2 \id k}{2\pi^2} P_\calR(k) \, W_{\mu}(k)
\\[1mm]
y_{\rm ac}&\approx  \int^\infty_{k_{\rm min}} \frac{k^2 \id k}{2\pi^2} P_\calR(k) \, W_{y}(k)
\end{empheq}
\esub
for adiabatic perturbations, where the $k$-space window functions are \cite{Chluba2013iso}
\bsub
\label{eq:window_def}
\beal
W_{\mu}(k)
&\approx  2.8 \,A^2 \left[
\exp\left(-\frac{\left[\frac{\hat{k}}{1360}\right]^2}{1+\left[\frac{\hat{k}}{260}\right]^{0.3}+\frac{\hat{k}}{340}}\right) 
- \exp\left(-\left[\frac{\hat{k}}{32}\right]^2\right)
\right]
\\[2mm]
W_{y}(k)
&\approx  \frac{A^2}{2}\,  \exp\left(-\left[\frac{\hat{k}}{32}\right]^2\right),
\end{align}
\esub
with $\hat{k}=k/[1\,\Mpc^{-1}]$ and cutoff scale, $k_{\rm min}\simeq 1\,\Mpc^{-1}$. The cut-off scale is introduced both because modes at $k<1\, \Mpc^{-1}$ are already tightly constrained by CMB measurements at large scales, implying only very low extra heating to create $y$-distortions, and because the analytic approximations for the photon transfer functions introduced above become inaccurate.
%

\subsubsection{Simple derivation based on energetics}
We can obtain the approximation for the effective heating rate in another way. Equation~\eqref{eq:Sac_full_TC_final_II_a} simply represents the redshift derivative of the total power integral, modulated by the damping function. From the discussion in Sect.~\ref{sec:sup_disc}, we can write the average energy density of the photon field as $\big<\rho_\gamma\big>\approx \rho^{\rm bb}_\gamma(\bar{T})[1+6\big<\Theta^2\big>]$, with the average CMB temperature $\bar{T}=\big<T\big>$, $\Theta=\Delta T(z, \vek{x}, \vek{n}) /\bar{T}$ and where $\big<...\big>$ denotes an universal average over location $\vek{x}$ and directions $\vek{n}$. Then we have the total energy density, $Q/\rho^{\rm bb}_\gamma(\bar{T})\approx 2 \big<\Theta^2\big>$, in the waves that can source distortions, and therefore $\partial_z (Q_{\rm ac}/\rho_\gamma)\simeq - 2 \partial_z \big<\Theta^2\big>$. If we rewrite $\Theta$ as spherical harmonic expansion and average over directions under the assumption of isotropy, we find $\big<\Theta^2\big>\approx \left<\Theta_{0}^2\right>+3\left<\Theta_1^2\right>\approx\left<|\Theta_0|^2\right>$. Here, we used that $\Theta_0$ and $\Theta_1$ are $\pi/2$ out of phase and that $3|\Theta_1|^2\approx |\Theta_0|^2$. In Fourier space, $|\Theta_0|_k\simeq A \expf{-k^2/\kD^2}$, and by comparing with Eq.~\eqref{eq:Sac_full_TC_final_II_a}, we can directly identify $\big<\Theta^2\big>\approx A^2 \int \frac{k^2\!\id k}{2\pi^2} P_\calR(k) \, \expf{-2 k^2/\kD^2}$. These simple arguments do not capture the full time-dependence of the damping process for a single mode (fast oscillations), but give the correct time-averaged heating rate. More carefully, one would find $\partial_z (Q_{\rm ac}/\rho_\gamma)\simeq - 2 \partial_z \big<\Theta^2\big>\equiv - 4 \big<\Theta  \partial_z\Theta \big>$, which using the expressions for $\partial_z\Theta$ to first order in perturbation theory again recovers the full time-dependence \citep[see Sect.~5.1 of,][]{Chluba2012}. Here, $\partial_z\Theta$ only includes scattering terms and the full gauge-independence is restored by also adding second order scattering corrections. For general scalar, vector and tensor perturbations, this was recently shown by \cite{Chluba2015}.

\subsection{Constraints on the small-scale power spectrum}
We now have all the ingredients for computing the spectral signal caused by the dissipation of small-scale perturbations. Before discussing possible constraints on features, we briefly highlight those on the power spectrum, using the standard parametrisation \citep{Kosowsky1995}
\beal
\label{eq:P_k_st}
P_\calR(k)&=2\pi^2 k^{-3} A_\zeta(k/k_0)^{\nS-1+\frac{1}{2} n_{\rm run} \ln(k/k_0)}.
\end{align}
Here, we have the power spectrum amplitude $A_\zeta$, spectral index $\nS$, its running $n_{\rm run}\equiv \id \nS/ \id \ln k$ and pivot scale $k_0$. Features are added as a perturbation around this background power spectrum and the important question is whether with spectral distortions we could tell the difference. As we argue below, CMB spectral distortions allow us to put interesting constraints on features caused by particle production models, while for oscillatory features the average signal remains too close to the signal produced by a featureless background power spectrum. 

\subsubsection{Standard power spectrum}
Constraints on the small-scale power spectrum for power law initial perturbations ($\nrun=0$) using CMB spectral distortions were first discussed by \cite{Daly1991}, \cite{Hu1994} and \cite{Hu1994isocurv}, both for adiabatic and baryon isocurvature modes. Using the early distortion limits from COBE/FIRAS and the amplitude of the CMB anisotropies measured with COBE/DMR at large scales, is was argued that $\nS<1.6$ for adiabatic modes \citep{Hu1994}. At that time, this was one of the tightest constraints available! 

Those early estimates were based on simple approximations for the energy release and it was argued that an improved treatment of the dissipation process was required to forecast experimental possibilities for PIXIE-like experiments \citep{Chluba2011therm}. The discussion was extended to cases with running by \cite{Khatri2011BE}, but without taking the detailed physics of the dissipation process into account developed later \citep{Chluba2012}. It was shown that a PIXIE-like experiment could detect the $\mu$-distortion at about $\simeq 1.5\sigma$, if the power spectrum is extrapolated from large to small scales assuming $\nS=0.96$ and $\nrun=0$ \citep{Chluba2012}. On the other hand, for an experiment similar to PRISM (basically 10 times the spectral sensitivity of PIXIE) the distortion would be detectable at the level of $\simeq 15\sigma$, which would provide an additional confirmation for the simplest inflation model all the way to $k\simeq 10^4\,\Mpc^{-1}$! Simple expressions to estimate constraints on $A_\zeta$, $\nS$ and $\nrun$ from $\mu$ and $y$-distortions are given in \cite{Chluba2013iso} for both adiabatic and isocurvature modes.

Generally, additional information from CMB distortions on the standard power spectrum (extrapolated all the way to small scales) does not improve the constraints on $A_\zeta$ and $\nS$ over the CMB anisotropy measurements \cite[e.g., see Fig.~11 of][]{Chluba2013PCA}, unless much larger sensitivity ($\gtrsim 100\times$PIXIE) is reached \citep{Chluba2013fore}. However, the long lever arm between large-scale CMB and the small scales relevant to spectral distortions can help improving the limits on running over CMB anisotropies alone \citep{Powell2012, Khatri2013forecast, Chluba2013PCA}. For $\nrun\simeq 0$, the constraint can be improved about $\simeq 1.7$ times when adding CMB spectral distortions to future anisotropy measurements expected from an imager similar to PRISM, while for $\nrun\simeq -0.02$ the improvement is $\simeq 20\%$ \citep{Chluba2013PCA, PRISM2013WPII}.

However, from the experimental point of view the small-scale power spectrum is much less constrained and a simple extrapolation from large to small scales involves (more or less well motivated) theoretical prejudice. At $3\,\Mpc^{-1}\lesssim k\lesssim 10^4\,\Mpc^{-1}$, from the experimental point of view \citep[e.g.,][]{BSA11} there are at least two orders of magnitude of wiggle-space relative to large-scale constraints. It is thus interesting to consider distortion bounds on the small-scale power spectrum independently \citep[e.g.,][]{Chluba2012inflaton, Chluba2013fore, Chluba2013PCA}. 
\begin{figure}[t]
\centering
\includegraphics[width=0.87\columnwidth]{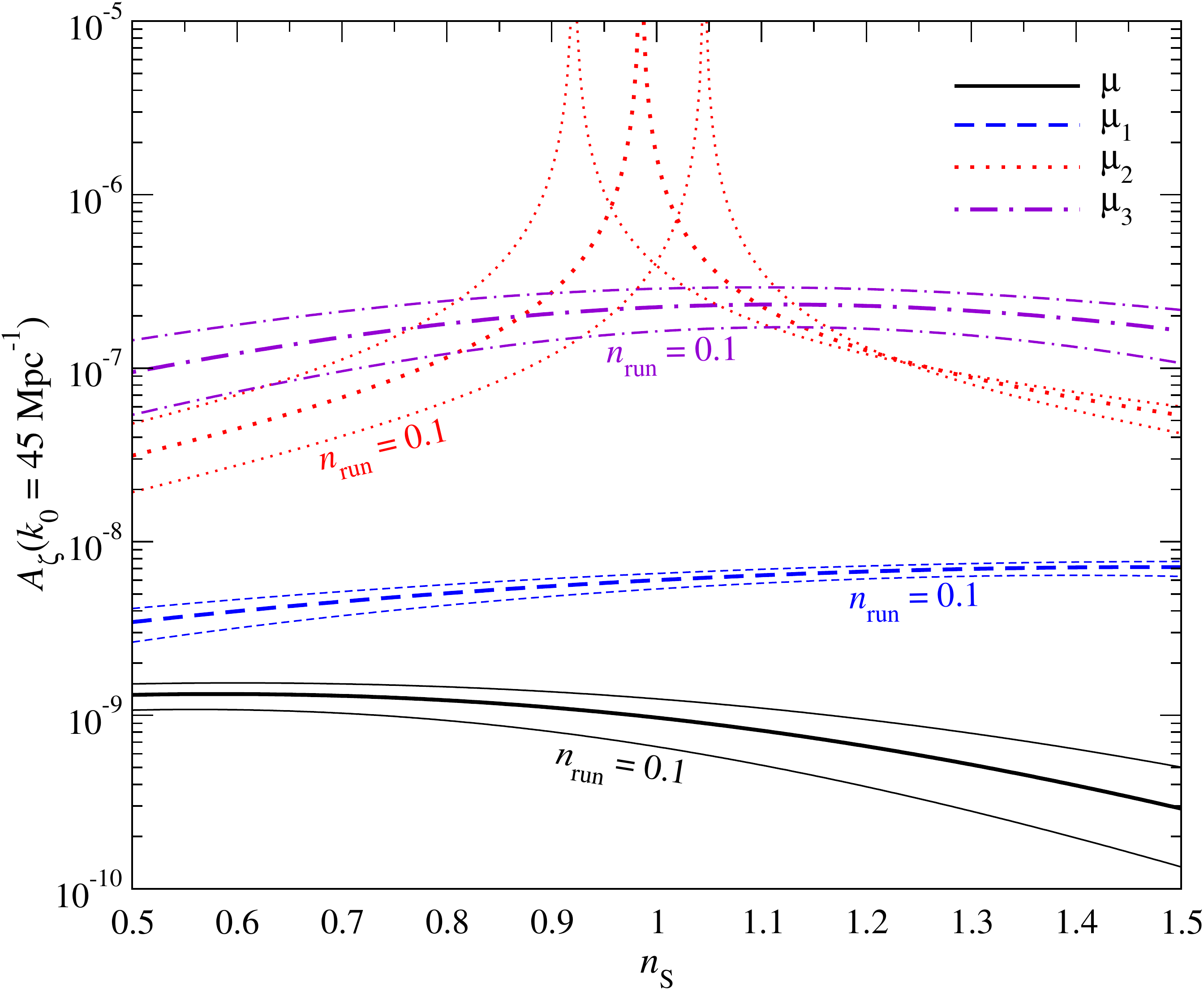}
\caption{$1\sigma$-detection limits for $\mu$, $\mu_1$, $\mu_2$, and $\mu_3$ caused by dissipation of small-scale acoustic modes for PIXIE-like settings. We used the standard parametrisation for the power spectrum with amplitude, $A_\zeta$, spectral index, $\nS$, and running $\nrun$ around pivot scale $k_0=45\,\Mpc^{-1}$. 
The heavy lines are for $\nrun=0$, while all other lines are for $\nrun=\{-0.1, 0.1\}$  in each group. For reference we marked the case $\nrun=0.1$. The figure is taken from \cite{Chluba2013PCA}.}
\label{fig:Limit_nS_nrun}
\end{figure}
Using the standard parametrisation, Eq.~\eqref{eq:P_k_st}, for the power spectrum around a pivot scale $k_0\simeq 45\,\Mpc$, the $1\sigma$ detection limits for $\mu$, $\mu_1$, $\mu_2$ and $\mu_3$ are given in Fig.~\ref{fig:Limit_nS_nrun} assuming PIXIE-like settings, idealised foreground removal and control of systematic effects. 
Around $\nS\simeq 1$, a detection of $\mu$ is possible for $A_\zeta\gtrsim 10^{-9}$, while $A_\zeta\gtrsim \pot{6}{-9}$ is necessary to also have a detection of $\mu_1$. In this case, two of the three model-parameters can in principle be constrained independently. To also access information from $\mu_2$ and $\mu_3$ one furthermore needs $A_\zeta\gtrsim 10^{-7}$. In this case, one could expect to break the degeneracy between all three parameters, $p=\{A_\zeta, \nS, \nrun\}$, with a PIXIE-type spectrometer. This implies that for excess small-scale power, CMB distortion measurements allow placing unique and complementary constraints on different early-universe models. A spectrometer similar to PRISM could further lower these detection limits, possibly by one order of magnitude \citep{Chluba2013PCA}.

\subsubsection{Features in the small-scale power spectrum}
Let us now turn to the question of whether distortions can help us to shed light on features in the small-scale power spectrum. A broad discussion of non-standard small-scale power spectra can be found in \cite{Chluba2012inflaton} and \cite{Clesse2014}. Here, we focus on features caused by {\it particle production} models and {\it changes in the sound speed}. The former scenario primarily leads to localised bumps in the power spectrum \citep{Barnaby:2009dd, Barnaby:2010ke}, whereas the latter primarily results in oscillatory modifications\footnote{Although in the limit of very slowly varying sound speeds, a net injection of power resulting in a bump is also possible (cf. Fig.~\ref{fig:long_cs}).} (Sect.~\ref{sec:theory}). 

For particle production models, \cite{Barnaby:2009dd} provided a simple parametrisation for the resulting bump in the primordial power spectrum which we will utilise in the discussion that follows: 
\beal
\label{eq:P_k_Neil}
\Delta P_{\calR, i}(k)&=2\pi^2 A_{\zeta, i} \left(\frac{\pi \expf{}}{3k_i^2}\right)^{3/2} \!\exp\left(-\frac{\pi}{2}\frac{k^2}{k^2_i}\right).
\end{align}
From Fig. \ref{particle_prod}, we see that this simple parametrisation accurately models the generated features calculated using heat kernel methods in the regime where back reaction on the background fields are negligible (cf. Sect.~\ref{sec:theory} and \ref{1l}). The amplitude of the feature, $A_{\zeta, i}$, is simply related to the value of coupling constant $g_i$ between the two fields;  $A_{\zeta, i}\simeq \pot{1.01}{-6} g^{15/4}_i$.  The derivation for this feature in the primordial power spectrum is only valid for $10^{-7}\lesssim g_i^2\lesssim 1$ \citep{Barnaby:2009mc}, so we are primarily interested in $A_{\zeta, i}$ values between $10^{-19}$ and $10^{-6}$. There are no restrictions on the location of the bump; $k_i$ is determined by the number of $e$-foldings between the moment when the field becomes massless ($\phi\simeq \phi_i$) and the end of inflation.

\begin{figure}[t]
\centering
\includegraphics[width=0.9\columnwidth]{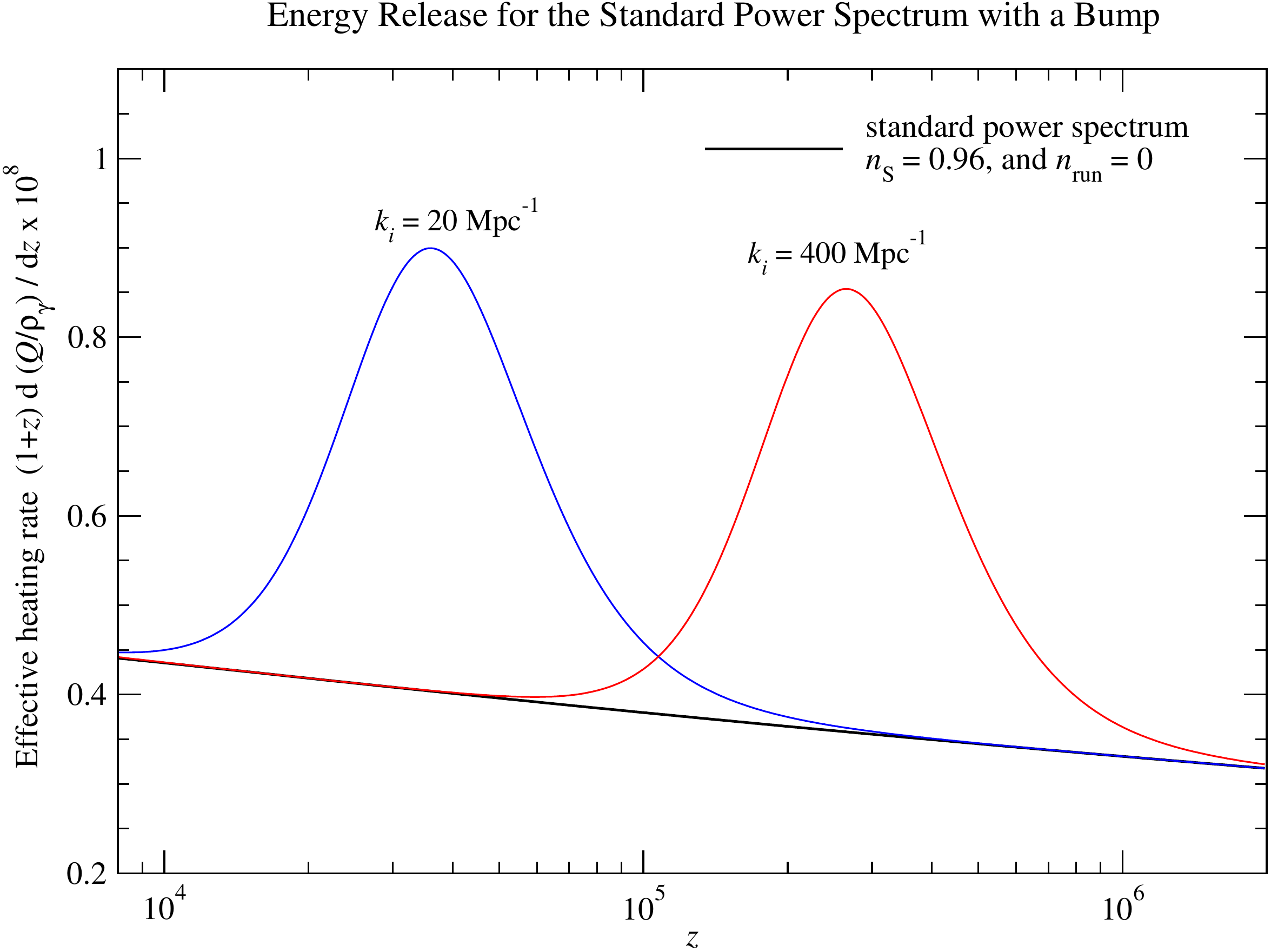}
\caption{Effective heating rate for the standard power spectrum with a feature caused by particle production during inflation at $k_i=\{20,400\}\Mpc^{-1}$.
For illustration we chose $(\nS,\nrun)=(0.96,0)$ for the background spectrum. Furthermore, we set $A_{\zeta, i}=\pot{3.5}{-9}$ in both shown cases. The figure is taken from \cite{Chluba2012inflaton}.
}
\label{fig:heating_rate_particles}
\end{figure}
Inserting Eq.~\eqref{eq:P_k_Neil} into Eq.~\eqref{eq:Sac_full_TC_final_II}, we find the approximation \cite{Chluba2012inflaton}
\beal
\label{eq:Qdot_Neil}
\boxed{\frac{\id (Q_{{\rm p},i}/\rho_\gamma)}{\id z}
\approx 
9.4 a \,A_{\zeta, i} \frac{\expf{3/2}}{\sqrt{6\pi}}\,\frac{(k_i/\kD)^2}{\left[1+\frac{4}{\pi}(k_i/\kD)^2\right]^{5/2}}}
\end{align}
for the effective heating rate.  Examples for two particle production scenarios are shown in Fig.~\ref{fig:heating_rate_particles}. The localised particle production feature in the small-scale power spectrum causes a burst of energy release at $z_{\rm p}\simeq \pot{5.1}{5}\left(k_i/10^3\Mpc^{-1}\right)^{2/3}$. This means that for models with $3\,\Mpc^{-1}\lesssim k_i \lesssim \pot{\rm few}{2}\,\Mpc^{-1}$ one can expect to be able to constrain both the position and the amplitude of the feature using information from the $r$-distortion. COBE/FIRAS constraints are too weak to limit the parameter space, since physical conditions already imply $A_{\zeta, i}\lesssim 10^{-6}$ \citep{Chluba2012inflaton}. However, for PIXIE-like settings one can expect to strongly improve the situation.

In Fig.~\ref{fig:detection_particles}, we show the expected $1\sigma$ detection limits for PIXIE-like settings.
\begin{figure}[t]
\centering
\includegraphics[width=0.9\columnwidth]{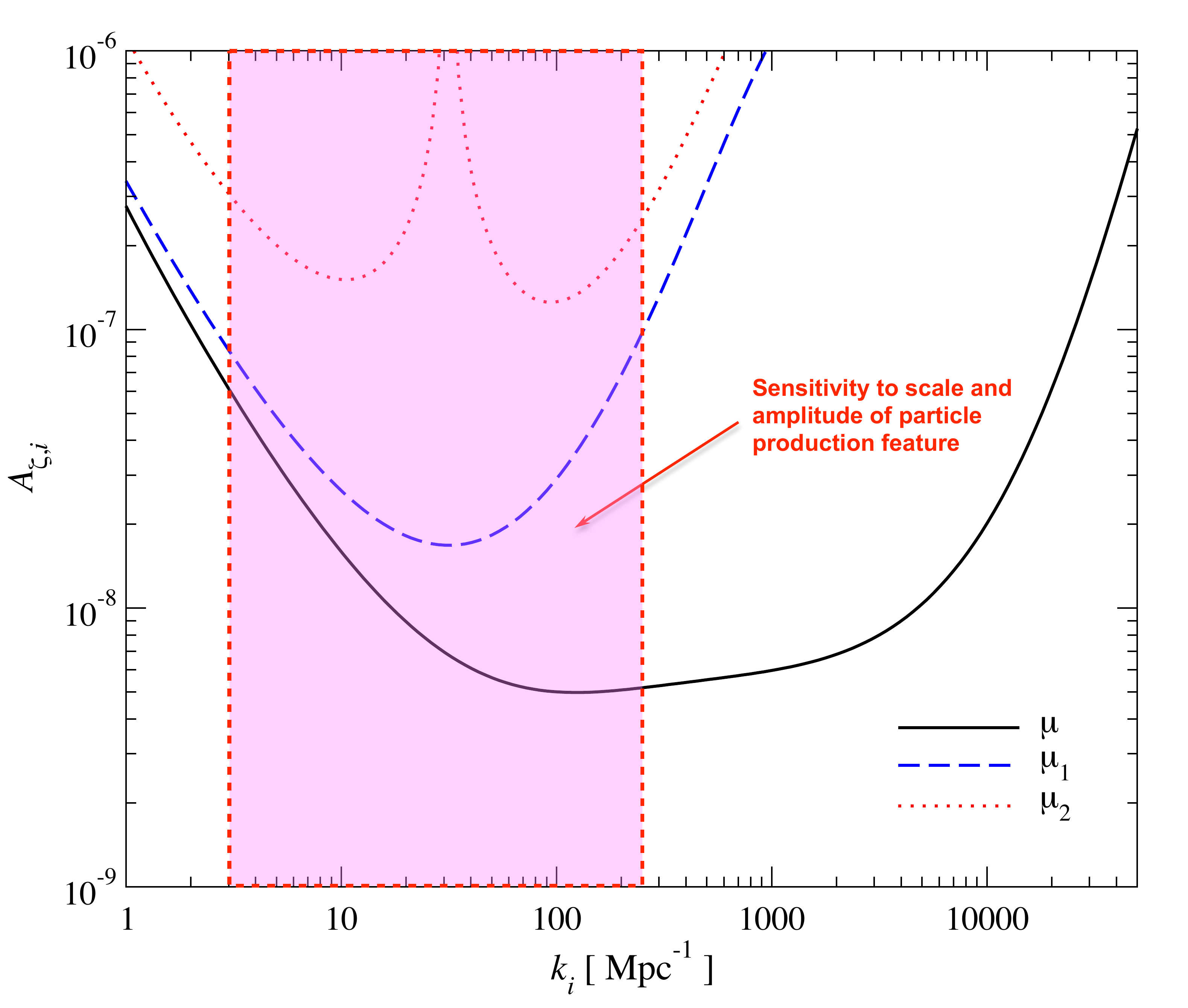}
\caption{$1\sigma$-detection limits for $\mu$, $\mu_1$ and $\mu_2$ on a power spectrum feature caused by particle production models for PIXIE-like settings. The figure was obtained with the Green's function method \citep{Chluba2013Green} of the {\sc CosmoTherm} package \citep{Chluba2011therm}.
}
\label{fig:detection_particles}
\end{figure}
For $80\,\Mpc^{-1}\lesssim k_i \lesssim 3000\,\Mpc^{-1}$, the detection limit from $\mu$ is $A_{\zeta, i}\simeq \pot{6}{-9}$, while in the range $3\,\Mpc^{-1}\lesssim k_i \lesssim 250\,\Mpc^{-1}$ one may be able to determine both, $A_{\zeta, i}$ and $k_i$, if $A_{\zeta, i}\gtrsim 10^{-7}$. At scales $k\gtrsim 10^3\,\Mpc^{-1}$, only $\mu$ may be detected if a particle production feature is present, but the detection limits become weak beyond $k\simeq \pot{\rm few}{4}\,\Mpc^{-1}$. A PRISM-like spectrometer could lower the detection limits by a factor of $\simeq 10$, thereby also widening the range over which both $A_{\zeta, i}$ and $k_i$ could be constrained individually.

Another question is whether a small-scale power spectrum with and without particle production feature can be distinguished. For this we have to consider the shape of the distortion, which is determined by the distortion shape parameters $\rho_i=(\mu_i/\Delta \mu_i)/(\mu/\Delta \mu)$. The ratios $r_i=\mu_i/\mu$ simply eliminate the dependence on the amplitude of the distortion, while weighting by the errors directly gives a sense of the sensitivities \citep{Chluba2013PCA}. Thus, combinations of $\rho_1$ and $\rho_2$ define a unique trajectory in parameter space which can be compared to the one of the standard power spectrum without a feature. 
\begin{figure}[t]
\centering
\includegraphics[width=0.9\columnwidth]{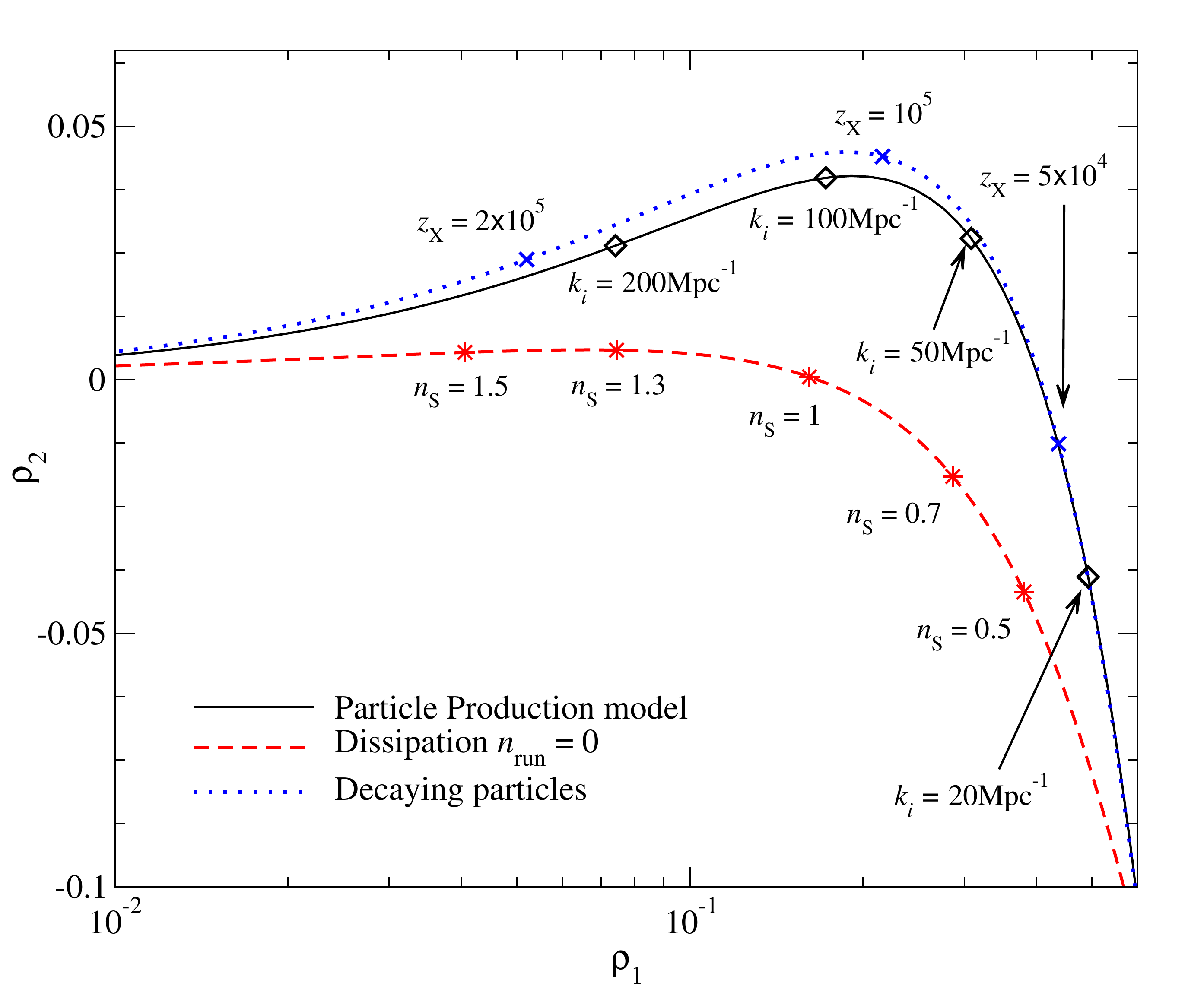}
\caption{Trajectories in the $\rho_1-\rho_2$ plane for different models. The black line shows the trajectory for a feature caused by a particle production scenario for varying $k_i$. For comparison, the red dashed line shows a power law spectrum (no running) for different spectral indices, $\nS$, while the blue dotted line gives the trajectory for decaying particle scenarios and varying particle lifetime $t_X\simeq [\pot{4.9}{9}/(1+z_X)]^2 \sec$. Particle production features produce a distortion that is very similar to the one of decaying particles but distinguishable from a normal featureless power spectrum.}
\label{fig:Model_com}
\end{figure}
This is illustrated in Fig.~\ref{fig:Model_com}, which compares these cases. The eigenspectra of particle production scenarios differ from the simple power law case. Thus, a particle production feature at scales $20\,\Mpc^{-1}\lesssim k_i \lesssim 200\,\Mpc^{-1}$ could be distinguishable for PIXIE-like settings, if the amplitude exceeds $A_{\zeta, i}\simeq 10^{-7}$. 

For comparison, we also show the signal caused by a decaying particle scenario. From Fig.~5 of \cite{Chluba2013fore} we can see that the shape of the energy release history is very similar to the one of a particle production feature, Fig.~\ref{fig:heating_rate_particles}. Consequently, the distortion signal is expected to be quite similar, a fact that is reflected in Fig.~\ref{fig:Model_com}. Thus, distinguishing a decaying particle scenario and a power spectrum feature due to particle production will be challenging. However, ultimately this is a question of sensitivity and how many distortion parameters $\mu_i$ can be measured.

Finally, we briefly consider oscillatory features in the small-scale power spectrum, for example, created by changes in the sound speed. As already mentioned in \cite{Chluba2012inflaton}, the energy release caused by oscillatory power spectrum features usually causes a small average effect. For the net energy release, power enhancements and deficits cancel out. In particular, for features with rapid oscillations only the average energy release history over several periods will cause a change of the distortion signal with respect to the standard featureless power spectrum. 

One rough estimate for the sensitivity to oscillations in the energy release history caused by oscillations in the small-scale power spectrum can be found from the typical width of the energy release eigenmode functions, determined in \cite{Chluba2013PCA}. Even if the first four $r$-distortion eigenmodes could be constrained, features in the energy release history that are narrower than $\Delta z/z\simeq 0.2-0.3$ in redshift, will not be resolved. This corresponds to $\Delta k/k \simeq 0.2-0.3$ around $k\simeq 45\,\Mpc^{-1}$ for oscillatory features in the power spectrum. For features with higher period, individual oscillations will leave an unobservable net distortion signal, unless power is modulated with a longer wavelength. However, given the large parameter space, a more detailed study will be left for the future.

}

\section{Statistical analysis \label{sec:statistics}}
{
Having introduced the most relevant observables, let us now turn to reviewing some basics of the statistical analysis of the data.   How can one actually tell whether a feature is ``required" by the data, and how would one go about looking for them?

The starting point of this discussion is the basic quantity delivered by an experiment measuring an observable~$\mathfrak{O}$ (e.g., the CMB angular power spectrum) -- the {\it likelihood function}~$\mathcal{L}(\mathcal{\mathfrak{O}_\mathrm{obs}}|\mathfrak{O}_\mathrm{th})$, i.e., the conditional probability of $\mathfrak{O}$ taking on the observed value $\mathfrak{O}_\mathrm{obs}$, under the premise that the true value of $\mathfrak{O}$ is given by $\mathfrak{O}_\mathrm{th}$.  In general, $\mathfrak{O}_\mathrm{th}$ is the prediction of a theoretical model~$\mathcal{M}$ for given values of the free parameters $\boldsymbol{\theta}$ of $\mathcal{M}$, and thus we can write the likelihood as a double conditional probability $\mathcal{L}(\mathfrak{O}_\mathrm{obs}|\mathfrak{O}_\mathrm{th}(\boldsymbol{\theta} | \mathcal{M}))$.  It should be noted that $\mathcal{L}$ is a function of the data~$\mathfrak{O}_\mathrm{obs}$, not of the theoretical prediction~$\mathfrak{O}_\mathrm{th}(\boldsymbol{\theta} | \mathcal{M})$, so taken just by itself, it cannot be used to distinguish between different models (or even different parameter values within the same model).

\subsection{Frequentist or Bayesian?}
Given the likelihood function, one now has two approaches to making quantitative statements about potential signatures of features in the data.   Their fundamental difference lies in what quantity is to be treated as a random variable.  One can either take the functional dependence of the likelihood at face value and treat the data as the outcome of a random process (we will refer to this as the Frequentist method), or one can take the actually measured data as a condition and relate the likelihood to a probability distribution where the theoretical prediction is a random variable (Bayesian method). 

Though the issue of which way to follow is sometimes the subject of ideological debates among practitioners in the field, we do not presume to take a side here.  In our opinion, both methods are valid and valuable items in the data analyst's toolbox, provided they are implemented correctly.

\subsubsection{Frequentist approach}
Applied to the problem at hand, the Frequentist approach can be motivated by the following line of thought:  even if the primordial power spectrum realised in Nature was smooth, realistic data will scatter around the theoretical expectation values.  It is therefore always possible to engineer a more complex (features-)model that will fit the data better than a smooth spectrum.  Assuming the true spectrum was smooth, one can now ask by how much a particular features-model would typically improve the fit.  And unless the improvement observed for the real data exceeds the expectation value by a sufficiently large amount, one should be rather careful about claiming a discovery of a feature.

More formally, the frequentist approach requires the formulation of a null hypothesis $\mathfrak{H}_0$ (e.g., ``the true power spectrum does not have any features"), and an alternative hypothesis $\mathfrak{H}_1$ (e.g., ``the true power spectrum has features, described by model $\mathcal{M}_1$"), along with the definition of a suitable test statistic $\mathfrak{S}$ (e.g., the likelihood ratio between the best fit featureless model and the best fit of features-model $\mathcal{M}_1$).  Then, random realisations of the data can be generated under the assumption that the null hypothesis is true.  On each of these random realisations, the test statistic will be evaluated, resulting in an estimate of the probability distribution $P(\mathfrak{S})$.  By comparing of $P(\mathfrak{S})$ with the actually observed value $\mathfrak{S}_\mathrm{obs}$, one can extract the fraction of simulations which yield a more extreme test statistic $\mathfrak{S} > \mathfrak{S}_\mathrm{obs}$ (or $\mathfrak{S} < \mathfrak{S}_\mathrm{obs}$), known as the $p$-value, and gives a qualitative indication whether or not the data prefer a feature.

Unfortunately, $p$-values are often misinterpreted (see for instance Ref.~\cite{goodman2008dirty} for commonly encountered misconceptions). Let us in particular emphasize that the $p$-value is {\it neither} the probability of the null hypothesis being wrong, {\it nor} is $1-p$ the probability of the alternative hypothesis being true.  Also, the widely-spread practice of calling $p < 0.05$ ``significant" evidence against the null hypothesis is rather dubious~\cite{Hurlbert2009}.  
In addition, great care needs to be taken when selecting the data and defining the test statistic: both choices can bias the $p$-value and ideally ought to be decided upon before having seen the actual data.  Focussing on a particularly anomalous subset of the data after having inspected them (``a posteriori reasoning") or considering several independent alternative hypotheses (``look-elsewhere effect"), for instance, can easily lead to artificially low $p$-values and premature claims of a feature detection.  

Issues like these, and the fact that one has to perform the already numerically demanding fitting/likelihood maximisation procedure not only once for the real data, but also for a large number of simulated data sets, have given the alternative Bayesian ansatz a slight edge in popularity in features searches.

\subsubsection{Bayesian approach}
The idea behind the Bayesian approach is to promote the likelihood function to a probability density over the parameter space of a model, or, going even further, over a space of different models (see~\cite{Trotta:2008qt} for a detailed discussion of Bayesian statistics in the context of cosmology).  This goal can be achieved with the help of the following relation for conditional probabilities, known as Bayes' theorem:
\begin{equation}
\boxed{P(B|A) = \frac{P(A|B) \cdot P(B)}{P(A)}.}
\end{equation}

In a first step, one can apply Bayes' theorem to the likelihood function and obtain the posterior probability density $\mathcal{P}$,
\begin{equation}\label{eq:bayes1}
\boxed{\mathcal{P}(\mathfrak{O}_\mathrm{th}(\boldsymbol{\theta} | \mathcal{M}) | \mathfrak{O}_\mathrm{obs}) \mathrm{d}\boldsymbol{\theta} = 
\frac{p(\boldsymbol{\theta} | \mathcal{M}) \mathrm{d}\boldsymbol{\theta} \cdot \mathcal{L}(\mathfrak{O}_\mathrm{obs}|\mathfrak{O}_\mathrm{th}(\boldsymbol{\theta} | \mathcal{M}))}{\mathcal{E}(\mathfrak{O}_\mathrm{obs} | \mathcal{M})}.}
\end{equation}
Here, the {\it prior probability density} $p(\boldsymbol{\theta} | \mathcal{M})$ represents our knowledge about the parameters independently of the data.\footnote{The prior probability density should encode, e.g., theoretical limitations on parameters, or knowledge about the parameters from previous independent measurements.  In most cases, however, there is no unique canonical prior on $\boldsymbol{\theta}$, and its choice will involve a fair amount of subjectivity.}  The {\it Bayesian evidence} $\mathcal{E}$ is basically the normalisation constant of $\mathcal{P}$ and given by an integral of the prior weighted by the likelihood over the full parameter space of $\mathcal{M}$:
\begin{equation}\label{eq:bayesianevidence}
\mathcal{E}(\mathfrak{O}_\mathrm{obs} | \mathcal{M}) = \int \mathrm{d}\boldsymbol{\theta} \; p(\boldsymbol{\theta} | \mathcal{M}) \mathcal{L}(\mathfrak{O}_\mathrm{obs}|\mathfrak{O}_\mathrm{th}(\boldsymbol{\theta} | \mathcal{M})).
\end{equation}
Invoking Bayes' theorem a second time and applying it to the evidence yields the probability of $\mathcal{M}$ given the data:
\begin{equation}\label{eq:bayes2}
\boxed{P(\mathcal{M} | \mathfrak{O}_\mathrm{obs}) = \frac{\tilde{p}(\mathcal{M}) \cdot \mathcal{E}(\mathfrak{O}_\mathrm{obs} | \mathcal{M})}{P(\mathfrak{O}_\mathrm{obs})}.}
\end{equation}
In order to compare two different models $\mathcal{M}_0$ and $\mathcal{M}_1$, it is useful to look at their model probability ratio, also known as the {\it Bayes factor},
\begin{equation}\label{eq:bayesfactor}
B_{01} \equiv \frac{P(\mathcal{M}_0 | \mathfrak{O}_\mathrm{obs})}{P(\mathcal{M}_1 | \mathfrak{O}_\mathrm{obs})}  = \frac{\mathcal{E}(\mathfrak{O}_\mathrm{obs} | \mathcal{M}_0)}{\mathcal{E}(\mathfrak{O}_\mathrm{obs} | \mathcal{M}_1)},
\end{equation}
where the second equality assumes that the two models have been assigned equal model prior probabilities.  The Bayes factor has a very straightforward interpretation: ``given the observed data and priors, $\mathcal{M}_0$ is $B_{01}$ times more probable than $\mathcal{M}_1$''.  Unlike a likelihood-ratio-based frequentist analysis, it also has the desirable property of naturally implementing Occam's razor principle, due to the averaging process in the evaluation of the evidence.  This rewards predictive models that yield a good fit over large parts of their parameter space and punishes models that are overly complex and ``can fit anything'' and thus provide a good fit over only a small fraction of their parameter space.

Clearly, any physical interpretation of Bayes factors must keep in mind the subjectivity of having to specify the prior probability distributions of the model parameters.  In particular, this also applies to the support of the priors.  Whereas in Bayesian parameter estimation, parameter constraints are often not very sensitive to the range of a parameter covered by the prior, this is not the case for Bayes factors.  Choosing too wide priors on the parameters of features models will typically disfavour them, but Bayes factors can also be pushed in the other direction by selecting (too) narrow priors around regions of parameter space with high posterior probability.  Now this would typically require knowledge of the data, and while it is legitimate in the Bayesian approach to adjust priors a posteriori, one should still be mindful to not wilfully bias the outcomes in this fashion.

Finally, the numerical evaluation of the Bayesian evidence can be a difficult problem in higher-dimensional parameter spaces.  In practice, the nested sampling algorithm~\cite{skilling2006} offers an efficient solution, and has been implemented and refined for applications in cosmology in the \texttt{MultiNest}~\cite{Feroz:2008xx,Feroz:2013hea} and \texttt{PolyChord}~\cite{Handley:2015fda} software packages.

\subsubsection{A note on the use of ``sigmas" \label{sec:sigma}}
It is not uncommon to find results of a search for features quoted in terms of a likelihood ratio or effective $\Delta \chi^2$ between the two models under consideration, $\Delta \chi^2_{\rm eff} \equiv 2 \mathcal{L}_1^\mathrm{max}/\mathcal{L}_0^\mathrm{max}$, sometimes even mapped to the ``number of sigmas", $\sqrt{\Delta \chi^2_{\rm eff}}$.  This is obviously motivated by the case of a 1-parameter Gaussian likelihood/posterior, in which case it can be straighforwardly related to a $p$-value or the Bayes factor, respectively.  In multi-dimensional parameter spaces, or when facing non-Gaussian likelihoods, $\sqrt{\Delta \chi^2_{\rm eff}}$ no longer has this straightforward interpretation and applying the common intuition formed by the simple case is likely to vastly overestimate evidence for the better-fitting model, as we shall demonstrate in the example scenario below.

\subsection{A simple example}
Let us illustrate the application of the Bayesian and Frequentist analysis techniques with the help of a simple, yet, from the point of view of features-searches, perhaps not entirely unrealistic example.

Consider a data set $\mathfrak{D}$, consisting of $N_\mathrm{p}$ data points $x_i$, each drawn randomly from a Gaussian normal distribution $G_{\mu,\sigma}(x)$ with mean $\mu = 0$ and standard deviation $\sigma = 1$.  Let us now compare the performance of two theoretical models, the ``smooth'' $\mathcal{M}_0$, and the ``features'' $\mathcal{M}_1$, defined as:
\begin{itemize}
\item[]{{\bf Model $\mathcal{M}_0$}: \\
Predicts $\hat{x}_i = 0$ for all $i \in \{ 1,..., N_\mathrm{p} \}$ and does not have any free parameters.\\}  
\item[]{{\bf Model $\mathcal{M}_1$}: \\
Predicts a discrete feature $\hat{x}(i) = x_\mathrm{f}$ at $i = i_\mathrm{f}$, and
$\hat{x}_i = 0$ for the remaining $N_\mathrm{p} - 1$ values of $i$.  This model has two free parameters, the position of the feature, $i_\mathrm{f}$, and its amplitude, $x_\mathrm{f}$. }
\end{itemize}
The likelihood functions are simply given by
\begin{align}
	-2 \ln \mathcal{L}_0 &= \sum_{i=1}^{N_\mathrm{p}}{x(i)}^2 \\
	-2 \ln \mathcal{L}_1 &= \sum_{i=1}^{N_\mathrm{p}}{x(i)}^2 - x(i_\mathrm{f})^2 + (x(i_\mathrm{f}) - x_\mathrm{f})^2.
\end{align}
The ratio of the likelihoods is maximised for $x_\mathrm{f} = x(i_\mathrm{max})$ and $i_\mathrm{f} = i_\mathrm{max}$, where $i_\mathrm{max}$ labels the $x$ with the largest absolute value, and thus we have $\Delta \chi^2_\mathrm{max} = x(i_\mathrm{max})^2$.

Of course we know from the construction of $\mathfrak{D}$ that $\mathcal{M}_0$ is the true underlying model and that $\mathcal{M}_1$ will achieve a better fit to the data only by overfitting the natural scatter.  Nonetheless, one might want to  analyse under what conditions $\mathcal{M}_1$ would actually be preferred.

\subsubsection{Frequentist analysis}
With the null hypothesis that $\mathcal{M}_0$ is the correct underlying model and $\Delta \chi^2_\mathrm{max}$ as the test statistic, we can generate a large number of simulated data sets $\mathfrak{D}_j$ and construct the probability distribution of the test statistic.  This gives us an idea of what the typical $\Delta \chi^2_\mathrm{max}$ is if there is no feature in the underlying model, and also allows us to quantify how large the $\Delta \chi^2_\mathrm{max}$ would have to be in order for us to be able to ``rule out'' the null hypothesis of the smooth model at a given level of confidence.  

In this particular case, the probability density function describing the distribution of $\Delta \chi^2_\mathrm{max}$ can be derived analytically (see Ref.~\cite{Fergusson:2014hya} for a similar example).  Noting that the $\Delta \chi^2$ for a single $x_i$ follows a chi-squared distribution with one degree of freedom, $f_{\chi^2_1}$, the probability of the maximum $\Delta \chi^2$ to not exceed a value of $y$ (i.e., the cumulative distribution function of $\Delta \chi^2_\mathrm{max}$) can be expressed as
\begin{align}
P(\Delta \chi^2_\mathrm{max} \leq y) &=  \left(P(\Delta \chi^2 \leq y)\right)^{N_\mathrm{p}}\\
&= \left(\int_0^{y} \mathrm{d}y' f_{\chi^2_1}(y')\right)^{N_\mathrm{p}}\\
&= \left(\mathrm{erf} (\sqrt{y/2})\right)^{N_\mathrm{p}},
\end{align}
and thus the probability distribution function of $\Delta \chi^2_\mathrm{max}$ is given by
\begin{equation}
\label{eq:pdf}
\mathrm{PDF}(y,N_\mathrm{p}) = \frac{N_\mathrm{p}}{\sqrt{2 \pi y}} \, e^{-y/2} \, \left(\mathrm{erf} (\sqrt{y/2})\right)^{N_\mathrm{p}}.
\end{equation}
In general however, the probability density function describing the distribution of $\Delta \chi^2_\mathrm{max}$ cannot be so easily derived analytically and will have to be established through simulations.  For independent random variables $x_i$ drawn from identical distributions, and $N_\mathrm{p} \gtrsim \mathcal{O}(10)$, it can be reasonably well approximated by a generalised extreme value distribution~\cite{Fisher:1928xx,Gnedenko:1948xx},
\begin{equation}
GEV_{\mu,\sigma,\xi}(\Delta \chi^2_{\rm max}) = \frac{1}{\sigma} \left( 1 + \xi \frac{\Delta \chi^2_{\rm max}-\mu}{\sigma} \right)^{-(\xi+1)/\xi} \, \exp\left[ - \left( 1+ \xi \frac{\Delta \chi^2_{\rm max}-\mu}{\sigma} \right)^{-1/\xi} \right],
\end{equation}
with $1 + \xi \left( \Delta \chi^2_\mathrm{eff} \right)/\sigma > 0$, but the free parameters of the distribution need to be obtained by fitting to the results of a simulation.

As an example, we plot in Fig.~\ref{fig:distribution} a histogram of $\Delta \chi^2_\mathrm{max}$ obtained from $10^6$ simulations, for $N_\mathrm{p} = 1000$.  Here, the expectation value is \mbox{$\langle \Delta \chi^2_\mathrm{max} \rangle = 11.9$} and the 97.725\%- and 99.865\%-quantiles\footnote{For a Gaussian normal distribution, these quantiles are at $x=\mu + 2\sigma$ and $x=\mu + 3\sigma$, so we will refer to them as the (2-tail) ``$2\sigma$"- and ``$3\sigma$" limits.  If one has to deal with a distribution that only has one recognisable tail, it may be more reasonable to use a 1-tail definition of these limits instead, with the $2\sigma$- and $3\sigma$-limits marked by the $95.45\%$- and $99.73\%$-quantiles. This choice is motivated by the half-normal distribution, where these quantiles are at $x=2\sigma$ and $x=3\sigma$.} are at $\Delta \chi^2_\mathrm{max} = 17.9$ and $23.3$, respectively -- much larger than the na\"ive attribution $2\sigma \to \Delta \chi^2_{\rm eff} = 4$ and $3\sigma \to \Delta \chi^2_{\rm eff} = 9$ (see the discussion in Section~\ref{sec:sigma}).  Following our procedure, this case can in fact be recovered, but only if $N_\mathrm{p} = 1$ (and using the 1-tail limit definition).  

Increasing the number of data points in the data set will, on average, lead to more extreme outliers; therefore it is not surprising that the larger $N_\mathrm{p}$, the higher the expectation value of the distribution of  $\Delta \chi^2_\mathrm{max}$, and the higher the threshold value of $\Delta \chi^2_\mathrm{max}$ required for rejecting the null hypothesis at a given level of significance.  As can be seen in Fig.~\ref{fig:threshold}, both the threshold values and the expectation value grow to approximately linearly with $\ln N_\mathrm{p}$.

The way we defined the test statistic above, i.e., $\Delta\chi_\mathrm{max}^2$ evaluated over the full data set, properly takes into account the look-elsewhere effect.  But we can also use this scenario to illustrate the danger of a posteriori reasoning:  imagine one were to first identify the most extreme $x_i$ and then ask how likely it is to have such a large deviation from the prediction of the null hypothesis {\it at this particular point}.  One would naturally find a $p$-value very close to zero, apparently disfavouring the null hypothesis.  However, choosing to look at the most extreme point in isolation is obviously motivated by the fact that one has already seen the data, whereas in a blind analysis there would no reason to single out this $x_i$.  So the fallacy here lies in ignoring all the other points; one should not be asking about how likely it is to observe such a feature at this specific point, but rather how likely it is to see any feature at all in the entire data set.

\begin{figure}[t]
\center
\includegraphics[height=.90\textwidth,angle=270]{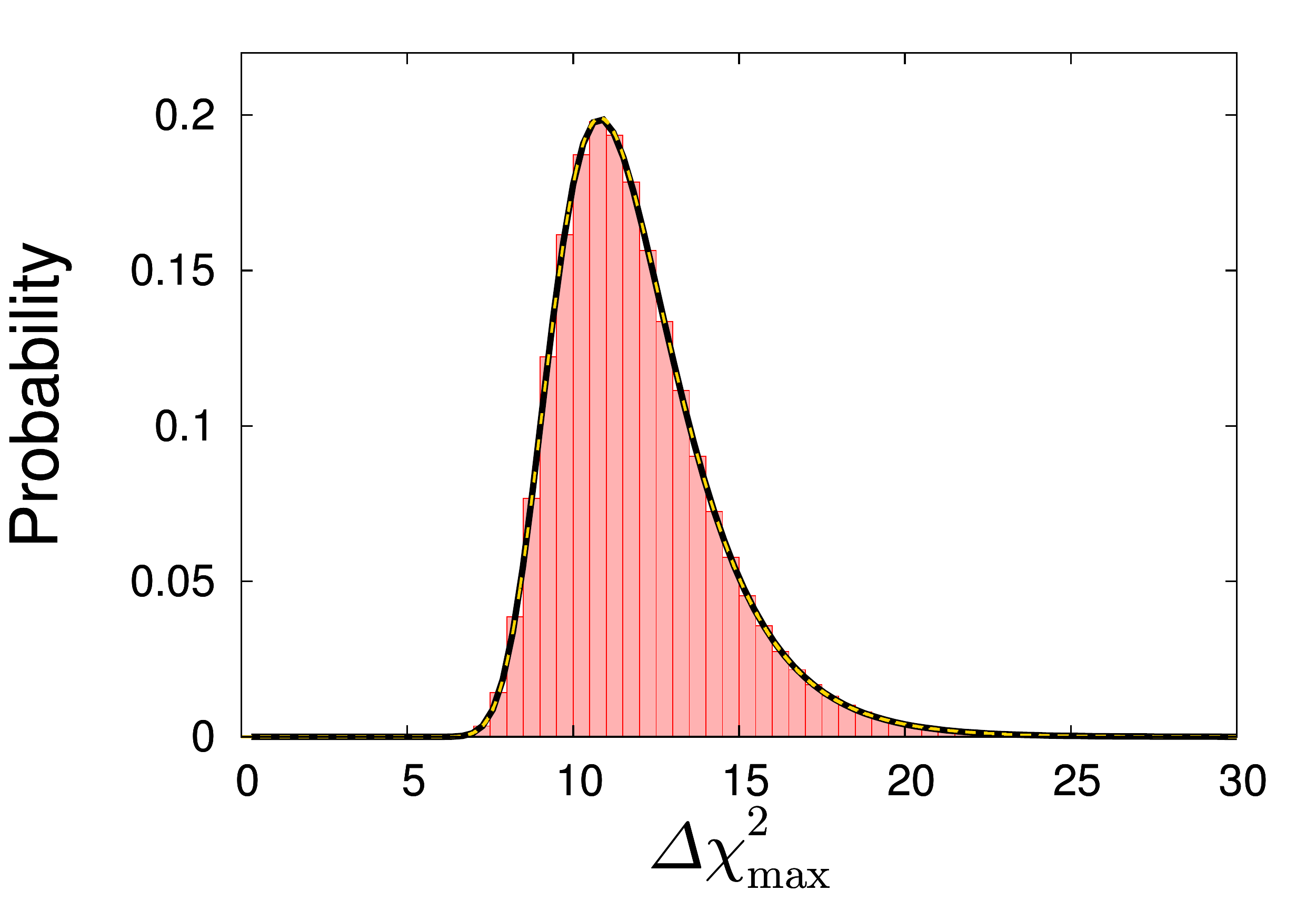}
\caption{Histogram of $\Delta \chi^2_\mathrm{max}$ for $N_\mathrm{data} = 10^6$ simulated data sets~$\mathfrak{D}_j$ with $N_\mathrm{p} = 1000$ data points each.  The thick black line is the theoretical probability distribution function (Eq.~\ref{eq:pdf}) and the dashed golden line is a fit of a generalised extreme value distribution with parameters $\mu = 10.83$, $\sigma = 1.85$ and $\xi = 0.0096$. \label{fig:distribution}}
\end{figure}

\begin{figure}[t]
\center
\includegraphics[height=.90\textwidth,angle=270]{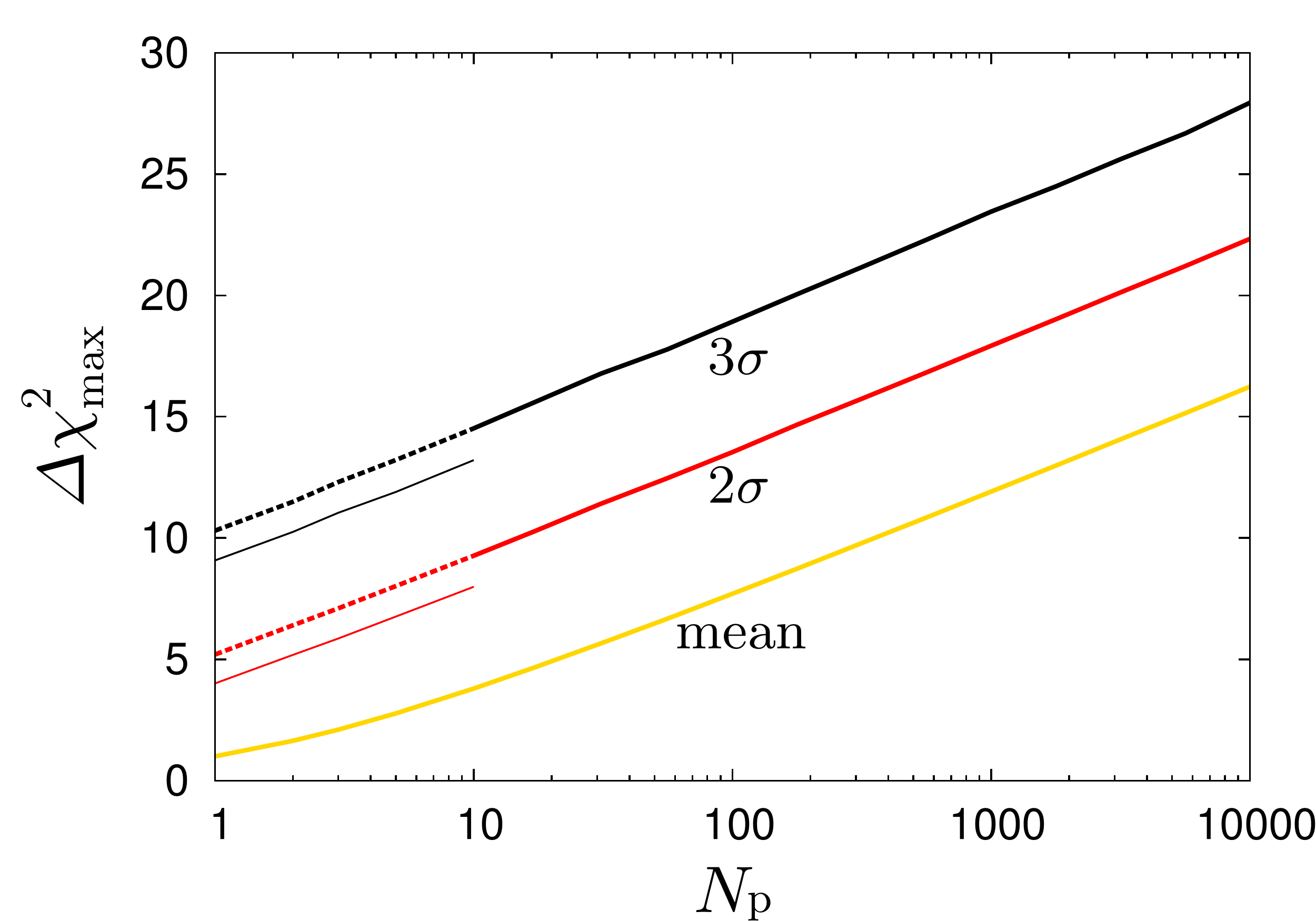}
\caption{Threshold values and mean of the distribution of $\Delta \chi^2_\mathrm{max}$ as a function of the number of data points $N_\mathrm{p}$ of the data set~$\mathfrak{D}$ for the frequentist analysis.  The golden line indicates the expectation value of $\Delta \chi^2_\mathrm{max}$.  The black and red lines mark the $\Delta \chi^2_\mathrm{max}$ at which the null hypothesis is ruled out at $3\sigma$ and $2\sigma$, respectively.  Thin lines denote the 1-tail definition, thick lines the 2-tail definition of $\sigma$.  If $N_\mathrm{p} \gtrsim 10$, the distribution of the $\Delta \chi^2_\mathrm{max}$ has negligible volume near $\Delta \chi^2_\mathrm{max} = 0$, and thus  the 2-tail definition of $\sigma$ should be used.  At $N_\mathrm{p} \lesssim 10$, one might want to prefer the 1-tail definition instead; in particular for $N_\mathrm{p} = 1$, this reproduces the na\"ive thresholds of $\Delta \chi^2_\mathrm{max} = 4$ for $2\sigma$ and $\Delta \chi^2_\mathrm{max} = 9$ for $3\sigma$.   \label{fig:threshold}}
\end{figure}

\subsubsection{Bayesian analysis}
Let us assign equal model prior probabilities to $\mathcal{M}_0$ and $\mathcal{M}_1$, and assume the following parameter prior probabilities for $\mathcal{M}_1$:  a uniform prior on  $i_\mathrm{f}$ (i.e., $P(i_\mathrm{f}) = 1/N_\mathrm{p}$ for all $i_\mathrm{f} \in \{1,...,N_\mathrm{p}]$\}, and a symmetric top hat prior of width $\beta$ on $x_\mathrm{f}$ (i.e., $P(i_\mathrm{f}) = 1/\beta$ for $i_\mathrm{f} \in [-\beta/2, \beta/2]$ and $P(x_\mathrm{f}) = 0$ if $|(x_\mathrm{f})| > \beta/2$. 
Then the Bayes factor between models 0 and 1 is given by
\begin{equation}\label{eq:bayesfactor_ex}
B_{01} = \frac{1}{2 \beta N_\mathrm{p}} \, \sum_{i=1}^{N_\mathrm{p}} \frac{ \int_{- \beta}^{\beta} \mathrm{d}x \,G_{x,1}(x_i)}{G_{0,1}(x_i)},
\end{equation}
which can be straightforwardly evaluated numerically.  

In figure~\ref{fig:bayesthreshold}, we plot the $\Delta \chi^2_\mathrm{max}$ at which the Bayes factor $B_{01}$ takes on values of $1$, $e^{-2}$ and $e^{-4.5}$.  These results depend on the choice of the prior width $\beta$; the Bayes factor would be maximised in favour of $\mathcal{M}_1$, if the prior on $x_\mathrm{f}$ was a Dirac delta distribution centred on the actually observed maximum of the $x_i$.  In this extreme (and, admittedly, not very realistic) case, a $B_{01}$ of $e^{-2}$ and $e^{-4.5}$ would correspond to the likelihood ratio at the 2-tail $2\sigma$- and $3\sigma$-limits of the normal distribution.  For our choice of a symmetric top-hat prior, the integral in equation~(\ref{eq:bayesfactor_ex}) is evaluated also over lower-likelihood regions of parameter space, leading to a suppression of the Bayes factor compared to the optimal case.  This averaging procedure is absent in the frequentist analysis, which relies solely on comparing the best-fit values of the likelihood between the two models.  As a consequence, the $\Delta \chi^2_\mathrm{max}$ required to reach a given Bayes factor is always greater than the one required to reach the corresponding frequentist $p$-value.

\begin{figure}[t]
\center
\includegraphics[height=.90\textwidth,angle=270]{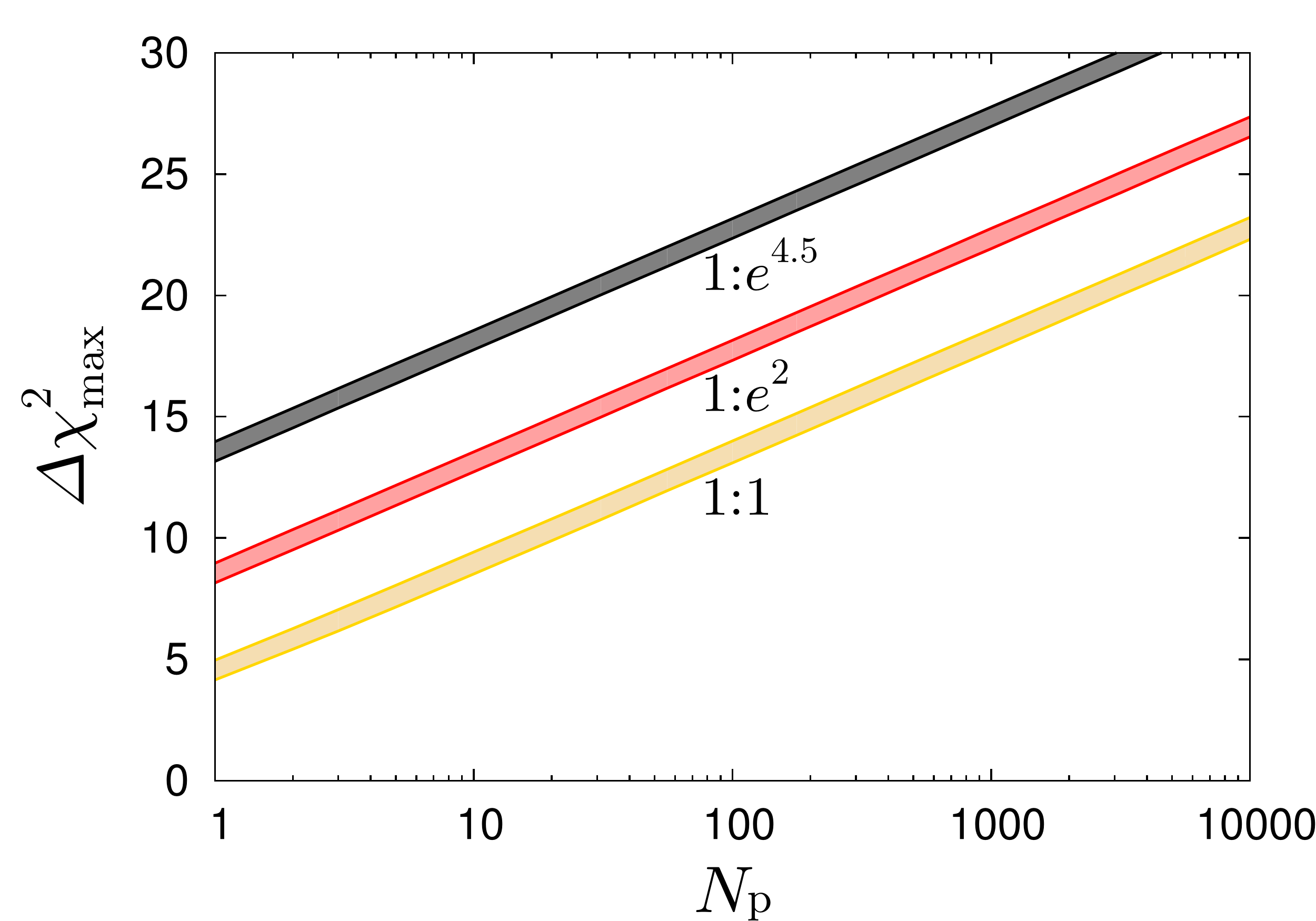}
\caption{Threshold values of $\Delta \chi^2_\mathrm{max}$ as a function of the number of data points $N_\mathrm{p}$ of the data set~$\mathfrak{D}$ for the Bayesian analysis.   The grey, pale red and pale golden bands mark the $\Delta \chi^2_\mathrm{max}$ at which the Bayes factor $\ln B_{01}$ reaches 4.5, 2, and 0, for a range of choices of the width of the prior probability distribution of $x_\mathrm{f}$.  Positive values of $\ln B_{01}$ indicate a preference for $\mathcal{M}_1$. The lower and upper edges of the bands correspond to choices of $\beta = 10$ and $\beta = 15$, respectively.  
\label{fig:bayesthreshold}}
\end{figure}

}

\section{Features searches \label{sec:searches}}
{
To a certain extent, the prospects of being able to detect possible features of the primordial power spectrum depend on how well the featureless reference model performs in relation to the data.  Since the release of Wilkinson Microwave Anisotropy Probe (WMAP) data~\cite{Spergel:2003cb}, the quality of fit of a power-law $\Lambda$CDM model to CMB anisotropy data has not been very good (but not too terrible either), with $p$-values hovering around the $5$-$10\%$ region.\footnote{For the $\ell \geq 30$ {\it Planck} temperature data alone the $p$-value is about $17\%$, and roughly $7\%$ for a combination of temperature and $E$-polarisation~\cite{Planck:2015xua}.  Adding the $\ell < 30$ data would likely lead to slightly lower numbers.}
This result certainly does not mean that power-law $\Lambda$CDM is incompatible with data, but it also leaves room for improvement and has inspired the hope of finding deviations from a smooth power-law spectrum.  Consequently, many groups have embarked on the quest for features, employing a large array of different methods.  This variety of methods can broadly be categorised into two different, but complementary, classes which we shall refer to as {\it bottom-up} and {\it top-down}.

The philosophy behind the top-down approach is to start from a theoretically motivated model and fit its predictions to data.  Top-down typically introduces at most a handful of additional free parameters and can be combined well with both Frequentist or Bayesian statistical methods.  While it is a great choice for the devout theorist, who wants to test their favourite model of inflation, it slightly suffers from the fact that its outcomes are very specific to one model, and so may not be the first choice for the agnostic phenomenologist.

This is where the bottom-up approach comes into play which aims to reconstruct ``ideal" primordial power spectra from the data.  Reconstruction methods often introduce a large number of extra parameters and, due to their great flexibility, a Bayesian evidence based analysis will generally favour the simplicity of a featureless spectrum.  Additionally, the reconstructed spectra do not typically correspond to any proper physical model and are thus by no means realistic.  However, they can nonetheless be very helpful in establishing whether the data indicate any significant deviations from smoothness (using frequentist methods), and, if so, in identifying what kind of features the data would prefer.  In this sense, the reconstruction can be seen as providing guidance and inspiration to help one find realistic models that reproduce the reconstructed spectra's most salient features.

\subsection{Bottom-up: Reconstruction of the primordial power spectrum}
Even though reconstruction is in principle meant to be a model-independent technique, one nonetheless needs to decide on a parameterisation of $\mathcal{P_R}(k)$.  At first glance, one might be tempted to expand $\mathcal{P_R}(k)$ over the observable region of wavenumbers in some functional basis, e.g., a Taylor series.  However, keeping in mind both feasibility of technical analysis and of physical interpretation, it is desirable that these parameters do not exhibit strong and complicated correlations with each other.  This requirement makes for instance a Taylor expansion ansatz fairly unpractical beyond the first few terms (though other more suitable basis functions, such as wavelets~\cite{Mukherjee:2003cz,Shafieloo:2006hs}, or a principal component analysis~\cite{Leach:2005av} are better behaved in this regard).

The majority of reconstruction methods in the literature are based on some sort of binning approach, where $\mathcal{P_R}(k)$ is parameterised in terms of a set of amplitudes $\left\{\mathcal{P_R}(k_i)\right\}$ at a given set of wavenumbers or {\it knots}, $\left\{k_i\right\}$, where the non-negligible entries of the correlation matrix are typically clustered around the diagonal.  In the most simplistic case, the $\left\{\mathcal{P_R}(k_i)\right\}$ are interpreted as bandpowers and the full power spectrum $\mathcal{P_R}(k)$ is a series of discontinuous steps~\cite{Wang:1998gb}.  Somewhat less unphysical spectra can be obtained by interpolating between the $\left\{\mathcal{P_R}(k_i)\right\}$, e.g., using linear interpolation~\cite{Bridle:2003sa,Hannestad:2003zs,Bridges:2008ta} or splines~\cite{Sealfon:2005em,Verde:2008zza,Ichiki:2009zz,Hu:2014aua}.  While the positions of the knots are often fixed before the analysis, it is possible to also treat the $\left\{k_i\right\}$ as variables~\cite{Vazquez:2012ux,Hazra:2013nca} and even use the Bayesian evidence to decide how many additional knots are actually preferred by the data \cite{Aslanyan:2014mqa,Ade:2015lrj}.

As long as the total number of free parameters does not exceed a few dozen, the problem is amenable to both maximum likelihood analysis or Bayesian methods.  However, with a thus limited number of degrees of freedom, the classes of spectra that can be obtained from reconstruction is limited as well: features finer than the bins defined by the  $\left\{k_i\right\}$ cannot be reconstructed.
There is no fundamental limit to how densely $\mathcal{P_R}(k)$ can be sampled though; the number of samples in $k$ can even exceed the number of data points (e.g., the $\mathcal{C}_\ell$s) if one is willing to forgo the possibility of an analysis based on exploring the likelihood/posterior and treat the issue as a deconvolution problem (i.e., an inversion of Equation~\ref{eq:cls}) instead.  Deconvolution techniques have been applied by several groups to CMB temperature data~\cite{Matsumiya:2002tx,Kogo:2003yb,Shafieloo:2003gf,TocchiniValentini:2004ht,Nagata:2008tk,Paykari:2014cna}, CMB temperature+polarisation data~\cite{Kogo:2004vt,Nicholson:2009pi,Nicholson:2009zj} and combinations of CMB data with large scale structure data~\cite{Hunt:2013bha}.  

In order to get rid of spurious high-frequency spikes that are likely to occur for noisy data when the primordial spectrum is oversampled, these methods typically involve a smoothing procedure. Effectively, the maximal possible resolution of the reconstruction is related to the width of the window functions of the data sets under consideration.
Note also that the behaviour at the boundaries of the reconstructed $k$-interval can easily be dominated by the first/last data point, potentially leading to drastic suppression/enhancement (depending on whether these data points lie above or below the featureless model's best-fit) of the reconstructed spectrum~\cite{Hamann:2009bz}.  If such a behaviour is undesired, it can be avoided by introducing a likelihood penalty, as for instance in Ref.~\cite{Gauthier:2012aq}.

A number of these reconstruction methods have been applied to {\it Planck} data~\cite{Hazra:2014jwa,Ade:2015ava} and find results in agreement with earlier analyses of WMAP data, e.g., those of Ref.~\cite{Hunt:2013bha}.  Several distinctive features have been identified in reconstructed spectra:
\begin{itemize}
\item{A cutoff-like suppression of power on the largest scales $k \lesssim 5\times10^{-4}$~$\mathrm{Mpc}^{-1}$.  This is driven by the fact that, as seen in Figure~\ref{fig:plonk_cls}, $\mathcal{C}_2, \mathcal{C}_3$ and $\mathcal{C}_4$ are all below the expectation value of the power-law model's best-fit.  While dramatic-looking, one should keep in mind that actually only these first few multipoles are sensitive to the scales near this cutoff (cf.\ Figure~\ref{fig:temp_delta}), and receive hardly any contribution from even larger scales, thus favouring as little power as possible at very small $k$.  Such a cutoff-like behaviour is therefore more of an extrapolation of the spectrum into areas where the data are not informative at all, and is also not uncommonly encountered in reconstructions from featureless simulated CMB data.}
\item{A dip around $k \approx 2\times10^{-3}$~$\mathrm{Mpc}^{-1}$ followed by a bump at $k \approx 4\times10^{-3}$~$\mathrm{Mpc}^{-1}$, corresponding to a similar pattern in the Sachs-Wolfe plateau of $\mathcal{C}_\ell^{TT}$ extending from $20 \lesssim \ell \lesssim 40$.  For the first-year WMAP data, this was shown to be the most important local deviation from a smooth spectrum~\cite{Shafieloo:2006hs}.}
\item{Later iterations of WMAP data as well as {\it Planck} data have revealed several additional local features, similar in shape to the previous one, but narrower in $\ln k$ (and therefore only found using methods with a high enough resolution), e.g., at wavenumbers $k \approx 3.5\times10^{-2}$~$\mathrm{Mpc}^{-1}$ and $k \approx 6\times10^{-2}$~$\mathrm{Mpc}^{-1}$, reproducing structures in the $\mathcal{C}_\ell^{TT}$ data at multipoles $\ell \sim 450$ and $\ell \sim 800$, respectively.}
\end{itemize}
While these are certainly interesting candidates for primordial features, none of them have been shown to be absolutely required by the data, neither in the frequentist nor in the Bayesian sense.  In Ref.~\cite{Ade:2015ava}, three methods are applied to the 2015 {\it Planck data}: the penalised likelihood reconstruction finds $p$-values no smaller than 0.045, and the spline-reconstruction in the same paper $p$-values obtains $p$-values of at least 0.11,\footnote{Quoted are the smallest $p$-values from among different data combinations and analysis parameters.} and the Bayesian evidence-based variable knot-method shows no preference for spectra more complicated than a power-law either.

Since the transfer functions are not independent of the late-time cosmology, it is also interesting to ask to what extent constraints on non-inflationary parameters depend on the assumption of a primordial power-law spectrum.  With present data, the constraints on the $\Lambda$CDM model's late-time cosmological parameters are in fact surprisingly stable when one allows a very general shape of the power spectrum compared to the power-law case~\cite{Hazra:2013eva,Ade:2015ava} (though in extended models this statement may not hold~\cite{dePutter:2014hza}).

\subsection{Top-down: Looking for specific features in the power spectrum}
In this approach, the data are tested for the presence of specific classes of theoretically motivated features.  They can take the form of a parameterisation of the primordial power spectrum, or, even more fundamentally, a parameterisation of the inflaton potential (which then requires $\mathcal{P_R}(k)$ to be calculated explicitly by integration of the mode equations).\footnote{It is also possible to combine a parameterisation of the spectrum and the potential; such a hybrid scheme has been employed to reduce dependence on the background inflaton potential when looking for features in $V(\phi)$.~\cite{Hamann:2007pa}}  
Focussing on a particular model, the number of additionally introduced parameters in this case is typicall very small.  This means that the scope of the analysis is not nearly as broad as in the bottom-up case and different classes of models would have to be tested case by case, but on the other hand, the higher degree of predictivity means that the Bayesian evidence is not doomed from the start to favour a smooth power spectrum.  Also, compared to binned reconstruction methods, there is no inherent limit on the resolution of features in this approach, so very sharp or high-frequency features become accessible (but the chances of detection are of course still subject to the inherent resolution of the data considered).

Just like for the reconstruction methods, the vast majority of parameterised features searches in the literature involve the CMB temperature angular power spectrum.  The models analysed can roughly be divided into three categories, based on their phenomenology: (i) cutoff-like models with a power-law behaviour at small and intermediate scales, but a suppression of power on large scales, (ii) models with local features joining up two power-law segments, and (iii) models with global features, i.e., modifications of the power-law on all scales.

\subsubsection{Models with a large scale power suppression}
The phenomenological attraction of these models obviously lies in the observed lack of power at the largest scales, but they are also interesting from a theoretical point of view, since the largest scales correspond to the earliest observable moments of inflation, potentially carrying information about the onset of a slow-roll phase of inflation, as discussed in Section~\ref{sec:preinf}.
Besides using ad-hoc parameterisations (e.g., piecewise power-laws, exponential cutoff, step-function cutoff), a range of physically motivated scenarios leading to a suppression of power have been explored: examples include a non-trivial topology of the universe~\cite{Jing:1994jw}, initial kinetic domination~\cite{Contaldi:2003zv}, initial radiation domination  \cite{Sinha:2005mn} or a sharp kink in the inflaton potential~\cite{Starobinsky:1992ts} have been explored in the literature.  

In the wake of the BICEP2 team's claimed detection of a primordial tensor contribution \cite{Ade:2014xna}, these types of models experienced a brief surge in popularity due to their ability to ameliorate the apparent tension of a primordial power-law spectrum with the large-scale CMB temperature data, which is exacerbated in the presence of tensors~\cite{Lesgourgues:1998mq}.  There was in fact even some indication of a preference over a power-law spectrum~\cite{Contaldi:2014zua,Abazajian:2014tqa,Hazra:2014aea,Hazra:2014jka,Hazra:2014goa}.  However, this conclusion had to be revised after the re-evaluation of the BICEP2 data using additional information from {\it Planck} and the {\it Keck array}~\cite{Ade:2015tva} which showed that the $B$-mode signal originally attributed to primordial tensor perturbations can be explained in terms of polarised galactic dust emission.  The initial kinetic domination cutoff model has been tested against the most recent {\it Planck} temperature and polarisation data, but both frequentist and Bayesian analyses favour a featureless spectrum~\cite{Ade:2015ava}.

\subsubsection{Models with a local feature}
Rather than massively deviating from an overall power-law shape of the spectrum, the general idea of these models is to improve the fit to the data by matching patterns confined to a small range of wavenumbers or multipoles.  Physically, such an effect can be obtained by the occurrence of transient phenomena between two stages of standard slow-roll inflation and realised by many of the scenarios presented in Section~\ref{sec:theory}.

Consequently there is a rich literature of models that have been confronted with data.  Resonant particle production (Section~\ref{sec:partprod}), for instance, was looked at in Ref.~\cite{Elgaroy:2003hp}.  

The mechanism of Section~\ref{sec:effpot} was discussed for a phase transition in a multiple inflation model where the effective inflaton mass undergoes a sudden change~\cite{Adams:1997de,Hunt:2004vt} (see also~\cite{Joy:2007na,Joy:2008qd}), with particular emphasis on exploring alternatives to late-time $\Lambda$CDM-cosmology~\cite{Hunt:2007dn,Hunt:2008wp}.  Another, more empirical case is a step in the inflaton potential parameterised by a $\tanh$-function~\cite{Adams:2001vc}, which has been found to match particularly well the $20 \lesssim \ell \lesssim 40$ feature of CMB temperature data~\cite{Peiris:2003ff} \cite{Covi:2006ci,Hamann:2007pa,Benetti:2012wu,Miranda:2013wxa,Miranda:2014wga}, with an effective $\Delta \chi^2 \approx 10$.  However, though in the eyes of {\it Planck} data this model compares favourably with other features models analysed in Ref.~\cite{Ade:2015ava}, it does not outperform the power-law spectrum.

The impact of heavy fields (Section~\ref{sec:heavyfields}) has been investigated in the context of a model with a temporary reduction in sound speed \cite{Achucarro:2013cva,Achucarro:2014msa,Hu:2014hra}.  Here, the largest improvement in the effective $\chi^2$ was found to come from fitting the $\ell \sim 800$ feature in the {\it Planck} temperature angular power spectrum, but the total improvement is relatively modest ($\sim 10$) for a model with three additional parameters.

\subsubsection{Models with global features}
If the mechanism causing the emergence of features is not transient, but recurring or constantly at work, one can expect the resulting power spectrum to deviate from a power-law at all scales.  A parameterisation of special interest here is the log-spaced sine-modulation of the primordial power spectrum,
\begin{equation} \label{eq:wiggles}
\mathcal{P_R}(k) = \mathcal{P}^0_\mathcal{R}(k) \left[ 1 + \alpha \cos \left( \omega \ln{k/k_*} + \varphi \right) \right],
\end{equation}
where $\mathcal{P}^0_\mathcal{R}(k)$ is a smooth power-law.  The modulation amplitude $\alpha$ and frequency $\omega$ can in principle be functions of $k$, but are often kept constant.  

From a technical perspective, the analysis of this model is somewhat challenging: it tends to display a very complex likelihood landscape with many isolated local maxima, especially in the $\omega$-direction, making it hard to properly sample with conventional Markov-chain Monte-Carlo methods.  And, while the numerical analysis of features models generally needs the the numerical precision settings of the Boltzmann code to be increased (hence slowing down computation of observables), this is particularly true for the high-frequency ($\omega \gtrsim 100$) part of this model's parameter space, where the primordial power spectrum starts to require a denser sampling than the window functions when integrating Eq.~(\ref{eq:cls}).

Eq.~(\ref{eq:wiggles}) can be used to describe the signatures of trans-Planckian physics (see Section~\ref{sec:initstate}, Figure~\ref{logktrans}) and was first confronted with CMB data in this regard~\cite{Martin:2003sg,Martin:2004iv,Martin:2004yi}.  Besides that, it is also a good approximation to the power spectrum predicted in the axion monodromy model discussed in Section~\ref{sec:effpot}, where the parameters $\alpha$ and $\omega$ can be related to the inflation potential.
Analyses of different incarnations of the WMAP data have yielded effective $\Delta \chi^2$ of up to $20$ for the monodromy model \cite{Flauger:2009ab,Meerburg:2011gd,Aich:2011qv,Peiris:2013opa}.  For {\it Planck} data, however, the $\Delta \chi^2$ is slightly lower~\cite{Planck:2013jfk,Easther:2013kla} and no compelling statistical evidence for a presence of oscillations has been found~\cite{Meerburg:2013cla,Meerburg:2013dla,Ade:2015lrj}.\footnote{This also holds true for linearly-spaced modulations of $\mathcal{P_R}(k)$, which were examined in Refs.~\cite{Meerburg:2013dla,Ade:2015lrj}.}

With the advent of {\it Planck} data, it has become possible to meaningfully go beyond the CMB power spectrum and extend the search for features to the bispectrum, as well as performing combined analyses~\cite{Fergusson:2014hya,Fergusson:2014tza}.   The {\it Planck} bispectrum data have been compared to a number of bispectrum features templates, including both logarithmically and linearly spaced oscillations, albeit with a reduced frequency range compared to the power spectrum ~\cite{Ade:2013ydc,Ade:2015ava}, and it was shown that no single of the analysed features was preferred over a Gaussian model after correcting for the look-elsewhere effect.

}

\section{Concluding remarks and outlook\label{sec:conclusions}}
{
As far as the prospects for cosmological observations are concerned, the present era represents a watershed. Not only might we be one mission approval away from an ultimate CMB measurement (e.g. \cite{Andre:2013afa}) targeting both polarization and spectral distortions, the Hubble volume in which we find ourselves constitutes a \textit{finite} region within which it is conceivable that a large fraction, if not all large scale gravitationally bound structures will be mapped one day-- possibly even within the lifetime of someone reading these words at the time of publication. Complementary 3-d (i.e. uncompressed) information about modes up to comoving scales commensurate with those seen in the CMB will be extracted from LSS surveys (up to $k = 10^{-1}\, {\rm Mpc}^{-1}$) and over hitherto unobserved scales through 21 cm observations (up to $k = 10^{2}\, {\rm Mpc}^{-1}$ ) and observations of CMB spectral distortions (up to $k = 10^{4}\, {\rm Mpc}^{-1}$)\footnote{In the context of inflationary cosmology, this corresponds to being granted a view of another 10 $e$-folds of inflation at work beyond the 6-7 that we see in the CMB anisotropies alone.}. We will thus potentially be privy to a vast amount of untapped information about primordial physics, provided it can be appropriately deconvolved from these observations, given we know what questions to ask of it.

As this review has attempted to highlight, features, if realized in nature, would offer a unique, discriminating window onto the microphysics that underlies the cosmological standard model. Although couched in terms of correlation functions, simple linear response theory allows us to extract new characteristic scales from features in certain contexts, enabling us to infer appropriately limited information about the parent theory (from the EFT point of view) in which the mechanism that generated structure is embedded. After surveying the various theoretically well grounded mechanisms through which features can be generated in correlation functions of the comoving curvature perturbation in the context of (adiabatic) inflationary cosmology, we reviewed the various ways these could be traced through different cosmological observables. These include the anisotropies and spectral distortions of the CMB as well as (at later times) the matter power spectrum and bispectrum inferred through LSS and conceivably, future 21cm surveys. For the latter observations, much remains to be understood in terms of extracting quantities from observations that can be compared to theoretical predictions (in addition to the theoretical predictions themselves) at the scales and redshifts of interest, further investigations into which this review hopes to spur.

}

\section*{Acknowledgments}
{
We wish to thank Ana Ach\'ucarro, Peter Adshead, Thorsten Battefeld, James Fergusson, Jinn-Ouk Gong, Dhiraj Kumar Hazra, Daniel Meerburg, Gonzalo Palma, Hiranya Peiris, Raquel Ribeiro, Subir Sarkar and Arman Shafieloo for valuable comments on the draft. In addition to all of the former, we further wish to thank Cliff Burgess, Vincent Desjacques, Ruth Durrer, Daniel Grin, Augusto Sagnotti, L. Sriramkumar and Michael Trott for many useful discussions over the course of preparing this review.

JC is supported by the Royal Society as a Royal Society University Research Fellow at the University of Cambridge, U.K. JH acknowledges the support of a Future Fellowship of the Australian Research Council, and would like to thank the Asia Pacific Center for Theoretical Physics, where part of this work was completed during the Focus Program ``Cosmology and Fundamental Physics 4''. SP was supported for part of his time at CERN by a Marie Curie Intra-European Fellowship of the European Community's 7'th Framework Program under contract number PIEF-GA-2011-302817.

\begin{center}
``{\textit{So wie es ist, bleibt es nicht.}}''\\ -- Berthold Brecht
\end{center}

}

\begin{appendix}
{
\section{A review of the `in-in' formalism}
\label{in-in}
In connecting fundamental theory with late time cosmological observations, a central object of interest is the finite time correlation function of a particular operator (or string of operators) of some primordial field-- $\langle\calO(\tau)\rangle$. The latter signifies a finite time correlator in a given quantum state (typically taken to be the Bunch-Davies vacuum state) at some initial or asymptotic time, unitarily evolved forward. This is to be contrasted with the matrix element of some time evolved initial state with some final state in the asymptotic future i.e. an S-matrix element, with which the reader might be more familiar. In the interaction picture, operator products of fields evolve by their free field equations of motion (and so admit an expansion in terms of the usual creation and annihilation operators), and states that are evolved forward in time by the unitary evolution operator:
\eq{dop}{U(\tau,\tau_0) = T\,{\rm exp}\left(-i\int^\tau_{\tau_0}H_I(\tau') \dint\tau' \right)}
where $T$ denotes time ordering, and $H_I$ denotes the totality of terms in the interaction Hamiltonian. Hence the operator expectation value $\langle\calO(\tau)\rangle$ is actually shorthand for
\eq{iidef}{\langle\calO(\tau)\rangle = {\langle 0_{\rm in}|\bar T\,\left[{\rm exp}\left(i\int^\tau_{\tau_0}H_I(\tau') \dint\tau' \right)\right] \calO(\tau)\,T\left[\,{\rm exp}\left(-i\int^\tau_{\tau_0}H_I(\tau') \dint\tau' \right)\right]|0_{\rm in}\rangle}}
where $\bar T$ denotes anti time ordering-- the result of taking the adjoint of a time ordered operator on the left hand side. One can view this as having time evolved from the initial time $\tau_0$ to time $\tau$, inserting the operator $\calO(\tau)$ at time $\tau$ and then evolving back to time $\tau_0$. Therefore one can equivalently consider (\ref{iidef}) as the product of $\calO(\tau)$ with the unitary operator:
\eq{iidefc}{\langle\calO(\tau)\rangle = {\langle 0_{\rm in}| T_C\,\left[{\rm exp}\left(-i\oint H_I(\tau') \dint\tau' \right)\right] \calO(\tau)|0_{\rm in}\rangle},}
where the contour integral now goes from $\tau_0 \to \tau$ and then back from $\tau\to \tau_0$, as illustrated below--
\begin{figure}[h!]
  \hfill
  \begin{minipage}[t]{.45\textwidth}
    \begin{center}
      \epsfig{file=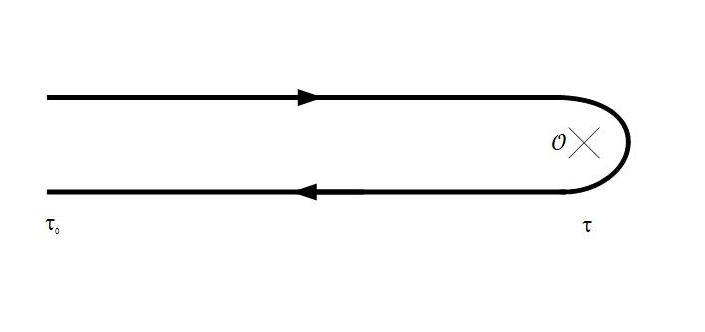, height=1.5in, width=2.5in}
      \caption{Contour used in evaluating (\ref{iidefc})}
      \label{cinin}
    \end{center}
  \end{minipage}
  \hfill
  \begin{minipage}[t]{.45\textwidth}
    \begin{center}
      \epsfig{file=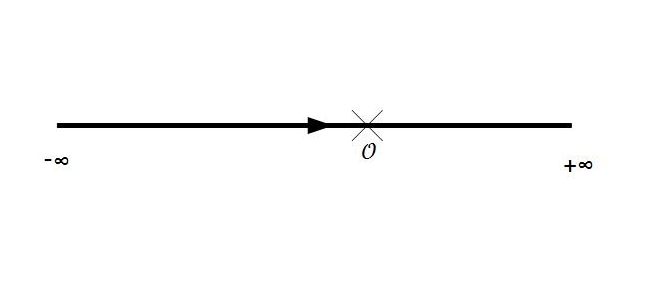, height=1.5in, width=2.5in}
      \caption{Contour for evaluating the S-matrix}
      \label{csm}
    \end{center}
  \end{minipage}
  \hfill
\end{figure}
The expression (\ref{iidefc}) is merely a fancy way of rewriting (\ref{iidef}), and each represents an equivalent way of doing the same computation. What is important to realize is that because of the anti-time ordered evolution operator on the left of $\calO(\tau)$ in the expectation value in (\ref{iidef}), any diagrammatic expansion of the above is rather more involved\footnote{For comparison, we recall that a typical S-matrix element is given by $\langle {\rm  out} |T\left[{\rm exp}\left(-i\int^\infty_{-\infty}H_I(\tau') \dint\tau' \right)\right]| {\rm in} \rangle$, where the comparable states in the expectation value (\ref{iidef}) would be the out and in the in vacua respectively (note that these do not have to be the same in an interacting theory). The contour one makes in evaluating the S-matrix is indicated in figure (\ref{csm}).}. Furthermore, were one to evaluate the above diagrammatically, great care would be needed, as not only are the associated Feynman rules more involved, one also has to account for the many cancellations among the relevant diagrams that are not immediately obvious from the above form. A more convenient representation of (\ref{iidef}) that avoids this potential issue is equivalently furnished by \cite{Weinberg:2005vy}:
\eq{inin}{\langle\calO(\tau)\rangle = \sum_{n=0}^\infty i^n \int^\tau_{\tau_0}\dint\tau_n\int^{\tau_n}_{\tau_0}\dint\tau_{n-1}...\int^{\tau_2}_{\tau_0}\dint\tau_1\langle[H_I(\tau_1),[H_I(\tau_2),...[H_I(\tau_n),\calO(\tau)]...]] \rangle}
One can prove that (\ref{inin}) is equivalent to (\ref{iidef}) inductively by showing that if up to order $N$, the time derivatives of (\ref{inin}) and (\ref{iidef}) up to $N^{th}$ order are equal, then they will also be so at $N+1^{th}$ order \cite{Weinberg:2005vy}. One is free to work with either a diagrammatic expansion of (\ref{iidef}) or the operator summation (\ref{inin}) as convenience dictates. For the purposes of this review, the operator evaluation (\ref{inin}) suffices. For a review of the diagrammatic approach-- which comes in handy when evaluating higher order or loop corrections to certain observables-- see \cite{calzetta2008nonequilibrium}.

\section{Effective actions and particle production}
\label{1l}
Consider a heavy field $\psi$ coupled to a light field $\phi$ that represents the inflaton:
\eq{start}{S = \int\sqrt{-g}\Bigl[\frac{1}{2}\phi\square\phi - V_\mathrm{inf}(\phi)\Bigr] + \int\sqrt{-g}\Bigl[\frac{1}{2}\psi\square\psi - \frac{1}{2}M^2(\phi)\psi^2\Bigr] + ...}
where for simplicity, we do not consider derivative couplings at quadratic order and ignore higher order terms in the heavy field\footnote{The latter can formally be accounted for by introducing sources and taking appropriate functional derivatives of the resulting generating functional.}. Integrating out the heavy field results in the effective action
\eq{eact}{e^{i W[\phi]} = e^{i S_\mathrm{inf}[\phi]}\int \mathcal D\psi~ e^{{-\frac{i}{2}}\int\psi(-\square + M^2[\phi])\psi} ~=~ e^{iS_\mathrm{inf}[\phi]}[\det(-\square + M^2[\phi])]^{-1/2},}
with $S_\mathrm{inf}[\phi]$ given by the first term in (\ref{start}). Thus
\eq{funcid}{W = \int\sqrt{-g}\Bigl[\frac{1}{2}\phi\square\phi - V_\mathrm{inf}(\phi)\Bigr] + \frac{i}{2} \mathrm{Tr} \ln(-\square + M^2[\phi]).}
As we demonstrate via the heat-kernel method in the next subsection, if $M^2(\phi)$ is independent of $\phi$, the functional determinant is straightforwardly evaluated resulting in an effective action with the effective potential
\eq{1lep}{V_\mathrm{eff}(\phi) = V_\mathrm{inf} + V_\mathrm{ct} + \frac{M^4}{64\pi^2}\ln\Bigl[M^2/\Lambda^2\Bigr],}
where $V_\mathrm{ct}$ are the usual infinite renormalizations that we absorb into the bare couplings of the action and $\Lambda$ is the mass scale introduced by regulating the loop integrals. This is the Coleman-Weinberg (CW) \cite{Coleman:1973jx} corrected effective potential\footnote{Allowing for derivative couplings between $\phi$ and $\psi$ and for $M^2(\phi)$ to depend arbitrarily on $\phi$ will in general result in corrections to (\ref{funcid}) that constitute the usual derivative expansion.}. More generally, we can rewrite (\ref{funcid}) recalling (\ref{eact}) as
\eq{efrw}{W = \int\sqrt{-g}\Bigl[\frac{1}{2}\phi\square\phi - V_\mathrm{inf}(\phi)\Bigr] - i~\ln~Z_{\psi}}     
with $Z_{\psi}$ defined as
\eq{zpdef}{Z_{\psi} = \int \mathcal D\psi~ e^{{-\frac{i}{2}}\int\psi(-\square + M^2[\phi])\psi}.}
The resulting quantum corrected equations of motion are obtained by varying the effective action functional $W$ with respect to $\phi$, which gives
\eq{qceom}{\square\phi - V_\mathrm{inf}'(\phi) = \frac{1}{2}M^{2'}[\phi]\langle \psi^2\rangle_{\phi},}
where it should be clear that
\eq{expdef}{\langle \psi^2\rangle_{\phi} := \frac{\int \mathcal D\psi~\psi^2~ e^{{-\frac{i}{2}}\int\psi(-\square + M^2[\phi])\psi}}{\int \mathcal D\psi~ e^{{-\frac{i}{2}}\int\psi(-\square + M^2[\phi])\psi}} }
and where it should be emphasized that the right hand side of (\ref{qceom}) is a correlation function of two coincident fields, evaluated at a fixed spacetime point and thus suitably regularized. The subscript on the expectation value denotes that it will in general be a functional of $\phi$ and its derivatives. If in the asymptotic past $M^2[\phi]\to M^2$, where $M$ characterizes some constant heavy mass scale, and if at any finite time $M^2[\phi]$ only ever differs from $M^2$ perturbatively, then the above can be straightforwardly evaluated via the Schwinger-Keldysh, or in-in formalism \cite{Schwinger:1960qe, Bakshi:1962dv, Bakshi:1963bn, Keldysh:1964ud} (see above). If the initial state of the system was in the adiabatic vacuum of the $\psi$ field, then the net effect of time evolution in the interaction picture (effected by the Dyson operator) will be to evolve the vacuum into a state with finite occupation number mode by mode, described by mode functions related to the initial vacuum via the Bogoliubov coefficients $\alpha_k$ and $\beta_k$: 
\eq{modmf}{v_k = \alpha_k u_k + \beta_k u^*_k} 
Given that such evolution will not excite modes of arbitrarily high energy, the Dyson operator corresponding to (\ref{modmf}) evaluated at that moment is given (up to a phase) by the equivalent unitary operator\footnote{Requiring that the Hilbert spaces spanned by the mode functions on either side of (\ref{modmf}) be unitarily equivalent implies that the $\Theta_k$ must tend to zero sufficiently fast for large $k$.}
\eq{UT}{U(\Theta) = e^{-\frac{1}{2}\int[\Theta_k a_k^2 - \Theta^*_ka^{\dag 2}_k]},}
where the $a^\dag_k$ and $a_k$ are the creation and annihilation operators associated with the vacuum of $\psi$ in the infinite past, and where $\Theta_k := \theta_k e^{i\delta_k}$ relates to the Bogoliubov coefficients of the transformation (\ref{modmf}) as $\alpha_k = \cosh\theta_k$, $\beta_k = e^{-i\delta_k}\sinh\theta_k$. From this, we straightforwardly evaluate $\langle\psi^2\rangle_\phi$, recalling that $\psi$ evolves as it would in the Heisenberg picture as
\eq{psB}{\langle\psi^2\rangle_\phi = \langle 0|U^\dag(\Theta)\psi^2U(\Theta)|0\rangle = \frac{1}{(2\pi)^3}\int \frac{\dint^3k}{2\omega_k}\Bigl[1 + 2|\alpha_k\beta_k|cos(2\delta_k) + 2|\beta_k|^2\Bigr]}
with $\omega_k^2 = k^2 + M^2$. The first term in the square brackets above results in the usual divergent contribution of evaluating a two point correlation function at coincident points. This term contributes the usual renormalizations of the couplings of our theory and the effective potential. That is to say, bringing this contribution over to the left hand side of (\ref{qceom}), when $M^2(\phi)$ varies slowly enough we find the usual CW correction\footnote{We drop all unphysical power law divergent terms in making this comparison.}:
\eq{CWcorr}{\frac{M^{2'}(\phi)}{4(2\pi)^3}\int^{\Lambda^2} \frac{\dint^3k}{\sqrt{k^2 + M^2}} = \frac{M^2(\phi)'}{32\pi^2}M^2(\phi)~\ln\Bigl[M^2/\Lambda^2\Bigr] \equiv V'_\mathrm{CW}(\phi),}
which evidently captures the variation of the vacuum energy density of the $\psi$ field along the inflaton trajectory. The second term in (\ref{psB}) corresponds to a phase associated with each excited wave number. The so called `random phase' states \cite{Kandrup:1988vg,Kandrup:1988sc} (such as thermal states or eigenstates of the number operator) contribute vanishingly. More generally, this contribution will be negligible compared to the last term. Thus we are left with 
\eq{ppans}{\langle\psi^2\rangle_\phi = \int \frac{\dint^3k}{\omega_k}\frac{|\beta_k|^2}{(2\pi)^3} \equiv \frac{1}{a^3}\int \frac{\dint^3k}{\omega_k}\frac{n_k}{(2\pi)^3}}
where $n_k$ is the number density of particles with comoving momenta indexed by $k$. Evidently then, our task in evaluating additional contributions to the one loop effective action on top of the usual CW contributions boils down to accounting for all relevant contributions to particle production of the heavy quanta. These contributions, which sum up to (\ref{ppans}), will in general not only depend on $\phi$, but also its velocity, acceleration, etc. In general these terms will resum to the usual derivative expansion we are accustomed to. We now discuss several ways by which one might calculate these contributions, beginning with a brief introduction to the method of the heat kernel. 

\subsection{The effective action via the heat kernel}

The (Euclidean) effective action for a light scalar $\phi$, coupled to a heavy scalar $\psi$ can be obtained by integrating out the field $\psi$. In the event that $\psi$ appears in the action quadratically, the contribution to the effective action is given by 
\eq{easf}{e^{-W[\phi]} = \int \mathcal D\psi~ e^{{-\frac{1}{2}}\int\psi(-\square + m^2[\phi])\psi} ~=~ [{\rm det}(-\square + m^2(\phi))]^{-1/2}.}
Thus 
\eq{fid}{W = \frac{1}{2}{\rm ln~det}[-\square + m^2(\phi)] = \frac{1}{2}{\rm  Tr~ln}[-\square + m^2(\phi)] = \frac{1}{2}\int^\infty_{0}\frac{\dint s}{s} {\rm Tr}~e^{- s[-\square + m^2(\phi)]},}
where the latter equality arises from the relation:
\eq{expint}{\lim_{\epsilon^2\to 0} \int^\infty_{\epsilon^2}\frac{\dint s}{s}~e^{- s x} = \lim_{\epsilon^2\to 0} - {\rm Ei}[- x\epsilon^2] \approx {\rm ln}\,[x],}
where irrelevant constant terms have been discarded in the above. At this point, there are several formal paths one could embark upon to actually compute the effective action, each with its own conceptual appeal. In this subsection, we summarize some standard results obtained via the rather versatile heat kernel methods (see \cite{Vassilevich:2003xt} for an excellent review, and the generalization to interacting fields).

To begin with, we first note that the matrix operator that we are tracing over in (\ref{fid}) can be written in position space basis as:
\eq{psb}{G(x,x';s) := \theta(s)\langle x| e^{- s[- \square + m^2(\phi)]} |x' \rangle.}
This function is clearly a solution to the p.d.e.:
\eq{pde}{[\partial_s - \square_x + m^2(\phi)]G(x,x';s) = \delta(s)\delta^4(x,x').}
which allow us to identify it as the heat kernel for a 5-d spacetime with `time' identified with the $s$ direction, with unit diffusion co-efficient, and the Euclidean dimensions identified as the spatial dimensions. In the case that $m^2$ is a constant (or if $\partial_\phi m^2(\phi)$ is smaller than any other mass scale in the problem), we can solve the above directly in Fourier space and transforming back, to obtain: 
\eq{gffp}{G(x,x'; s) = \theta(s)\frac{e^{-m^2(\phi)  s}}{16\pi^2 s^2}e^{-\frac{(x-x')^2}{4 s}}.}
Therefore, if the characteristic time scales associated with the dynamics of $\phi$ are much longer than that of $\psi$ (i.e. it is a much lighter degree of freedom), we can approximate $m^2(\phi)$ above as a constant and by using (\ref{gffp}), compute the effective action, regulating the divergent lower limit of the integral over $s$ via the regulator $\epsilon^2 = 1/\Lambda^2$ (where $\Lambda$ is some UV mass scale):
\begin{eqnarray}
\label{eea}W &=&  \frac{1}{2}\int^\infty_{\frac{1}{\Lambda^2}}\frac{\dint s}{s} \int d^4x~\langle x|e^{- s[-\square + m^2(\phi)]}|x\rangle\\ &=& \nonumber \frac{1}{2}\int^\infty_{\frac{1}{\Lambda^2}}\frac{\dint s}{s} \int \dint^4x~G(x,x; s) = \frac{1}{32\pi^2}\int \dint^4x~\int^\infty_{\frac{1}{\Lambda^2}} \frac{\dint s}{ s^3} e^{-m^2(\phi) s}
\end{eqnarray}
The integral over $ s$ evaluates (to leading divergences) as:
\eq{sintev}{\int^\infty_{\frac{1}{\Lambda^2}} \frac{\dint s}{ s^3} e^{-m^2(\phi) s} = \frac{1}{2}\Bigl[\Lambda^4 - m^4(\phi){\rm ln}\Bigl(m^2(\phi)/\Lambda^2\Bigr) \Bigr],}and so the one loop correction to the potential for the $\phi$ field $V_{\rm inf}(\phi)$ results in the effective potential:
\eq{1lc}{V_{\rm eff}(\phi) = V_{\rm inf} + V_{\rm ct} + \frac{m^4(\phi)}{64\pi^2}{\rm ln}\Bigl[m^2(\phi)/\Lambda^2\Bigr],}
where $V_{\rm ct}$ contains the standard counter terms one has to add to the inflaton potential $V_{\rm inf}$ in this renormalization scheme. This is the standard Coleman-Weinberg effective potential, and the consequences of this correction for the predictions of inflation have been explored in detail in \cite{Burgess:2003zw}. The derivative corrections to the above that would result if $m^2(\phi)$ is no longer assumed to be a slowly varying function of $\phi$, we result in the standard effective field theory (derivative) expansion that is discussed in the context of inflation in \cite{Weinberg:2008hq}. In certain special cases as exploited in this paper, it is possible to compute the leading order contributions to the functional determinant for non-constant  $m^2(\phi)$  exactly. 

\subsection{Particle production via the heat kernel}

As an illustration of the heat kernel method, we discuss a canonical example that has as a limit, an example that has previously been studied by other methods \cite{Kofman:2004yc}\cite{Barnaby:2009mc}. We consider a light field $\phi$ coupled to a massive scalar field $\psi$:
\eq{isoc}{\mathcal L = -\frac{1}{2}\partial_\mu\phi\partial^\mu\phi -\frac{1}{2}\partial_\mu\psi\partial^\mu\psi - \frac{M^2}{2}\psi^2 - \frac{\lambda^2}{2}(\phi-\phi_*)^2\psi^2.}
Our task is to compute the effective action (\ref{easf}) around the background trajectory of the light field. If $\phi$ corresponds to a slowly rolling inflaton, the potential (\ref{isoc}) results in the following \textit{Euclidean} mass term (tachyonic, for small enough $M^2$) for the $\psi$ field\footnote{Since particle production, if it occurs at all, will happen in a very brief window around $\phi = \phi_*$ we adopt the common \cite{Kofman:2004yc}\cite{Barnaby:2009mc} approach of approximating de Sitter space with Minkowski space in what follows.}:
\eq{isocp}{m^2(\phi) = M^2 -\lambda^2\dot\phi^2_0(\tau -\tau_*)^2,}
where $\phi_0$ denotes the spatially homogeneous solution, and $\dot\phi_0$ is the field velocity so that the background solution when $\psi\equiv 0$ is given by $\phi_0(t) = \dot\phi_0t = -i\dot\phi_0\tau$. According to (\ref{easf}) and (\ref{fid}), we are then required to evaluate
\eq{fdet}{W[\phi] = \frac{1}{2}\int^\infty_{0}\frac{\dint s}{s} {\rm Tr}~e^{- s[-\square - \lambda^2\dot\phi^2_0\tau^2 + M^2]},}
where we have set $\tau_* = 0$ for simplicity. In spite of the non-trivial spacetime profile of the inflaton field, the operator trace we have to evaluate turns out to have a closed analytic expression \textit{in the limit we can consistently treat $\phi_0$ as an external background field\footnote{i.e. when we can neglect backreaction on the background trajectory that may arise, for example from particle production.}}. An analogous operator trace arises when comptuing the effective action for Dirac fermions in a constant background electric field \cite{Schwinger:1951nm}\cite{Dunne:1998ni}. 

In proceeding, we first take note of an alternative prescription to the Schwinger-Keldysh, or in-in approach towards computing (\ref{qceom}), where the two point correlation function of interest is evaluated by evolving initial states from the asymptotic past to some finite time. If we were instead only willing to compute the time evolution operator along the contour from the asymptotic past to the asymptotic future (as we do when we compute S-matrix elements), we could still calculate the contribution to the effective action due to particle production from the imaginary part of the effective action:
\begin{eqnarray}
\label{inout}\langle 0_{\rm out}|0_{\rm in}\rangle &\equiv& e^{iW[\phi]}\\
\nonumber &=& e^{i[ {\rm Re}(W) + i {\rm Im}(W) ],}    
\end{eqnarray}
and so therefore
\eq{vacexp}{|\langle 0_{\rm out}|0_{\rm in}\rangle|^2 = e^{-2 {\rm Im}(W)}.}
In words, if the adiabatic conditions are met throughout, the in vacuum will evolve to the out vacuum. If these conditions are not met even at localized events, the in vacuum will have evolved into a state which is populated as far as the out vacuum is concerned, and the two states will have an overlap that differs form unity. The difference of this overlap from unity tells us exactly how many excited quanta have been created per spacetime unit volume, and is evidently given by the imaginary component of the effective action (\ref{fdet}) thus computed. To evaluate this in the context we are interested in, consider the relevant functional trace:
\eq{ftrace}{{\rm Tr}~e^{- s[- \square - \lambda^2\dot\phi^2_0 \tau^2 + M^2]}= {\rm Tr}~e^{- s[-\partial_\tau^2 - \lambda^2\dot\phi_0^2\tau^2 - \nabla^2 + M^2]},}
we recognize in the above that the first two terms group into the Hamiltonian of a harmonic oscillator in an inverted potential \cite{Dunne:1998ni}:
\eq{harmtr}{{\rm tr}~e^{-\sqrt{s}[-\sqrt{s}\partial_\tau^2 - \sqrt{s}\lambda^2\dot\phi_0^2\tau^2]},}
with the identification
\eq{hamid}{H = -\frac{\partial_\tau^2}{2m} + \frac{k}{2}\tau^2 \equiv -\sqrt{s}\partial_\tau^2 - \sqrt{s}\lambda^2\dot\phi_0^2\tau^2}
so that the mass of the oscillator is to be read off as $m = 1/2\sqrt{s}$ and the spring constant as $k = -2\sqrt{s}\lambda^2\dot\phi_0^2$, resulting in the imaginary frequency $\omega = 2i\lambda\dot\phi_0\sqrt{s}$. We can immediately evaluate this sub-trace to yield:
\eq{harmtr20}{{\rm tr}~e^{-\sqrt{s}[-\sqrt{s}\partial_\tau^2 - \sqrt{s}\lambda^2\dot\phi_0^2\tau^2]} = {\rm tr}~e^{-\sqrt{s} H}= -i~{\rm tr}~e^{-\sqrt{s}\omega[a^\dag a + \frac{1}{2}]} = -i \sum_{n=0}^{\infty} e^{-2i\lambda\dot\phi_0 s(n + \frac{1}{2})},}
where the sum is straightforwardly evaluated so that 
\eq{harmtr2}{{\rm tr}~e^{-\sqrt{s}[-\sqrt{s}\partial_\tau^2 - \sqrt{s}\lambda^2\dot\phi_0^2\tau^2]} = -i \frac{e^{-i\lambda\dot\phi_0 s}}{1-e^{-2i\lambda\dot\phi_0 s}}}
where the factor of $-i$ appears from the Jacobian in transforming to the ladder operator basis. The remaining factor in the operator trace is easily evaluated:
\eq{remtr}{{\rm tr}~e^{s{\nabla^2}} = \int \frac{\dint^3p}{(2\pi)^{3/2}}e^{- s p^2}= \frac{1}{(2s)^{3/2}},}
so that we are left evaluating:
\eq{immeff}{{\rm Im}~W = {\rm Im}~\frac{i}{2}\int^\infty_0 \frac{\dint s}{s}~ \frac{e^{-sM^2}}{(2s)^{3/2}} \frac{e^{i\lambda\dot\phi_0 s}}{1-e^{2i\lambda\dot\phi_0 s}}= \frac{1}{2}\sum_{n=1}^{\infty}\frac{(-1)^{n + 1}}{2 n}e^{\frac{-n\pi M^2}{\lambda\dot\phi_0}}\Bigl(\frac{\lambda\dot\phi_0}{2n\pi}\Bigr)^{3/2},}
where the latter follows from the standard principal value evaluation of the integral which has poles along the positive s axis \cite{Schwinger:1951nm}:
\eq{pval}{\lim_{\epsilon\to +0} \frac{1}{x - i \epsilon} = P\frac{1}{x} + \pi i \delta(x).}
It is interesting to leave the functional trace in (\ref{remtr}) undone to obtain the equivalent expression:
\begin{eqnarray}
\label{equivexp} {\rm Im}\, W &=& {\rm Im}\, \frac{i}{2}\int\frac{\dint^3p}{(2\pi)^{3/2}}\int\frac{\dint s}{s}e^{-s(p^2+M^2)}  \frac{e^{i\lambda\dot\phi_0 s}}{1 - e^{2i\lambda\dot\phi_0 s}}\\
\nonumber &=& \frac{1}{2}\int\frac{\dint^3p}{(2\pi)^{3/2}}\sum_{n=1}^{\infty}e^{- n\pi (p^2+M^2)/\lambda\dot\phi_0}\frac{(-1)^{n+1}}{2n},
\end{eqnarray}
the truncation to $n=1$ of which, is to be compared to (A.8) of \cite{Kofman:2004yc}\footnote{Also, to (A.10) in the same reference, as originally derived in \cite{Bachas:1995kx}.}. The form above permits an easy interpretation in terms of the spectrum of the $\psi$ quanta created. We start by noting that the pair production probability per unit spacetime is given by
\eq{ppp}{P/V \simeq \int \frac{\dint^3p}{(2\pi)^{3/2}}|\beta(p)|^2,}
where $\beta(p)$ is the Bogobliubov coefficient of the created mode $\vec p$. However we note from (\ref{vacexp}) that the quantity (\ref{equivexp}) has an identical interpretation:
\eq{effect}{|\langle 0_{\rm out}|0_{\rm in}\rangle|^2 = e^{-2 {\rm Im}(W)} \equiv 1 - 2\, {\rm Im}(W) \to P = 2 {\rm Im}\,(W) }
Therefore we can read of the spectrum of created particles as:
\eq{creatspect}{|\beta(p)|^2 = \sum_{n=1}^{\infty}e^{- n\pi (p^2+M^2)/\lambda\dot\phi_0}\frac{(-1)^{n+1}}{2n}.}
We immediately see that when $M^2 \gg \lambda\dot\phi_0$, particle production is heavily suppressed. However when $M^2\to 0$, we see from (\ref{isoc}) that $\psi$ has a effective mass everywhere except when $\phi= \phi_*$, at which point $\psi$ is effectively massless resulting in copious particle production at that instant, quantified as:
\eq{creatspect2}{|\beta(p)|^2 = \sum_{n=1}^{\infty}e^{- n\pi p^2/\lambda\dot\phi_0}\frac{(-1)^{n+1}}{2n},}
the $n=1$ truncation of which was studied in \cite{Barnaby:2009mc, Barnaby:2009dd, Barnaby:2010sq, Barnaby:2010ke} as a means to generate features in the power spectrum as well as to generate non-trivial shapes for the CMB bispectrum. In \cite{Kofman:2004yc} particle production at such `enhanced symmetry points' was suggested as a means to trap rolling moduli that approach such points\footnote{See \cite{Patil:2004zp} for how massless string states consistently stabilize moduli at such points.}. For completeness, one could evaluate the sum in (\ref{creatspect2}) and perform the subsequent momentum integral for small velocities
\eq{svans}{P/V = \frac{1}{16}\Bigl(\frac{\lambda\dot\phi_0}{\pi}\Bigr)^{3/2}[2^{5/2}-1]\zeta(5/2).}  
We note that our derivation is strictly only valid in the limit when particle production does not appreciably backreact onto the trajectory of $\phi_0$. Were it to do so, we would have invoke an iterative procedure in accurately estimating the imaginary part of the effective action. However at weak coupling, the first order treatment compares very accurately to lattice simulations \cite{Barnaby:2009mc}. 

\subsection{An analytic example}
\label{aexample}
We consider the potential (\ref{ep2}) in which a heavy direction suddenly becomes fractionally lighter for a brief window
\eq{ep}{V(\phi,\psi)= V_{\rm inf}(\phi) + M^2\psi^2\Bigl[1 - \lambda^2 {\rm sech}^2[(\phi - \phi_*)/\mu]\Bigr],}
In the above, $\mu$ is some mass scale that characterizes the skew field gradient of $V(\phi,\psi)$, $V_{\rm inf}$ is a potential which admits slow roll, and $\lambda$ is a dimensionless parameter.

If $M^2$ is greater than any other scale in $V_{\rm inf}$, then classically (i.e. at tree level), after the inflaton $\phi$ has settled on it's attractor, it will roll along the trough of these potentials uninterrupted as $\psi$ will be in its ground state. If there are no bends in the trajectory, $\psi$ remains in its ground state throughout. However at one loop we know that this will no longer be the case-- the zero point energies of the heavy modes change as the effective mass of $\psi$ changes. This change manifests adiabatically through the variations of the parameters of the CW potential and non-adiabatically through the particle production contribution to the effective potential (respectively, the first and the third terms of (\ref{psB})).   

We are interested in computing the contribution (\ref{ppans}) for (\ref{ep}). From this, we can readily compute corrections to CMB observables that result from contributions to the effective potential due to particle production. We first note that the equation of motion for the $\psi$ field that results from (\ref{ep}) is given by:
\eq{peom}{\ddot \psi + M^2\Bigl[1 + \frac{k^2}{M^2} - \frac{\lambda^2}{\cosh^2[\frac{\phi-\phi_*}{\mu}]}\Bigr]\psi = 0,}
where since we are only interested in the behaviour of the field in a brief window around $\phi_*$ in computing the spectrum of produced particles, we can safely approximate the background as Minkowski in what follows. Furthermore, assuming that $\phi$ is slow-rolling, to first order we consider $\phi - \phi_*$ simply to be $\dot\phi_0(t-t_*)$ in the above. Certainly, higher order corrections can be entertained, but as we shall see, if the backreaction of the produced particles on the background inflaton trajectory is negligible, this is a consistent approximation. Therefore, the equation of motion for the $\psi$ quanta 
\eq{peom2}{\ddot \psi + M^2\Bigl[1 + \frac{k^2}{M^2} - \frac{\lambda^2}{{\rm cosh}^2[\frac{t-t_*}{\Delta}]}\Bigr]\psi = 0,}
where $\Delta := \mu/\dot\phi_0$. We note that far enough away from $\phi_*$, the $\psi$ quanta propagate as free particles. Particle production would be indicated by a non-trivial overlap between the exact mode functions of (\ref{peom}) with those of the corresponding negative frequency free particles at late times. Computing the Bogoliubov coefficients $\beta_k$  becomes computationally identical to determining the ratio of the reflection and transmission coefficients ($R/T$) of the wave function defined by (\ref{peom}) when considered as a Schr\"odinger equation in the presence of the barrier defined by the inverted potential in the above. Exact mode functions that satisfy (\ref{peom}) are in fact available, and the precise problem of barrier transmission in the potential defined by (\ref{peom}) has been studied previously by Eckart \cite{Eckart:1930zza}. With the appropriate boundary conditions (the normalization $T=1$, and setting $t_* = 0$), we find the exact solutions
\eq{peomsol}{\psi_k(t) = e^{iM\tilde\omega_k t} F\Bigl[\frac{1}{2} - \frac{\kappa}{2},-\frac{1}{2} + \frac{\kappa}{2}, 1 - i \frac{\tilde\omega_k}{\sqrt{\beta_{(1)}}}; \frac{1}{1 + e^{2Mt \sqrt{\beta_{(1)}}}}\Bigr],}
with 
\eq{b1d}{\beta_{(1)} := \frac{\dot\phi_0^2}{\mu^2M^2} = \frac{1}{\Delta^2 M^2}}
and where $\tilde\omega_k = \sqrt{1 + k^2/M^2}$, $F(x)$ is the hypergeometric function $_2F_1(x)$, and:
\eq{kappa}{\kappa := \Bigl(1 - 4\frac{\lambda^2}{\beta_{(1)}}\Bigr)^{1/2}.}
For large $t$ given that $F(0) = 1$, we see that we recover an outgoing plane wave with transmission co-efficient unity, as per our normalization:
\eq{asympf}{\psi(t\to \infty) \sim e^{iM\tilde\omega_k t}.}
For $t \to -\infty$, one can exploit the transformation formula (with $\xi:= -e^{2Mt\sqrt{\beta_{(1)}}}$) \cite{Eckart:1930zza}:
\begin{eqnarray}
\label{transform} &&(-\xi)^{i\tilde\omega_k/2\sqrt{\beta_{(1)}}}F\Bigl[\frac{1}{2} - \frac{\kappa}{2}, -\frac{1}{2} +\frac{\kappa}{2}, 1 - i\frac{\tilde\omega_k}{\sqrt{\beta_{(1)}}}; \frac{1}{1 - \xi}\Bigr]\\ \nonumber
&=& a_1(-\xi)^{i\tilde\omega_k/2\sqrt{\beta_{(1)}}}F\Bigl[\frac{1}{2}- \frac{\kappa}{2}, -\frac{1}{2}+ \frac{\kappa}{2}, 1 + i\frac{\tilde\omega_k}{\sqrt{\beta_{(1)}}}, \frac{\xi}{\xi - 1}\Bigr]\\
\nonumber
&+& a_2 \Bigl(\frac{-\xi}{(1 - \xi)^2}\Bigl)^{-i\tilde\omega_k/2\sqrt{\beta_{(1)}}}F\Bigl[\frac{1}{2}- \frac{\kappa}{2} - i\frac{\tilde\omega_k}{\sqrt{\beta_{(1)}}}, -\frac{1}{2}+ \frac{\kappa}{2} - i\frac{\tilde\omega_k}{\beta_{(1)}}, 1 - i\frac{\tilde\omega_k}{\sqrt{\beta_{(1)}}}, \frac{\xi}{\xi - 1}  \Bigr]
\end{eqnarray}
to obtain the past asymptotic form (with $\xi = -e^{2tM\sqrt{\beta_{(1)}}} \to 0$):
\eq{asympf2}{\psi(t\to \infty) \sim a_1 e^{iM\tilde\omega_k t} + a_2e^{-iM\tilde\omega_k t},}
and with
\begin{eqnarray}
\label{aadef}
a_1 &=& \frac{\Gamma[1 - i\tilde\omega_k/\sqrt{\beta_{(1)}}]\Gamma[-i\tilde\omega_k/\sqrt{\beta_{(1)}}]}{\Gamma[1/2 + \kappa/2 -i\tilde\omega_k/\sqrt{\beta_{(1)}}]\Gamma[1/2 - \kappa/2 -i\tilde\omega_k/\sqrt{\beta_{(1)}}]}\\
\nonumber
a_2 &=& \frac{\Gamma[1 - i\tilde\omega_k/\sqrt{\beta_{(1)}}]\Gamma[i\tilde\omega_k/\sqrt{\beta_{(1)}}]}{\Gamma[1/2 + \kappa/2]\Gamma[1/2 - \kappa/2]}.
\end{eqnarray}
From here, it is straightforward to compute the Bogoliubov coefficients $|\beta_k|^2 = |R/T| = |a_2/a_1|^2$ as:
\eq{reflcof}{|\beta_k|^2 = \frac{1 + \cos[\pi\kappa]}{\cosh[2\pi\tilde\omega_k/\sqrt{\beta_{(1)}}] + \cos[\pi\kappa]}.}
We can immediately infer some basic properties of the above by taking some limits. Clearly, when $\lambda \to 0$ in (\ref{ep}), $\kappa = \sqrt{1 - 4\lambda^2/\beta_{(1)}}\to 1$, and so $\beta_k \to 0$ for all $k$, as expected. Next, when taking the $\Delta = \mu/\dot\phi_0 \to 0$ limit, we find:
\eq{bl2}{|\beta_k^2| = \frac{\lambda^4M^4\Delta^2}{32(k^2 + M^2)},}
which demonstrates as expected the preferential production of long wavelength modes, which disappears once the window $\Delta$ drops significantly below the characteristic time scale $M^{-1}$ of the $\psi$ field. We note that in the above $\kappa$ can be imaginary as well, where in this case the identical expression would result except with $|\kappa|$ in place of $\kappa$ \cite{Eckart:1930zza}. A particularly transparent special case is when $\kappa = 0$, i.e. when $\Delta = 1/2\lambda M$:
\eq{specb}{|\beta_k|^2 = \frac{2}{1 + \cosh\Bigl[\frac{\pi \sqrt{(1 + k^2/M^2)}}{\lambda}\Bigr]},}
Where again $\beta_k\to 0$ as $\lambda \to 0$, and modes with wavelengths shorter than $M^{-1}$ are preferentially produced. Here and in its other guises above, we notice how particle production appears as a non-perturbative result (vanishing faster than any power of the the perturbative coupling, hence inaccessible to any perturbative expansion \cite{Coleman:1973jx})-- as had been remarked upon previously \cite{Brezin:1970xf}\cite{Dunne:1998ni}. We plot the number density of produced $\psi$ quanta at $\phi_*$ in fig \ref{particle_prod}, given through (\ref{ppans}). 
\begin{figure}[t]
\begin{center}
\label{particle_prod2}
\epsfig{file=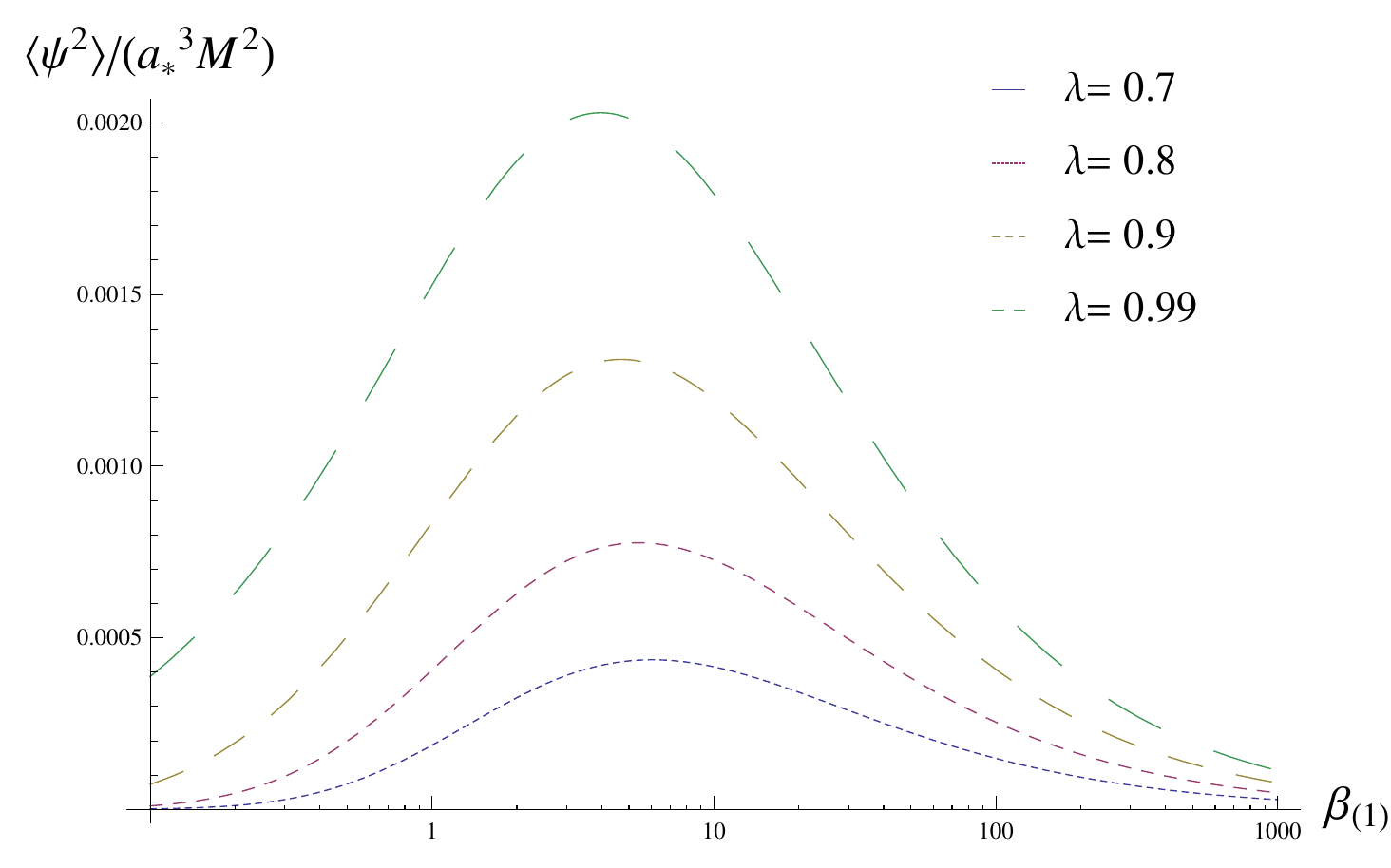, height=2.5in, width=4in}
\caption{Particle production induced at $\phi = \phi_*$ as a function of $\beta_{(1)}$ for differing values of $\lambda$.}
\end{center}
\end{figure}

}
\end{appendix}


\bibliographystyle{JHEP}
\bibliography{jan_refs,subodh_refs,jens_refs}

\end{document}